\documentclass[aps,prd,reprint,noshowpacs,nofootinbib,superscriptaddress,notitlepage,amsmath,amssymb,twocolumn,floatfix]{revtex4-2}

\usepackage{xspace}
\usepackage{graphicx}
\usepackage[top=2cm,bottom=2cm,left=3cm,right=3cm]{geometry}
\usepackage[utf8]{inputenc}

\newcommand{\savefootnote}{\edef\tmpfn{\the\value{footnote}}}

\usepackage{todonotes}
\usepackage{lineno}
\usepackage{float}
\usepackage{xcolor}
\usepackage{placeins}
\usepackage{isotope}
\usepackage{gensymb} 
\usepackage{blindtext}
\usepackage[colorlinks = true,
            linkcolor = red,
            urlcolor  = blue,
            citecolor = red,
            anchorcolor = blue]{hyperref}
\usepackage[capitalize]{cleveref}
\usepackage{multirow}
\usepackage{siunitx}
\usepackage[nowatermark]{fixmetodonotes}
\usepackage[normalem]{ulem}

\newcommand{\dd}{\mathrm{d}}
\newcommand{\numu}{$\nu_\mu$\xspace}

\newcommand{\mb}{MiniBooNE\xspace}

\newcommand{\minerva}{MINERvA\xspace}

\newcommand{\pt}{\ensuremath{p_T}\xspace}
\newcommand{\pz}{\ensuremath{p_{||}}\xspace}

\newcommand{\mares}{\ensuremath{M^{\mbox{\scriptsize{RES}}}_{\textrm{A}}}\xspace}
\newcommand{\maqe}{\ensuremath{M_{\textrm{A}}^{\mbox{\scriptsize{QE}}}}\xspace}

\definecolor{SGColor}{rgb}{0.133,0.545,0.133}

\newcommand{\Genietwoa}{G18\_02a\xspace}
\newcommand{\Genietena}{G18\_10a\xspace}
\newcommand{\Genietenb}{G18\_10b\xspace}

\begin{document}

\renewcommand{\tableautorefname}{Tab.}
\renewcommand{\figureautorefname}{Fig.}
\renewcommand{\equationautorefname}{Eq.}
\renewcommand{\sectionautorefname}{Sec.}
\renewcommand{\subsectionautorefname}{Sec.}
\renewcommand{\subsubsectionautorefname}{Sec.}



\author{M. Buizza Avanzini}
\affiliation{Laboratoire Leprince-Ringuet, CNRS, Ecole polytechnique, Institut Polytechnique de Paris, Palaiseau, France}

\author{M. Betancourt}
\affiliation{Fermi National Accelerator Laboratory, Batavia, IL, USA}

\author{D. Cherdack}
\affiliation{University of Houston, Houston, TX, USA}

\author{M. Del Tutto}
\affiliation{Fermi National Accelerator Laboratory, Batavia, IL, USA}
\affiliation{Harvard University, Cambridge, MA, USA}

\author{S. Dytman\thanks{Corresponding author}}
\affiliation{University of Pittsburgh, Pittsburgh, PA, USA}

\author{A.P. Furmanski}
\affiliation{University of Manchester, Manchester, UK}
\affiliation{University of Minnesota, Minneapolis, MN, USA}

\author{S. Gardiner}
\affiliation{Fermi National Accelerator Laboratory, Batavia, IL, USA}

\author{Y. Hayato}
\affiliation{University of Tokyo, Kamioka, Japan}

\author{L. Koch}
\affiliation{University of Oxford, Oxford, UK}

\author{K. Mahn}
\affiliation{Michigan State University, East Lansing, MI, USA}

\author{A. Mastbaum}
\affiliation{Rutgers University, Piscataway, NJ, USA}

\author{B. Messerly}
\affiliation{University of Pittsburgh, Pittsburgh, PA, USA}
\affiliation{University of Minnesota, Minneapolis, MN, USA}

\author{C. Riccio}
\affiliation{INFN Sezione di Napoli and Università di Napoli, Napoli, Italy}
\affiliation{State University of New York at Stony Brook, Stony Brook, New York, USA}

\author{D. Ruterbories}
\affiliation{University of Rochester, Rochester, NY, USA}

\author{J. Sobczyk}
\affiliation{University of Wroc\l{}aw, Wroc\l{}aw, Poland}

\author{C. Wilkinson}
\affiliation{Lawrence Berkeley National Laboratory, Berkeley, CA, USA}

\author{C. Wret}
\affiliation{University of Rochester, Rochester, NY, USA}

\title{Comparisons and challenges of modern neutrino-scattering experiments (TENSIONS 2019 report)}
\date{\today}

\begin{abstract}
A set of comparisons among neutrino interaction experiments (MiniBooNE, MINERvA, T2K, and MicroBooNE) is presented.  This gives a broad view of the field of neutrino-nucleus interactions.  The emphasis is on charged current inclusive, quasielastic-like, and pion production experiments. Measurements are compared in new ways.  Comparisons of recent data with available event generator codes are made more comprehensively than is regularly found in most previous publications.  Generator studies show sensitivities for experimental model dependence.  Efficiencies calculated with different generators are presented in a novel way.  A comparison of different forward folding techniques is also presented.
\end{abstract}

\maketitle

\section{Introduction}
\label{sec:intro}
Neutrino oscillations are a rich subject which sees a variety of long baseline neutrino oscillation experiments in the $0.1-10$ GeV range currently running~\cite{Abe:2011ks,nova,sbn}, studying the parameters of the PNMS matrix~\cite{10.1143/PTP.28.870,Bilenky:1978nj} and other effects beyond the Standard Model~\cite{DUNE:fd2020,DUNE:nd2021,DUNE:bsm2020}. 
Future experiments~\cite{DUNE:2020ypp, Abe:2011ts} aim to be dominated by systematic uncertainties and will depend on critically on the details of modeling of neutrino-nucleus interactions.
These experiments aim to reconstruct the neutrino energy of a particular flavor using the products of the neutrinos charged-current interaction, at a specific distance from the neutrino source.
Although most running experiments employ the use of a near-detector to constrain some aspect of the neutrino flux and interaction uncertainties, the study of neutrino interactions are important to understand the effects which cause the neutrino energy and flavour to be poorly reconstructed, as these do not always cancel in a near-far extrapolation. An experimental and theoretical review can be found in Ref.~\cite{nustec-review}.

There have been a variety of reviews of neutrino interactions~\cite{nustec-review} with the main goal of describing the physics content.  Many types of interaction are possible, all denoted by the principal interaction, i.e. what the neutrino does microscopically.  The dominant interactions are quasielastic (QE, where the neutrino interacts with a single nucleon and only a single nucleon is emitted), multinucleon processes (usually referred to as $2p2h$ because the interaction is predominantly with 2 nucleons), resonant (RES, where the struck nucleon is excited to one of the broad nucleon resonances which decay to a nucleon and various mesons), and deep inelastic scattering (DIS, where the neutrino interacts with either a nucleon or predominantly a quark and a variety of mesons are emitted).  Neutrinos can interact via charged current (CC) or neutral current (NC).  For CC interactions, a charged lepton is emitted and its flavor provides a way to infer the neutrino flavor.

Many neutrino cross-section measurements have moved away from publishing data according to interaction mode---which requires heavily model-dependent corrections to be made---and now publish data of a more objective nature.
Emphasis is now on topological content of the final state observed in the detector.  For example, CCQE cross-section measurements have generally been replaced with measurements that specify a final state of one muon and no pions. These measurements are referred to as CC0$\pi$ or CCQE-like. Thus cross-section measurements tend to have a mix of contributions to their signal; for instance, CC$0\pi$ without any restriction on the number of nucleons will see contributions from CCQE, $2p2h$, and pion production processes where the pion is absorbed.  CC1$\pi$ measurements have different meanings depending on the neutrino energy.  At low energies ($E_\nu < 1.5~\text{GeV}$), the $\Delta$(1232) resonance dominates.  However, higher energy neutrinos can excite a variety of higher mass resonances or interact directly with the quarks (DIS). 

Neutrino beams are wide-band, i.e. the flux distribution width is a large fraction of the peak energy.  In addition, the beam contains neutrinos of different flavors.  All these properties must be measured or calculated by each experiment.  Although most neutrino detectors have good acceptance over a broad range of kinematics, there are still kinematic holes, e.g. very low energy hadrons, and neutral particles are often hard to detect.  These aspects make the experiments difficult and Monte Carlo (MC) calculations are often used to fill the holes.
Hence, the established method for interpreting the results is to generate events and make predictions for interaction generators using the same signal definition, neutrino flux, and target as the experiment, and comparing the calculations to data. We follow this method throughout this publication.

The Tensions workshop series attempts to examine how measurements are defined/carried out and make comparisons with a variety of Monte Carlo calculations.
The first Tensions (2016) workshop~\cite{tensions2016} discussed difficulties in comparing results with different signal definitions, published comparisons of data against a variety of models, and discussed their model dependence. All cross-section experiments were represented, and generator experts were present. General issues of experiments and modeling were discussed from different points of view, and adjustments to the methods were suggested to the experiments.

One of the interesting studies in the first Tensions workshop was pion production where signal definitions are especially complicated.
Each experiment chooses analysis methods and signal definition to best utilize their detector and their view of what measurements are needed.
The result is that comparing measurements is a difficult job that must be done via generators which are capable of reproducing a wide range of signal definitions.
Event generator codes were able to accurately simulate the signal definitions of the MiniBooNE~\cite{Wilking} and first \minerva results~\cite{Eberly:2014mra} and as a result compare the same models to each data set.
One of the main conclusions of Tensions 2016 was that the two data sets were largely incompatible with our models as a result of comparison with a variety of calculations.

This article comes from the second Tensions workshop which was held in the summer of 2019 at University of Pittsburgh, USA, as well as further investigation which followed this workshop. For this article, we discuss new measurements from T2K, MINERvA, and MicroBooNE, which have been published with significant improvements in methodology. We use three recent versions of GENIE~\cite{Andreopoulos:2009rq}, one recent version of NuWro~\cite{Golan:2012wx}, and one recent version of NEUT~\cite{neut}. To compare the generators to each other and to experiments' cross-section data, we use NUISANCE~\cite{Stowell:2016jfr}. 

The emphasis is on CC inclusive (\autoref{sec:incl}), CC0$\pi$ (\autoref{sec:cc0pi}), and CC1$\pi$ (\autoref{sec:pion}) measurements. 
We make comparisons of the event generators against recent cross-section data, and present novel comparisons beyond model-to-data comparisons.  In the last few years, collaboration between theorists and generator authors has significantly increased.  The result is improved models and the plots contained in this paper shows the results. 
We compare similar measurements in the same kinematics, and pseudo-efficiencies from generator predictions were compared with actual efficiencies for each class of measurement to look for model dependence. 
We give a retrospective on the MiniBooNE results and their relevance to modern measurements is provided (\autoref{sec:minib} and \autoref{sec:minibapp}).
In \autoref{sec:gen_studies} we compare generator predictions for quantities that are of great interest to experiments, such as neutral particle energy content and the dependence of efficiency on detection threshold.
Finally, some important observations on forward-folding techniques are made (\autoref{sec:forwardfold}).

The workshop attendees were experts from each of the experiments' cross-section programs, and generator experts were present throughout, all of whom are contributing authors to this document.

\section{Event generator overview}

\label{sec:generator}

This section briefly introduces the generator models that are compared to data later in this work. Generator models are essential to experiments, providing a means by which to estimate efficiencies and background contributions, develop selection cuts and corrections, and assess systematic uncertainties due to interaction modeling. Understanding interaction-model dependence in neutrino-oscillation experiments is a topic of particular interest~\cite{Abe:2018wpn,Acero:2019ksn,DUNE:2020ypp}. The generators used in existing accelerator experiments, GENIE~\cite{Andreopoulos:2009rq} and NEUT~\cite{neut,Hayato:2021heg}, have been developed primarily within the experimental community, drawing on published theoretical work or direct collaborations with theorists to develop models. While in many cases models are similar with respect to the underlying theory, differences in implementation and parameter choices can lead to important differences in generator predictions; this is evident in the comparisons shown in later sections. The use of generators for interpretation of experimental data introduces additional challenges. Complete coverage of phase space is necessary for use in a full detector simulation, typically leading to inconsistency in regions which are poorly understood or where models overlap. Mechanisms for assessment of systematic uncertainties are also necessary and challenging; these must incorporate all relevant degrees of freedom within models while remaining computationally tractable. An additional generator, NuWro~\cite{Golan:2012wx}, is frequently used as a benchmark. With rapid integration of improved theoretical models and a more consistent handling of certain interaction modes, NuWro has provided both a point of comparison and an avenue for developing new models which are later integrated into GENIE and NEUT. Detailed discussion of the models implemented in each of these generators is provided in the following subsections. First, the general structure and common components are introduced.

In all generators considered here, neutrino-nucleus interactions are modeled as a two-step process in an impulse approximation. Here, the first step is the primary interaction, where interactions occur on individual bound and moving nucleons. The second step is final state interactions (FSI), where interactions of particles from the primary vertex with the residual nucleus are considered. A major difficulty in the interaction modeling is to consistently describe the nucleus at three different scales. For quasielastic scattering, meson exchange current or two-particle two-hole  interactions where the neutrino scattering is with a correlated nucleon pair, and resonance production, the nucleus is modeled as an ensemble of nucleons. In deep inelastic scattering (DIS), nucleon substructure in the form of quarks becomes most relevant. Finally, in coherent neutrino-nucleus scattering (COH), the nucleus is essentially a single composite object. These three classes of models tend to be developed separately, and must be merged into a consistent picture within the generators. The primary focus of this work is to study processes where the neutrino interacts with one or two nucleons at a time: QE, MEC or  $2p2h$, and RES.

Model choices for QE-like nucleon-level interactions are summarized in Table~\ref{tb:qemodels}. For modeling of the initial state bound nucleon momentum distributions, two types of models are typically used. The relativistic Fermi gas (RFG) model is the traditional approach, with implementations of the Smith-Moniz~\cite{Smith:1972xh}
and Bodek-Ritchie~\cite{BodekRitchie} versions available in generators. The more recent local Fermi gas (LFG) model provides a more realistic distribution based on the position-dependent local nucleon density. For describing quasielastic scatters, generators have historically used the RFG and Llewellyn-Smith~\cite{LlewellynSmith:1971uhs} model; in GENIE v3.00.06 (v3), the G18\_02a model set uses these models, to provide a point of comparison to these choices. In current versions, generators have shifted to use the LFG and Valencia group's self-consistent QE model. This model by Nieves {\it et al.}~\cite{Nieves:2004wx,nieves_2011} includes long-range nucleon-nucleon (RPA) correlations and Coulomb effects for the outgoing charged lepton on single-nucleon ($1p1h$ or true QE) and multi-nucleon ($2p2h$) interactions. These effects can modify interactions significantly at energies near or below 1 GeV, so these models are more applicable to these lower neutrino energies while predictions are identical to Llewellyn-Smith for higher energies.
For multi-nucleon interactions, the Valencia $2p2h$ model~\cite{Nieves_2p2h_14} has been widely used, particularly since Gran and Sanchez~\cite{Gran:2013kda} studied its features and application in collaboration with the theory authors. It is included as a distinct interaction channel which explicitly incorporates additional nucleons in the final state.  Broader applicability is gained by suppressing events for which $q_{3} > 1.2\mbox{ GeV/c}$, where $q_3$ is the magnitude of the three-momentum transfer. An alternative model, based on an empirical enhancement of the total cross section in a region of energy-momentum transfer space between QE and RES due to multi-nucleon interactions~\cite{Katori:2013eoa}, is used in GENIE v3 G18\_02a which is similar to code commonly used in experiments discussed here. Final state nucleons from $2p2h$ processes are generally distributed via phase space~\cite{Sobczyk:2012ms}. Consistency in the isospin decomposition of emitted nucleons is an ongoing problem. While some advanced theoretical models~\cite{Lovato:2014eva} have explored interference between one- and two-body currents, the implementation of these in event generators is just beginning to be explored~\cite{GENIESTA}.

\begin{table*}
\begin{ruledtabular}
\caption{QE-like models implemented in each generator.}
\label{tb:qemodels}
\centering
\renewcommand{\arraystretch}{1.2}
\begin{tabular}{cccccc}
          & Nuclear & QE & \maqe & $2p2h$ NN  & Long-range NN\\
Generator & model & model & (GeV) & correlations & correlations \\
\hline
GENIE v3 G18\_02a & RFG~\cite{BodekRitchie} & Llewellyn-Smith~\cite{LlewellynSmith:1971uhs} & 0.99 & Empirical~\cite{Katori:2013eoa} & None\\
GENIE v3 G18\_10a & LFG & Nieves~\cite{Nieves:2004wx} & 1.05 & Nieves~\cite{nieves_2011,Gran:2013kda} & RPA~\cite{Nieves:2004wx}\\
GENIE v3 G18\_10b & LFG & Nieves~\cite{Nieves:2004wx} & 1.05 & Nieves~\cite{nieves_2011,Gran:2013kda} & RPA~\cite{Nieves:2004wx}\\
\hline
NuWro 19.02     & LFG & Llewellyn-Smith~\cite{LlewellynSmith:1971uhs} & 1.03 & Nieves~\cite{nieves_2011,Gran:2013kda} & RPA~\cite{Graczyk:2003ru} \\
\hline
NEUT v5.4.0.1    & LFG & Nieves & 1.05 & Nieves~\cite{nieves_2011,Gran:2013kda} & RPA~\cite{nieves_2011} \\
\end{tabular}
\end{ruledtabular}
\end{table*}

\begin{table*}[h]
\begin{ruledtabular}
\caption{Models for pion production in the $\Delta$(1232) resonance region. R-S refers to Rein-Sehgal~\cite{Rein-Sehgal}, B-Y to Bodek-Yang~\cite{Yang:2009zx}, B-S to Berger-Sehgal~\cite{Berger-Sehgal}, and S-O to Salcedo-Oset~\cite{Salcedo-Oset}.}
\label{tb:1pimodels}
\centering
\renewcommand{\arraystretch}{1.2}
\begin{tabular}{ccccccc}
          & Resonance & \mares & Nonresonant & Form & $\pi$ FSI \\
Generator & model & (GeV) & model & factor & model \\
\hline
GENIE v3 G18\_02a & B-S & 1.23 & Scaled B-Y~\cite{bodek-yang} & Ref. \cite{Graczyk:2009qm} & Empirical~\cite{Andreopoulos:2009rq} & \\
GENIE v3 G18\_10a & B-S  & 1.23 & Scaled B-Y & Ref. \cite{Graczyk:2009qm} & Empirical~\cite{Andreopoulos:2009rq} & \\
GENIE v3 G18\_10b & B-S  & 1.23 & Scaled B-Y & Ref. \cite{Graczyk:2009qm} & S-O  \\
\hline
NuWro 19.02& $\Delta$-only~\cite{Graczyk:2009qm} &  0.94 & Scaled B-Y & Ref. \cite{Graczyk:2009qm} & S-O  \\
\hline
NEUT v5.4.0.1 & B-S       & 0.95 & Scaled B-Y & Ref. \cite{Graczyk:2009qm} & S-O  \\
\end{tabular}
\end{ruledtabular}
\end{table*}

\begin{table*}[h]
\begin{ruledtabular}
\caption{Models for meson production in the kinematic region with resonances of mass larger than $\Delta$(1232) resonance. R-S refers to Rein-Sehgal~\cite{Rein-Sehgal}, B-Y to Bodek-Yang~\cite{Yang:2009zx}, B-S to Berger-Sehgal~\cite{Berger-Sehgal}, and S-O to Salcedo-Oset~\cite{Salcedo-Oset}.}
\label{tb:npimodels}
\centering
\renewcommand{\arraystretch}{1.2}
\begin{tabular}{ccccc}
& Resonance & DIS & RES/DIS & Coherent \\
Generator & model & model & boundary & model \\
\hline
GENIE v3 G18\_02a & B-S &  B-Y~\cite{bodek-yang}/Pythia~\cite{Sjostrand:2006za} & 1.93 & B-S~\cite{Berger:2008xs} \\
GENIE v3 G18\_10a & B-S  &  B-Y/Pythia~\cite{Sjostrand:2006za} & 1.93 & B-S~\cite{Berger:2008xs}\\
GENIE v3 G18\_10b & B-S  &  B-Y/Pythia~\cite{Sjostrand:2006za} & 1.93 & B-S\\
\hline
NuWro 19.02& None &  B-Y/Pythia~\cite{Sjostrand:2006za, Juszczak:2005zs} & 1.3-1.6 & B-S\\
\hline
NEUT v5.4.0.1 & B-S         & B-Y with Custom~\cite{Hayato:2021heg}(W$<$2.0) & 1.4--2.0 & B-S \\
  &            & and PYTHIA (W$>$2.0)                           &          &   \\
\end{tabular}
\end{ruledtabular}
\end{table*}

The models for production of pions in the $\Delta(1232)~P_{33}$ resonance are summarized in Tab.~\ref{tb:1pimodels}. At the core of most generator resonance models is the Rein-Sehgal~\cite{Rein-Sehgal} (R-S) model. The R-S model uses a non-relativistic quark model~\cite{FKR} to derive helicity amplitudes to produce resonances, and then describes the subsequent decay of those resonances. Berger and Sehgal~\cite{Berger-Sehgal} updated the R-S model to include effects due to lepton mass. 

The resonance parameters such as masses, decay widths, and form factors have changed significantly as the data improved since the development of the R-S model in 1981. All generator groups have implemented these updates, and in certain models have incorporated updated tuning to neutrino scattering data. There are several ways to describe non-resonant pion production~\cite{nustec-review}. Strength can come from the tail of DIS processes, referred to here as a scaled Bodek-Yang~\cite{bodek-yang} (B-Y) model, or via low order diagrams~\cite{nieves_2011}. The scaled B-Y choice uses a factor that decreases model strength to achieve agreement with data, and necessarily includes both resonant and nonresonant contributions. 

For the kinematic region for values of $W$ (invariant mass) greater than 1.4 GeV, model choices are given in Tab.~\ref{tb:npimodels}.  This is the kinematic region that is critical to successful interpretation of DUNE~\cite{DUNE:2020ypp} data.  Berger and Sehgal~\cite{Berger-Sehgal} provide models for all highly-rated nucleon resonances~\cite{ParticleDataGroup:2020ssz} as well as coherent pion production~\cite{Berger:2008xs}, both of which are updated versions of the corresponding Rein-Sehgal~\cite{Rein-Sehgal,Rein:1982pf} models. Unlike $\Delta$ production, these resonances have weaker excitation strength from neutrino interactions and are poorly known as a result. The relevant couplings have not been fit to modern electron-nucleon data and the decay distributions are considered to be isotropic. A major challenge is finding the optimal way to describe the soft, or shallow, inelastic scattering (SIS) kinematic region, which covers values of $W$ between 1.4 and 2.0 GeV. There exists no hard boundary and theoretical guidance is meager.  As a result, modeling at the boundary between RES and DIS is empirical and event generator groups have adopted different strategies, using both RES and DIS models to describe the data in this region. All generators have a transition between resonance- and DIS-dominated kinematic regimes, though they differ in the location and treatment of this transition.  GENIE, for example, has a sharp boundary at $W\sim$1.9 GeV (depends on model) while NuWro linearly interpolates over a range just above the $\Delta$ peak.  Lacking true non-resonant models, generator codes use scaled versions of the B-Y model. 

For $W>2.2$~GeV, true DIS processes are dominant and the strategy of using Pythia~\cite{Sjostrand:2006za} is optimal.  At low values of $W$, resonance decays to baryon and meson states are reasonably well understood and are the appropriate description.  However, treatments that extend DIS models into the SIS region at lower $W$ rely on empirical models~\cite{Yang:2009zx,GENIEtuneAGKY} for hadronization processes. 

Final state interactions occurring during the propagation of produced particles through the remnant nucleus are described using intranuclear cascade models based on free hadron-nucleon cross sections. In models simulating the full intranuclear cascade (NEUT, NuWro, and GENIE G18\_10b), nuclear medium modifications are added in a local density approximation. These have been derived by Salcedo-Oset~\cite{Salcedo-Oset} for pions and by Pandharipande-Pieper~\cite{Pandharipande:1992zz} for nucleons. The GENIE G18\_02a and G18\_10a configurations have a data-driven model~\cite{Andreopoulos:2009rq} which has partial inclusion of medium dependence effects.

\subsection{GENIE overview} 
\label{sec:genie}

The GENIE~\cite{Andreopoulos:2009rq} generator evolved from NEUGEN, the primary event generator for the MINOS experiment~\cite{Gallagher:2002sf}. The current major version, GENIE v3, was released in October 2018. GENIE includes a variety of model sets, which are user-selectable via configuration files. The work described here uses three GENIE configurations: \Genietwoa, \Genietena, and \Genietenb. These alphanumeric codes represent the choice of GENIE physics models: the G18\_02a model set is an updated version of the historical default model from GENIE v2, and the G18\_10a and G18\_10b configurations have newer models for almost all processes relevant here \cite{GENIEv3Highlights}. For all G18 model sets, the last character defines the FSI treatment used. The letter `a' refers to the $hA$ effective cascade model and `b' to the $hN$ full cascade model, each described below.
The official labels for the GENIE configurations studied here all include an additional suffix \_02\_11a, which is omitted for simplicity. This suffix indicates that several model parameters were tuned to neutrino scattering data on hydrogen and deuterium targets~\cite{GENIEFreeNucleonTune}.

The G18\_10 model sets use a local Fermi gas (LFG) nuclear model and the Nieves models for CCQE~\cite{nieves_2011} and $2p2h$~\cite{nieves_2011, Gran:2013kda}. The $2p2h$ implementation in GENIE is fully described in Ref.~\cite{Schwehr:2016pvn}. The \Genietwoa~model set provides an updated version of models used within the community for many years: a Bodek-Ritchie~\cite{BodekRitchie} relativistic Fermi gas (RFG) nuclear model including a high-momentum tail due to short range nucleon-nucleon correlations, the Llewellyn-Smith model~\cite{LlewellynSmith:1971uhs} for primary CCQE processes, and an empirical $2p2h$ model based on fits to MiniBooNE data~\cite{Katori:2013eoa}. All model sets use a dipole axial form factor and BBBA07 vector form factors~\cite{bbba07}, and for nuclear targets they apply Pauli blocking requiring that the momentum of the outgoing nucleon exceeds the Fermi momentum $k_F$ for the nucleus in question. 

Although all GENIE resonance models are based on the Rein-Sehgal treatment~\cite{Rein-Sehgal}, a variety of changes have been implemented, e.g. regular updates for new resonance masses and widths. For all versions, the effect of the lepton masses on the allowed region of phase space is taken into account. G18\_10a/b fully include Berger-Sehgal lepton-mass corrections~\cite{Rein:2006di} and the pion-pole diagram~\cite{Berger-Sehgal}. In the G18\_02a model set, the axial and vector form factors are the modified dipole forms as in the Berger-Sehgal model. In GENIE v3 G18\_10a/b, the $\Delta$ form factors have been updated from fits to \mb~data~\cite{AguilarArevalo:2010bm}. While the $\Delta\rightarrow\pi$ decay is isotropic for the G18\_02a model set, G18\_10a/b use the angular distribution from Rein-Sehgal~\cite{Rein-Sehgal} which was fit to ANL data~\cite{Radecky:1981fn}. All GENIE models neglect interference between resonances. The non-resonant contribution to pion production comes from scaled versions of the Bodek-Yang~\cite{bodek-yang} model, with hadronization described by the custom AGKY model~\cite{Yang:2009zx,GENIEtuneAGKY} and Pythia~\cite{Sjostrand:2006za}.

DIS processes are handled in GENIE with a mix of Bodek-Yang~\cite{Bodek:2003wc}, a special fragmentation model~\cite{Andreopoulos:2009rq}, and Pythia~\cite{Sjostrand:2006za}. The transition between resonance and DIS processes comes at a cutoff value of $W$ which is part of the single-nucleon fit~\cite{GENIEFreeNucleonTune}; the value for model sets used here is 1.93 GeV. However, the Bodek-Yang model is valid for all energies above the $\pi N$ threshold and is used (scaled to neutrino-hydrogen and neutrino-deuterium scattering data) for nonresonant processes below the cutoff value.

GENIE has a unique FSI model~\cite{Andreopoulos:2009rq,Dytman:2021ohr} called $hA$ which uses a single interaction to approximate the multiple steps in traditional cascade models. This has been tuned to hadron-nucleus scattering data for a wide range of nuclei and energies. This model is denoted with an `a' in the configuration name and is used in G18\_02a and G18\_10a.  A multi-step cascade model called $hN$, which includes medium corrections for pions~\cite{Salcedo-Oset} and nucleons~\cite{Pandharipande:1992zz}, is denoted with a `b' and used in G18\_10b.
Both models use SAID hadron-nucleon fits to data~\cite{said} in calculations of mean free path and various angular distributions.

\subsection{NEUT overview} 
\label{sec:neut}

The NEUT Monte Carlo generator has been developed for Super-Kamiokande, T2K, and other experiments, and simulates neutrino-nucleus interactions from $\sim$100~MeV to $\sim$100~GeV. Simulations shown in this paper were performed using NEUT version 5.4.0.1.
 
The QE and pion production models are as described in Tables~\ref{tb:qemodels}--\ref{tb:npimodels}. NEUT takes into account inter-resonance interference, consistently using the Rein-Sehgal~\cite{Rein-Sehgal} model. Multi-pion production events are generated with the custom code that assumes the Koba-Nielsen-Olsen (KNO) scaling~\cite{Koba:1972mr} and the measured multiplicity of pions as a function of $W$. DIS events are generated with Pythia5/JETSET~\cite{Sjostrand:2006za}. For the multi-pion and DIS channels, the GRV98 parton distribution functions, including Bodek-Yang corrections~\cite{Bodek:2005de}, are used. $2p2h$ events are simulated with the Valencia model~\cite{nieves_2011}, and coherent pion production events are simulated with the Berger-Sehgal model~\cite{Berger:2008xs}. To reflect the idea of the formation zone, hadrons and mesons produced by interactions other than the (quasi-)elastic scatterings or coherent scattering have their production positions shifted toward the direction of the outgoing particles.

For pion FSI, the mean free paths (MFP) of absorption and inelastic scattering are calculated with the model developed by Oset {\it et al.}~\cite{oset_1987} below 400~MeV/c. Above 500~MeV/c, the MFP are extracted from pion-nucleon scattering data. In the transition region in between, the fraction of low energy model decreases linearly from 1 at 400~MeV/c to 0 at 500~MeV/c. The normalizations of the mean free paths were tuned using pion-nucleus scattering data \cite{PinzonGuerra:2016uae}. The kinematics are determined using the results of a phase shift analysis with the medium correction suggested by Seki {\it et al.}~\cite{Seki:1983si}. If a pion is absorbed in the nucleus, multiple nucleons are emitted. Kaon and eta FSI is handled similarly to pions by the cascade model, with the MFP deduced using kaon-nucleon or eta-nucleon scattering data. The nucleon FSI model is based on the work by Bertini et al.~\cite{Bertini:1972vz}.
Finally, nuclear de-excitation is considered in the case of an oxygen target, and includes production of additional low energy $\gamma$ or nucleons following the neutrino interaction.

\subsection{NuWro overview}
\label{sec:nuwro}

The NuWro~\cite{Golan:2012wx} generator has become an important `sandbox' for other generators since its inception at University of Wroc\l{}aw around 2005, introducing new theoretical models which are used for testing before being adopted by NEUT and GENIE. It covers a neutrino energy range from $\sim$100~MeV to $\sim$100~GeV. For neutrino scattering on a free nucleon, NuWro includes contributions from the three different regions discussed above.

Quasielastic interactions are described using the Llewellyn-Smith~\cite{LlewellynSmith:1971uhs} model with BBBA05~\cite{bbba05} vector and dipole axial vector form factors. For nuclear effects, NuWro offers many options: global and local Fermi gas, hole spectral function~\cite{Benhar:1994hw}, effective spectral function~\cite{Ankowski:2005wi}, and a density- and nucleon momentum--dependent potential~\cite{Juszczak:2005wk}. In the case of LFG, long range correlations calculated with an RPA technique can be included~\cite{Graczyk:2003ru}. For the simulations used in this paper, the LFG model including RPA effects has been used.

Resonance production is described with a model optimized for the $\Delta$ resonance peak region. The $\Delta$ resonance is explicitly included, with nucleon-$\Delta$ form factors taken from Ref.~\cite{Graczyk:2009qm} with parameters obtained as a simultaneous fit to ANL~\cite{Radecky:1981fn} and BNL~\cite{Barish:1978pj} single pion production data. Non-resonant backgrounds are taken as a fraction of the DIS contribution, extrapolated down to the pion threshold.  This is added incoherently to the $\Delta$ contribution as described in Ref.~\cite{Juszczak:2005zs}. In the region $W\in (1.3, 1.6)$~GeV, NuWro employs a linear interpolation between the described model and the DIS pion production cross sections. For nuclear target reactions the $\Delta$ self-energy is included in an approximation based on Ref.~\cite{Sobczyk.:2012zj}. The effect of the finite $\Delta$ lifetime is included as described in Ref.~\cite{Golan:2012wx}. The angular distribution of pions resulting from $\Delta$ decays is described based on values of density matrix elements informed by ANL and BNL experimental studies~\cite{Radecky:1981fn, Barish:1978pj}.

In the DIS region, NuWro uses the Bodek-Yang prescription~\cite{Bodek:2002vp}. Hadronic final states are generated using Pythia~\cite{Sjostrand:2006za} fragmentation routines~\cite{Sjostrand:2006za}, with modifications described in Ref.~\cite{Nowak:2006sx}. NuWro performance is optimized to reproduce charged hadron multiplicities reported in Ref.~\cite{Nowak:2006xv}. For DIS events, formation zone effects are included \cite{Golan:2012wx}.

Simulation of $2p2h$ events can be done using a variety of models for the overall contribution to the cross section and distribution of final state leptons. For the charged current reaction, the default is the Valencia model~\cite{nieves_2011, Gran:2013kda}. Other options include a transverse enhancement model (which can be applied to neutral current reactions)~\cite{Bodek:2011ps}, the Marteau-Martini model~\cite{Marteau:1999jp, Sobczyk:2003nx, Martini:2009} and the SuSAv2 model~\cite{Megias:2016fjk}. Simulations in this work use the Valencia $2p2h$ model. For the hadronic part, NuWro uses a model proposed in Ref.~\cite{Sobczyk:2012ms}. The basic assumption is that the distribution of outgoing nucleons in the hadronic center of mass frame is uniform. However, this assumption can be relaxed with a suitable parameter. Implementation of the Valencia and SuSAv2 $2p2h$ models is done with five tabulated nuclear response functions for carbon and oxygen. Extrapolation to heavier targets is done using methods similar to those proposed in Ref.~\cite{Schwehr:2016pvn}.
 
A custom cascade model for pions and nucleons propagating through the nucleus with a realistic density profile~\cite{Golan:2012wx, Niewczas:2019fro} is employed. The key inputs are microscopic pion-nucleon and nucleon-nucleon in-medium cross sections, with Pauli blocking effects implemented locally. In the $\Delta$ region, pion-nucleon cross sections are described with the Salcedo-Oset~\cite{Salcedo:1987md} model. At larger pion energies, free pion-nucleon total cross sections are taken from experimental data and differential cross sections are provided by the SAID model~\cite{said}. For nucleon-nucleon elastic interactions, in-medium modifications are taken from Ref.~\cite{Pandharipande:1992zz}, while for inelastic reactions the model from Ref.~\cite{Klakow:1993dj} is adopted. Short-range correlation effects are included by reducing the nuclear density near every hadron-nucleon interaction point~\cite{ANLdensity, Carlson:2014vla} and introducing a compensating factor at larger distances~\cite{Niewczas:2019fro}. 

\section{Cross-section extraction methods}
\label{sec:xsec_methods}
In recent years, there have been significant developments in our understanding of the potential for bias in neutrino cross-section extraction techniques. The bias of particular concern is towards the input Monte Carlo used when developing an analysis and extracting the cross section; the assumptions made about the channel to be measured can affect the analysis.
The bias can enter in a number of ways, including signal definitions and strategies for handling backgrounds.
The data utilized in this work has been produced over a number of years, as these techniques have evolved, and as such the results do not consistently use the latest techniques or follow the measurement strategies that each experiment would use today. In this work, we consider all results as they were published or released, without systematically evaluating the techniques used to extract them. However, we briefly comment here regarding the potential issues in the extraction of neutrino cross-section measurements, as these issues are critically important and their treatment will determine the long-term utility of measurements from currently-operating experiments.

Crucial areas of potential model dependence are: (1) the choice of signal definition and selection variables; (2) the treatment of unmeasured phase space and efficiency corrections; and (3) unfolding methods and techniques~\cite{Mahn:2018mai}.

\begin{enumerate}
\item Extracting cross-section measurements as a function of variables other than those which can be directly measured in the detector, or which are a convolution of them, requires Monte Carlo corrections to translate to them from the measured variables, which necessarily introduces some model dependence. More subtly, selection cuts which are as a function of variables which are not accessible in the detector also requires similar model-dependent corrections. Typically safe variables are kinematic variables of final-state particles.

\item A related issue is the treatment of unmeasured regions of phase space and efficiency corrections for the final state particles. If a selection excludes certain regions of phase space (e.g., low proton momenta), then a na\"ive efficiency correction would simply add the missing strength based on the Monte Carlo prediction. A now widespread technique to mitigate this problem is to explicitly remove such regions of low or no efficiency from the signal definition. More subtle issues relate to regions of rapidly changing efficiency in binned cross-section measurements, where the model may be implicitly relied upon when integrating the efficiency across each bin. Despite increasing awareness of issues relating to efficiency corrections across the field, this remains a very challenging problem with no easy solution, particularly for signal definitions which allow event topologies with many final-state particles. Reconstruction algorithms tend to perform badly for very high multiplicity events, and correlations between particles in the final state may have a significant impact on the efficiency, in ways that are very challenging to capture in an analysis.

\item Unfolding is a general term for removing the smearing of resolution due to properties of a measuring device used for a measurement~\cite{Prosper:2011zz, kuusela_master, kuusela_phd}. It describes the process for producing a result as a function of a true variable (e.g., the true muon momentum) from a reconstructed variable (e.g., the reconstructed muon momentum). A common issue when unfolding is that statistical fluctuations in reconstructed space can cause large fluctuations between bins in true space. Various methods exist for regularizing, or smoothing, unfolded results, by preferring results in which the result fits some prior expectation, which necessarily adds some bias~\cite{Prosper:2011zz, kuusela_master, kuusela_phd}. The challenge is to tune the strength of the regularization to balance the bias in the result with the variance in each bin (known as the bias--variance trade-off). The most popular unfolding method used in the field is D'Agostini unfolding~\cite{DAgostini:1994zf, dagostini2nd}, which can be characterized as an algorithm for maximum-likelihood estimation with early stopping~\cite{kuusela_master, kuusela_phd}. In the D'Agostini method, the input Monte Carlo is used as a reference for regularizing the result. Each iteration of the algorithm reduces the strength of the regularization, and the size of the bias, by allowing the bin-to-bin variance to increase. The main problem for the D'Agostini method is that the stopping criterion is generally set by Monte Carlo studies of the potential bias. If the simulations used for those studies are substantially different from data, it is likely that the result is strongly affected by the bias towards the input Monte Carlo. Similar issues exist for other unfolding and regularization methods. One solution would be to present regularized and unregularized results or to avoid regularization altogether, which in the D'Agostini case would mean iterating until convergence. Unregularized results can sometimes look unphysical, but are statistically correct. It is also possible to avoid unfolding altogether, and smear the model to match the data, rather than unsmear the data to match the model~\cite{Cousins:2016ksu}, as is discussed in \autoref{sec:forwardfold}. Even in this case, there remain complications when resolutions and efficiencies depend on more than the variable being measured; experiments must ensure the smearing functions are as complete as possible.
\end{enumerate}

The extent to which any of the above issues can bias a result depends on many aspects of the detector design and analysis methodology, so cannot be assessed outside the collaboration reporting the result, and therefore cannot be reliably quantified for historical purposes. As a consequence, studies here only based on Monte Carlo studies and are then necessarily incomplete.  They can this indicate sensitivity, but not fully assess the consequences. 

Adopting methods that minimize the model dependence of neutrino cross-section results will help to ensure their continued reliability when the models currently used in neutrino interaction simulation packages become obsolete. Heavily model-dependent results can only be judged in the context of the models used to extract them, and will have a limited utility as a result.

Furthermore, one always needs to take care not to over-interpret fluctuations in plots  of differential cross-section measurements, especially for unfolded results.
The unfolding procedure causes correlations in the data points' uncertainties, making it difficult to impossible to judge how well a given prediction fits the data by just looking at the plots.
Instead it is necessary to calculate measures for the goodness of fit that take the correlations into account and then judge the model-data-agreement based on these.
The most commonly used measure for this purpose is the Mahalanobis distance, which is often called ``the chi-square'' in the field of physics. This is largely objective but interpreting a $\chi^2$ value is not always easy as the effect of correlations is not shown in the plot and the value can be dominated by single bins which are not easily visible.
Comparing the generator predictions without any bin-to-bin correlations directly to one another is a more straightforward exercise.

\section{Efficiencies}
\label{sec:eff}
Determination of efficiencies is a very important part of every experiment.
Although they should be based on data as much as possible, Monte Carlo is often used to supplement or replace data.
To study one aspect of the resultant model dependence, selection efficiencies were provided by participating experiments for the measurements discussed in this work.  These make it possible to see potential sources of difference between measurements.  For example, although two experiments may measure the same process, they might be sensitive to different regions of phase space.  As well as the providing efficiencies in terms of measured variables presented by each analysis (e.g., $p_{\mu}$--$\cos\theta_{\mu}$), they were also provided in terms of various true kinematic quantities of interest, for example true $q_{0}$--$q_{3}$, or $Q^{2}$--$W$. There are well documented problems with making measurements in these variables, but they allow the qualitative comparison of experiments in terms of the variables of interest from a theoretical point of view.

\begin{figure}[htbp]
    \centering
    \includegraphics[width=\linewidth]{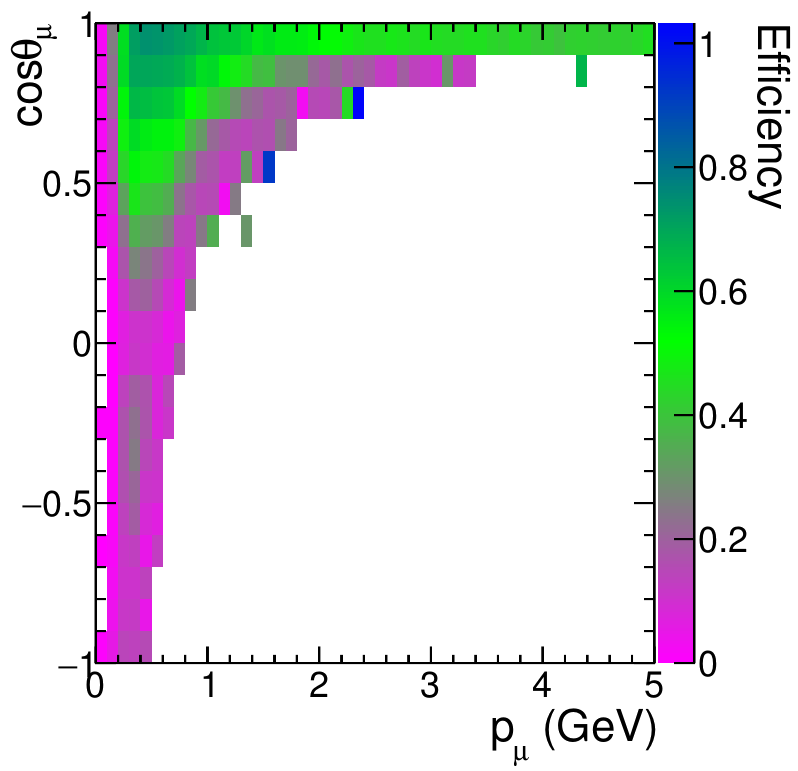}
    \caption{T2K selection efficiency for the $\nu_{\mu}$--hydrocarbon CC0$\pi$ measurement presented as a function of $p_{\mu}$--$\cos\theta_{\mu}$, corresponding to the analysis presented in Ref.~\cite{Abe:2020jbf}.}
    \label{fig:t2k_eff_example}
\end{figure}
\begin{figure*}[htbp]
    \centering
    \includegraphics[width=0.49\linewidth]{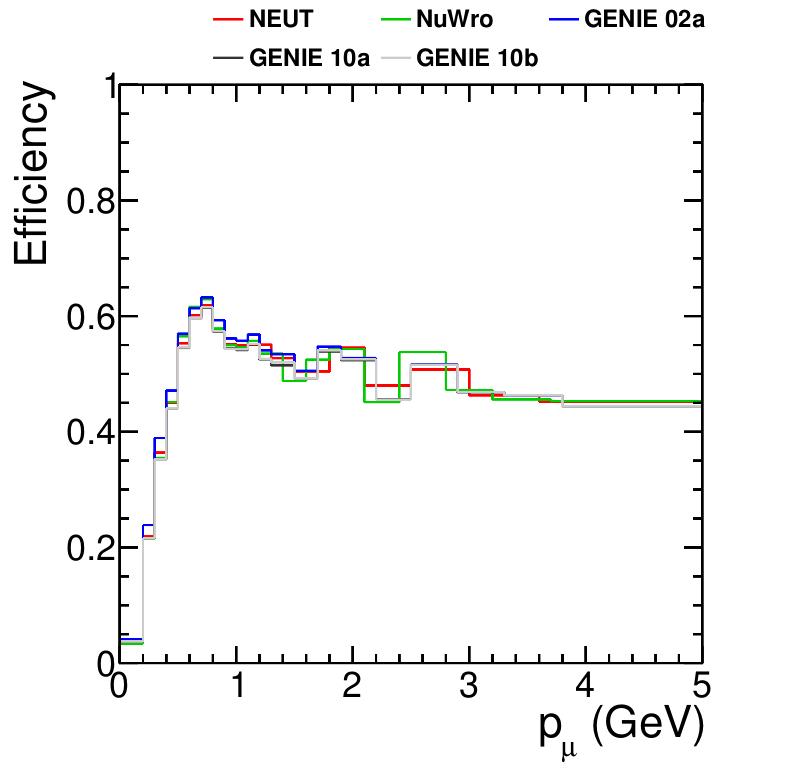}
    \includegraphics[width=0.49\linewidth]{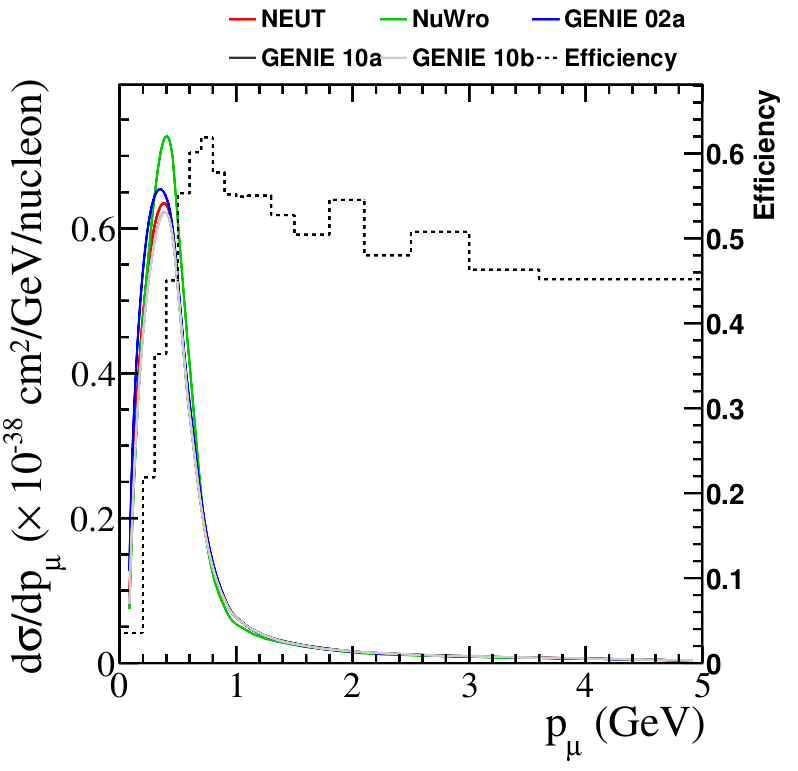}
    \caption{Left: a comparison of the one-dimensional selection efficiencies when different generator models are used to collapse from two to one dimension. Right: various generator predictions are shown and compared to the NEUT selection efficiency.}
    \label{fig:t2k_eff_comp}
\end{figure*}

Efficiencies were typically provided in two-dimensional phase space, with all other degrees of freedom implicitly integrated out. For example, Fig.~\ref{fig:t2k_eff_example} shows the selection efficiency as a function of $p_{\mu}$--$\cos\theta_{\mu}$ for T2K's $\nu_{\mu}$--hydrocarbon CC0$\pi$ measurement from Ref.~\cite{Abe:2020jbf}.  For ease of presentation, in this paper we compare single-dimensional efficiencies with generator predictions, for different measurements.  In order to collapse a two-dimensional efficiency as provided by the experiments to the single-dimensional efficiencies shown here, it is necessary to go through an intermediate, model-dependent step. 
This can be understood by considering Fig.~\ref{fig:t2k_eff_example} and asking what the efficiency is for a $p_{\mu}$ bin. Clearly it depends on the distribution of events within that bin in $\cos\theta_{\mu}$, as the efficiency also varies as a function of $\cos\theta_{\mu}$. Therefore we have to multiply the efficiency in two-dimensions by the predicted rate given the relevant flux, incident neutrino species and target material, and then collapse that distribution onto the axis of interest, before dividing by the total number of simulated events in that bin. This {\it pseudo-efficiency} is in principle dependent on the model used to transform from two to one dimensions. The resulting efficiencies as a function of $p_{\mu}$ when following this procedure from the $p_{\mu}$--$\cos\theta_{\mu}$ in Fig.~\ref{fig:t2k_eff_example} is shown in Fig.~\ref{fig:t2k_eff_comp} for all generators used in this work. The differences between the one-dimensional efficiencies produced using different model assumptions are small in this case, but are present. Figure~\ref{fig:t2k_eff_comp} also shows an example of the sort of efficiency plots found later in this work, where different generator predictions of the cross section are compared with the efficiency. In this figure, and all others in the work, the NEUT generator prediction is used to calculate the overlaid efficiency.

We note that this procedure, integrating over a model prediction for other kinematic variables, is explicitly done by experiments when they present results as a function of a single variable, and has the potential to introduce severe model dependence in some cases.

\section{Experiment overview} 

\subsection{Neutrino fluxes} \label{sec:expfluxes}
\begin{figure} 
  \centering
  \includegraphics[width=\linewidth]{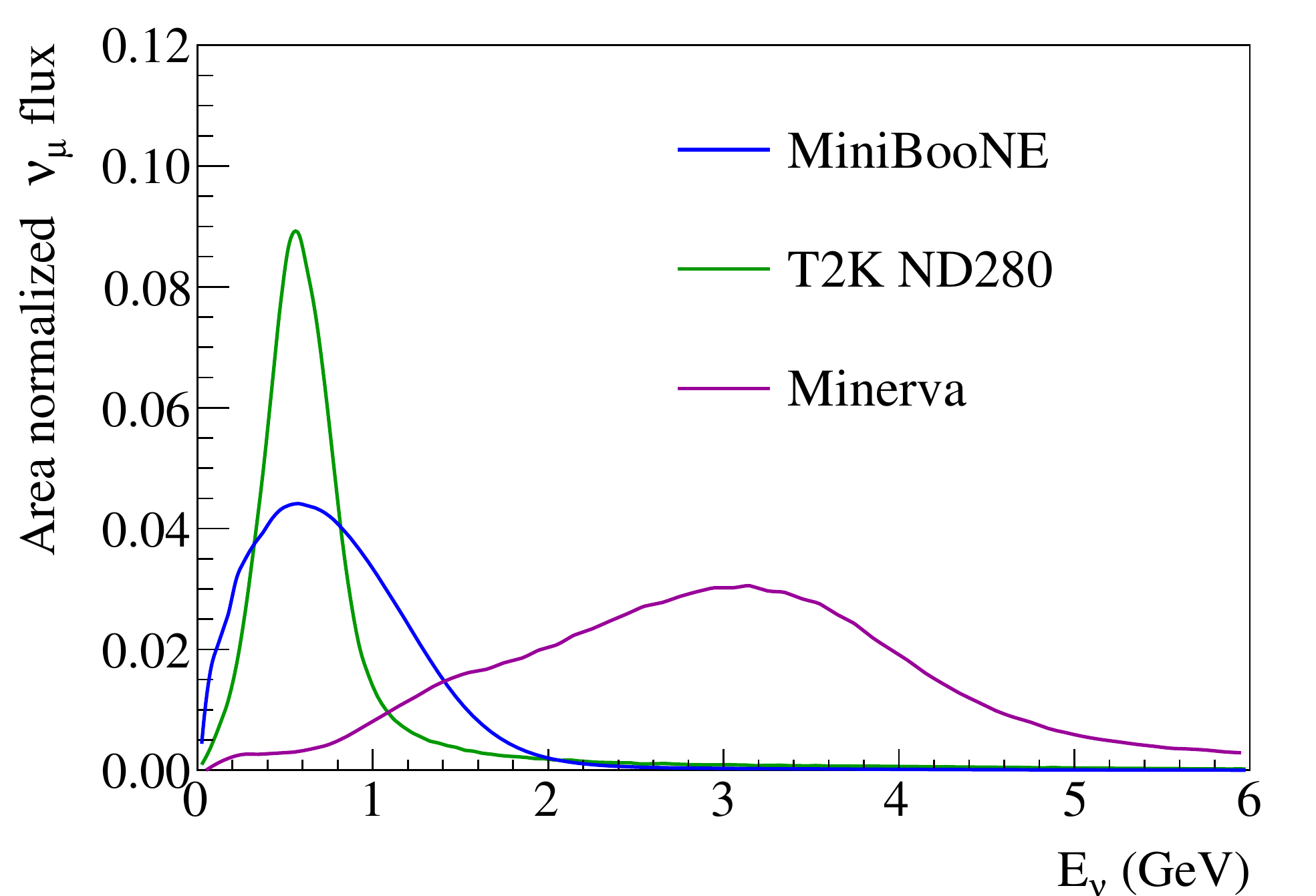}
  \caption{Area normalized neutrino flux distributions for the T2K, MINER$\nu$A Low Energy run, and MiniBooNE experiments.  Only the muon neutrino component is considered, the wrong-sign background contributions are not presented.  The T2K flux is that at the off-axis near-detector, ND280.  The MicroBooNE flux shape is almost identical to the MiniBooNE flux shape.}
  \label{fig:expfluxes}
\end{figure}

{Neutrino interaction measurements are intrinsically produced averaged over an incoming neutrino energy spectrum.
Figure \ref{fig:expfluxes} shows the neutrino flux spectra for the three experiments considered in this paper.
The different distributions of neutrino energies impact the interpretation of measurements and their relationships to one another.
All beams start with a primary proton beam which impinges on a target to produce pions and kaons.  These secondary particles are focused using magnetic horns and subsequently decay into neutrinos.  Table \ref{tab:fluxes} shows the proton energy, target material, decay pipe length, and resultant peak neutrino energy for the three neutrino beams used by the running experiment discussed in this work.
It should be noted that the T2K experiment uses the off-axis strategy, which results in a lower and narrower energy spectrum than the on-axis flux in the same beam.
}

\begin{table*}\label{tab:fluxes}
  \caption{Various parameters that control the neutrino beam energy distributions for the beams considered in this paper.}
  \label{tb:fluxes}
\begin{ruledtabular}
{\renewcommand{\arraystretch}{1.2}
  \begin{tabular}{c|c|c|c|c}

  Experiment & Proton energy & Target material & Decay pipe length & Peak energy \\
  \hline 
  T2K & 30~GeV & Graphite & 90~m & 0.6~GeV \\
  \minerva & 120~GeV & Graphite & 500~m & 3~GeV \\
  Mini/Micro-BooNE & 8~GeV & Beryllium & 50~m & 0.6~GeV \\

  \end{tabular}}
  \end{ruledtabular}
\end{table*}

\subsection{The T2K experiment} \label{sec:t2kexp}
The Tokai-to-Kamioka (T2K) experiment has detectors both on- and off-axis. The data considered in this paper are from the off-axis near detector, ND280~\cite{nd280}, which sits $2.5\degree$ from the beam center.
The off-axis angle provides a narrow-band neutrino beam peaked at $E_\nu\sim600\text{ MeV}$ with a suppressed high energy tail. FLUKA is used to model the proton-graphite target interaction and GEANT3/GCALOR is used to propagate the particles through the horns and decay volume. The simulation uses proton beam monitor measurements as inputs, and the modeling of hadronic interactions in the target is constrained using thin target hadron production measurements from the NA61/SHINE experiment at $\sim30\text{ GeV}$ \cite{Abe:2012av}. Downstream of the decay volume and absorbers, there are two muon monitors (MUMON) \cite{SUZUKI2012453,MATSUOKA2010591} monitoring the muon direction for muons with $p_\mu>5\text{ GeV}$. The neutrino rate and direction is monitored by the INGRID detector, which sits near ND280 but is centered on the neutrino beam, spanning $\pm 5\text{ m}$ in both dimensions \cite{ABE2012211}.

ND280 is composed of several subdetector systems, all enclosed in a 0.2 T magnetic field. The Pi-Zero Detector (P0D) \cite{Assylbekov:2011sh}  is composed of orthogonal scintillator tracking planes interleaved with refillable water layers, and sheets of brass. The two Fine-Grained Detectors (FGDs) \cite{Amaudruz:2012esa} are composed of orthogonal scintillator tracking planes (CH), one of which also contains alternating planes filled with water (FGD2). The FGD's scintillator bars are composed of $86.1\%$ carbon, $7.4\%$ hydrogen, $3.7\%$ oxygen, $1.7\%$ titanium, $1\%$ silicon, and $0.1\%$ nitrogen by mass. {Interleaved between} each of the three sub-detector modules (P0D, FGD1, FGD2), as well as downstream of FGD2, are gaseous argon Time Projection Chambers (TPCs) \cite{Abgrall:2010hi} which measure track characteristics at high resolution, providing sign selection and momentum measurements of tracks.  All of the sub-detector modules are surrounded by electromagnetic calorimeters (ECals)\cite{Allan:2013ofa}. The ECals surround the tracker and consists of 13 modules made up of plastic scintillator bars alternating with lead sheets. Finally, surrounding the ECals, burrowed in slats in the magnet, is the Side Muon Range Detector (SMRD) \cite{Aoki:2012mf}, used to tag escaping particles and particles entering from outside the detector, e.g. cosmic muons.

The neutrino event generator NEUT \cite{neut} is used to simulate neutrino interactions in the detector, more details are reported in \autoref{sec:neut} and GEANT4 version 4.9.4
~\cite{geant} is used to simulate the detector response and passage of particles through materials. 

\subsection{The MINERvA experiment}
MINERvA is located {on-axis in} the NuMI beamline at Fermilab.
{The on-axis beam peaks at 3~GeV and contains 95\% $\nu_\mu$, with the remainder consisting of $\bar{\nu}_\mu$, $\nu_e$, and $\bar{\nu}_e$~\cite{Aliaga:2016oaz}. The data presented here is from the Low Energy run, the MINERvA collaboration has started to produce several cross section measurements with the NuMI Medium energy flux~\cite{MINERVAME, MINERvA:2021owq}.
The neutrino beam is simulated with GEANT4 9.2.p03~\cite{geant}, and constrained with thin-target hadron production measurements and an in-situ neutrino electron scattering constraint~\cite{Park:2015eqa}. }

The \minerva detector uses plastic scintillator bars with a triangular cross section, arranged in 3 directions each 60 degrees from the other two. 
The \minerva detector~\cite{minervanim} is segmented longitudinally into several regions: nuclear targets, the scintillator tracker, and downstream electromagnetic and hadronic calorimeters. 
The nuclear target region contains five solid passive targets of carbon (C), iron (Fe), and lead (Pb), separated from each other by 4 or 8 scintillator planes for vertex and particle reconstruction. 
Targets 1, 2 and 3 contain distinct segments of Fe and Pb planes that are 2.6 cm thick; target 3 also has a C segment which is 7.6 cm thick and target 5 has Fe and Pb segments which are 1.3 cm thick.  
The tracker is made solely of scintillator planes, the fiducial volume contains 106 planes. The target mass of the fiducial volume is a mix of carbon in 88.51\%, hydrogen in 8.18\%, oxygen in 2.5\%, titanium in 0.47\%, chlorine in 0.2\%, aluminum in 0.07\%, and silicon in 0.07\%.
The MINOS Near Detector is two meters downstream of the \minerva detector and serves as a magnetized muon spectrometer \cite{Michael:2008bc}.

The neutrino event generator GENIE 2.12.6 with some additions is
used to simulate neutrino interactions in the detector. 
$2p2h$ interactions and long range correlations estimated using the Random Phase Approximation (RPA) from Valencia model are included. The interactions and decays of particles produced in the neutrino interactions of the final-state particles that exit the nucleus are simulated by Geant4 9.4.2~\cite{geant}.

\subsection{The MiniBooNE experiment}

{MiniBooNE used neutrinos from the Booster Neutrino Beam. 
The beam has an average energy of 800 MeV, and is $93.6\%$ $\nu_{\mu}$ with 5.9$\%$ (0.5\%) contamination of $\bar{\nu}_{\mu}(\nu_e,\bar{\nu_e})$.
The beam simulation was tuned to external hadron production measurements from HARP experiment}~\cite{HARPP}.

The detector is composed of 800 tons of mineral oil (CH$_{2}$) that serves as both the target for neutrino interactions and the medium in which charged particles produced in neutrino interactions radiate Cherenkov and scintillation photons. The photons are detected on an array of 1520 photomultipliers, and the resulting spatial and temporal patterns of light are used to identify and reconstruct the interactions. 
For particles above Cherenkov threshold, the scintillation light is a minor component, however the scintillation light is important for interactions that do not produce any particles above threshold. Scintillation light provides only position and energy information, whereas Cherenkov light additionally provides direction information.

The NUANCE event generator was used to simulate neutrino interactions and GEANT3-based program to simulate the response of the detector to neutrino interactions. The NUANCE generator main components includes: a relativistic Fermi gas model for CCQE and NC elastic scattering, a baryonic resonance model for CC/NC single pion production model, a deep inelastic scattering model and a final-state interaction model to simulate re-interaction of final state hadrons in nuclear medium. The simulation did not include long range correlation such Random Phase Approximation (RPA) or $2p2h$ nuclear effects.

\subsection{The MicroBooNE experiment}

MicroBooNE sits in the booster neutrino beam (BNB) upstream of MiniBooNE.  As a smaller detector than MiniBooNE it subtends a smaller angle relative to the neutrino beam direction, and as such has a slightly different energy spectrum, but the difference is very small.
MicroBooNE uses the same flux simulation chain and data constraints as MiniBooNE.

The MicroBooNE detector is a Liquid Argon Time Projection Chamber (LArTPC) with 85 tons of active mass.
In the MicroBooNE detector, charged particles leave trails of ionization electrons as they traverse the argon and also create prompt ultraviolet scintillation photons. The electrons drift in an electric field to one side of the detector, where they are detected by a series of sensing wires on three separate planes. The scintillation photons are instead detected by 32 photomultipliers. The liquid argon acts both as target material and as detector for charged particles.

For the analyses covered here, neutrino interactions are simulated
using the GENIE v2.12.2 version with the addition of the empirical
MEC, while cosmogenic particles (which constitute a significant
background in many MicroBooNE analyses) are simulated with
CORSIKA~\cite{corsika}. Particles are then propagated by GEANT4, while
the simulation of the MicroBooNE detector is performed in the LArSoft
framework~\cite{larsoft}.

\section{Inclusive interactions}
\label{sec:incl}
\subsection{Introduction} 

Inclusive interactions include all the neutrino interactions with the nucleus, without any particular requirement on the number or type of final state particles. Inclusive measurements are important because they often allow selecting a large sample of neutrino interactions with high efficiency and purity, they are mildly sensitive to hadron uncertainties and finally, allow testing multiple contributing processes at once. In addition, some calculations require integration over the hadronic final state and are available only for the inclusive cross sections. 
In this section, a comparison between charged-current inclusive measurements from T2K, MicroBooNE, and MINERvA is presented. The measurements from T2K and MicroBooNE use similar fluxes (see \autoref{sec:expfluxes}) and the same observables from the muons---$\cos\theta_\mu$ and $p_{\mu}$.  The measurement from MINERvA uses a neutrino flux with a higher mean of 3~GeV, and 
is made with different observables---both muon and hadronic information is measured to test more information about the models.
Additionally, the signal definition is different so MINERvA sees different contributions from interaction channels, such as CCQE and CC DIS.  MINERvA's charged-current inclusive measurement reports events with low three-momentum transfer, $|q_3|<0.8~\text{GeV}$, which does not include the multi-$\pi$ and DIS events, with more details in \autoref{sec:minerva_ccinc}. Meanwhile, T2K and MicroBooNE include all the events in the inclusive sample. The next sections outline details about the T2K, MicroBooNE and MINERvA measurements.

\subsection{T2K results}

The T2K collaboration produced a muon-neutrino CC-inclusive double-differential cross section on a carbon target using a beam of muon neutrinos with a peak energy of 0.6~GeV~\cite{Abe2018}.
The cross section was extracted as a function of the unfolded muon momentum $p_\mu$ and $\cos\theta_\mu$, where $\theta_\mu$ is the angle between the muon and the average incoming neutrino direction.

The main selection consists of four samples of $\nu_{\mu}$ charged-current interactions inside a $\sim \SI{1}{m^3}$ FV in FGD1, based on the angle of the muon with respect to the detector axis: forward going (FWD), backward going (BWD), high angle forward going (HAFWD), and high angle backward going (HABWD). 
The aim of the selection is to find events with at least a muon in the final state.
Depending on the angle of the muon inside the detector, different selection criteria are used.
Forward going and backward scattered muons are identified by the energy deposited in the gaseous argon TPCs and their track curvature in the magnetic field, i.e. their momentum.
Muons that are scattered close to perpendicular to the neutrino direction do not cross a TPC, and are instead identified in the electromagnetic calorimeters by a multivariate discriminator which separates muon and pion tracks from showering particles.
Additionally, the Side Muon Range Detector is used to tag forward going high angle muons, as well as veto seemingly backward going cosmic muon background events.
The composition of the selected signal events depends on the sample and is shown in \autoref{tab:t2ksamples}.
The contribution of $2p2h$ is between $5.5\%$ in the backward-going samples and $7.5\%$ in the forward-going sample.

\begin{table}[ht]
    \caption{Relative composition (\%) of the charged-current signal in the four samples of the T2K measurement according to NEUT.~\cite{Abe2018}}
    \label{tab:t2ksamples}.
    \centering
    \begin{tabular}{l|SSSS}
         & {FWD} &  {BWD} &  {HAFWD} &  {HABWD}  \\
        \hline
        QE & 44.7 & 82.0 & 67.3 & 83.2 \\
        $2p2h$ & 7.5 & 5.5 & 7.2 & 5.5 \\
        RES & 25.4 & 8.6 & 17.6 & 8.0 \\
        DIS & 19.9 & 3.8 & 7.2 & 3.4 \\
        COH & 2.5 & 0.0 & 0.7 & 0.0
    \end{tabular}
\end{table}

Migration of events between the kinematic bins is handled by an unfolding procedure (see ~\cite{Abe2018} for more details) producing a spectrum of events in ``true'' kinematic variables.
The number of background events is constrained by the selection of dedicated control regions, and fitted to the data in a simultaneous fit.
That means both the signal and background event rates are determined together and correlations and migrations between samples are handled naturally.
The resulting signal event distributions are then scaled by a bin-dependent efficiency correction to account for detection and reconstruction inefficiencies, and converted into a flux-integrated cross section using the known neutrino flux profile and number of target nuclei.

To judge the model dependence of the unfolding procedure, the result is extracted with two different models as the nominal assumptions.
Despite best efforts to make the unfolding procedure as model-independent as possible, the extracted cross-sections differ slightly but noticeably between the two.
This is due to a differing efficiency under the two model assumptions, especially at low-momentum, forward-going bins in muon kinematics.
Since the muon kinematics should not differ much within a single bin, these deviations probably come from differences in the distributions of hadronic particles.
Although they are integrated over for the inclusive cross-section, they still affect the overall reconstruction efficiency.
This underlines the need to understand and investigate the efficiency performance of an analysis for ``hidden'' variables that are not included in the signal definition.

\subsection{MicroBooNE results} 

The MicroBooNE collaboration produced a muon-neutrino CC-inclusive double-differential cross section using a beam of muon neutrinos with a mean energy of 0.8~GeV~\cite{Abratenko2019}. The cross section was extracted as a function of the reconstructed muon momentum, $p_\mu^{\textrm{reco}}$, and the muon direction, $\cos\theta_\mu^{\textrm{reco}}$, where $\theta$ is the angle between the muon and the beamline.

The inclusive sample of $\nu_{\mu}$ charged-current interactions is selected~\cite{Abratenko:2020sga,ub_publicnote_1045} inside a 44 t fiducial volume (FV), requiring one muon with or without the presence of other particles in the final state. The muon can be either contained inside, or can exit the detector, and can have any direction. The muon momentum is calculated using multiple Coulomb scattering, by fitting an argon-tuned Highland formula along the candidate muon trajectory~\cite{microboone_mcs}. This method is equally applicable to muons that are fully contained, and those that exit, but suffers from relatively poor (10-20\%) resolution. No angular or energy cut is applied. The deposited charge per unit length ($dQ/dx$) is used to discriminate muons from protons. Several algorithms ensure the quality of the fitted track by limiting the allowed spatial dispersion of the reconstructed hits with respect to the track hypothesis. Since the MicroBooNE detector is on the Earth's surface and takes several milliseconds to read out data, cosmic rays are the dominant background for an inclusive muon neutrino analysis. A series of algorithms is used to identify these background event, by looking at tracks that traverse the detector from top to bottom, that do not match with the light activity arriving in time with the neutrino beam, and by looking at the Bragg peak and Michel electrons to identify stopping muons, which overall reduce the cosmic rate by more than 3 orders of magnitude.

The analysis follows a so-called forward-folding technique and the measurement is presented in terms of reconstructed variables instead of true, unfolded, ones. More details on the forward-folding method and its limits of applicability are given later in \autoref{sec:forwardfold}. The analysis reported in the following uses data collected between February to July 2016, and corresponds to $1.6 \times 10^{20}$ protons on target.

\subsection{MINERvA results} 
\label{sec:minerva_ccinc}
MINERvA reported the first inclusive charged-current double-differential cross section as a function of three momentum transfer and available energy. The three momentum transfer is obtained using the four momentum transfer $Q^2$ and the energy transfer $q_0$
\begin{equation}
q_3=\sqrt{Q^2 + q_0^2}
\end{equation}
where the $Q^2$ is obtained using the energy of the neutrino, muon angle and momentum ($Q^2=2E_{\nu}(E_\mu-p_\mu\cos\theta_\mu)-M_{\mu}$). The muon momentum is calculated by using the ionization energy loss for a muon traversing the material in the MINERvA detector in conjunction with the momentum reconstructed from MINOS experiment. The neutrino energy is reconstructed using ($E_{\nu}=E_{\mu}+q_0$).
The energy transfer, $q_0$ is estimated by summing the visible hadronic energy and applying 
model-dependent corrections for unobserved neutrons and nucleon removal energy ~\cite{Rodrigues:2015hik}. 
A new variable called available energy ($E_\text{avail}$)  was defined to unfold and report the cross section. 
This variable is close to the true energy transfer, but does not include energy of the neutrons (because they leave very small energy in the detector), or other forms of missing energy (nuclear recoil, binding energy, etc). The resolution of $E_\text{avail}$ varies from $55\%$ to $38\%$.
The true $E_\text{avail}$ is defined as:

\begin{equation}
    E_\text{avail}=\sum T_p+\sum T_{\pi^{\pm}}+\sum E_{\text{particles}}
\end{equation}
where $\sum T_p$ is the proton kinetic energy, $\sum T_{\pi^{\pm}}$ is the pion kinetic energy
\begin{equation}
    \begin{aligned}
    \sum & E_{\text{particles}} = \\ 
         & \sum E_{K^{\pm}}+\sum E_{e^{\pm}}+\sum E_{\pi^{0}}+\sum E_{\gamma}
    \end{aligned}
\end{equation}
 and $\sum E_{\text{particles}}$ is the total energy of other particles except neutrons. In the reconstruction $E_\text{avail}$ is estimated using the calorimetric sum of the visible energy not associated with the muon.

The inclusive sample of $\nu_{\mu}$ charged current interactions is selected using events in MINERvA's 5.3 ton active-tracker FV, the sample includes muon tracks that are matched to a track in the MINOS detector and $\theta_{\mu}<20\degree$ and $p_{\mu}>1.5$~GeV. The signal definition is charged-current $\nu_\mu$ with $2$~GeV$<E_{\nu}<6$~GeV in the true neutrino energy, $p_{\mu}>1.5$ GeV and $\theta_{\mu}<20\degree$. The measurement is reported for low three-momentum transfer ($q_3<0.8$~GeV). Selection on neutrino energy might introduce model dependence, a better signal definition should avoid cuts on neutrino energy or any other observable with model dependence.

An unfolding procedure~\cite{DAgostini:1994zf} with four iterations was applied in two dimensions to translate the data from reconstructed quantities to true ($E_\text{avail}$, $q_3$). GENIE 2.8.4 was used to correct for the acceptance of the FV, the efficiency of the MINOS muon match, and the subtraction of small ($3\%$) neutral-current and $\mu^+$ backgrounds~\cite{Rodrigues:2015hik}.

\subsection{Comparisons of event generator predictions}

To explore in more detail the contributions from each experiment, \autoref{tb:t2k_gen_predictions}, \autoref{tb:ub_gen_predictions} and \autoref{tb:minerva_gen_predictions} shows the event generator predictions broken down by true interaction channels, including the signal definition cuts for each measurement. Concerning the QE predictions of the generators at the three experiments, we see by far the lowest prediction from {\Genietwoa} at MicroBooNE, which then is largest prediction at \minerva and T2K has similar prediction for {\Genietwoa} and \Genietena. The {\Genietena} prediction roughly agrees with NEUT and NuWro at MicroBooNE, but is 10\% lower at \minerva energies, whereas NEUT and NuWro seem to scale similarly to each other.  At T2K energies, NuWro is 13\% higher compared with NEUT and both GENIE versions.  This is interesting because many similar choices have been made according to Tab.~\ref{tb:qemodels}, making implementation important for understanding these differences.  Looking at the $2p2h$ prediction, NEUT, NuWro and {\Genietena} all utilize the Valencia $2p2h$ model, but we see up to 20\% difference due to implementation choices. GENIE {\Genietwoa} uses Empirical MEC for the $2p2h$ model, yet still produces a similar $2p2h$ predictions for MicroBooNE and T2K, but smaller prediction for MINERvA.  This is likely due to both models being tuned to MiniBooNE CCQE-like data, but handling the neutrino energy scaling differently. This suggests that different data sets could be used to constrain the energy dependence of the models. In T2K, $2p2h$ predictions are similar for each generator except for {\Genietwoa}, which is 27\%  higher because it uses a different model than the others. 

Focusing on the MicroBooNE and T2K breakdown, the single-$\pi$ production, multi-$\pi$ production, single $\eta$ production, and DIS contributions are all different. Some of this is due to nomenclature of defining interaction modes, e.g. summing the resonant, multi-$\pi$, $\eta$ and DIS contributions is required to get directly comparable  contributions from the generators. There are differences in strategy; for instance  NuWro favors the CC1$\pi$+1p for its tuning whereas NEUT tries to tune to all the data and inflates uncertainties accordingly.  GENIE~\cite{GENIEFreeNucleonTune} also  does a fit to all available data with different choices.  NuWro, GENIE, and NEUT are tuned to ANL and BNL bubble chamber data \cite{Graczyk:2009qm, Abe112008}.  All have similar predictions for MicroBooNE where $\Delta(1232)$ excitation dominates due to the neutrino energy range. When scaling up the same interaction mode to \minerva energies, the predictions separate with NuWro largest; NEUT is somewhat smaller and GENIE smaller yet.  This again displays the different choices for energy dependence of different interaction modes.  Although the choices allowed given the large uncertainties in the single nucleon data are important, nuclear modeling can also cause differences.  This is especially important for higher energy long baseline neutrino oscillation experiments such as NOvA and DUNE.

\begin{table*}
\begin{ruledtabular}
    \centering
        \caption{Event generator cross section predictions for the different true interaction channels for T2K charged-current inclusive cross section ($\times 10^{-38} \text{cm}^2$/GeV/($\cos(\theta$)) / nucleon).}
    \label{tb:t2k_gen_predictions}
    {\renewcommand{\arraystretch}{1.2}
    \begin{tabular} {l|cc||cccccc|c}
        
        Generator& QE& $2p2h$& $1\pi^+1p$ & $1\pi^0$ & $1\pi^+1n$ & $N\pi$ & $1\eta$ & DIS & 
        \begin{tabular}{c} 
            $1\pi^{+,0}+N\pi$ \\ 
            $+ 1\eta+\text{DIS}$
        \end{tabular} \\
        \hline
        GENIE v3 18\_02a   &2.80  &0.82 & 1.04 & 0.31 & 0.32 & 0.53 & 0.03 &0.42 & 2.65\\
        GENIE v3 18\_10a   &2.79  &0.55 & 1.12 & 0.33 & 0.34 & 0.57 & 0.03 &0.46 & 2.84\\
        NuWro 19.02         &3.19  &0.55 & 1.16 & 0.28 & 0.23 & 0.12 & 0.03 &0.79 & 2.61\\
        NEUT v5.4.0.1       &2.84  &0.56 & 1.19 & 0.36 & 0.37 & 0.56 & 0.07 &0.50 & 3.05 \\

    \end{tabular}}
    \end{ruledtabular}
\end{table*}

\begin{table*}
\begin{ruledtabular}
    \centering
        \caption{Event generator cross section predictions for the different true interaction channels for MicroBooNE charged-current inclusive cross section ($\times 10^{-38} \text{cm}^2$/GeV/($\cos(\theta$)) / nucleon).}
    \label{tb:ub_gen_predictions}
    {\renewcommand{\arraystretch}{1.2}
    \begin{tabular} {l|cc||cccccc|c}
        
        Generator & QE& $2p2h$& $1\pi^+1p$ & $1\pi^0$ & $1\pi^+1n$ & $N\pi$ & $1\eta$ & DIS & 
        \begin{tabular}{c} 
            $1\pi^{+,0}+N\pi$ \\ 
            $+ 1\eta+\text{DIS}$
        \end{tabular} \\
        \hline
        GENIE v3 18\_02a   &2.88  &0.75 & 1.03 & 0.40 & 0.39 & 0.31 & 0.03 &0.05 & 2.21\\
        GENIE v3 18\_10a   &3.30  &0.73 & 1.10 & 0.43 & 0.41 & 0.34 & 0.03 &0.05 & 2.36\\
        NuWro 19.02         &3.42  &0.58 & 1.07 & 0.38 & 0.30 & 0.14 & 0.00 &0.23 & 2.13\\
        NEUT v5.4.0.1       &3.33  &0.61 & 1.14 & 0.48 & 0.49 & 0.42 & 0.08 &0.06 & 2.68 \\
        
    \end{tabular}}
    \end{ruledtabular}
\end{table*}

\begin{table*}
\begin{ruledtabular}
    \centering
        \caption{Event generators predictions for the different true interaction channels for MINERvA charged-current inclusive cross section ($\times 10^{-38}  \text{cm}^2$/ GeV$^2$ / nucleon) }
    \label{tb:minerva_gen_predictions}
    {\renewcommand{\arraystretch}{1.2}

    \begin{tabular} {l|cc||cccccc|c}
        
        Generator& QE& $2p2h$& $1\pi^+1p$ & $1\pi^0$ & $1\pi^+1n$ & $N\pi$ & $1\eta$ & DIS & 
        \begin{tabular}{c} 
            $1\pi^{+,0}+N\pi$ \\ 
            $+ 1\eta+\text{DIS}$
        \end{tabular} \\
        \hline
        GENIE v3 18\_02a   &7.99  &1.62 & 1.88 & 0.33 & 0.43 & 0.05 & 0.00 &0.00 &2.69 \\
        GENIE v3 18\_10a   &6.76  &1.49 & 1.93 & 0.33 & 0.44 & 0.05 & 0.00 &0.00 & 2.75\\
        NuWro 19.02         &7.85  &1.49 & 2.60 & 0.39 & 0.46 & 0.05 & 0.00 &0.00  & 3.51\\
        NEUT v5.4.0.1       &7.29  &1.40 & 2.30 & 0.48 & 0.81  & 0.07 &0.02 &0.00 & 3.68 \\
      
    \end{tabular}}
    \end{ruledtabular}  
\end{table*}

\subsection{Comparisons of generators with  T2K, MicroBooNE and MINERvA inclusive data}
\begin{figure*}[ht!]
    \centering
    \includegraphics[width=\linewidth]{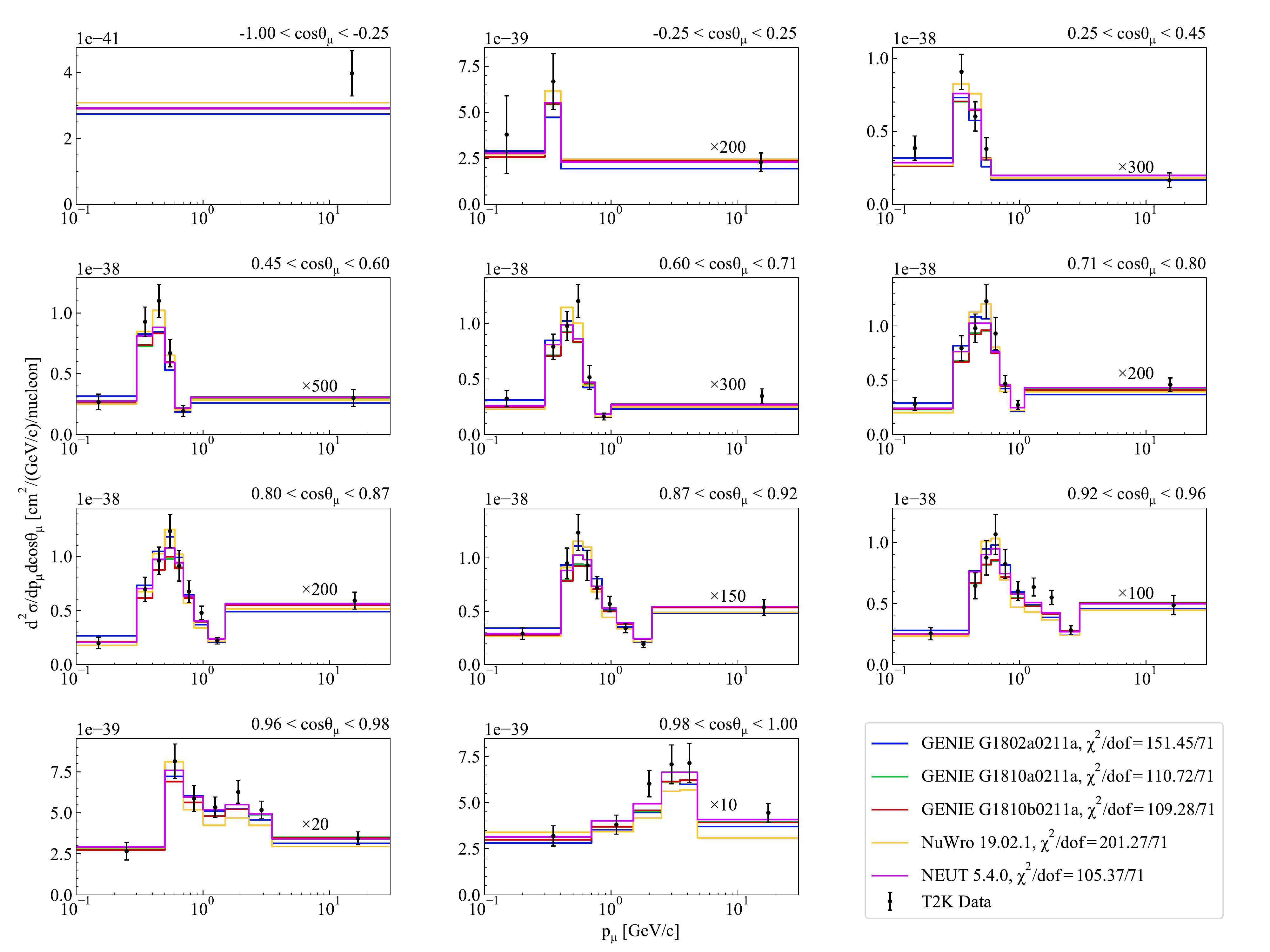}
    \caption{T2K double-differential cross-section $\textrm{d}^2\sigma/(\textrm{d}p_\mu^{\textrm{reco}}\textrm{d}\cos\theta_\mu^{\textrm{reco}})$ in nine regions of $\cos\theta_\mu^{\textrm{reco}}$ is compared to NuWro, NEUT, and three versions of GENIE.}
    \label{fig:inclusivet2k}
\end{figure*}

\begin{figure*}[ht!]
    \centering
   \includegraphics[width=\linewidth]{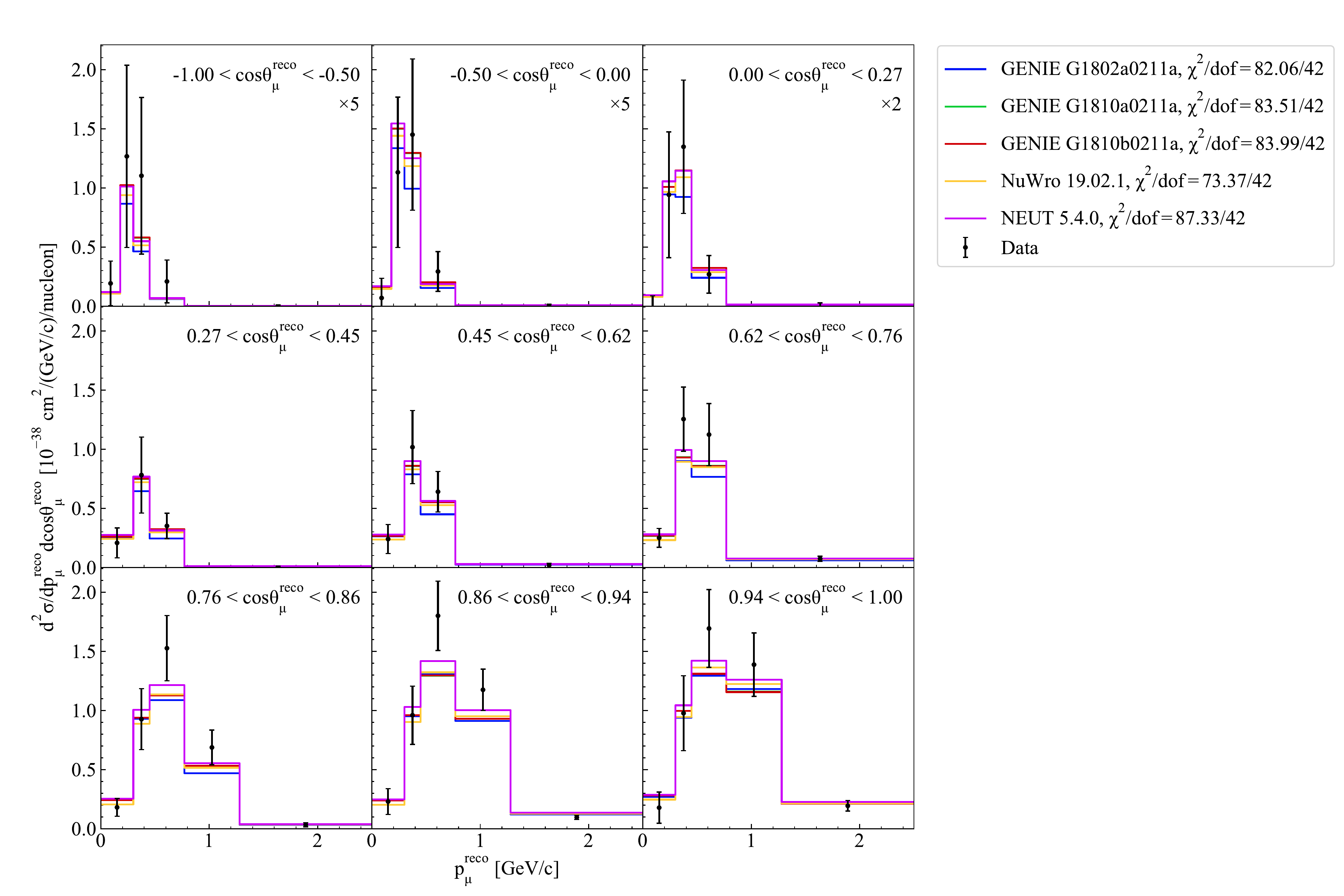}
    \caption{MicroBooNE double-differential cross-section $\textrm{d}^2\sigma/(\textrm{d}p_\mu^{\textrm{reco}}\textrm{d}\cos\theta_\mu^{\textrm{reco}})$ in nine regions of $\cos\theta_\mu^{\textrm{reco}}$ is compared to NuWro, and three versions of GENIE.}
    \label{fig:inclusivemicroboone}
\end{figure*}

\begin{figure*}[ht!]
    \centering
    \includegraphics[width=\linewidth]{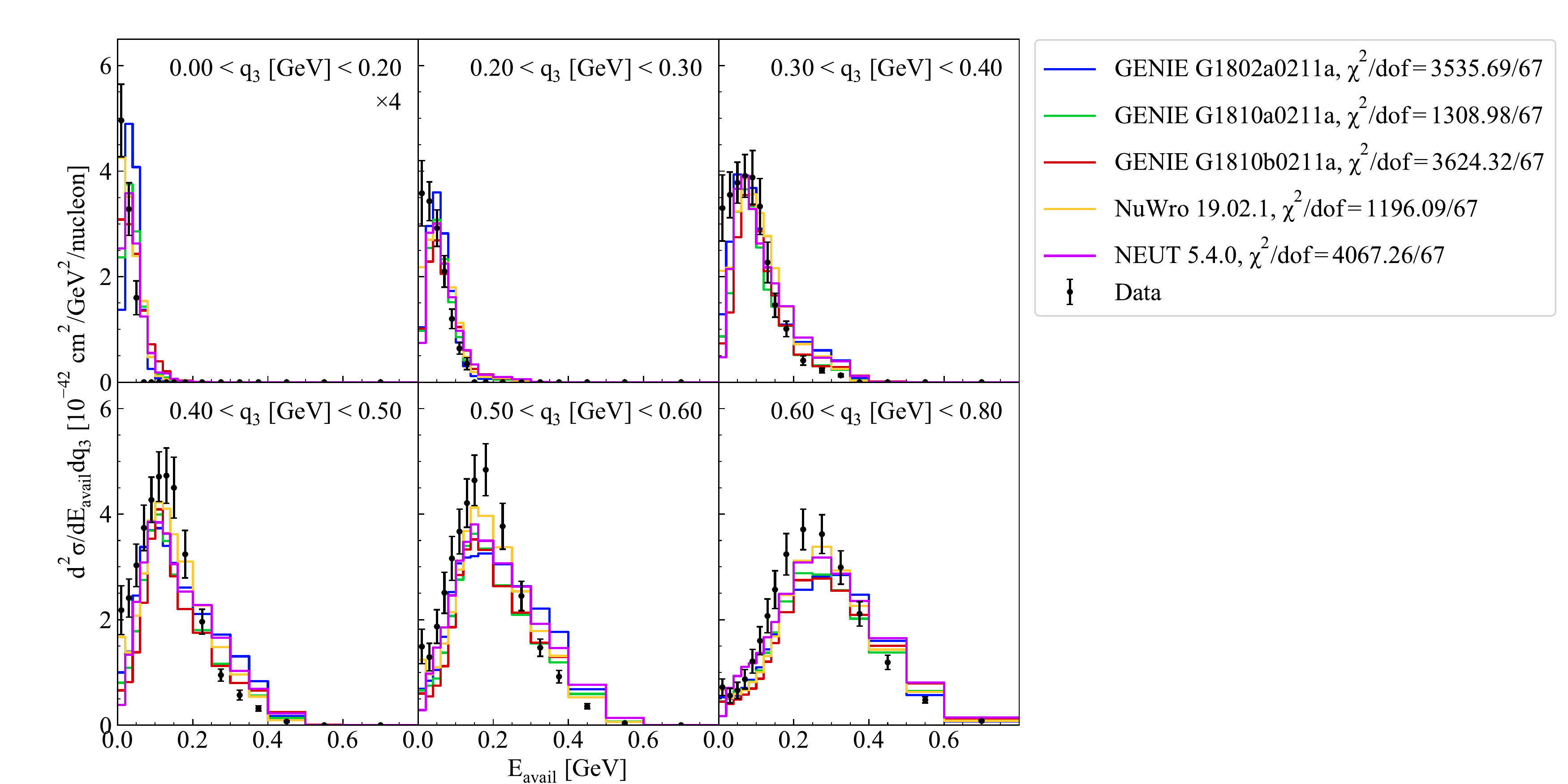}
    \caption{MINERvA double-differential cross-section $\textrm{d}^2\sigma/\textrm{d}E_{\textrm{avail}}\textrm{d}q_3$ in six regions of $q_3$ is compared to NuWro,  and three versions of GENIE.}
    \label{fig:inclusiveminerva}
\end{figure*}

This paper reports comparisons of the T2K, MicroBooNE and MINERvA's double-differential cross-section measurements with the different event generator predictions in Fig.~\ref{fig:inclusivet2k}, ~\ref{fig:inclusivemicroboone} and~\ref{fig:inclusiveminerva}. T2K and MicroBooNE have similar kinematic variables and use similar fluxes, but different targets.  MINERvA's target is  the same as T2K's, but the measurement variables, acceptance and fluxes are different from T2K and MicroBooNE. Therefore, the comparisons between these experiments is not straightforward. 
Importantly, while T2K and MicroBooNE exclusively measure the muon in these data, MINERvA's measurement uses kinematic variables that contain the muon and hadronic information from the event which is a further test of the models used to reconstruct the neutrino energy in oscillation experiments, noting that the NOvA and DUNE neutrino experiments both use the full kinematics of the event to reconstruct the neutrino energy at the cost of model-dependent corrections. 

Figure~\ref{fig:inclusivet2k} shows the inclusive T2K result compared to some model predictions and 
\autoref{fig:inclusivemicroboone} shows the same model comparisons with the MicroBooNE data.
Based on the $\chi^2$ values, the models describe the T2K data poorly. The best $\chi^2 / \mathrm{dof} = 105/71$ is given by the NEUT event generator, while NuWro shows the worst agreement among the compared generators, with a $\chi^2 / \mathrm{dof} = 201/71$.
Most of the difference between NuWro and the other generators seems to be located as a lower prediction in the high-momentum, very forward-going data points ($\cos\theta > 0.92$ and $p_\mu > \SI{1}{GeV}$), as well as a higher prediction in the peak of the cross section at around $p_\mu = \SI{500}{MeV/c}$.

In MicroBooNE's muon-neutrino double-differential cross-section measurement, about half of the events are quasi-elastic processes, and the remaining half are $2p2h$ and resonance processes, with a small contribution from deep inelastic scattering.
The data likewise is described poorly by the generators especially in the forward-going region where there is the largest tension between the data and the generators. Here, NuWro has the best agreement with $\chi^2 / \mathrm{dof} = 73/42$, and NEUT has the worst score of $\chi^2 / \mathrm{dof} = 87/42$.
The difference is much smaller than in the T2K case, though.
Also, as the difference between NuWro and the other generator predictions is less pronounced, it is more difficult to say what kinematic region is actually causing the difference. Based on separate $\chi^2$ analysis for each measurement, the generators have equivalent ability to describe general characteristics at neutrino energies of $\sim$1 GeV in Carbon and Argon. 

In MINERvA's double-differential cross-section measurements, the region at low available energy below 0.15~GeV is dominated by QE processes, the region at high available energy above 0.2~GeV is dominated by delta resonance events, and events in the intermediate region contains $2p2h$ contributions. Discrepancies between data and the different generator predictions are visible. None of the generators correctly predict the first bin of available energy for the momentum transfer region $0$~GeV$<q_3<0.4$~GeV --- dominated by QE events. In the region between the QE and delta process, where $2p2h$ events are expected, all generators underestimate the data. The best prediction is from the NuWro event generator with a $\chi^2 / \mathrm{dof}=1196$ and the GENIE $G18\_10a$ with a $\chi^2 / \mathrm{dof}=1308$, both simulations contain the same nuclear model (LFG), the same $2p2h$ model from Nieves and different long range correlations (RPA) models. However, the $\chi^2/dof$ are usually big values, most likely due to the strong correlations in the uncertainties among the data points.
Overall, none of the event generators predict the inclusive data well, data is under predicted in different regions of QE, MEC and RES. The main sources of disagreement is for low values of available energy and the middle region where the MEC and RES events are located. 

In the publication~\cite{Rodrigues:2015hik}, the data were compared with GENIE 2.8.4 with reduced pion production.  In this publication a discrepancy was reported, specifically in the region between QE and delta. In addition, the data were compared to a simulation that contained RPA~\cite{Nieves:2004wx, Gran:2017psn} and $2p2h$ contributions from the Nieves group. The simulation with the addition of RPA showed good agreement at the lowest $E_\text{avail}$ bins. The new version of GENIE has different predictions at the lowest $E_\text{avail}$ bins due to a different treatment of the binding energy of the protons, new nuclear model and has adopted Valencia QE and $2p2h$, see more details at \autoref{sec:generator}.  The new versions of GENIE are consistent with other generators predictions at the lowest $E_\text{avail}$ bins. 
However, all event generator underestimate the data in the lowest $E_\text{avail}$ bins.

\subsection{Comparisons of efficiency generator predictions}

\begin{figure}
    \centering
     \includegraphics[width=\linewidth]{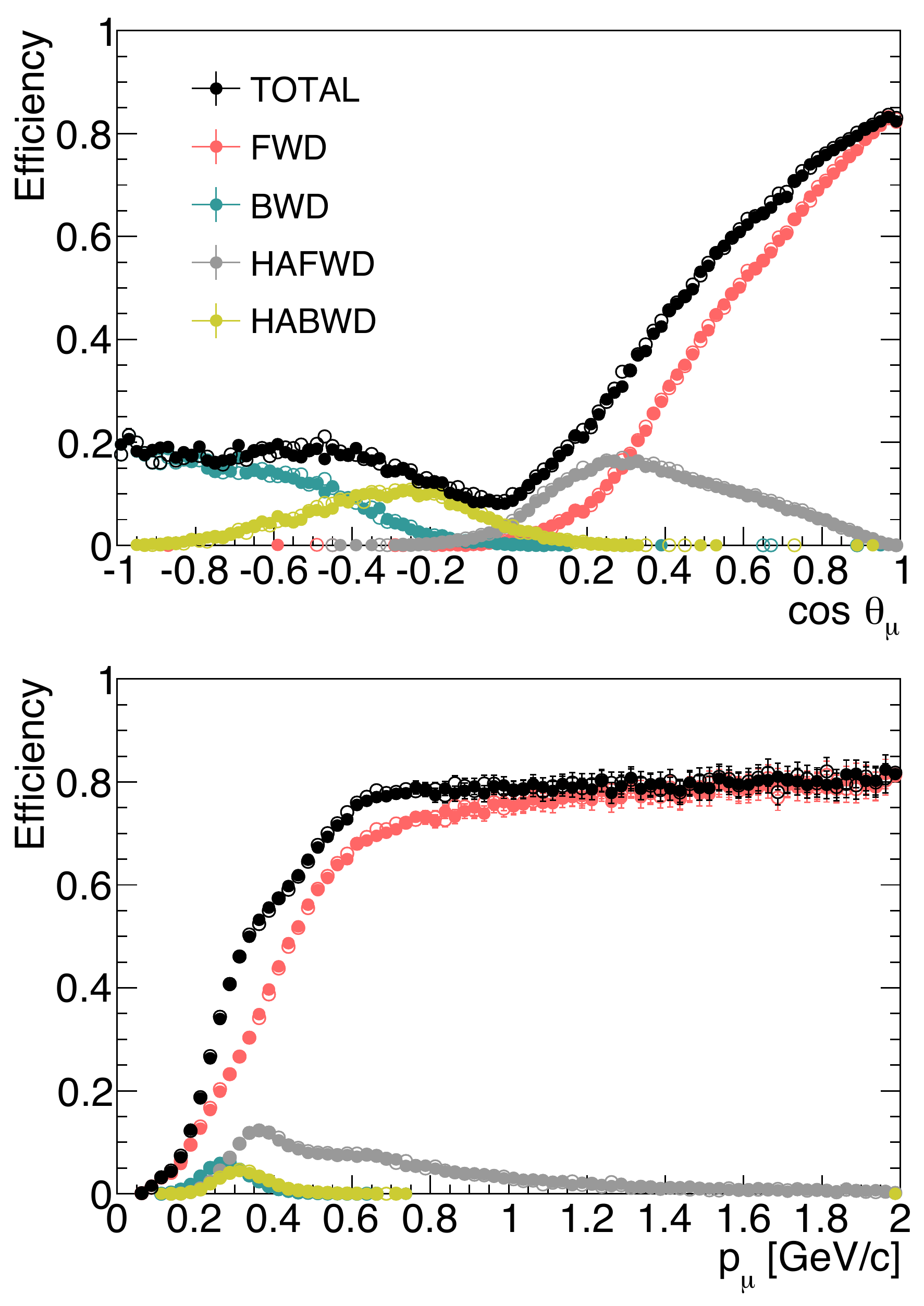}
    \caption{T2K efficiency for the GENIE (empty dots) and NEUT (filled dots) generators as function of the muon momentum (top) and the angle (bottom)\cite{Abe2018}. The different colors show the percentage of true signal events being reconstructed in the 4 different selection samples, split by muon direction (forward, backward, high-angle forward, and high-angle backward). While the efficiency seems to be identical for both generators in these projections, the 2D efficiency map presented in the original paper actually shows some significant differences in the low-momentum, forward-going bins.}
    \label{fig:inclusivet2keff}
\end{figure}

\begin{figure*}[ht!]
    \centering
    \includegraphics[width=0.32\linewidth]{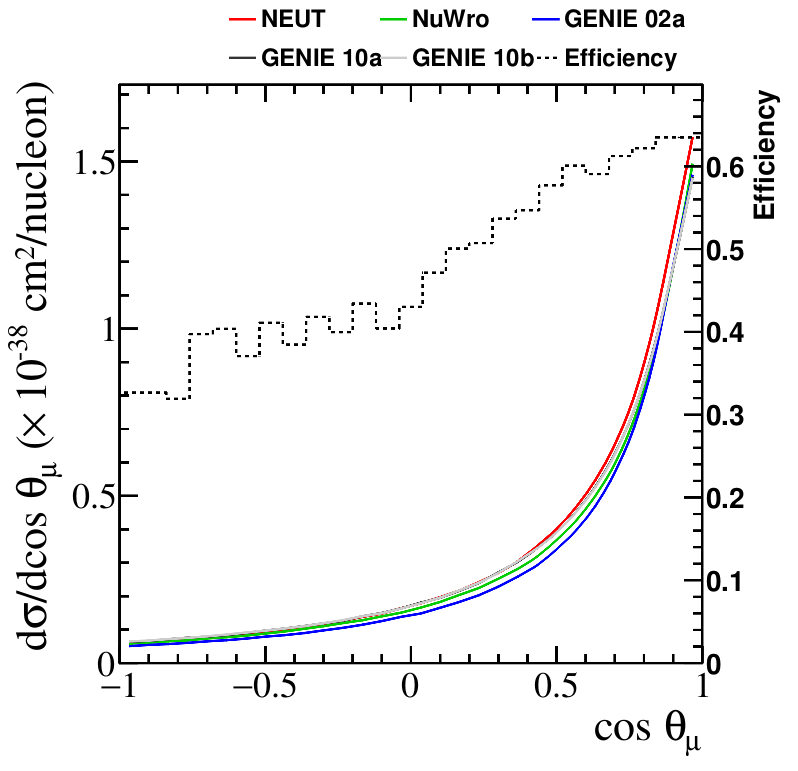}
    \includegraphics[width=0.32\linewidth]{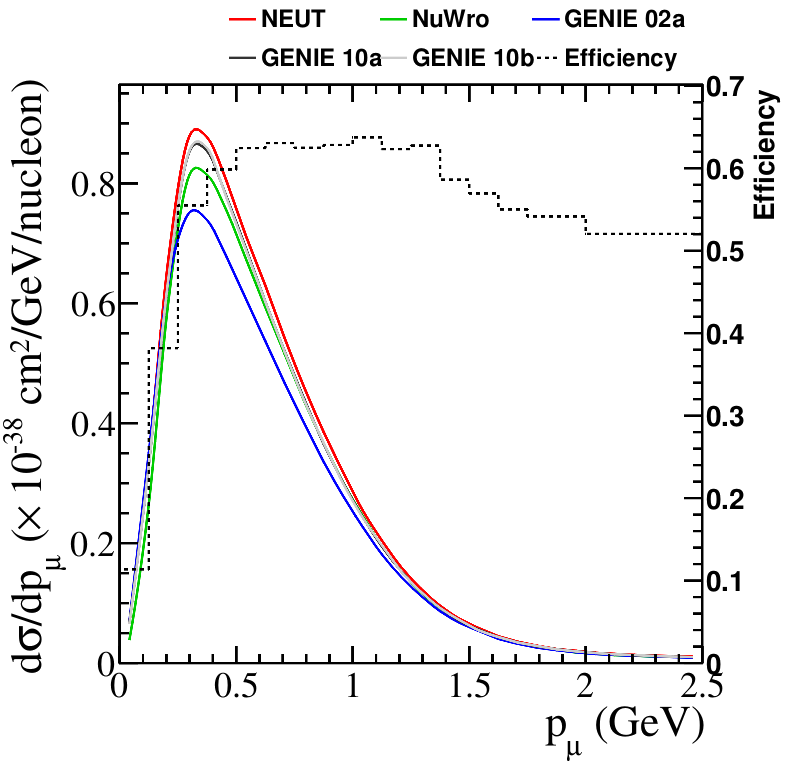}
    \includegraphics[width=0.32\linewidth]{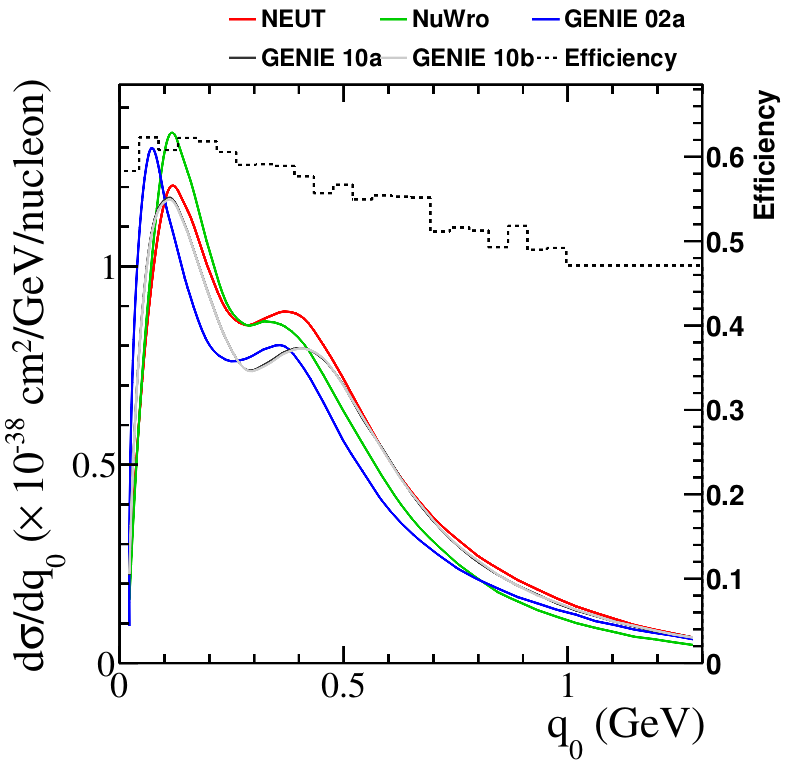}
    \caption{MicroBooNE efficiency (dotted line) and cross-section predictions (solid line) for the different generators as function of the muon angle (left), muon momentum (middle), and the hadronic energy $q_0$ (right).}
    \label{fig:inclusivemicrobooneeff_q3}
\end{figure*}

\begin{figure}[ht!]
    \centering
    \includegraphics[width=\linewidth]{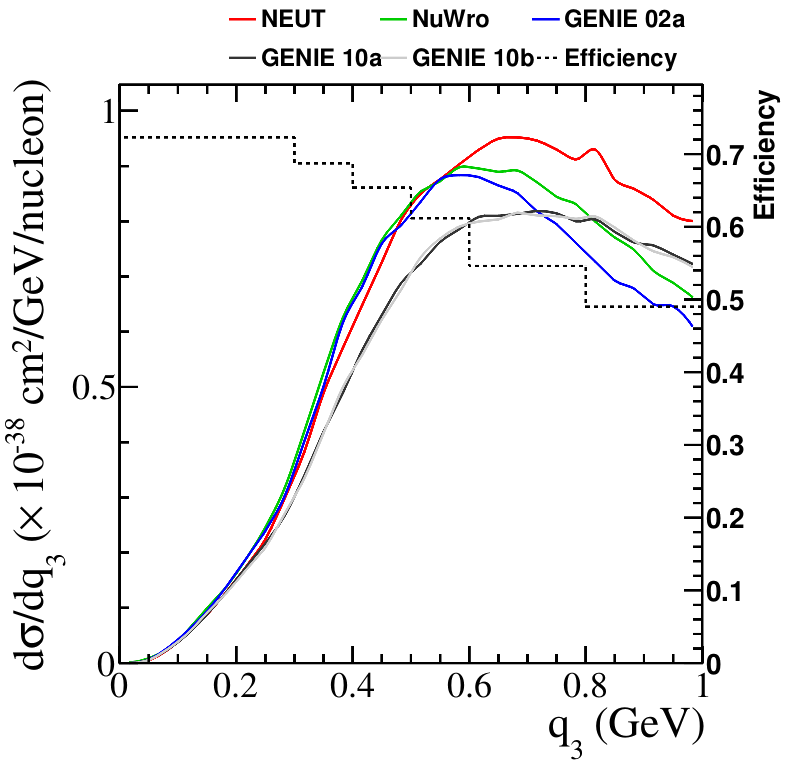}\\
    \includegraphics[width=\linewidth]{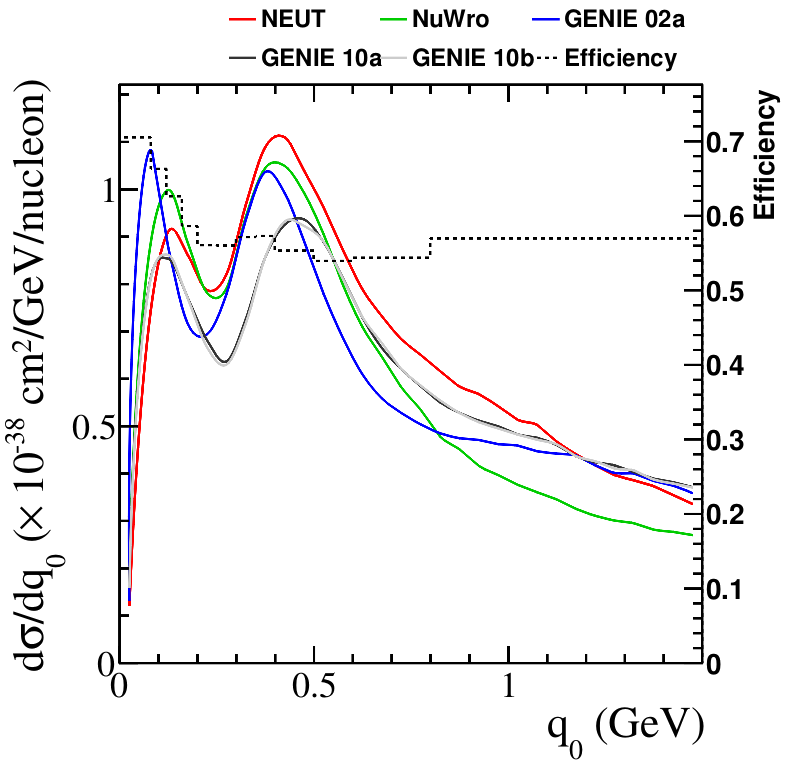}
    \caption{MINERvA efficiency (dotted line) and cross-section predictions (solid line) for the different generators as function of three momentum transfer (top) and the hadronic energy $q_0$ (bottom).}
    \label{fig:inclusiveminervaeff_q3}
\end{figure}

Figures~\ref{fig:inclusivet2keff} through~\ref{fig:inclusiveminervaeff_q3} show the event selection efficiency as a function of the muon angle (T2K and MicroBooNE), muon momentum (T2K and MicroBooNE), hadronic energy (MicroBooNE and MINERvA), and three-momentum transfer (MINERvA).
For MicroBooNE and MINERvA, the plots also show the cross section predicted by the generators studied in this paper as a function of the same variables.

The T2K measurement is sensitive to muon momenta above $\sim\SI{200}{MeV/c}$, with a flat efficiency above $\SI{600}{MeV/c}$. This is shown in  Fig.~\ref{fig:inclusivet2keff}.
The efficiency for back-scattered muons is lower than for forward-going muons, mostly due to those muons also being of lower momentum.
Muons produced at high angles are reconstructed without the help of the TPCs and can travel along the length of the FGD scintillator bars, making them harder to reconstruct.
It is important to note that the 2D efficiency shows model dependence in some of the measurement bins.
In low-momentum, forward-going bins the efficiencies evaluated with the NEUT and GENIE event generators differ significantly.
This is caused by event properties that are not part of the analysis binning, but which influence the event reconstruction.
In this case, it seems to be caused by the different handling of DIS events by NEUT and GENIE.
The difference is not visible in the 1D projections of the efficiencies as shown in Fig.~\ref{fig:inclusivet2keff}.
It is very important to check the efficiencies not only for dependence on the single measured variables, but also on their multidimensional combinations, as well as any other implicit assumptions about nuisance event property distributions.

In MicroBooNE, the efficiency in muon-momentum increases from 0 to 0.5~GeV/c due to the effect of detector and reconstruction thresholds; it is more constant above 0.5~GeV/c.  Fig.~\ref{fig:inclusivemicrobooneeff_q3} shows that the biggest tension between different cross-section models happens just before 0.5~GeV, right at the place where the efficiency changes rapidly. On the other hand, the efficiency in $q_0$ is quite constant with a slight decrease moving towards higher $q_0$ values. At higher $q_0$ the events become more complicated to reconstruct and hence more difficult to select.  The different models shown in the figure present a quite different behaviour as function of $q_0$.
Since the efficiency depends strongly on other event properties (like the muon momentum), it must be be assumed that the overall efficiency difference between the considered models is stronger than the flat efficiency in $q_0$ suggests.

In MINERvA, the efficiency as a function of the three momentum transfer and the hadronic energy, $q_0$, are shown in Fig.~\ref{fig:inclusiveminervaeff_q3}. The efficiency at low three momentum transfer is $70\%$ and decreases for values of $q_3>0.3$~GeV to $50\%$, where the models show different predictions. Event generators show different predictions in normalization, for example at $q_3=1$~GeV NEUT and GENIE v3 $G18\_02a$  event generators differ by $20\%$. In the region $0.4 < q_3 < 1$~GeV, the event generators have different predictions in shape and normalization. The efficiency as a function of $q_0$ is $60\%$ for values $q_0>0.2$~GeV and efficiency is higher at low $q_0<0.2$~GeV. Model predictions are different in shape and normalization for all values of the hadronic energy.  

\subsection{Discussion}

The $q_0$ efficiency for MicroBooNE and MINERvA, shown in Figs.~\ref{fig:inclusivemicrobooneeff_q3} and~\ref{fig:inclusiveminervaeff_q3}, is similar for the two experiments, about 80\% efficiency which slightly decreases with increasing $q_0$. Having a constant efficiency for the MINERvA measurement is extremely important giving the fast changing cross sections and also the differences among the models shown in the same figures.

The MicroBooNE cross section, shown in  Fig.~\ref{fig:inclusivemicroboone}, shows a poor agreement with the different predictions, with tension especially visible in the forward going bins,
where the MC shows some deficits compared to the data in some momentum bins. From the $\chi^2$, calculated with the full covariance matrix, NuWro appears to give the best prediction, though the other generators are comparable.

The comparisons of MINERvA's inclusive cross section with different event generator predictions in Fig.~\ref{fig:inclusiveminerva} show disagreement in different regions of $q_3$ and available energy. A deficit at low available energy is observed from all generators and strength is missing for the regions with MEC and RES, around $0.4\text{~GeV}<q_3<0.8\text{~GeV}$.

Comparisons of T2K and MicroBooNE with MINERvA can't be directly made. The inclusive charged current measurements were done as a function of different variables and with different neutrino energy spectra.  A common outcome from the comparisons for MicroBooNE and MINERvA inclusive charged current measurements is that NuWro provides better predictions for some but not all kinematics, although no generator is able to successfully describe the MINERvA data at low available energy.
Conversely, for the T2K measurement NEUT has better $\chi^2/dof$ compared with NuWro. This apparent tension is hard to resolve.  The GENIE \Genietena and \Genietenb predictions have $\chi^2$ values between NEUT and NuWro for MicroBooNE, but have the best $\chi^2$ values for T2K.  It is interesting that for \minerva \Genietenb has $\chi^2$ twice as large as \Genietena even though the only difference is choice of FSI model.  This shows an interesting sensitivity to FSI in these data.  Although GENIE \Genietwoa has older models than \Genietena and \Genietenb, the $\chi^2$ values are not significantly worse overall. 

\newpage

\FloatBarrier

\section{CC-mesonless interactions}
\label{sec:cc0pi}
\subsection{Introduction}

Quasielastic-like, also referred to as CC-0$\pi$, refer to a topological classification of neutrino-nucleus interactions where the resulting particles exiting the nucleus contain only nucleons and no mesons. This interaction is a critical process providing a dominant channel for neutrino oscillation experiments operating in the few-GeV region \cite{sbn, Abe:2011ks, nova, DUNE:fd2020, Rodrigues:2015hik}. Appropriate estimators of neutrino energy and the ability to simulate these kinematics is of the utmost importance to this experimental program.

The CC-0$\pi$ topology is mainly composed of CCQE events, where indeed only one nucleon is expected to exit the interaction vertex. Recent measurements have shown the importance of the $2p2h$ process, 
although the details remain uncertain. Other effects due to nucleon-nucleon correlations in the nucleus, like RPA, are now included in simulations, but also remain uncertain. Finally, FSI can impact the CC-0$\pi$ channel, by altering the nucleon final state kinematics or by re-absorbing final state pions before they exit the nucleus.

In this section we will review several CC meson-less measurements from T2K and \minerva. We will start in \autoref{subsec:t2kcc0pi} and \autoref{subsec:minervacc0pi} with a comparison between the cross section measurements in muon kinematics from both experiments and the predictions from generators described in this paper, using samples which include protons below tracking thresholds. With the same samples, in \autoref{subsec:cc0piefficiency} and \autoref{subsec:t2kmincc0picomp} we will focus on a more direct comparison between T2K and \minerva results, trying to select a region of the phase space common to both experiments. A comparison of the $q_0$-$q_3$ phase space for the two experiments is also discussed.

In order to focus more on the ability of current generators to describe nuclear effects, in \autoref{subsec:stvgeneratorcomparison}, we will review measurements of a number of variables, known as Single Transverse Variables, which have specific sensitivity to a variety of nuclear effects.

Measurement of QE-like interactions in $Q^{2}_{QE}$, using muon kinematics, from \minerva is also compared to MC predictions in \autoref{subsec:minervaq2}, while in \autoref{subsec:minervaAscale} generator predictions for different targets (CH, Fe and Pb) will be compared to \minerva results in $Q^{2}_{QE}$ variable, using proton kinematics. We will finally discuss the presented comparisons in \autoref{subsec:cc0pidiscussion}.

\subsection{T2K results and comparisons with generators}
\label{subsec:t2kcc0pi}

The T2K collaboration has published two CC-0$\pi$ measurements combining different targets and flavors, and using the data taken at the off-axis near detector (ND280): the first simultaneous extraction of the $\nu_\mu$ and $\bar{\nu}_\mu$ CC-0$\pi$ cross-sections on hydrocarbon employing the data taken with a neutrino and antineutrino beam~\cite{Abe:2020jbf} and the first simultaneous extraction of the $\nu_\mu$ CC-0$\pi$ cross-sections over oxygen and carbon~\cite{Abe:2020uub}. The cross-sections have been extracted as function of the muon momentum and cos$\theta$ without any phase space restriction. 

A simultaneous measurement has many advantages. The knowledge of the correlation between the measured cross-sections allows further information (cross-section ratio, asymmetry, sum, difference) to be obtained through a proper combination, often reducing common systematics uncertainties. In addition, a joint measurement further allows a less model dependent background subtraction, as is the case for the combined neutrino-antineutrino CC-0$\pi$ cross-section, where the neutrino background in antineutrino beam is relatively large. By fitting $\nu$ and $\overline{\nu}$ samples at the same time, it is possible to simultaneously extract both cross sections, thus disentangling the neutrino and antineutrino contributions, without needing a bare background subtraction. A similar approach is exploited for the combined carbon-oxygen cross-section measurement. In this case, the two cross sections are extracted using the interactions occurring in the ND280 FGDs. The first FGD (FGD1) is completely made of plastic scintillator bars, while in the second one (FGD2) the scintillator bars are interleaved with water targets~\cite{Abe:2011ks}. The sample of CC-0$\pi$ interactions on carbon is a background for the oxygen measurement, since the water modules are passive and all the interactions are reconstructed in the scintillator layers. Based on the starting position of the muon track, it is possible to construct carbon- and oxygen-enhanced samples to be used in a simultaneous fit, thus allowing the oxygen and carbon components to be extracted at the same time.

In the following, we will only consider the $\nu_\mu$ CC-0$\pi$ cross-sections on carbon and hydrocarbon, in order to allow a more direct comparison with \minerva $\nu_\mu$ CC-0$\pi$ cross-section described in \autoref{subsec:minervacc0pi}, which uses the same target material.

The two T2K measurements use the same event selection for the $\nu_\mu$ CC-0$\pi$ interactions; two control samples are also employed to constrain the background in the signal sample, mainly made up of interactions from CC resonant pion production and CC deep inelastic scattering. Events are selected exploiting the particle identification capabilities of the FGDs and the TPCs, and the timing between sub-detectors to distinguish between forward- and backward-going (w.r.t. beam direction) tracks. Five different signal samples with at least a negatively charged muon entering in TPC or fully contained in FGD, and, eventually, one or more protons, have been selected. For the control samples, events with one negatively charged muon and one positively charged pion (CC-1$\pi^+$) or more than one (CC-Other) entering in the TPC are selected.

The two measurements also share the same cross-section extraction method. An extended binned likelihood fit is used to extract the true number of CC-0$\pi$ events in bins of muon kinematics (momentum and cos$\theta$) that are subsequently corrected by the signal efficiency, the integrated flux, the number of targets and the bin width. Uncertainties are taken into account by adding a penalty term to the likelihood and are then propagated when estimating the cross section.

The differences between the two measurements are related to: 
\begin{itemize}
    \item the combined cross section: in one case, the complementary measurement is the $\overline{\nu}_\mu$ CC-0$\pi$ on CH and in the other case is the $\nu_\mu$ CC-0$\pi$ on O;
    \item the target: in one case the detector target is the FGD1 and the cross section is extracted per  hydrocarbon nucleons, while in the second case both FGD1 and FGD2 are used and the cross section is extracted per carbon nucleon\footnote{It should be underlined that when quoting the CH cross section, the full detector mass of the FGD1 is considered and it thus includes also small percentage of non CH elements (as detailed in \autoref{sec:t2kexp}). On the other hand, when quoting the cross section on carbon, the contribution from all the other elements is removed.};
    \item the binning: for the measurement on CH the $p_\mu - \cos\theta_\mu$  binning is finer than for the measurement on C. However the latter better matches the MINER$\nu$A phase space.
\end{itemize}

Due to the number of common points between the two measurements, results on carbon or hydrocarbon should in principle give similar information. However, since they are partial results of more complex and different analyses, we decided to report both.

\begin{figure*}
	\centering
	\includegraphics[width=1.\linewidth]{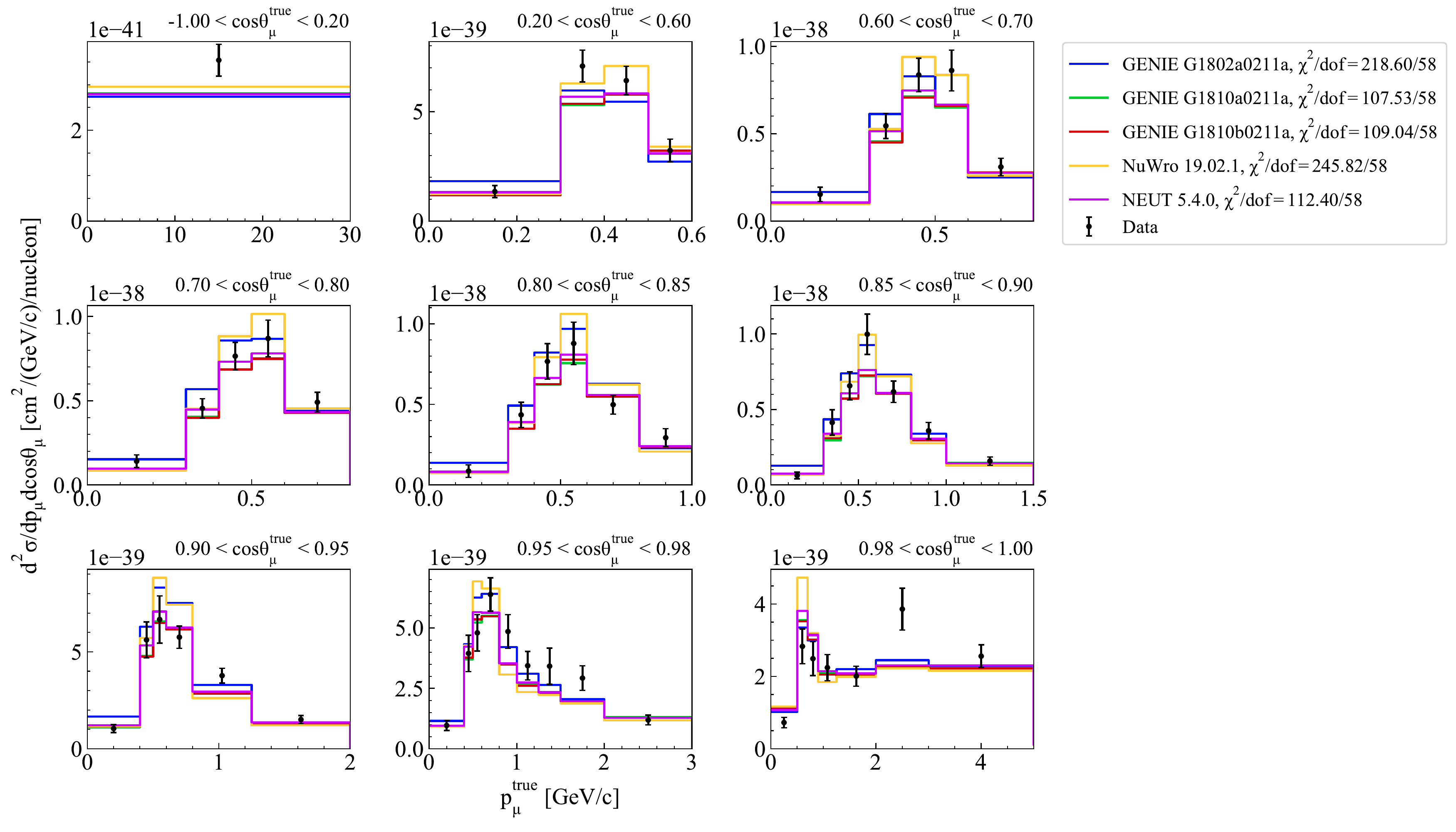}
	\caption{Measured T2K $\nu_\mu$ CC-0$\pi$ double-differential cross sections on hydrocarbon in bins of true muon kinematics. The  results  are  compared  to GENIE v3 G18\_02a (blue), G18\_10a (green) and G18\_10b (red), NuWro 19.02.1 (orange) and NEUT 5.4.0 (violet). The last bin in momentum is not displayed for readability.}
	\label{fig:t2knumucc0pich2dpcos}
\end{figure*}

\begin{figure*}
	\centering
	\includegraphics[width=1.\linewidth]{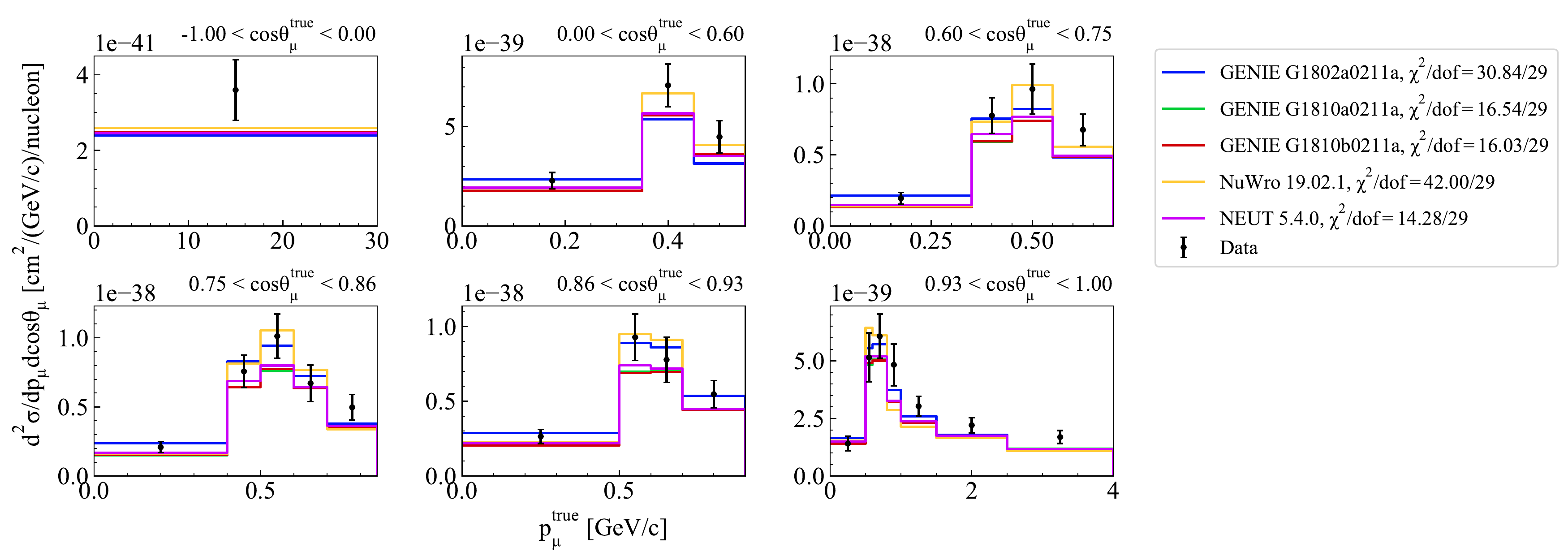}
	\caption{Unregularised T2K $\nu_\mu$ CC-0$\pi$ double-differential cross sections on carbon in bins of true muon kinematics.  The results are compared to GENIE v3 G18\_02a (blue), G18\_10a (green) and G18\_10b (red), NuWro 19.02.1 (orange) and NEUT 5.4.0 (violet). The last bin in momentum is not displayed for readability.}
	\label{fig:t2knumucc0pic2dpcos}
\end{figure*}

\Cref{fig:t2knumucc0pich2dpcos,fig:t2knumucc0pic2dpcos} show the measured $\nu_\mu$ CC-0$\pi$ double-differential cross sections on hydrocarbon and carbon, respectively, in bins of true muon kinematics, compared with the MC predictions. Although the two measurements share several data samples and the cross-section extraction method, some differences can be noticed when compared to generator predictions. 

In the legends of \cref{fig:t2knumucc0pich2dpcos,fig:t2knumucc0pic2dpcos}, we report the $\chi^2$ obtained using a reduced covariance matrix for the neutrino-only and the carbon-only part of the measurements. Although the two results show similar preferences, it should be noticed that $\chi^2$ values are in general smaller for the CC-0$\pi$ measurement on carbon than on hydrocarbon. This can be partially explained by the fact that the two measurements use different binning, different statistics and are sub-set of two more complex measurements, one including antineutrino and the other one including oxygen. Looking at the $\chi^2$, carbon data seems to clearly prefer NEUT, GENIE {\Genietena} and {\Genietenb}. We should notice that the only difference between GENIE {\Genietena} and {\Genietenb} is related to the pion FSI model, and CC0$\pi$ measurements in muon kinematics are not very sensitive to these model differences. On the other hand, the measurement on hydrocarbon is clearly overestimated in the most forward bin for momenta below 1 GeV. This can be due to incorrect $1p1h$ predictions in the region of small energy transfer to the nucleus, where the treatment of various nuclear effects, like RPA, is not well understood.

\subsection{MINERvA result and comparison with generators} 
\label{subsec:minervacc0pi}

\begin{figure*}
	\centering
	\includegraphics[width=\linewidth]{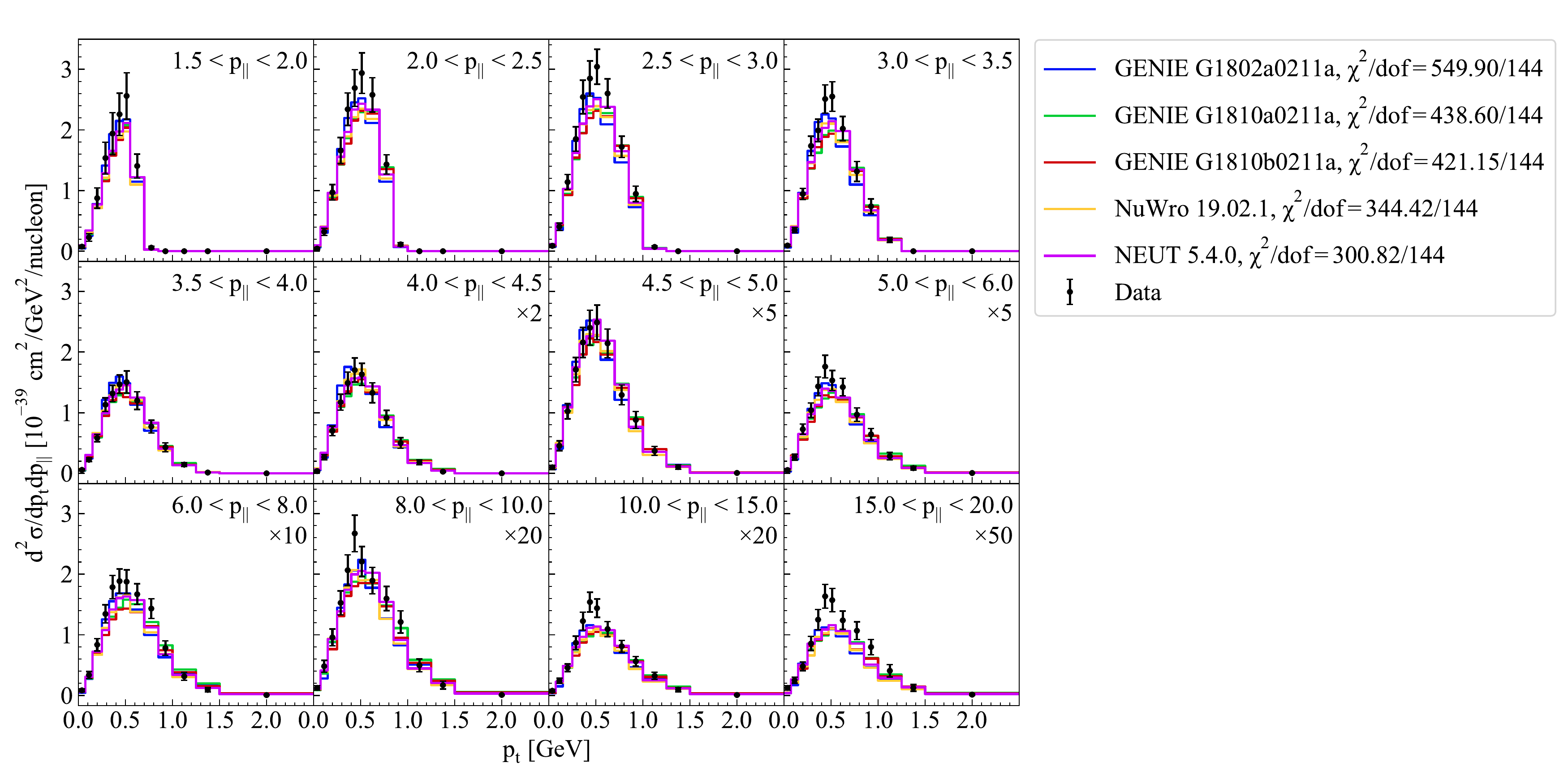}
	\caption{\minerva~CCQE-like cross-section measurement on hydrocarbon compared to various models in regions of $p_{||}$ [GeV].}
	\label{fig:cc0piminerva2Dplot}
\end{figure*}

The \minerva~collaboration produced a CCQE-like double differential cross-section, shown in Fig. \ref{fig:cc0piminerva2Dplot}, using a beam of primarily muon neutrinos at a mean energy of 3.5 GeV. The measurement uses as observables the transverse (\pt) and longitudinal (\pz) muon momentum \cite{Ruterbories:2018gub}. This variable combination was chosen because at high neutrino energy, as is the case for \minerva, $p_T^2$ is correlated to $Q^{2}_{QE}$ and \pz is correlated to the neutrino energy.

A selection of CCQE-like events employs a combination of selection criteria using particle identification to remove different sub-samples of backgrounds. A Michel electron tagging algorithm is used to identify late in time electrons from pion decay near the interaction vertex, and all track end points. An isolated cluster algorithm is used to count the multiplicity of showers in the interaction. In addition, a 500 MeV restriction on the visible energy not associated with tracked particles is imposed to remove the DIS and neutral resonant pion production with large pion energies.

The selected sample requires no Michel electrons and no more than one isolated cluster. Three control samples are populated using a combination of the interactions failing these cuts. A charged pion dominated sample is constructed by requiring a single Michel electron and no more than one isolated cluster. A multi-pion dominated sample is constructed by requiring more than one Michel electron and more than two isolated clusters. The third control region is a mix of single charged, neutral, or multi-pion and is constructed by selecting interactions with no Michel electrons and two isolated showers. 

The cross section is extracted using a data-constrained background subtracted sample which is unfolded, using four iterations, using D'Agostini unfolding via the RooUnfold package \cite{Adye:2011gm}. The sample is efficiency corrected using the GENIE prediction. The result is then corrected for the number of nucleons in the FV and the integral of the NuMI flux between 0 and 100 GeV.

Each prediction's $\chi^{2}$ values using the full covariance matrix are reported in the legend of Fig. \ref{fig:cc0piminerva2Dplot}. We note that none of the generators is able to well reproduce the data because the $\chi^2$ values are all larger than 300 for 144 bins.  
The smallest $\chi^2$ values are shown for NEUT and NuWro, while GENIE {\Genietwoa} shows the largest disagreement with data. The region where models struggle the most in reproducing data is in the highest \pz region. Unlike what was shown in \ref{subsec:t2kcc0pi}, NuWro and NEUT predictions are very similar at higher beam energies. In general, predictions from generators using LFG as nuclear models are similar and show differences, as expected, from {\Genietwoa}.

\subsection{T2K/\minerva phase-space comparison}
\label{subsec:cc0piefficiency}
In this section we compare true experimental efficiencies with projected calculations of 2D cross sections calculated by the generators (see \autoref{sec:eff} for details), specifically for the T2K and MINERvA measurements described in \autoref{subsec:t2kcc0pi} and in \autoref{subsec:minervacc0pi}, respectively.  
We remind here that, although both T2K and MINERvA  performed 2D measurements and thus provided two-dimensional efficiency maps, for display purposes 2D efficiencies were projected in each dimension and compared with generator predictions. \\

\begin{figure*}
	\centering
	\includegraphics[width=0.48\linewidth]{T2K_CC0pi_mom_13_eff}
	\includegraphics[width=0.48\linewidth]{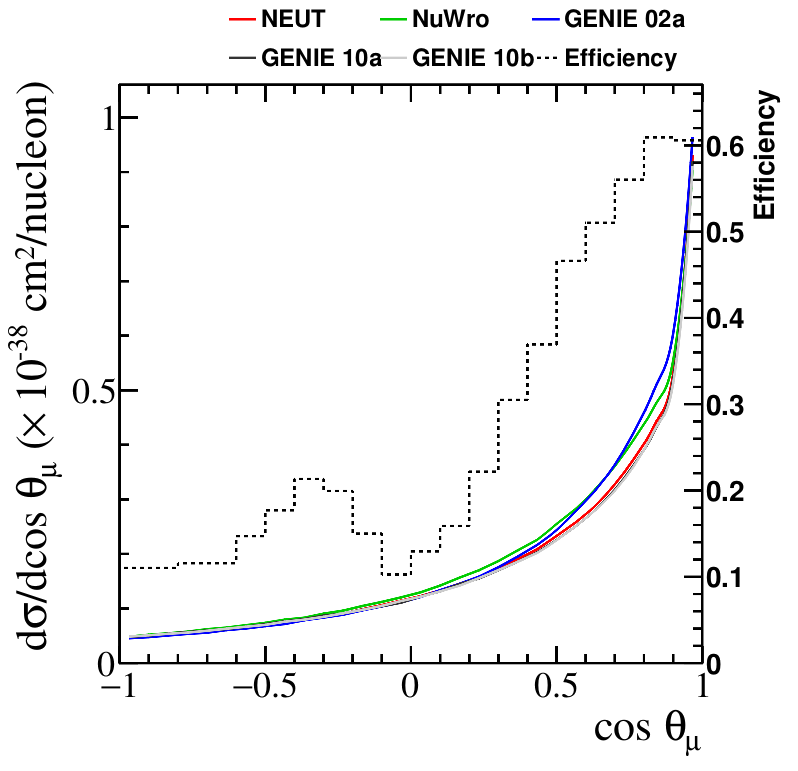}
	\caption{Efficiency for the T2K CC0$\pi$ measurement (dotted line using the right vertical scale) and cross-section predictions (solid line using the left vertical scale) for the different generators employed in this paper as function of true muon momentum (left) and cosine of the muon scattering angle (right).}
	\label{fig:t2kcc0picosth13eff}
\end{figure*}

The efficiencies of the T2K CC-0$\pi$ selected sample as a function of true muon momentum and cosine of the scattering angle is shown in \autoref{fig:t2kcc0picosth13eff} (dotted line). On the same plot, the cross sections (solid lines) predicted by the different generators studied in this paper are compared. The dip for $\cos\theta_{\mu} = 0$ is a result of the intrinsic inefficiency of the detector to track particles perpendicular to the beam direction.  One goal is to examine model dependence in situations where the efficiency is rapidly changing.  The efficiency is rapidly rising for muon momentum values up to about 0.5 GeV.  Although the cross section predictions are also changing rapidly in that region, no significant disagreements among models are seen.

\begin{figure*}
	\centering
	\includegraphics[width=0.48\linewidth]{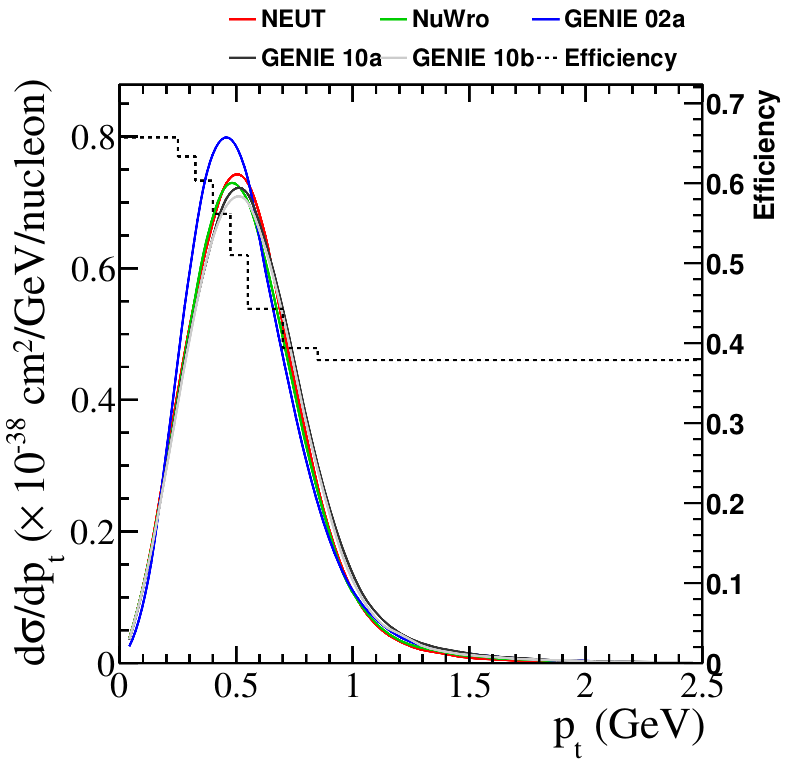}
	\includegraphics[width=0.48\linewidth]{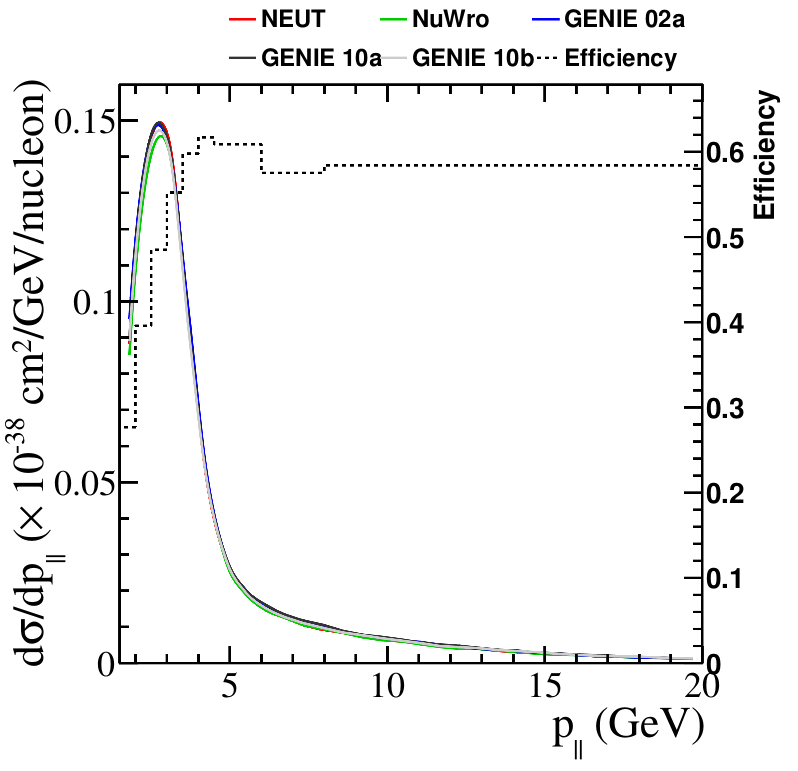}
	\caption{Projected efficiency in \pt (left) and \pz (right) for the \minerva CCQE-like measurement. The dotted line is the efficiency using the right vertical scale while the colored lines are cross section predictions using the left vertical scale for the generators used in this paper.}
	\label{fig:cc0pi_minerva_2D_efficiencyplot}
\end{figure*}

The efficiencies of the \minerva CCQE-like sample as a function of transverse and longitudinal momentum are shown in Fig. \ref{fig:cc0pi_minerva_2D_efficiencyplot}, as a dotted line. The cross section predictions from the generators used in this paper are also shown. Also in this case, all the generators show a similar behavior, although we can notice that {\Genietwoa} predicts a slightly higher cross section in \pt. The projected efficiency ranges from 40\% at large \pt~to 70\% at small \pt and ranges from 30\% at small \pz~to a plateau of about 60\% for \pz greater than 4 GeV. The decrease in efficiency at low \pz is due to the requirement that the muon is reconstructed in MINOS, which puts a threshold of about 2 GeV depending on the vertex location in the scintillator tracker. The efficiency drop off at large \pt is due to the same \minerva-MINOS track requirement.

\begin{figure*} [ht!]
	\centering
	\includegraphics[width=0.4\linewidth]{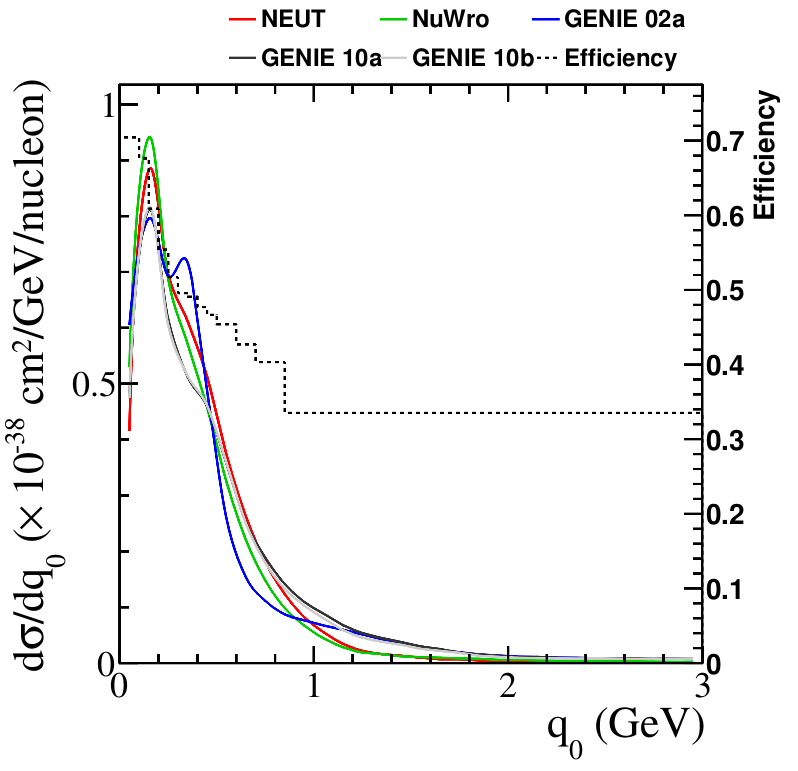}
	\includegraphics[width=0.4\linewidth]{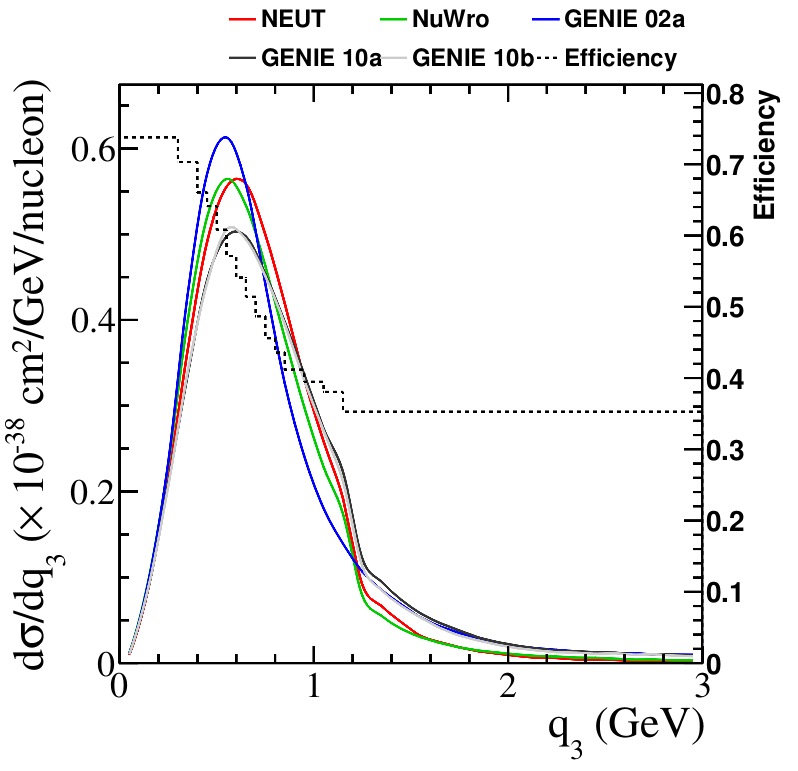}
	\includegraphics[width=0.4\linewidth]{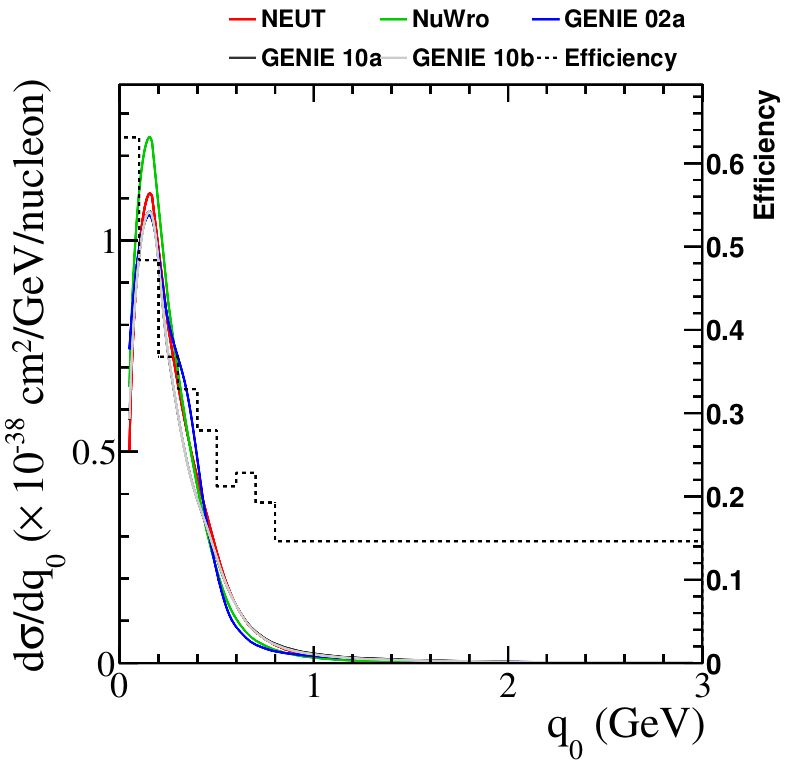}
	\includegraphics[width=0.4\linewidth]{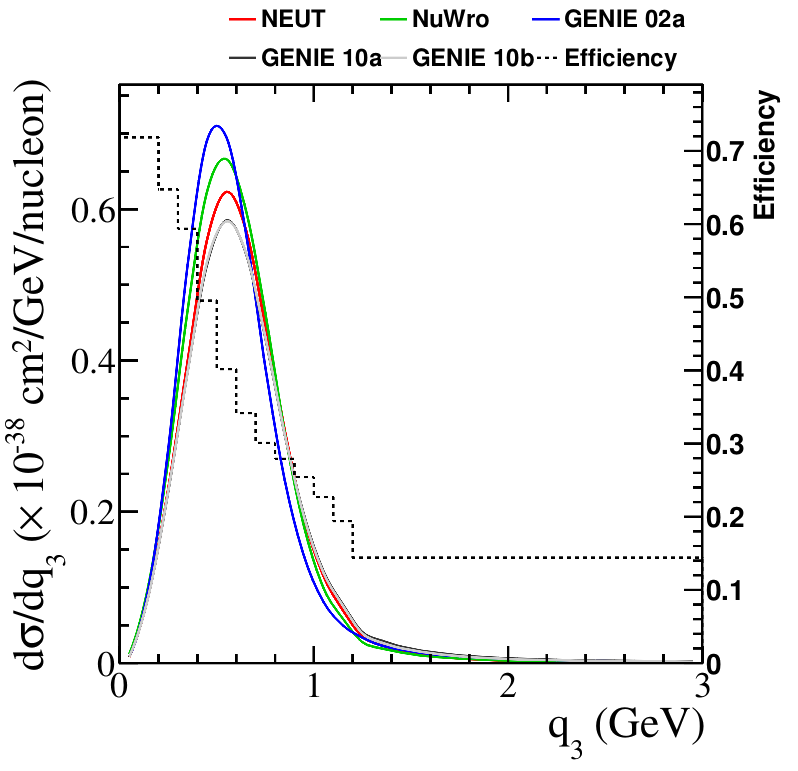}
	\caption{Comparison of the projected efficiencies and predicted cross sections for $q_0$ (left column) and $q_3$ (right column) for \minerva (top) and T2K (bottom).}
	\label{fig:cc0piq0q3eff}
\end{figure*}

In addition to the efficiencies described above, in order to facilitate a direct comparison between T2K and MINERvA results,  the $q_0-q_3$ projected phase spaces and relative efficiencies are shown in \cref{fig:cc0piq0q3eff}, using the method described in \autoref{sec:eff}.
Due to the higher beam energy, the accessible phase space in both $q_0$ and $q_3$ is wider in \minerva than in T2K. For these plots, the same kinematic ranges were used.  \minerva and T2K have efficiencies which are large at low values of $q_0$ and $q_3$, about 70\% for \minerva and 65-70\% for T2K, which decrease with increasing $q_0$ and $q_3$. The lowest efficiency \minerva has is about 35\%, while for T2K it is about 15\%.

\subsection{Comparisons between T2K and \minerva data sets}
\label{subsec:t2kmincc0picomp} 

\begin{figure}[h!]
    \centering
    \includegraphics[width=0.97\linewidth]{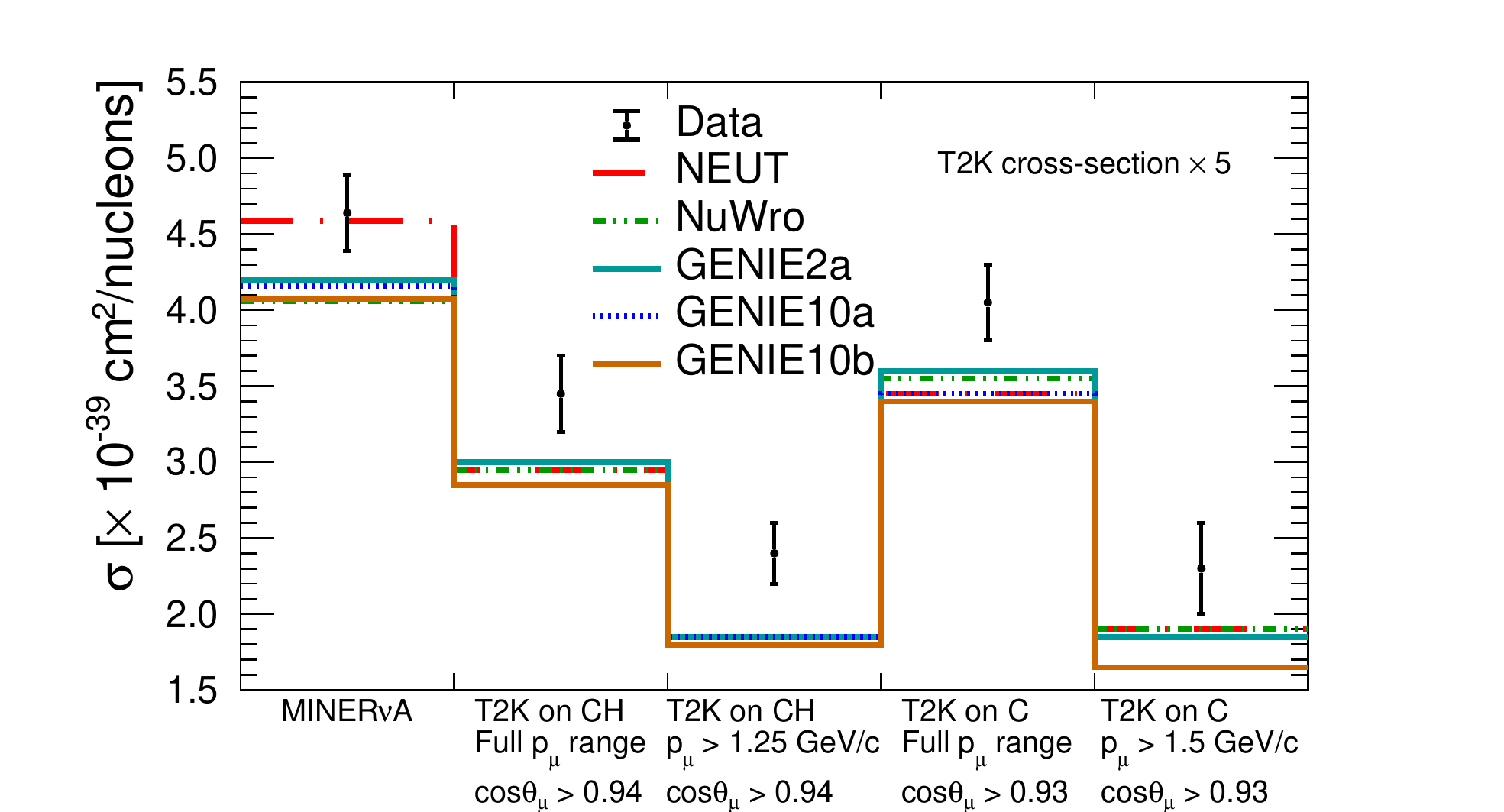}\\
    \includegraphics[width=0.97\linewidth]{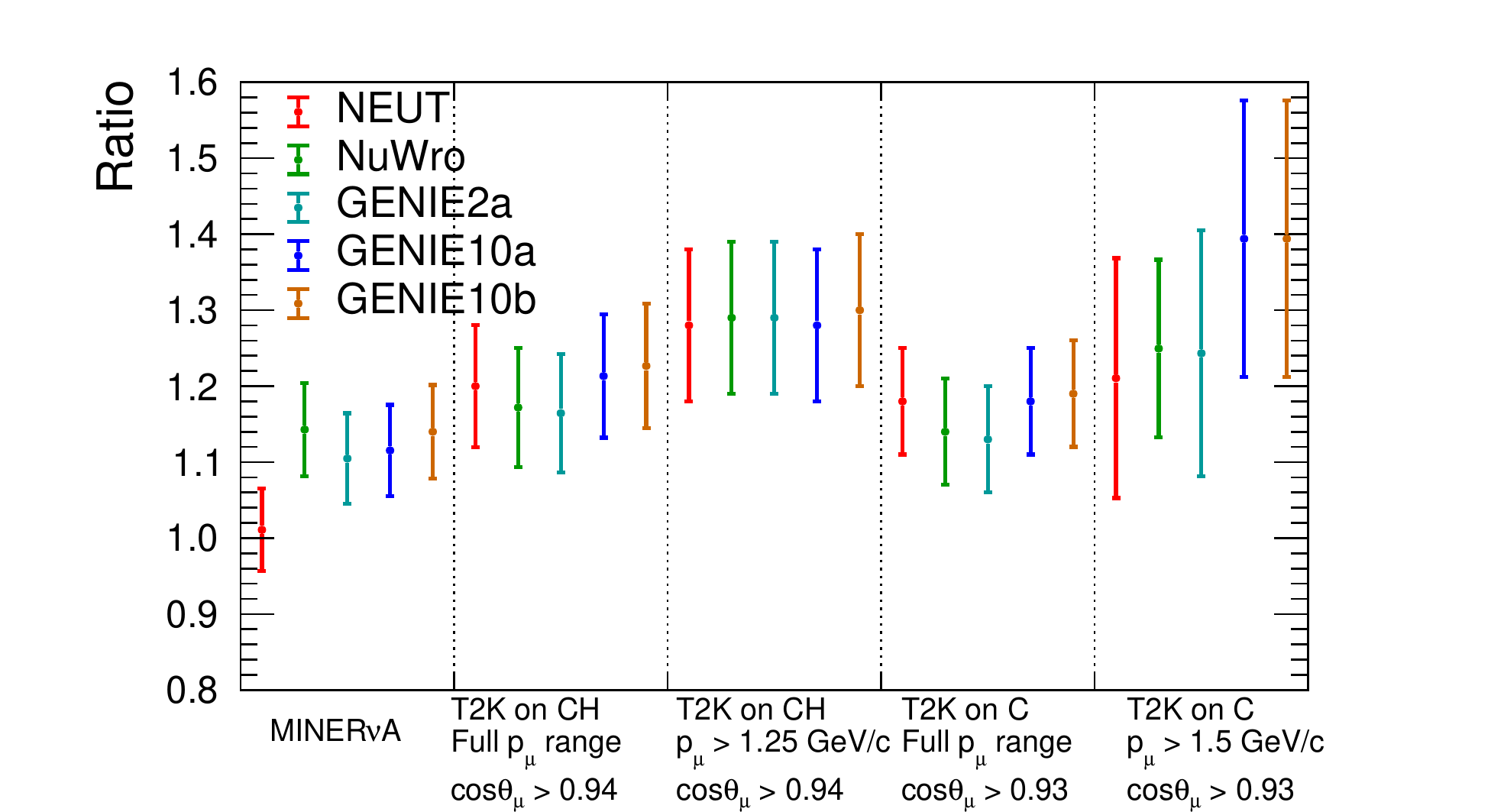}
    \caption{Top figure: \minerva and T2K CC0$\pi$ cross-section measurements compared with the MC employed in this paper. The phase space of the T2K measurements is restricted to match \minerva and the obtained values of the cross section are multiplied by a factor 5 for display purposes. Bottom figure: Ratio between \minerva and T2K CC0$\pi$ cross-section measurements and the MC.}
    \label{fig:cc0pit2kvsminerva}
\end{figure}

In the upper panel of \autoref{fig:cc0pit2kvsminerva} the \minerva~and T2K $\nu_\mu$ CC-0$\pi$ neutrino cross-section measurements are compared with the MC samples studied in this paper and described in \cref{sec:nuwro,sec:genie,sec:neut}. Even if \minerva reports the cross section as function of \pz and \pt, given the restricted phase space in which the cross section has been measured, such variables can be associated to p$_\mu$ and $\cos\theta_{\mu}$ respectively.  Although MINERvA has a higher beam energy and therefore larger range of kinematics among the final state particles, the muon must be at less than 20$^\circ$. The phase space of the T2K measurements is restricted for the purposes of this comparison to match MINER$\nu$A phase space (p$_\mu >$ 1.5\,GeV and $\theta_\mu <$ 20$^{\circ}$). For T2K both measurements over hydrocarbon and carbon are included, and two different sets of phase space cuts, are compared:
\begin{itemize}
    \item In one case, the full muon momentum phase space is exploited, while the cosine of the muon scattering angle is required to be greater than 0.94 for the cross section on hydrocarbon and 0.93 for the cross section on carbon; those cuts correspond to require $\theta_\mu$ smaller than about 20$^{\circ}$.
    \item In the other case, the momentum is restricted to be greater than 1.25 GeV/c for hydrocarbon and 1.5 GeV/c for carbon, while the cut on the muon $\cos\theta$ remain the same.
\end{itemize}  

With respect to the binning used for the T2K measurement on hydrocarbon, the binning used for the measurement on carbon allows to better match the MINERvA $p_\mu$ phase space.  
The values plotted in Fig. \ref{fig:cc0pit2kvsminerva} represent the cross section per nucleon assuming pure CH or C targets, respectively.

In order to mitigate the energy dependence due to the different neutrino fluxes at which \minerva and T2K detectors are exposed, and thus to allow a clearer comparison between the two experiments, 
the ratio between the measured cross-sections and the MC predictions has been computed and is shown in the lower panel of Fig. \ref{fig:cc0pit2kvsminerva}. In the case of \minerva, the ratio between the measured cross-section  and NEUT is consistent with one, while T2K measures a cross section higher than what was predicted by NEUT. The other ratios are compatible across the different predictions and both experiments show that the MC underestimate the data. It is important to stress that for T2K this effect is only true for high momentum bins, while for low momentum the MC overestimate the data, as shown in the last angular bins in \cref{fig:t2knumucc0pich2dpcos,fig:t2knumucc0pic2dpcos}. Indeed, when using the full momentum phase space for this forward region, the MC underestimation is less visible.

In summary, most generators seem to systematically underestimate the data with one exception: NEUT is in excellent agreement with the \minerva data. Concerning the T2K restricted phase space, this trend is confirmed, even if the low statistics in this region results in larger error bars.

\subsection{Transverse kinematic imbalance variables and comparisons with generators} 
\label{subsec:stvgeneratorcomparison}

As explained in \cref{subsec:t2kcc0pi,subsec:minervacc0pi}, both T2K and \minerva include events with outgoing protons detected in their CC-0$\pi$ selections. By using the sub-samples where one or two protons are reconstructed, T2K and \minerva have measured the CC-0$\pi$Np cross-section as a function of Transverse kinematic imbalance variables (TKI), as reported in \cite{Abe:2018pwo} and \cite{Lu:2018stk} respectively. In this section, published results have been compared with MC predictions studied in this work.

TKI variables quantify the imbalance between the outgoing lepton and proton kinematics in the plane transverse to the incoming neutrino, and are thus able to offer a probe of nuclear effects~\cite{Lu:2018stk,Cai:2019hpx}. They are defined as follow: 
\begin{eqnarray}
\delta p_T &=& |\mathbf{\delta p}_T| = \left| \mathbf{p}_T^\mu + \mathbf{p}_T^p \right|,   \\
\delta \alpha_T &=& {\rm arccos}\left(- \frac{\mathbf{p}_T^\mu \cdot \mathbf{\delta p}_T}{p_T^\mu \delta p_T} \right), \ \\
\delta \phi_T &=& {\rm arccos}\left( - \frac{\mathbf{p}_T^\mu \cdot \mathbf{p}_T^p}{p_T^\mu p_T^p}\right)
\end{eqnarray}
where $\mathbf{p}_T^\mu$ and $\mathbf{p}_T^p$ are the momentum of the outgoing muon and  the highest momentum proton, respectively, projected on the plane transverse to the incoming neutrino. In the case of an interaction on a free nucleon, $\delta p_T$ and $\delta \phi_T$ are expected to be zero (while $\delta \alpha_T$ is undefined) and any difference from zero is an indication of nuclear effects. In particular, $\delta p_T$ is most sensitive to the nuclear structure, specifically the momentum distribution of the struck nucleon, while $\delta \alpha_T$ is most sensitive to final state interactions (FSI).\\

Both T2K and \minerva measure cross sections as a function of TKI variables over a restricted phase space where their detectors are sensitive, as summarized in Tab.~\ref{tab:phaseSpace}.
\begin{table*}
\begin{ruledtabular}
\centering
{\renewcommand{\arraystretch}{1.2}
\begin{tabular}{ l|c|c|c|c } 

Analysis & $p_p$ & $\cos\theta_p$ & $p_\mu$ & $\cos\theta_\mu$  \\
 \hline
T2K &  $0.45-1.0$~GeV & $>0.4$ & $>250$~MeV & $>-0.6$ \\
\minerva &  $0.45-1.2$~GeV & $>0.342$ & $1.5-10$~GeV & $>0.940$ \\

\end{tabular}}
\caption{Signal phase space restrictions for the T2K and MINERvA results in TKI variables.}
 \label{tab:phaseSpace}
 \end{ruledtabular}
\end{table*}

\begin{figure*}[ht!]
	\centering
	\includegraphics[width=\linewidth]{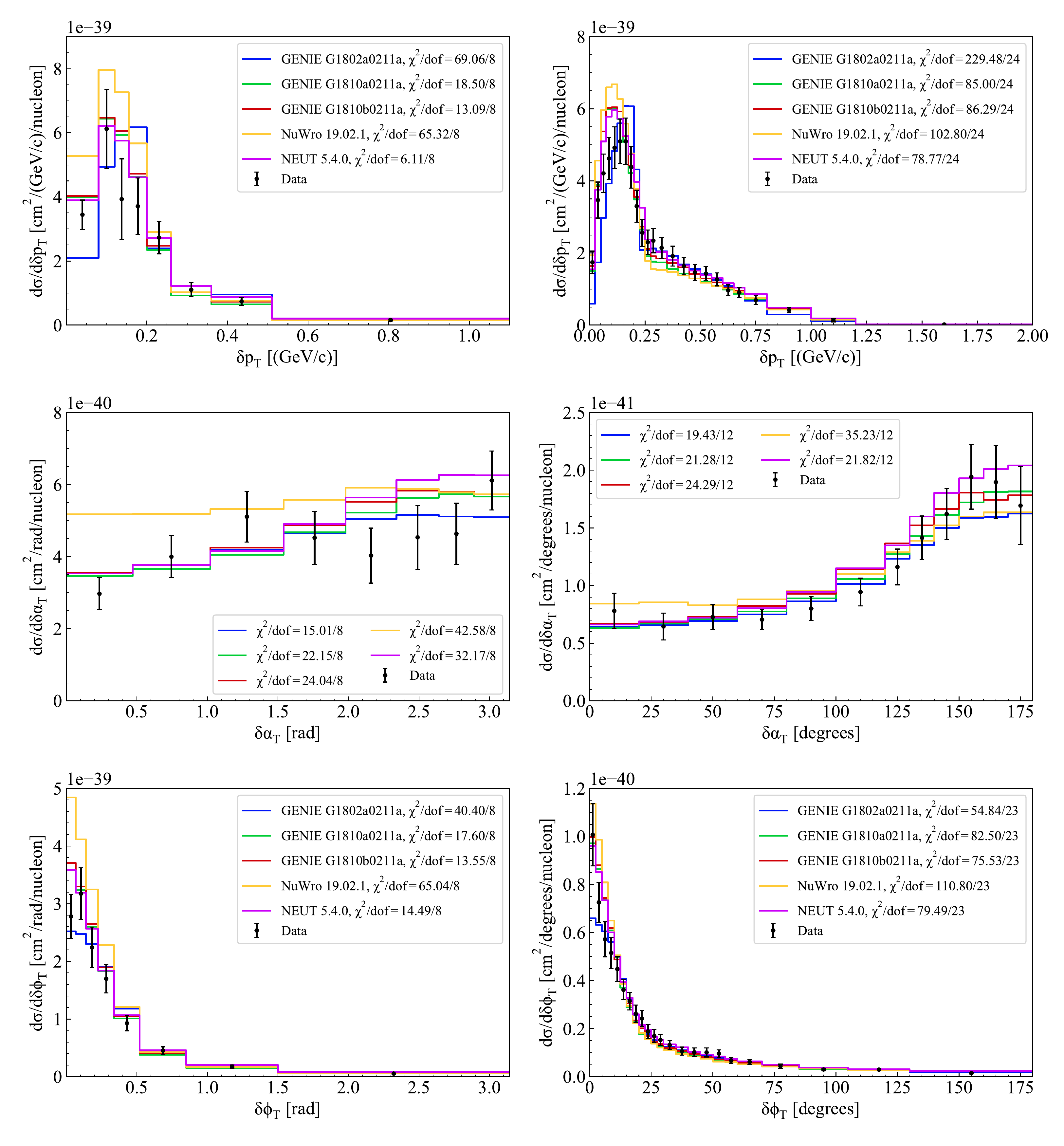}
	\caption{T2K (left) and \minerva (right) CC0$\pi$ TKI variable measurements on hydrocarbon compared to various models. Generator names, suppressed in the  middle row, follow the same colors as above and below.}
	\label{fig:t2kcc0pinpstv}
\end{figure*}

In the left panels of \cref{fig:t2kcc0pinpstv}, T2K TKI variables are shown and compared with generator predictions. All calculations use the LFG nuclear model except {\Genietwoa} which uses RFG. Considering the cross section as a function of $\delta p_T$, models using LFG generally have better agreement with the data, while  {\Genietwoa} has the worst agreement.  It is surprising that NuWro has poor agreement with the data, although it also uses LFG as nuclear model.  The high momentum tail where $2p2h$ and FSI processes are more relevant, is well described by all models.

Concerning $\delta \alpha_T$, most of the generator predictions have similar shapes, and show a rise at high angles, as is expected in the presence of FSI effects. While {\Genietwoa} shows the smallest rise at high $\delta \alpha_T$, NuWro has an almost flat shape in this variable. Data has a slight preference for {\Genietwoa}. Finally, {\Genietwoa} and NuWro show a different shape with respect to the other generators  also in $\delta \phi_T$.

Although none of the generators correctly reproduces all the data in the three variables, the largest disagreement is shown by NuWro and {\Genietwoa}, that have $\chi^2/$ndf values between 5 and 8 and between 2 and 8, respectively. On the other hand, NEUT and GENIE with LFG show similar and lower $\chi^2$/ndf values, between 0.8 and 3, depending on the variable; NEUT and both {\Genietena} and G18\_10b are fairly good in reproducing $\delta p_T$ and $\delta \phi_T$ distributions. Focusing on G18\_10, we do not see a particular preference for one of the two FSI models used.\\

\begin{figure*}[ht!]
	\centering
	\includegraphics[width=0.49\linewidth]{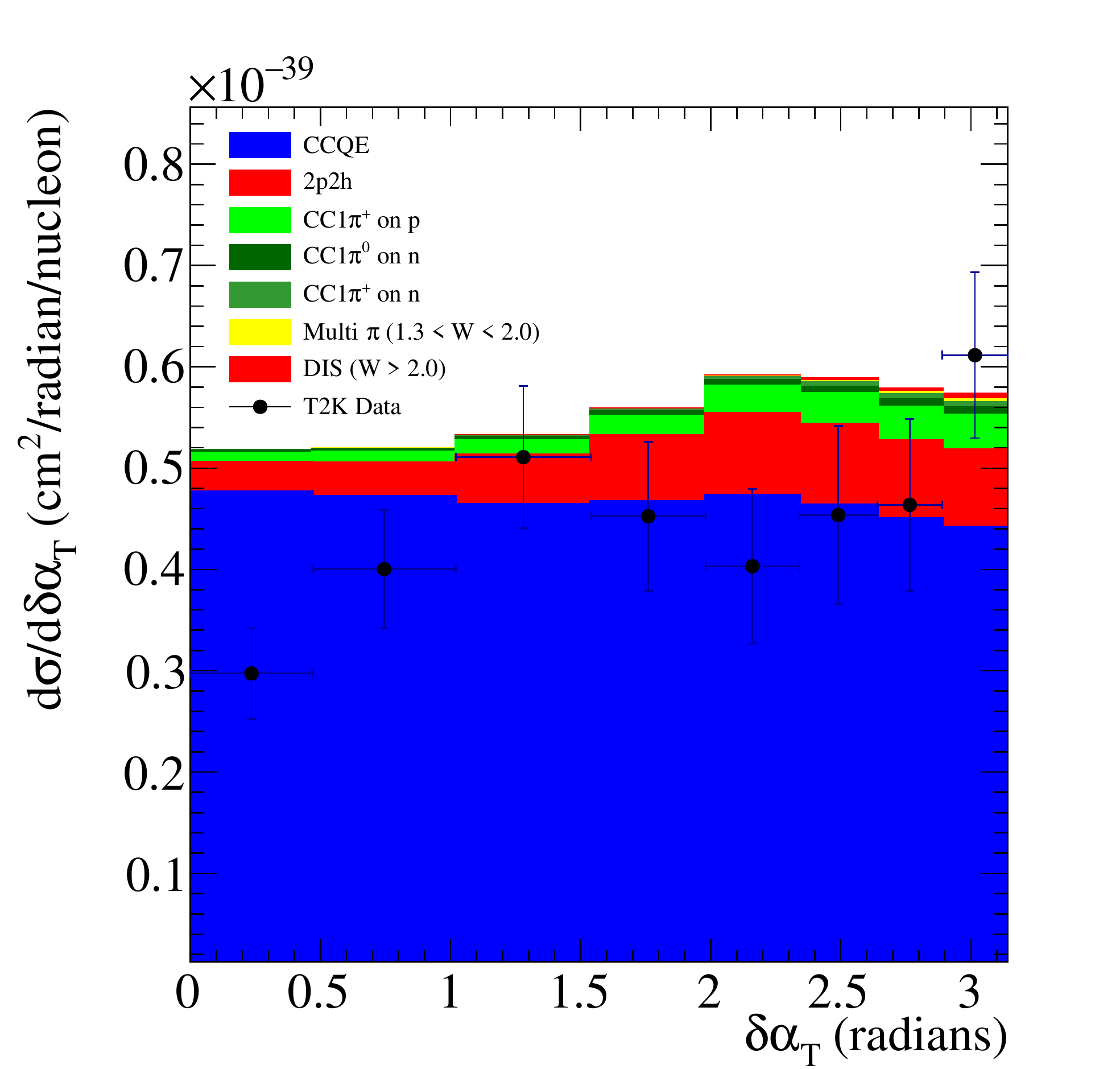}
	\includegraphics[width=0.49\linewidth]{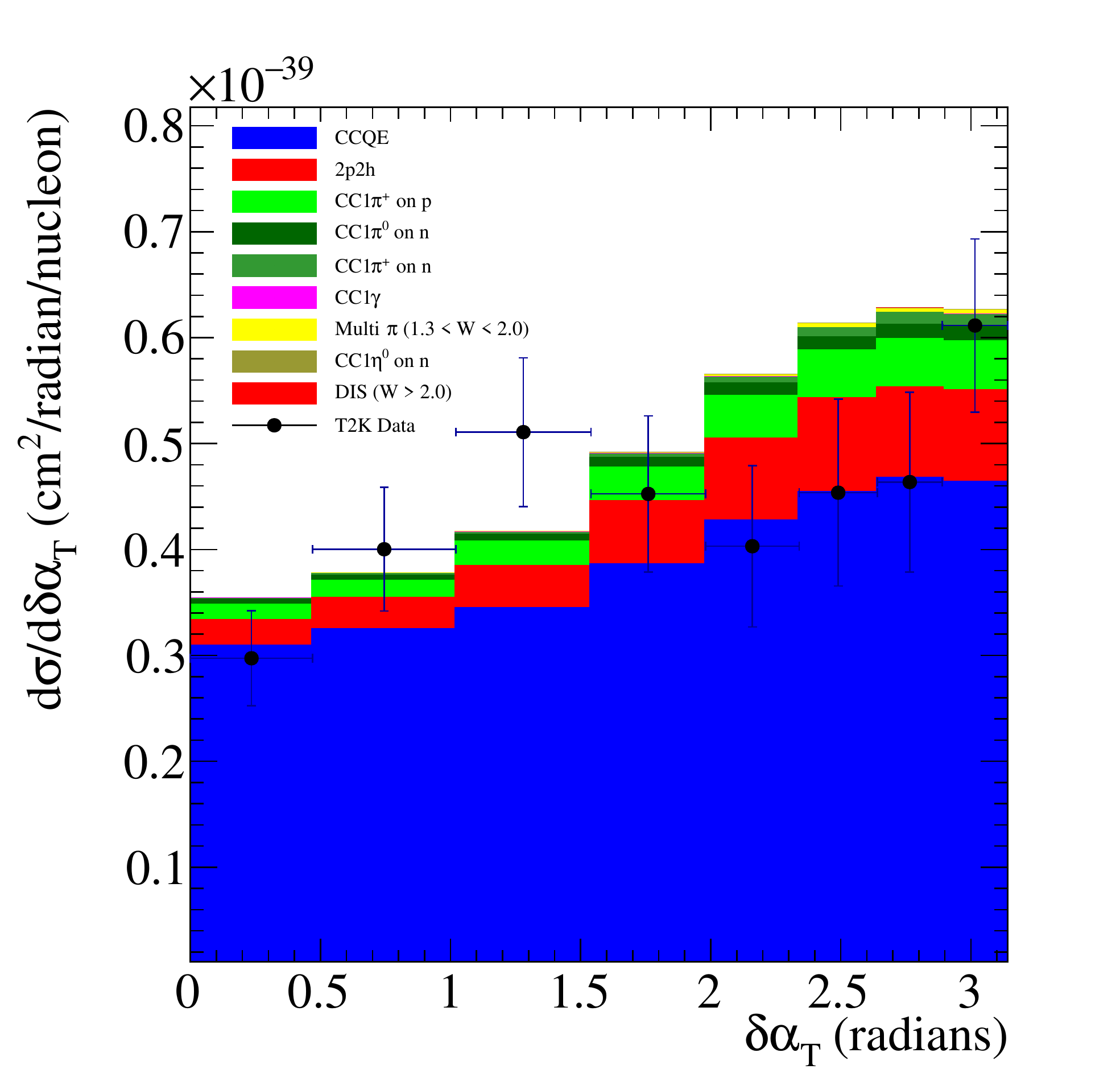}\\
	\includegraphics[width=0.49\linewidth]{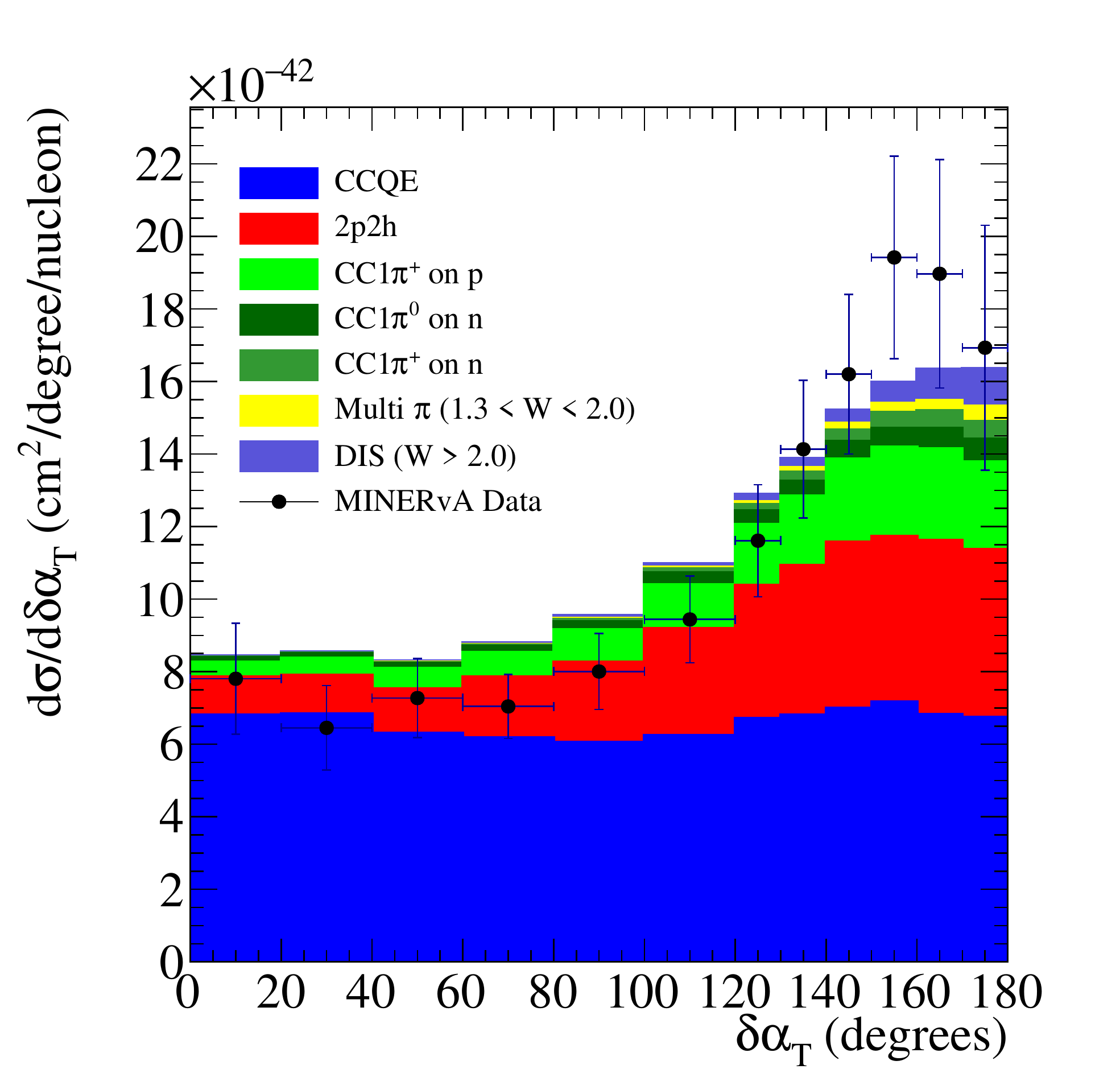}
	\includegraphics[width=0.49\linewidth]{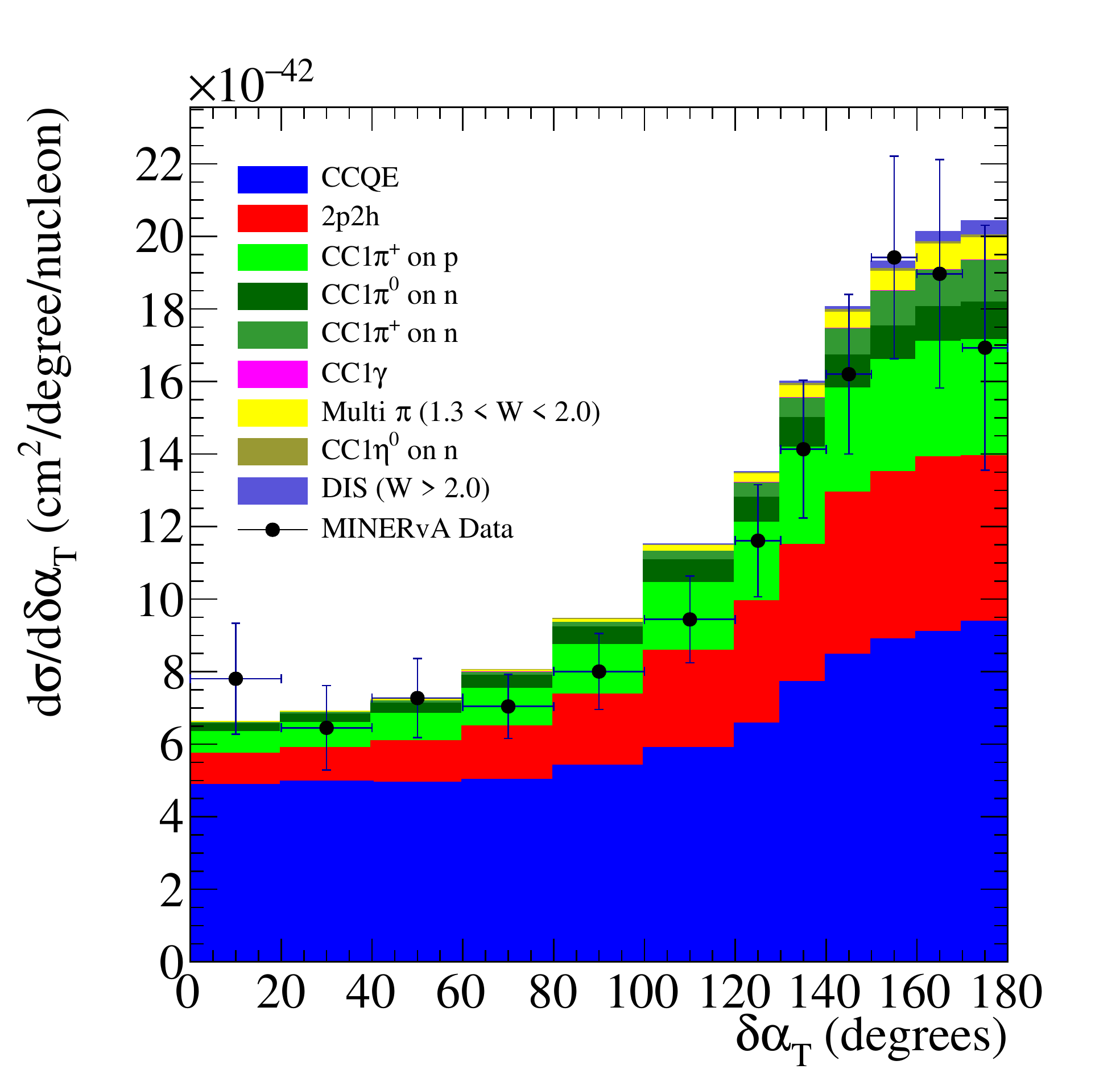}\\

	\caption{T2K CC0$\pi$ (top) and \minerva~(bottom) TKI variable measurements on hydrocarbon compared to NuWro (left) and NEUT (right) predictions, split by neutrino interactions.}
	\label{fig:stvt2kneutnuwro}
\end{figure*}

\minerva TKI variable results are compared to the generator predictions in the right panels of  \cref{fig:t2kcc0pinpstv}. Comparisons indicate varying degrees of data-MC agreement. In particular, generator  predictions for $\delta p_T$ show different peak positions to one another, indicating a difference in the description of the nucleus; {\Genietwoa} shows the largest disagreement with data ($\chi^2/$ndf $\simeq$ 9). In general, predictions from generators that use LFG as nuclear model, show similar shapes in $\delta p_T$, although $\chi^2$ values are smaller for NEUT and GENIE ($\chi^2/$ndf $\simeq$ 3.3) w.r.t. NuWro ($\chi^2/$ndf $\simeq$ 4.3). Overall, none of considered generators is able to well reproduce the $\delta p_T$ distribution. 

Concerning $\delta \alpha_T$, all the generators show similar predictions and thus have a similar fairly good agreement with the data ($\chi^2/$ndf $\simeq$ 2), with the only exception of NuWro that shows a disagreement ($\chi^2/$ndf $\simeq$ 3) and a flatter distribution as already observed for T2K. Despite $\delta \alpha_T$ being in principle sensitive to FSI effects, there is no clear preference for either of the two FSI models used in G18\_10. 

Concerning $\delta \phi_T$, none of the generator shows agreement with \minerva data, although {\Genietwoa} predictions shows the smaller $\chi^2$ value ($\chi^2/$ndf $\simeq$ 2.4), while NuWro the largest ($\chi^2/$ndf $\simeq$ 5); $\chi^2$ values for NEUT and {\Genietena}/b are similar ($\chi^2/$ndf $\simeq$ 3.5). \\

To better understand the difference between NuWro and others generators that use LFG in the prediction of $\delta\alpha_T$, we report in Fig. \ref{fig:stvt2kneutnuwro} the Nuwro and NEUT TKI variable distributions split by neutrino interactions for T2K and \minerva.

For T2K, by looking at first plots of Fig. \ref{fig:stvt2kneutnuwro}, it is evident that the remarkable difference between the predicted $\delta \alpha_T$ shapes is due to CCQE interactions: NuWro predicts an almost flat shape for CCQE, while NEUT shows a dip in CCQE at low $\delta \alpha_T$. Contributions from $2p2h$ and CC1$\pi$ are instead largely similar. We can also notice that the normalisation difference visible in the three variables is still due to different predictions of CCQE, that has a higher cross section in NuWro than in NEUT.

For \minerva, see \cref{fig:stvt2kneutnuwro}, the differences between NuWro and NEUT in TKI variables prediction are less relevant, but are again due to the different contribution from CCQE interactions, that also in this case have a flat distribution in $\delta \alpha_T$ and a higher cross section. Also, NEUT predicts a higher contribution from CC1$\pi$, mainly visible at $\delta p_T \simeq$ 0.3-0.4 GeV.\\

In general, none of the generators investigated here are able to correctly reproduce the data in the three variables; in particular, both T2K and \minerva data disfavor NuWro predictions in all the three TKI variables. Generators have difficulties in reproducing the \minerva $\delta p_T$ distribution, while the agreement is better with T2K data, especially for NEUT and GENIE LFG predictions. Conversely, generators seem to well reproduce \minerva $\delta \alpha_T$ distribution, while they have more difficulties with the T2K $\delta \alpha_T$ distribution. Finally, concerning $\delta \phi_T$, T2K data seems to favor NEUT and GENIE LFG calculations, while this is not evident for \minerva data.

\subsection{Comparison of generators using lepton derived four momentum transfer}
\label{subsec:minervaq2}

Although recent measurements tend to report cross sections as a function of variables related to detectable muon observables, in this and following  sections we focus on CCQE-like measurements that have been performed as a function of the momentum transfer, $Q^2$. 

\begin{figure}[htbp]
	\centering
	\includegraphics[width=0.96\linewidth]{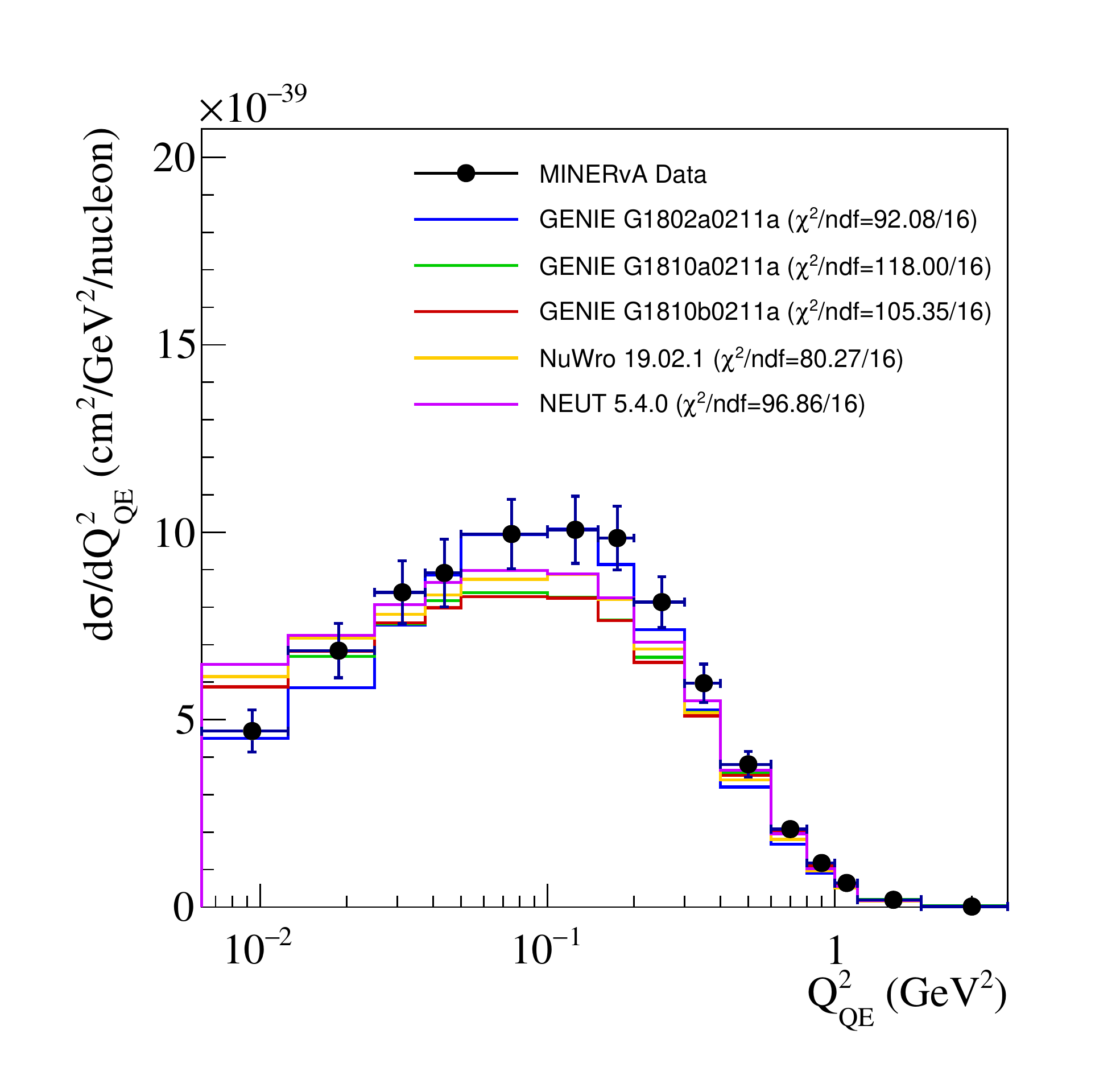}
	\caption{\minerva~CCQE-like $Q^{2}_{\mathrm{QE}}$ cross-section measurement on hydrocarbon compared to various models.}
	\label{fig:cc0piminervaq2qetrackerl}
\end{figure}

An estimator of the four-momentum transfer can be derived for QE-like events under the assumption of a stationary target, as shown in Eqs. \ref{eq:enuqe} and \ref{eq:qqqe}.

\begin{equation}
\label{eq:enuqe}
E_{\nu,QE} = \frac{M_p^2-(M_n-E_b)^2-M_\mu^2+2(M_n-E_b)E_\mu}{2(M_n-E_b-E_\mu+P_\mu cos\theta_\mu)}
\end{equation}

\begin{equation}
\label{eq:qqqe}
Q^{2}_{QE} = 2E_{\nu,QE}(E_\mu-P_\mu cos\theta_\mu) - M_\mu^2
\end{equation}
where $M_p$ and $M_n$ are the masses of proton and nucleon, respectively, $E_b$ is the binding energy, $E_\mu$ is the energy of the outgoing muon and $P_\mu$ and cos($\theta_\mu$) are the muon momentum and direction.\\
The MINERvA experiment measures QE-like interactions in the scintillator tracker region between 0 and 4 GeV$^2$/c$^2$ using an on-axis beam with an average energy of ~3.5 GeV \cite{Ruterbories:2018gub}. Figure \ref{fig:cc0piminervaq2qetrackerl} shows a comparison of this result with a set of predictions from NuWro, NEUT and GENIE. 

The NuWro prediction shows the lower $\chi^2$ while {\Genietwoa} and {\Genietenb} show the highest disagreement with data. However, $\chi^2/$ndf values are between 5 and 7, thus indicating that none of the generators properly describes the data.

\subsection{Comparisons of generators with data from different targets (CH, Fe and Pb)} 
\label{subsec:minervaAscale}

The \minerva experiment measured QE-like interactions on different targets using the same neutrino beam \cite{Betancourt.119.082001}. Although this result was not published recently, it has a unique standing as it spans various nuclei. The signal is defined as an event with one muon, no pions and at least one proton with momentum greater than 450 MeV/c exiting the nucleus. The measurements have been performed using the four momentum transfer $(Q^2_p)$ reconstructed using the proton kinetic energy $(T_p)$. Under the assumption of CCQE scattering from a neutrino at rest the $Q^2_p$ value is reconstructed using:

\begin{multline}
Q^2_p=(M_n-\epsilon_B)^2-M^2_p+\\ 2(M_n-\epsilon_B)(T_p+M_p-M_n+\epsilon_B),
\end{multline}
where $M_{n,p}$ is the appropriate nucleon mass, and $\epsilon_B$ is the effective binding energy of 34 MeV$/c^2$.

\begin{figure*}[ht!]
	\centering
	\includegraphics[width=0.49\linewidth]{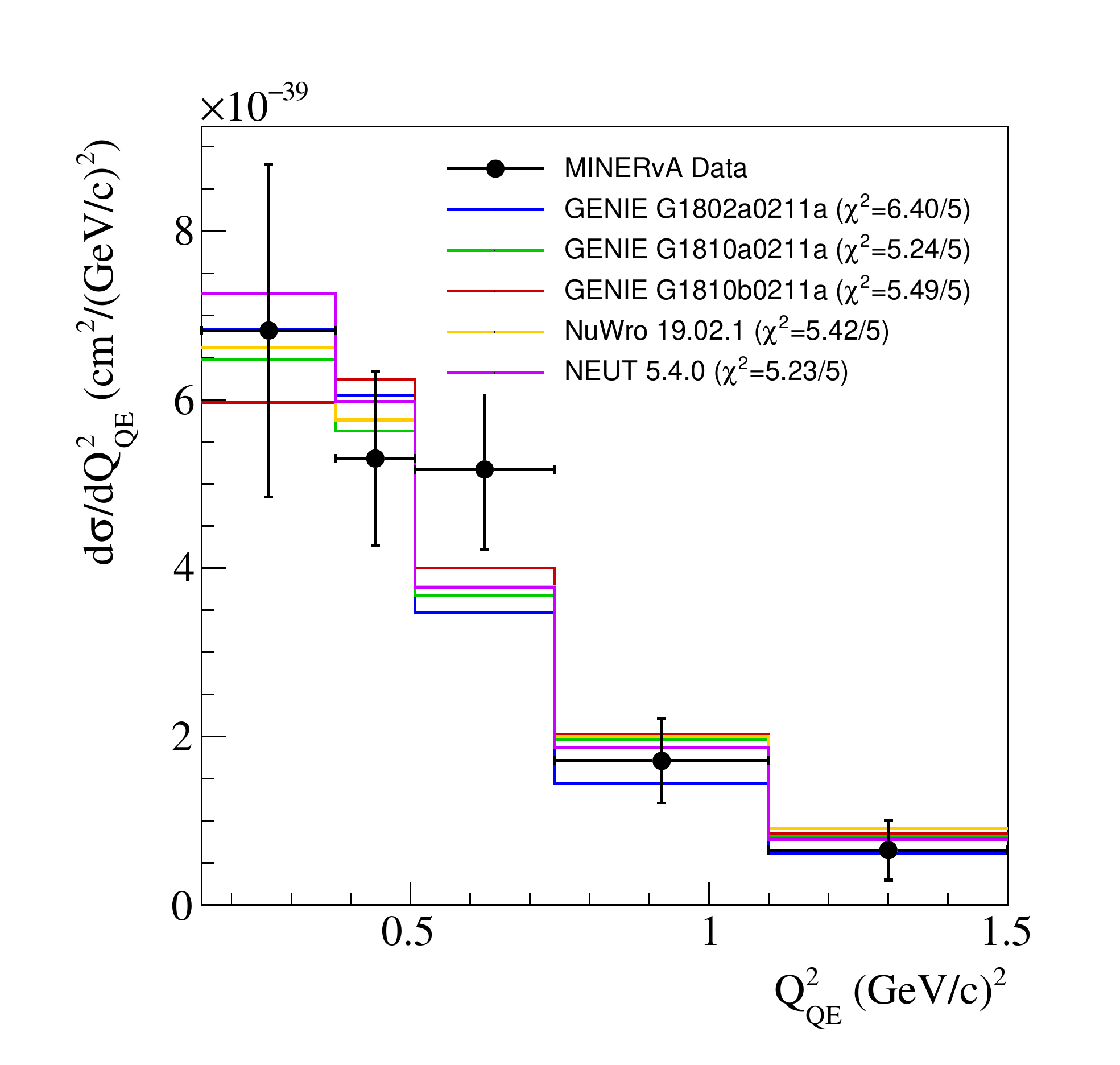}
	\includegraphics[width=0.49\linewidth]{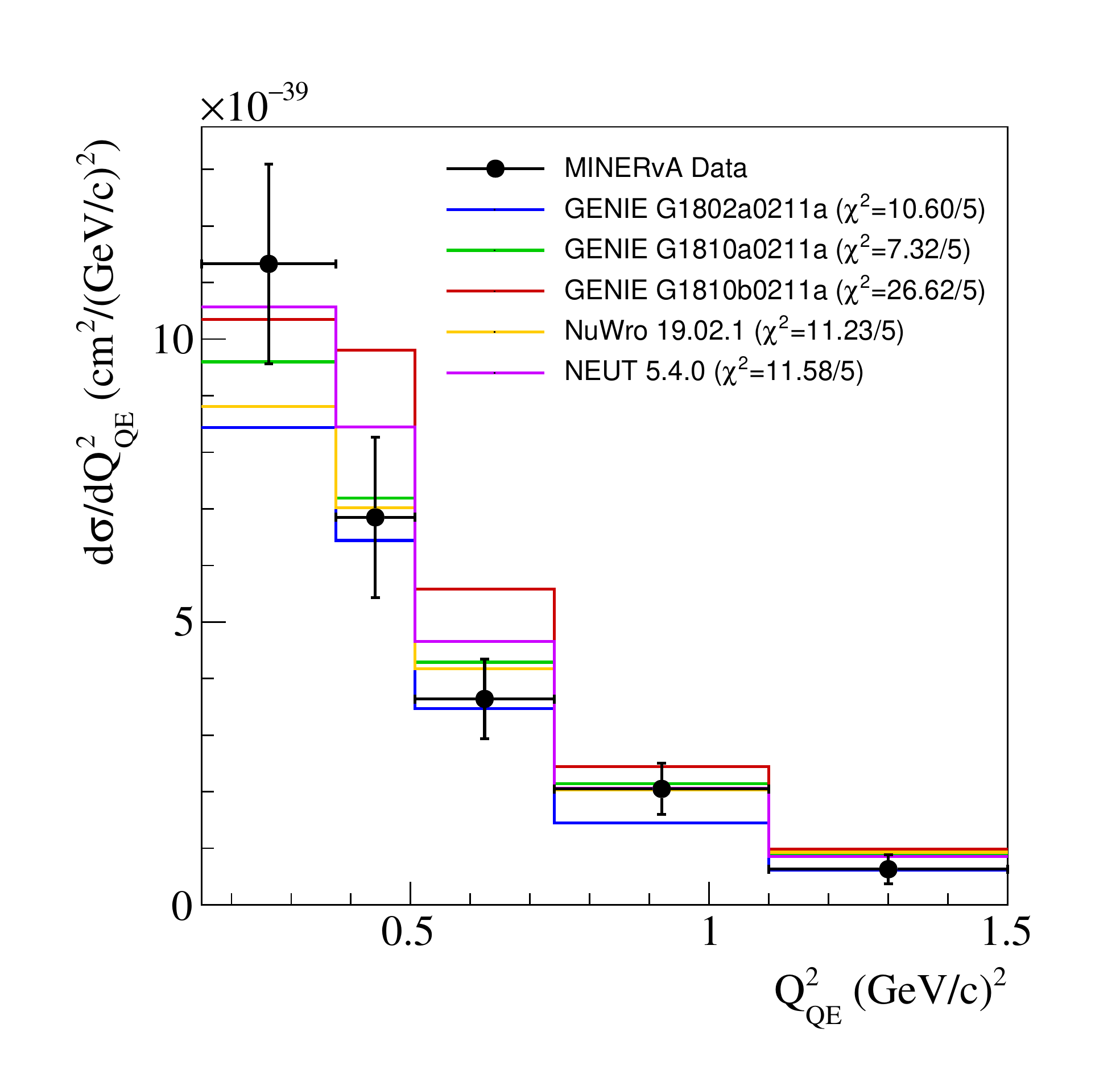}
	\includegraphics[width=0.49\linewidth]{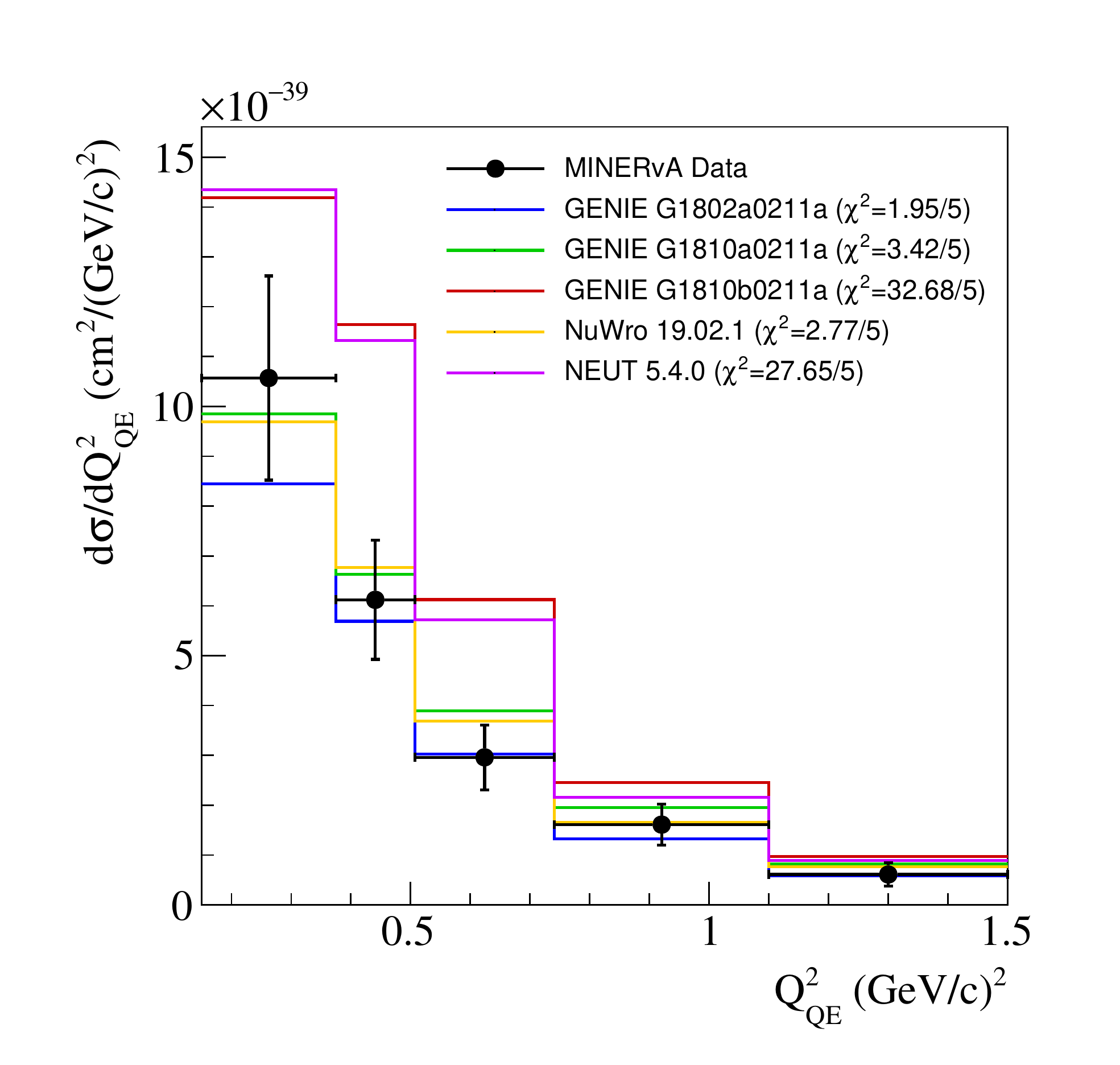}
	
	\caption{Differential cross sections as a function of $Q^2_p$ for CH  (top-left), Fe (top-right) and Pb (bottom) compared with NuWro, NEUT and three versions of GENIE.}
	\label{fig:minervaccqeliketargets}
\end{figure*}

Figure \ref{fig:minervaccqeliketargets} shows comparisons of data in different materials (CH, Fe and Pb) with the NuWro, NEUT and GENIE predictions. 

Since proton kinematics in the final state are very sensitive to final state interactions, these measurements test the atomic mass (A) dependence of models. For the CH target measurement, all generators show similarly good agreement with data and have a $\chi^2$/ndf of about 1.0, although no tuning was done for any of the models. The heavier targets provide a more interesting test of models, since we expect larger effects from FSI.  The spread of distributions predicted by generators and the spread of corresponding $\chi^2$ both increase significantly for these nuclei. For Fe the best prediction comes from {\Genietena} and for Pb the best predictions are from {\Genietwoa}, G18\_10a and NuWro. NEUT 5.4.0 and {\Genietenb} have large disagreement with the Pb data with $\chi^2$ per data point of about 5. The most significant deviations are at low $Q^2_p$ where nuclear effects tend to be more important. Perhaps, the most interesting comparison is between G18\_10a and G18\_10b because only the FSI model changes between them. Although G18\_10a describes the A dependence well, G18\_10b (as well as NEUT) grows faster at low $Q^2_p$ than the data.

\subsection{Discussion}
\label{subsec:cc0pidiscussion}

A number of interesting comparisons against three results from T2K~\cite{Abe:2020jbf,Abe:2020uub,Abe:2018pwo} and \minerva~\cite{Ruterbories:2018gub,Betancourt.119.082001,Lu:2018stk} have been shown in this section. All use $\nu_\mu$ beams and all use C or CH targets except Ref.~\cite{Betancourt.119.082001}.

Recent cross-sections measurements are presented as a function of muon kinematics, in either $p_\mu - cos\theta_\mu$ (T2K, \cref{fig:t2knumucc0pich2dpcos,fig:t2knumucc0pic2dpcos}) or \pt-\pz (\minerva, \cref{fig:cc0piminerva2Dplot}), that are variables directly measured, thus ensuring a minimization of model dependence. In order to help the comparison between the two mentioned experiments, in \cref{fig:cc0piq0q3eff}, model predictions for $q_0-q_3$ are also reported, together with the corresponding detection efficiencies. 

When comparing data and generator predictions, it is interesting to note that the older and less sophisticated model ({\Genietwoa}) is the worst model overall in describing the data but is not significantly worse. Also, NuWro seems unable to reproduce T2K data well, while it behaves slightly better for \minerva. GENIE {\Genietwoa} and {\Genietenb} show similar behaviors and are very close to NEUT predictions in the case of T2K, while they slightly differ when looking at \minerva results.

It should be stressed that $\chi^2$ presented in this section are obtained using the covariance matrices as provided by the two collaborations. This means that normalization and shape errors are accounted at the same time, thus implying significant correlations between bins, due for instance to the flux normalization uncertainty. Therefore, the $\chi^2$ calculation could be affected by Peelle's Pertinent Puzzle (PPP)~\cite{ppp}, although according to T2K publications~\cite{Abe:2020jbf,Abe:2020uub}, where also shape-only $\chi^2$ are provided, the conclusions about data-MC agreement do not change.

A comparison of T2K and \minerva data in a similar phase space region in muon kinematics was made.  To take out the known energy dependence (larger cross section for the higher beam energies of \minerva), the ratio data$/$Monte Carlo is also shown.  In general, in this particular phase space region, MC predictions underestimate the data. The only exception is represented by NEUT that matches very well the \minerva results. For all the other generators, the underestimation of the data seems to be equivalent in T2K and \minerva, thus suggesting that the energy dependence of the models is approximately correct for the CC-0$\pi$ interaction.  \\

The TKI variables (Fig.~\ref{fig:t2kcc0pinpstv}) provide a more detailed way to explore nuclear and FSI models. $\delta p_T$ distributions, which are more sensitive to the nuclear model, disfavor {\Genietwoa} for both T2K and \minerva. The T2K preference is clearly for LFG models as implemented in GENIE and NEUT, while NuWro is disfavored. Concerning FSI models, that can be in principle tested by looking at $\delta \alpha_T$ distributions, neither T2K nor \minerva show a particular preference for {\Genietena} or {\Genietenb}; all $\chi^2$ values are similar, except for NuWro that is the most disfavored.\\

Two different \minerva analyses measuring $Q^2$ have been considered: one where the $Q^2$ variable is estimated starting from muon kinematics and the other one where the proton in the final state is used to evaluate the $Q^2$; this second, and older, analysis used data of neutrino interactions on three different nuclei (see Fig.~\ref{fig:minervaccqeliketargets}).  It has been re-examined because of its uniqueness as a way to probe the A dependence. 
Agreement between data and MC is best for the lightest target (CH) that does not show a particular preference for one generator. NuWro and {\Genietena} are also able to nicely describe cross section on heavier nuclei, while  NEUT and {\Genietwoa} show a more rapid increase in the low $Q^2$ region and a significantly higher normalization than the Pb data. Both Fe and Pb data seems to prefer {\Genietena} over {\Genietenb}, thus preferring the data-driven FSI model.

\FloatBarrier

\section{Pion production interactions}
\label{sec:pion}
Pion production data are crucial for neutrino oscillation experiments at higher beam energies such as NOvA and DUNE, where final states with one or several pions are dominant. Models for neutrino-induced single pion production in NEUT and GENIE are based on modifications to the Rein-Sehgal model, whereas NuWro uses a $\Delta$-only model. These models are generally tuned to low statistics bubble chamber data, where the neutrino scatters on deuterium and/or hydrogen.  Effective models for the nuclear medium are added, but none of the models contain effects on the $\Delta$ production amplitude shown to be important in descriptions of pion-nucleus scattering data~\cite{Freedman:1982yp}.  
In addition, the effects of long-range NN correlations are unknown.
NN correlations and the nucleon density distributions~\cite{Benhar:1994hw} are known to be important for CC-mesonless interactions (see \autoref{sec:cc0pi}) but have not been applied to pion production interactions.  
Another major challenge in using neutrino-\emph{nucleus} scattering data to tune the underlying nucleon production model is the presence of pion FSI.
Because of the difficulties in describing hadron propagation through nuclei with a quantum mechanical model, generators use intra-nuclear cascade (INC) models.
These are often semi-classical models with corrections accounting for nuclear effects via some effective approach. As described in \autoref{tb:1pimodels}, GENIE v3, NEUT, and NuWro all have implementations of the nuclear effects in the Salcedo-Oset~\cite{Salcedo-Oset} cascade model, although the details differ slightly. GENIE additionally offers an effective ``single-step'' cascade model with its {\it hA} model which is tuned to hadron-nucleus data.
FSI models can additionally be informed by pion-nucleus scattering data~\cite{,Salcedo-Oset,PinzonGuerra:2018rju}, although the relationship between pion-nucleus scattering and pions being produced in-medium and propagating out is non-trivial~\cite{Niewczas:2019fro,Dytman:2021ohr}.

In this section, we investigate recent charged pion production publications from T2K and MINERvA which serves as an update on the previous TENSIONS2016 work~\cite{tensions2016}, where tensions between the MiniBooNE CC$1\pi^+$ result~\cite{Wilking} and the first \minerva CC$1\pi^\pm$ result~\cite{Eberly:2014mra} were investigated.

\subsection{T2K results}
The most recent CC1$\pi^{+}$ analysis from T2K~\cite{t2k_cc1pi_fgd1} used data corresponding to $1.51\times10^{21}$ POT and used FGD1, a plastic scintillator detector (C$_8$H$_8$), as its target. The measurement is dominated by single pion production via a resonant interaction, most prominently the $\Delta(1232)$ resonance due to T2K's beam energy. Coherent pion production and soft inelastic interactions also make small contributions to the single pion topology, with very few deep inelastic events. The analysis used NEUT 5.1.4.2 as its neutrino interaction simulation (\autoref{tb:1pimodels}) during the development of the analysis.

\subsubsection{Selection}
The analysis selects events with a forward-going muon and a charged pion in the final state. The pion is tagged either by the presence of a pion-like track in a TPC, that shares a common vertex in FGD1 with the muon track, or by a time-delayed Michel electron in FGD1 implying a pion with momentum below tracking threshold. Proton-like tracks were allowed but not required. The selection provides a pure sample with minimal muon-pion confusion, primarily due to the charge as determined by track curvature from the magnetic field in the TPC, with good momentum and angle resolution. The selection limits the ranges in track momentum and angle, since muon tracks must have enough momentum to escape FGD1 and they must have relatively forward angles with respect to the beam direction to enter the TPC. The signal definition includes restrictions on both muon and pion kinematics to reflect this.

\subsubsection{Variable definitions}
The differential cross-section is analysed in seven distributions: the two dimensional $d^{2}\sigma/dp_{\mu}d\cos\theta_{\mu}$, and the one dimensional $d\sigma/dQ^{2}_{\mathrm{rec}}$, $d\sigma/dp_{\pi}$, $d\sigma/d\theta_{\pi}$, $d\sigma/d\theta_{\pi\mu}$, $d\sigma/d\phi_{\text{Adler}}$, and  $d\sigma/d\cos\theta_{\text{Adler}}$.

Importantly, the ``theory variables'' of $Q^{2}_{\mathrm{rec}}$, $\cos\theta_\text{Adler}$ and $\phi_\text{Adler}$ used in the measurement do not correspond to the usual theory variables which require exact knowledge about the neutrino energy and direction, which would involve model-dependent corrections to the observed event. Instead they are proxy variables, derived using the lepton and pion kinematics only, assuming that there were three outgoing particles. The variables require both a muon and a pion track, and $Q^{2}_{\mathrm{rec}}$ is reconstructed as $Q^{2}_{\mathrm{rec}} = -q^2 = -(p_\nu-p_\mu)^2=-\left(m^2_\mu - 2E_\nu^{\mathrm{rec}}\left(E_\mu - |\vec{p}_\mu| \cos\theta_\mu\right)\right)$, where the neutrino energy is \emph{reconstructed} using detector observables to reduce model dependence
\begin{equation}
    E_\nu^{\mathrm{rec}} = \frac{m^2_p-\left(m_p-E_b-E_\mu-E_\pi\right)^2 +\left|\vec{p}_\mu+\vec{p}_\pi\right|^2}{2\left(m_p-E_b-E_\mu-E_\pi+\vec{d}_\nu \cdot  \left(\vec{p}_\mu+\vec{p}_\pi\right)\right)}
    \label{eq:t2k_enu}
\end{equation}
with $E_b=25\text{ MeV}$ and $\vec{d}_\nu$ being the average predicted neutrino direction. 
Although this means $Q^{2}_{\mathrm{rec}} \neq Q^2_{\mathrm{true}}$ for the T2K result, it was preferred in the publication as it avoids model dependence, and still maps well to $Q^2_{\mathrm{true}}$.

\begin{figure}[h]
    \centering
    \includegraphics[width=0.98\linewidth]{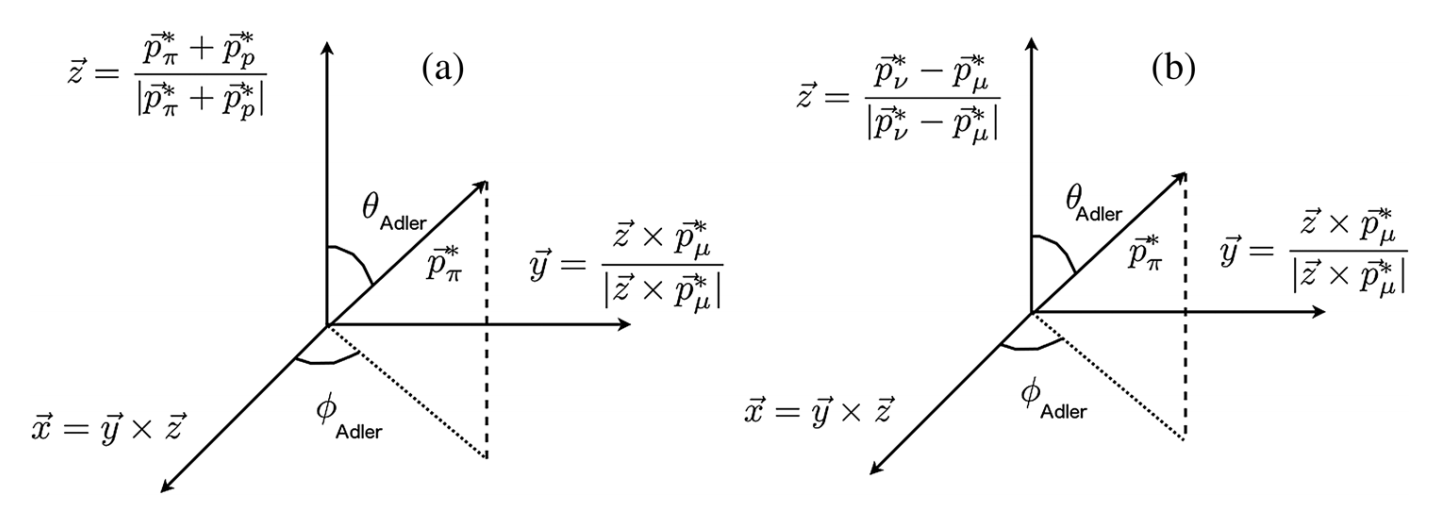}
    \caption{Definition of the Adler angles in the resonant rest frame, using a) pion and nucleon kinematics, b) using neutrino and muon kinematics. Reproduced from Ref.~\cite{t2k_cc1pi_fgd1}.}
    \label{fig:t2k_adler_def}
\end{figure}

The Adler angles are defined in \autoref{fig:t2k_adler_def} which, when combined with \autoref{eq:t2k_enu} to estimate $E_\nu^{\mathrm{rec}}$, provide a more model independent proxy variable for the usual Adler angles, which are defined on the nucleon level in the resonance rest frame. Hence the Adler angle measurements produced by T2K do not perfectly correspond to the more common $\Delta$ decay angles, since producing them involves significant dependence on the nuclear model~\cite{Sanchez:2015yvw}. Furthermore, the publication uses the Adler angles defined by the neutrino and muon kinematics instead of the more commonplace pion and nucleon kinematics, since the ability of ND280 to reconstruct the outgoing nucleon is relatively poor. The Adler angles can be used to infer the strength of interference between resonant and non-resonant pion production, which polarizes the resonance~\cite{Sanchez:2015yvw,PhysRevD.93.093015}.

\subsubsection{Signal definitions}
The cross-section measurements' signal definitions are summarized in \autoref{tab:t2k_1pi_cuts}, and differ slightly from each other due to the different efficiencies in the variables, and the kinematics that the variables require. For instance, the $d^2\sigma/dp_\mu d\cos\theta_\mu$ distribution only requires a measurement of the muon and a tag for the pion; hence T2K can tag the pion via a delayed Michel electron in addition to the usual track criteria, so the restrictions on $p_\pi$ and $\cos\theta_\pi$ are removed. The statistical power of the measurement is therefore the highest out of all the variables in the publication.

The cuts are always on the outgoing observed particle kinematics according to truth (MC), and not on auxiliary variables, such as the hadronic mass, $W$.

\begin{table*}
\caption{Phase space cuts on particle kinematics included in the signal definition in the various T2K CC$1\pi^+$ measurements.}
\label{tab:t2k_1pi_cuts}
\begin{ruledtabular}
\centering
{\renewcommand{\arraystretch}{1.2}   
\begin{tabular}{c||c| c| c| c}

     Measurement & $\cos\theta_\mu$ & $p_\mu \text{ (GeV/c)}$ & $\cos\theta_\pi$ & $p_\pi \text{ (GeV/c)}$ \\
\hline
     $d^2\sigma/dp_\mu d\cos\theta_{\mu,\nu}$ & $>0.0$  &         &         & \\
     $d\sigma/dp_{\pi}$                 & $>0.2$ & $>0.2$ & $>0.2$ &   \\
     $d\sigma/d\theta_{\pi,\nu}$        & $>0.2$ & $>0.2$ & $>0.0$    &  $>0.2$ \\
     \hline
     \begin{tabular}{c} $d\sigma/dQ^{2}_{\mathrm{rec}}$\\ $d\sigma/d\theta_{\pi,\mu}$\\ $d\sigma/d\phi_{Adler}$\\ $d\sigma/d \cos\theta_{Adler}$\end{tabular}             & \multicolumn{4}{c}{$>0.2$} \\

\end{tabular}}
\end{ruledtabular}
\end{table*}

\subsubsection{Cross-section extraction}
The measurement uses three sideband samples looking for a right-signed muon, further enriched in CC0$\pi$1$p$, CC2$\pi^{+}$, and CC$Ne^\pm0\pi^{+}$ events to constrain pion/proton confusion, missing charged pions, and neutral pion backgrounds, respectively. The event rate extractions were performed separately for each differential cross section, and the number of events for overlapping regions of phase space were not required to match. However all results were found to be consistent with the model within statistical and systematic uncertainties.

After background-subtraction, the number of events in each bin of the true kinematic variable was estimated from the data with a single iteration of the D'Agostini~\cite{DAgostini:1994zf} unfolding procedure, and was repeated for each of the aforementioned kinematic variables. The single iteration unfolding was chosen by studying full NEUT and GENIE productions on ND280 with different alternate parameter sets, balancing the statistical error size with the unfolding robustness~\cite{CastilloFernandez:2015wbg}. However, the use of a single iteration of D'Agostini, and the features of the efficiency function (see \autoref{sec:t2k_minerva_eff_comps}), opens potential for bias toward the input signal MC in this analysis. 

\subsubsection{Results}
\begin{figure*}[ht!]
    \centering
    \includegraphics[width=0.98\linewidth]{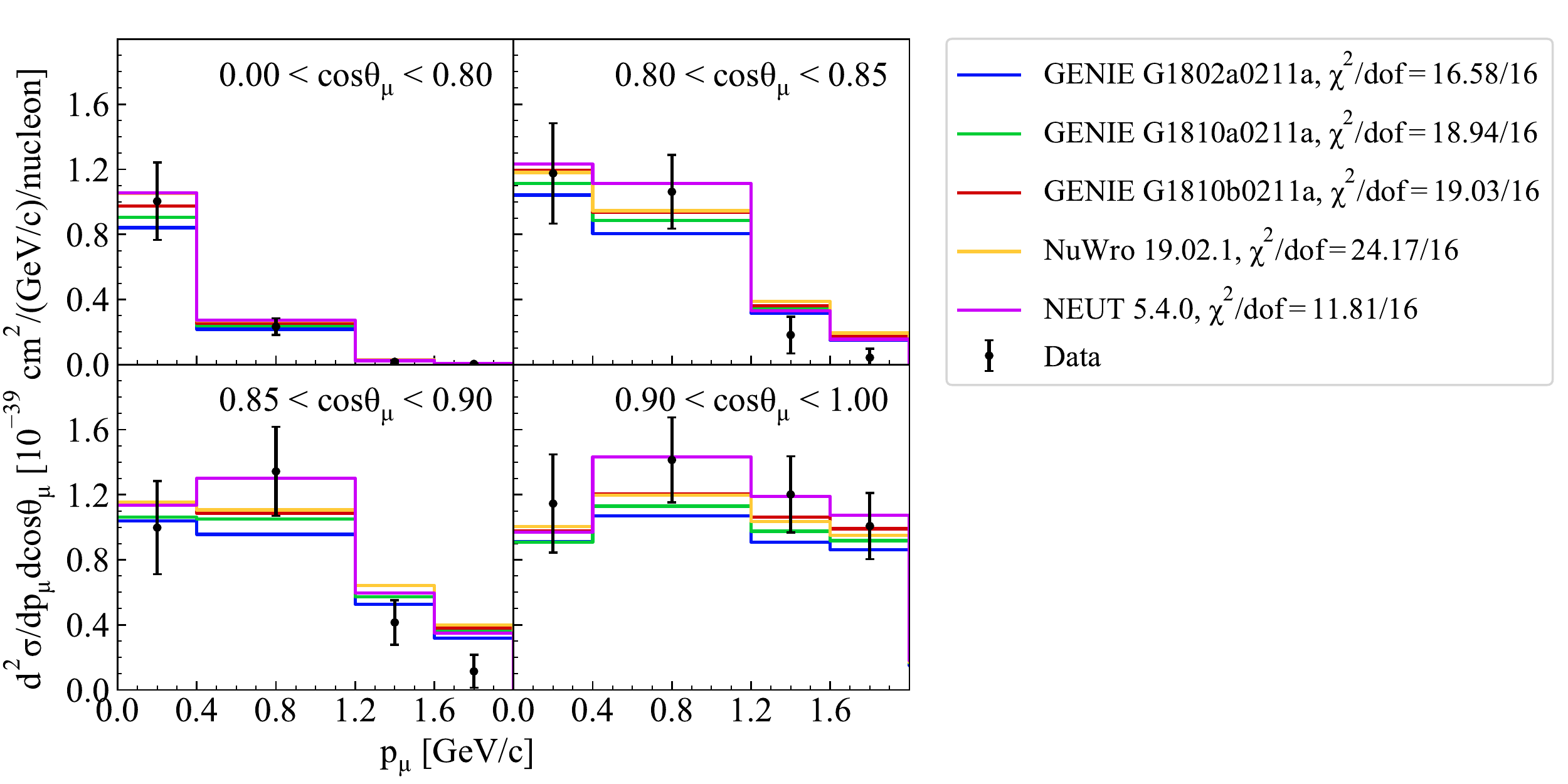}
    \caption{Double-differential cross section as a function of $p_\mu$ and $\cos\theta_{\mu}$ from T2K, showing the predictions of various neutrino interaction generators.}
    \label{fig:t2k_cc1pi_mu}
\end{figure*}

\begin{figure*}[ht!]
    \centering    
    \includegraphics[width=0.48\linewidth]{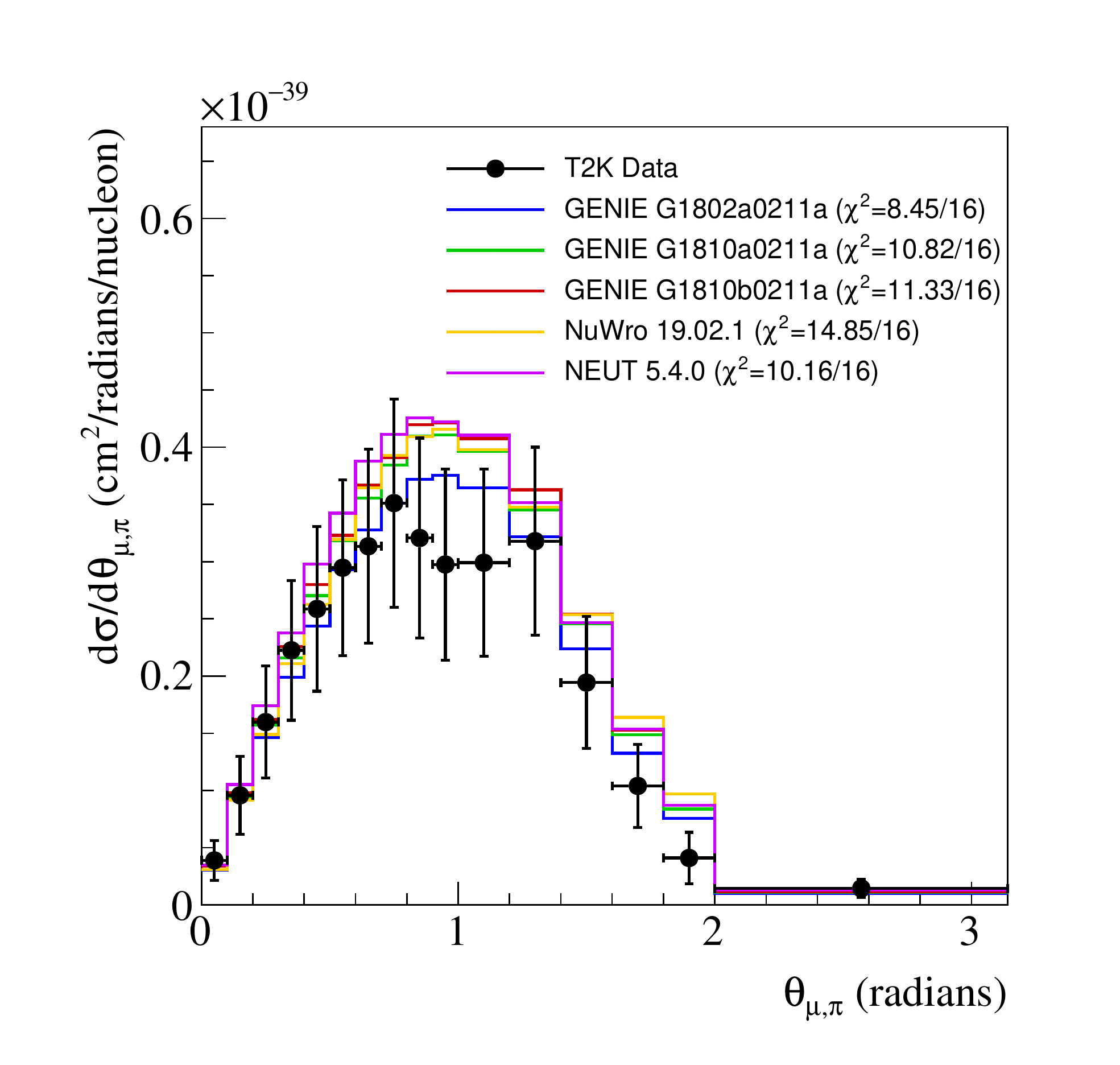}
    \includegraphics[width=0.48\linewidth]{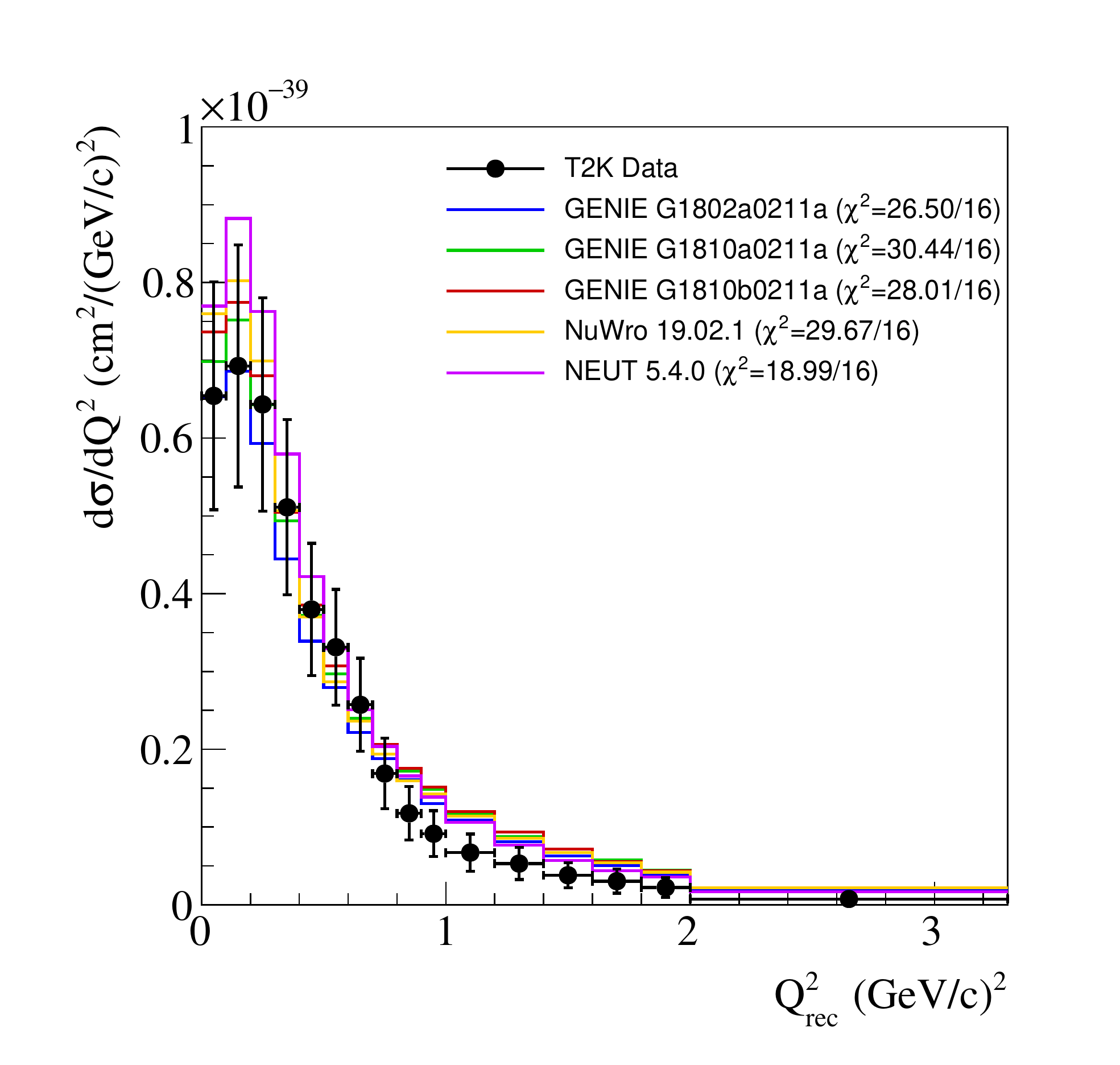}
    \caption{Single-differential cross section in $\theta_{\mu,\pi}$ and $Q^{2}_{\mathrm{rec}}$ from T2K, showing the predictions of various neutrino interaction generators. The $Q^{2}_{\mathrm{rec}}$ definition does not correspond to the fundamental interaction vertex $Q^2_{\mathrm{true}}$, but is instead reconstructed from the visible particles observed in the detector after they exit the nucleus.}
    \label{fig:t2k_cc1pi_other}
\end{figure*}

\begin{figure*}[ht!]
    \centering
    \includegraphics[width=0.48\linewidth]{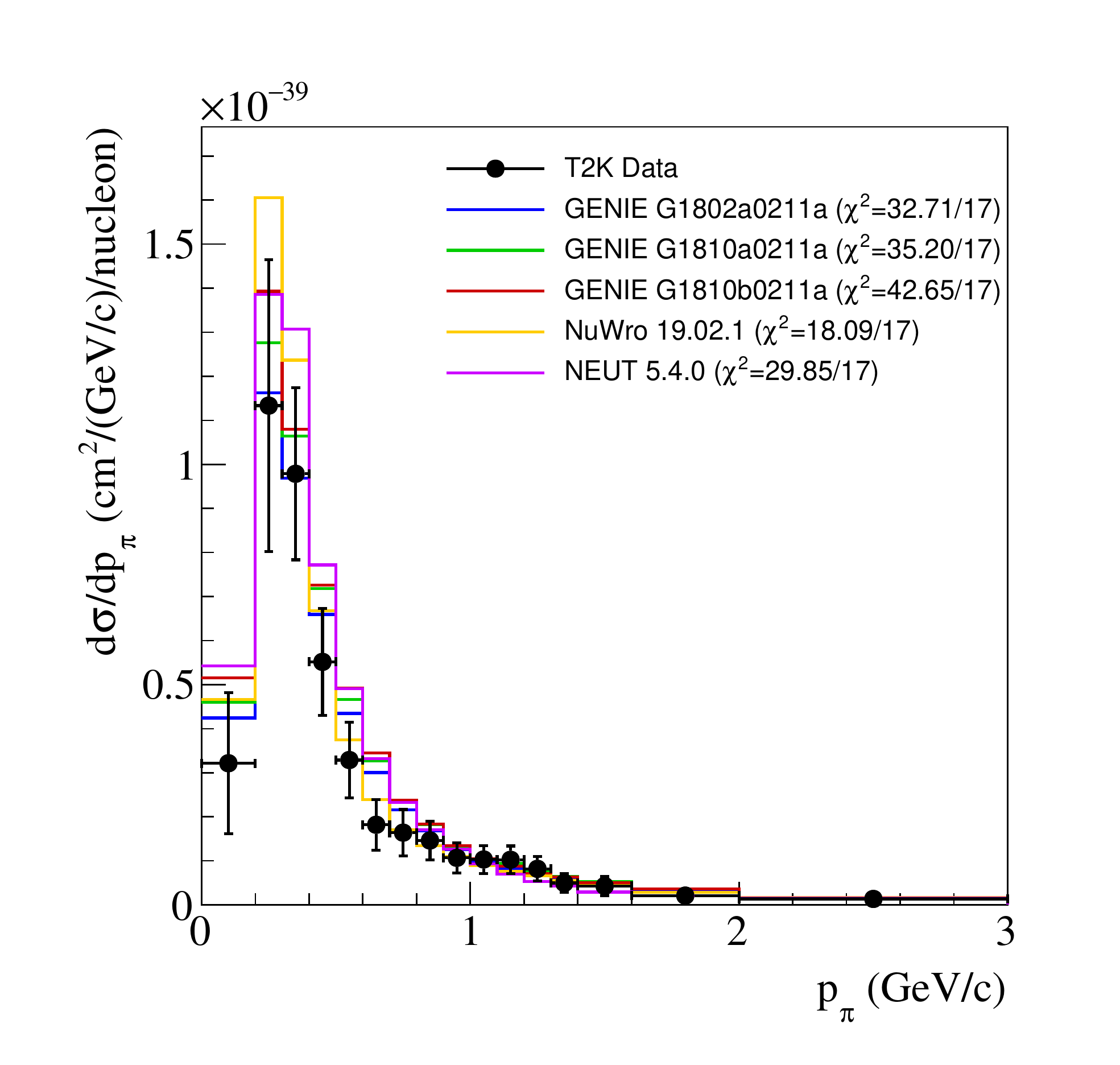}
    \includegraphics[width=0.48\linewidth]{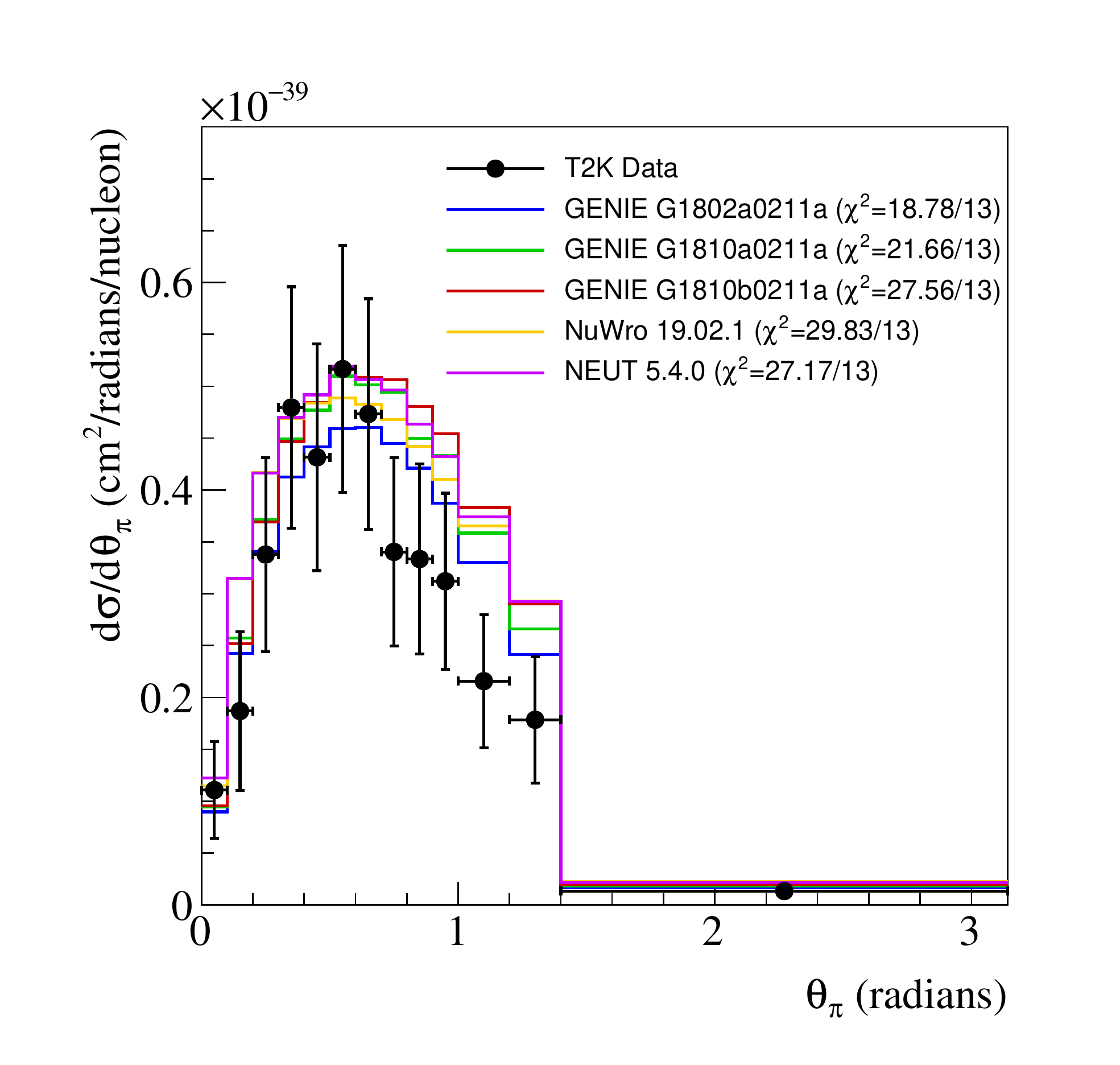}
    \caption{Single differential cross-section in $p_\pi$ and $\theta_{\pi,\nu}$ from T2K, showing the predictions of various neutrino interaction generators}
    \label{fig:t2k_cc1pi_pi}
\end{figure*}

\begin{figure*}[ht!]
    \centering
    \includegraphics[width=0.48\linewidth]{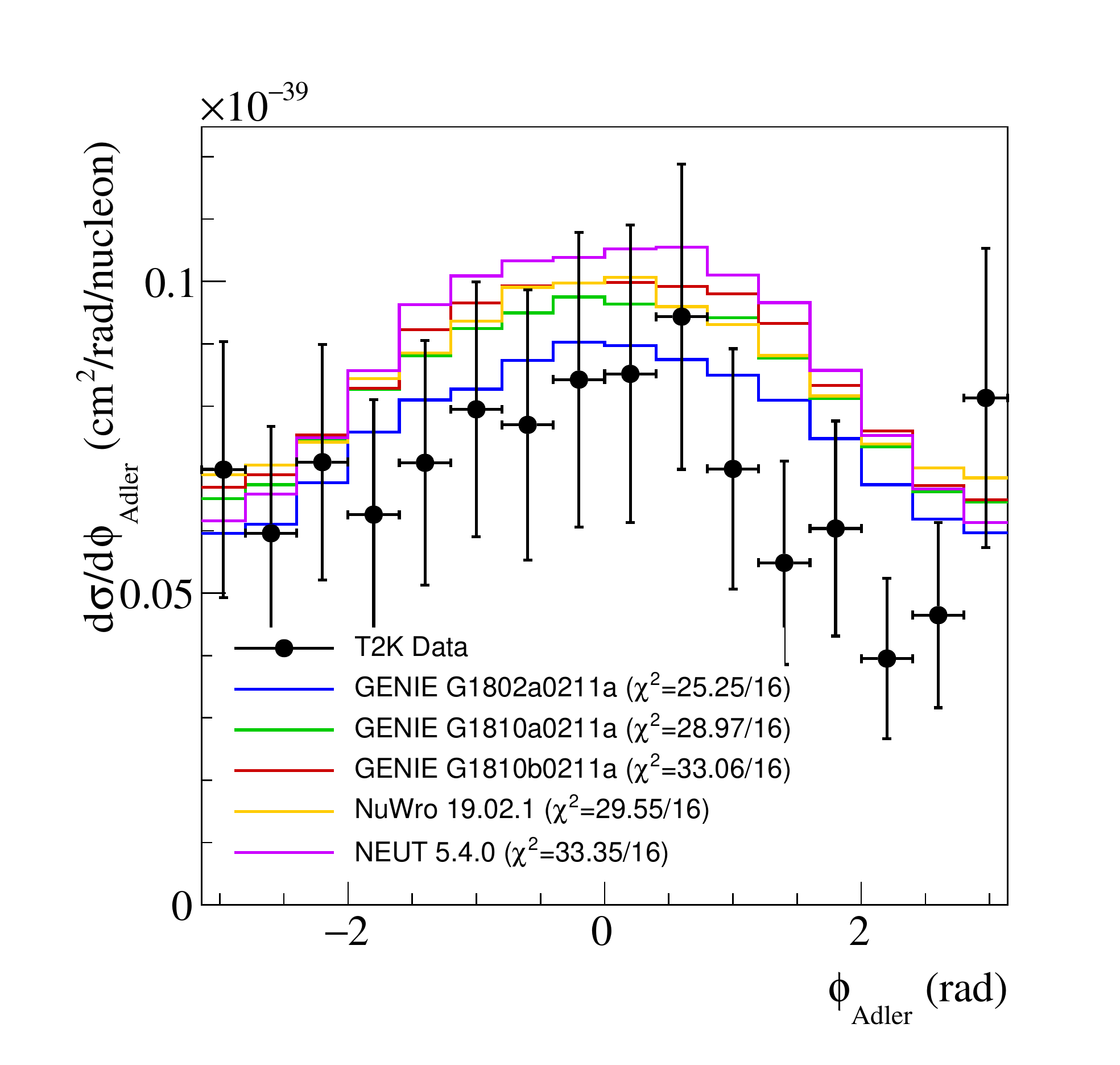}
    \includegraphics[width=0.48\linewidth]{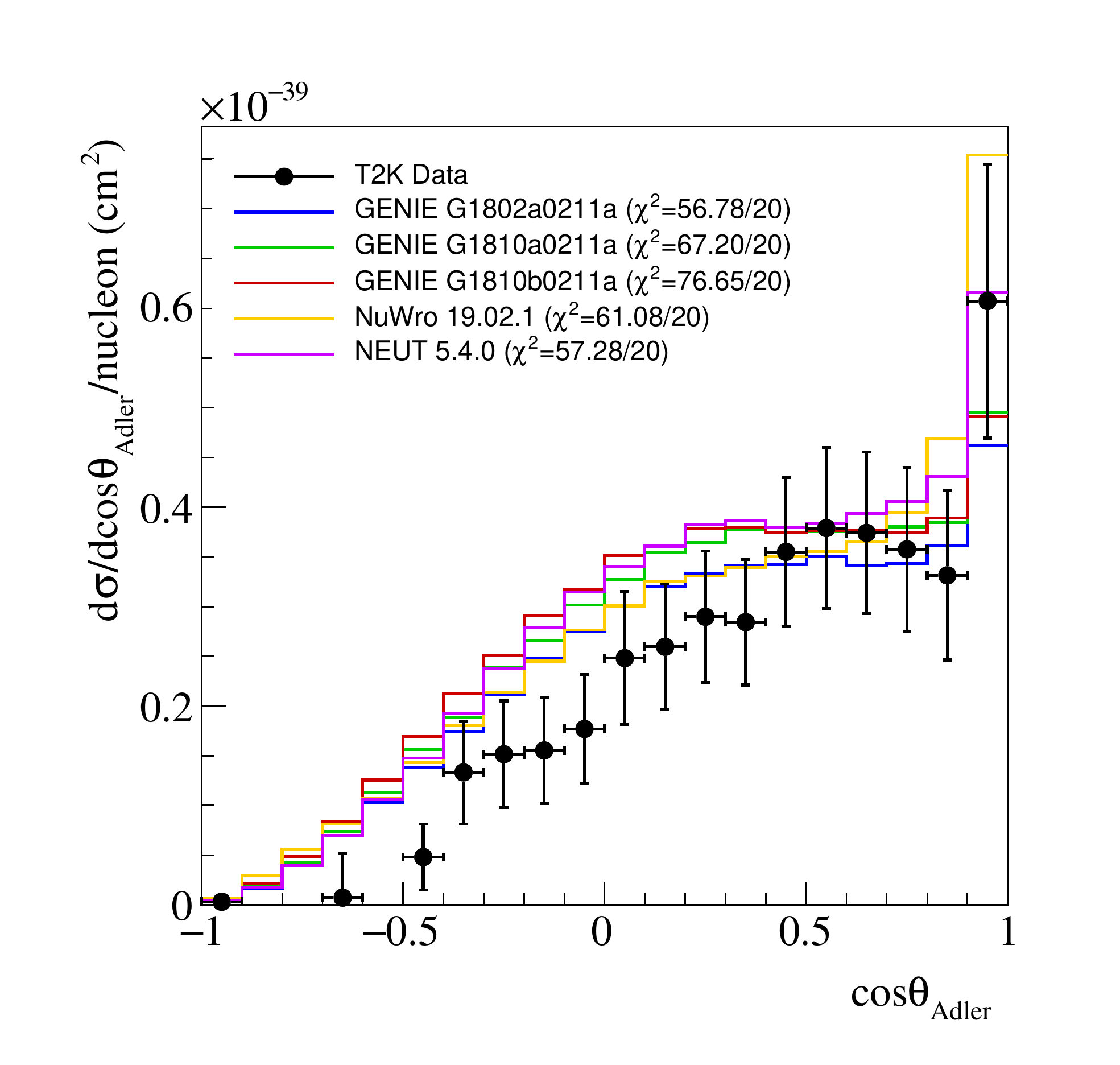}
    \caption{Single differential cross-section in the Adler angles $\phi_{\mathrm{Adler}}$ and $\cos\theta_{\mathrm{Adler}}$ from T2K, showing the predictions of various neutrino interaction generators. The Adler angles do not correspond to the fundamental interaction vertex Adler angles, but is instead reconstructed from the visible particles observed in the detector after they exit the nucleus.}
    \label{fig:t2k_cc1pi_adler}
\end{figure*}

The differential cross section in muon momentum in the four muon angular bins is shown in \autoref{fig:t2k_cc1pi_mu}.  
The overall variation in predictions among the generators is relatively small, and they are generally good against data. NuWro has the largest $\chi^2$, coming predominantly from the two highest momentum bins in the $0.80 < \cos\theta_\mu<0.90$ region. NEUT has the lowest $\chi^2$, and a notably different prediction in the 0.4-1.2 GeV bin; however, no particular region drives the lower $\chi^2$. Little variations is seen between the various GENIE models, which use different form factors and have a different FSI treatment.
The six single differential measurements shown in \autoref{fig:t2k_cc1pi_other}, \autoref{fig:t2k_cc1pi_pi}, and \autoref{fig:t2k_cc1pi_adler} provide more detailed information, and exhibit larger differences between the generators.

For the $Q^{2}_{\mathrm{rec}}$ distribution in \autoref{fig:t2k_cc1pi_other}, we note that all generators exhibit a turn-over behaviour at low $Q^{2}_{\mathrm{rec}}$, also observed in the data. The publication~\cite{t2k_cc1pi_fgd1} used older versions of GENIE (2.8.4) and NEUT (5.1.4.2), and is saw an over-prediction at low $Q^{2}_{\mathrm{rec}}$, which does not appear with our recent generator versions. NEUT shows the best overall agreement, although it consistently over-predicts at $Q^{2}_{\mathrm{rec}}<0.5\text{ GeV}^2$, possible due to the strong bin-by-bin correlations in the data. Looking only at the shape of the distribution, GENIE under-predicts the low and over-predicts the high $Q^2_\mathrm{rec}$ region, whereas NEUT and NuWro do better at low $Q^2_\mathrm{rec}$, but also over-predict the high $Q^2_\mathrm{rec}$ region although to a lesser degree. NuWro and the various GENIE versions receive largest $\Delta\chi^2\sim4$ contributions from the bins around $Q^2_\textrm{rec}=1-1.5~\text{GeV/c}^2$ and all generators do poorly in the last bin, with $\Delta\chi^2\sim$2--3.

The muon-pion opening angle distribution in \autoref{fig:t2k_cc1pi_other} is sensitive to a variety of effects related to the initial state and the resonance decay. The generators all have good $\chi^2/N_\text{bins}$, even though the predictions are notably different, due to 
the weak statistical power of the measurement. Generally, NEUT has the largest prediction, and \Genietena, \Genietenb and NuWro all predict similar distributions, with \Genietwoa differing to the other GENIE versions mostly by normalisation.

\autoref{fig:t2k_cc1pi_pi} shows the pion momentum distribution, and all generators tend to be higher than the data for $p_\pi<0.8~\text{GeV/c}$. Although the generators predict the same peak at $0.2-0.3$ GeV/c, NuWro has a more pronounced peak which appears to over-predict the data considerably. However, NuWro also has the best $\chi^2$ by factor 2, coming predominantly from the $0.5<p_\pi<1~\text{GeV/c}$ region, where the other generators consistently over-predict. All GENIE versions have the worst $\chi^2$, coming from the last pion momentum bin, which contributes 9 units of $\chi^2$, compared to less than 1 unit of $\chi^2$ for NEUT and NuWro. NuWro's dominant bin contribution is $1.3-1.4$ GeV/c with $\Delta\chi^2=4$, and NEUT's is $0.6-0.7$ GeV/c with $\Delta\chi^2=5$.

The changes to the GENIE pion FSI model are most pronounced around the pion momentum peak, although this region does not particularly affect the $\chi^2$ against the data. Comparing the shapes, all three GENIE versions produce similar predictions. We theorize that $p_\pi$ is more sensitive to the details of the resonance decay in the nuclear medium and secondarily to pion FSI, so the larger deviations in $p_\pi$ compared to $p_\mu-\cos\theta_\mu$ and $Q^{2}_{\mathrm{rec}}$, likely come from these treatments.

Conversely, the situation is reversed for the distribution in the lab pion angle, $\theta_{\pi}$, shown in \autoref{fig:t2k_cc1pi_pi}. Here GENIE agrees best with the data, with NEUT, NuWro and \Genietenb having almost double the $\chi^2$ of \Genietwoa, indicating that no simple picture is possible. \Genietwoa has the best $\chi^2$ due to capturing the aggressive drop-off in cross-section at higher pion angles. We also note that the pion FSI model in GENIE has barely any effect until $\theta_\pi>0.6\text{ rad}$. All generators predict somewhat larger cross-sections than data at larger pion angles ($\theta_{\pi}>0.8$ rad), which is also the region where we observe largest difference between the generators. At lower angles, NEUT is the outlier, predicting a larger cross-section than the other generators and the data.

The Adler angles are traditionally considered as the angles of the outgoing pion in the rest frame of the decaying resonance, before final state interactions.  As previously discussed, the T2K measurement's ``Adler angles'' instead concern a proxy variable of the ``true'' Adler angles, using only the outgoing muon and pion kinematics to form the variable, avoiding corrections for nuclear effects such as FSI. As such, they do not perfectly map to the traditional Adler angles, but also do not contain the strong model dependence that deriving the true Adler angles would. The true Adler angle variables were measured in bubble chamber experiments~\cite{Radecky:1981fn} and is the source of the event generators' decay models; GENIE, NEUT, and NuWro all use parameterizations of the  $\Delta(1232) \rightarrow \pi N$ system based on neutrino H/D data. The Adler angles may also have some sensitivity to resonance/non-resonance interference~\cite{Sanchez:2015yvw}, which the generators also implement differently: GENIE ignores interference, NuWro only models a single non-interfering $\Delta(1232)$ resonance, and NEUT has 17 interfering resonances and one non-interfering non-resonant $I_{1/2}$-only background.

The Adler angles for the T2K data in \autoref{fig:t2k_cc1pi_adler} are very poorly predicted by all generators, although the predictions differ. In $\phi_\text{Adler}$ NEUT shows the most peaked distribution and has the worst $\chi^2$, whereas the other generators show similar behaviour up to a normalization, with the \Genietwoa prediction having the lowest cross-section and the best $\chi^2$ with $\chi^2/N_{\mathrm{bins}}\sim1.5$. All three GENIE versions have the same decay implementation, and the small differences show the effect of other aspects of the dynamics. The effect of pion FSI between \Genietena and \Genietenb appears mostly flat up to a normalization. 
In the $\cos\theta_\text{Adler}$ distribution, we note a consistent over-estimation in $\cos\theta_\text{Adler}<0.5$ for all generators.
The $\chi^2$ are about 3 per bin for $\cos\theta_\text{Adler}$; the problems are with both shape and magnitude. Comparing the generators, we note significantly different behaviour in the most forward $\cos\theta_{\text{Adler}}$ angles, where NuWro rapidly rises, with almost 50\% larger cross-section the \Genietwoa. NEUT also rises rapidly, but to a lesser extent, possibly a reflection of the similar non-isotropic $\Delta$ decay the two generators have in common. 

\begin{table*}
    \caption{$\chi^2$ contributions for the generators against T2K CC1$\pi^+$ data, over a total of 114 bins.  The $\chi^2$ values for individual distributions are calculated using the provided covariance matrices. The summed $\chi^2$ for each generators does not account for correlations between measurements in different variables, so do not represent a proper global $\chi^2$.}
    \label{tab:t2k_chi2}
\begin{ruledtabular}
    \centering
{\renewcommand{\arraystretch}{1.2}
    \begin{tabular}{l | c | c c c c c }
  
          Measurement & $N_\text{bins}$ & {\Genietwoa} &  {\Genietena} & {\Genietenb} & NuWro & NEUT \\
          \hline
         $p_\mu-\cos\theta_\mu$ & 16 & 16.58 & 18.94 & 19.03 & 24.17 & 11.81 \\
         $p_\pi$ & 17               & 32.71 & 35.20 & 42.65 & 18.09 & 29.85 \\
         $\cos\theta_\pi$ & 13      & 18.78 & 21.66 & 27.56 & 29.83 & 27.17 \\
         $\theta_{\mu,\pi} $ & 16   & 8.45  & 10.82 & 11.33 & 14.85 & 10.16 \\
         $Q^2_{\mathrm{rec}}$ & 16           & 26.50 & 30.44 & 28.01 & 29.67 & 18.99 \\
         $\phi_{\text{Adler}} $ & 16& 25.25 & 28.97 & 33.06 & 29.55 & 33.35 \\
         $\cos\theta_{\text{Adler}} $& 20 & 56.78 & 67.20 & 76.65 & 61.08 & 57.28 \\
         \hline
         All & 114 & 185.05 & 213.23 & 238.29 & 207.24 & 188.61 \\
        
    \end{tabular}}
    \end{ruledtabular}
\end{table*}

A summary of the $\chi^2$ contributions for all variables is provided in \autoref{tab:t2k_chi2}. Overall all the generators appear to be able to describe the data reasonably well in muon kinematics and $\theta_{\pi,\mu}$. The $Q^2_{rec}$ distribution is moderately well described, with all generators having $p_{\chi^2}>0.01$ in common. NuWro is notably best at describing the pion momentum distribution, {\Genietwoa} at describing the pion and pion-muon angular distributions, and NEUT at describing the muon kinematic distributions and $Q^{2}_{\mathrm{rec}}$.
A general trend is that the generators all agree better with data in muon kinematic variables than pion kinematic variables. The worst agreement is seen for the variables derived from a number of measured quantities, $Q^{2}_{\mathrm{rec}}$ and the Adler angles.
Amongst the GENIE flavors,
\Genietwoa performs best in every distribution. The impact of pion FSI in \Genietena vs \Genietenb has very little effect on the predictions for $p_\mu-\cos\theta_\mu$, $Q^{2}_{\mathrm{rec}}$ and $\theta_{\mu,\pi}$. For the pion kinematic variables and Adler angles, \Genietena is consistently preferred, although the overall agreement with data in those variables is still not good.

\subsection{\minerva results}
\minerva has published pion production measurements on CH with varying final states---\numu CC$1\pi^+$~\cite{Eberly:2014mra}, $\bar{\nu}_\mu \text{CC}1\pi^0$~\cite{McGivern:2016bwh}, $\nu_\mu \text{CC}1\pi^0$~\cite{Altinok:2017xua}, and $\bar{\nu}_\mu \text{CC}1\pi^-$~\cite{Le:2019jfy}. For this paper, we have selected $\nu_\mu$ CC$1\pi^\pm$ data to allow optimal comparison with T2K~\cite{t2k_cc1pi_fgd1} and MiniBooNE~\cite{Wilking}. The analysis has been updated since Ref.~\cite{Eberly:2014mra} to use the improved signal definitions of Ref.~\cite{McGivern:2016bwh}, and is available as a public data release~\cite{minerva_datarelease}. The updated results for this measurement include an improved flux estimate~\cite{Aliaga:2016oaz} and a modified signal definition. As with the CC-mesonless measurement discussed in \autoref{subsec:minervacc0pi}, the measurement uses a muon neutrino beam with mean energy of 3.5 GeV, and it selects events in the \minerva central tracker, composed mainly of CH.

\subsubsection{Signal definition}
\label{sec:minerva_pion_signaldef}

The analysis focuses on $\pi^\pm$ production in the $\Delta$ (P$_{33}$(1232)) resonance region which is the most prominent baryon resonance for neutrino interactions.  It is the successor to Ref.~\cite{Eberly:2014mra} with some significant improvements that were developed for Ref.~\cite{McGivern:2016bwh} including a new flux calculation~\cite{Aliaga:2016oaz} and an updated signal definition. It has no signal definition restriction corresponding to the acceptance limitations for pion kinetic energy, pion angle, muon momentum, or muon angle\footnote{The original analysis~\cite{Eberly:2014mra} did include a version with a muon angle constraint in its appendix, but this constraint was removed for the updated analysis, to better align with \minerva's later pion measurements, specifically Ref.~\cite{McGivern:2016bwh}.}.

There are kinematic limits on the signal in terms of quantities that are not directly observable in the detector.  The true neutrino energy is required to be between 1.5~GeV and 10~GeV.  Additionally, a key kinematic constraint in the analysis is on the invariant hadronic mass, $W$.  The purpose is to suppress higher mass resonances and allow comparison with measurements and calculations that emphasize $\Delta$(1232) resonance.
In the original version of the analysis~\cite{Eberly:2014mra}, a signal constraint was placed on the true, generator invariant hadronic mass, requiring $W < 1.4$ GeV.
In the new version (method in Ref.~\cite{McGivern:2016bwh} and data in  Ref.~\cite{minerva_datarelease}), the signal constraint was placed on the event's true $W_{\mathrm{exp}}$ calculated with true muon kinematics and true $E_{\nu}$, as defined in \autoref{eq:w}.

\begin{eqnarray}
\nu &=& E_{\nu} - E_{\mu}
\label{eq:nu}\\
Q^2 &=& 2E_\nu\left(E_\mu - \left|\vec{p}_\mu\right|\cos\theta_{\mu\nu}\right) - m_\mu^2
\label{eq:q2}\\
W^2_{\mathrm{exp}} &=& m_p^2 - Q^2 + 2m_p\nu
\label{eq:w}
\end{eqnarray}

The signal definition requires a single charged pion ($\pi^\pm$), though the Michel requirement selects $>99\%$ $\pi^+$. The CC$1\pi^-$ contribution to the CC$1\pi^\pm$ cross-section is calculated to be less than 2\% at \minerva energies. Any additional baryons and mesons (including $\pi^0$) are allowed.

\subsubsection{Selection}
This analysis uses the selection of Ref.~\cite{McGivern:2016bwh} as applied to the signal definition defined above. It selects events with a muon track and charged pion track that share a common vertex. The muon track must exit the back of \minerva and enter MINOS, effectively restricting its momentum to about $p_{\mu} > 1.5$ GeV/c and angle to $\theta_{\mu} < 20\degree$. All $\pi^+$ in this analysis are identified by a Michel electron in \minerva, and are required to produce a track with an energy deposition signature consistent with a charged pion. This imposes limitations on the kinetic energy of the pion, $T_\pi$: there is a $T_\pi > 35\text{ MeV}$ tracking threshold in \minerva, and pions with $T_\pi > 350\text{ MeV}$ typically exit the back of \minerva, and therefore cannot be fully reconstructed.  Due to \minerva's planar design, pion tracking efficiency drops to zero between $80\degree < \theta < 110\degree$. Since the $p_\mu,\theta_\mu,T_\pi,\theta_\pi$ restrictions exist in the selection but not in the signal definition, the model is relied upon in these regions at the efficiency correction stage of the cross-section calculation which necessarily implies model dependence. 

The reconstructed energy transfer, $\nu^{\mathrm{reco}}$, referred to as $E_{\mathrm{recoil}}$ and $E_{\mathrm{hadronic}}$ in the publications, is estimated by summing the visible energy not associated with the muon, and applying corrections determined from simulation that correct for energy missing in passive material, energy escaping the detector, energy below threshold, and energy lost to the nucleus (i.e. binding energy). 
The reconstructed neutrino energy, $E_{\nu}^{\mathrm{reco}}$, is calculated as $E_{\mu}^{\mathrm{reco}} + \nu^{\mathrm{reco}}$.
The reconstructed $Q^2$ and $W_{\mathrm{exp}}$ are calculated as in \autoref{eq:q2} and \autoref{eq:w} respectively, but using reconstructed muon kinematics and the reconstructed energy transfer.

\subsubsection{Cross-section extraction}
The signal selection efficiency is approximately 8\%, with the largest losses coming from the Michel selection. The high-$W_{\mathrm{exp}}$ background is constrained to side-bands as in Ref.~\cite{Eberly:2014mra} with the new signal definition.  The cross section is extracted using D'Agostini unfolding~\cite{DAgostini:1994zf} with four iterations. The number of iterations was determined by examining MC simulations and the value chosen had the best balance between correctly unfolding and problems with statistics.
The second major improvement in the updated result is an improved neutrino flux~\cite{Aliaga:2016oaz} which resulted in a increase in cross section which was flat across pion kinetic energy.   The redefined signal definition resulted in a decrease in cross section largely independent of pion energy.

\subsubsection{Results}
\begin{figure*}[ht!]
    \centering
    \includegraphics[width=0.48\linewidth]{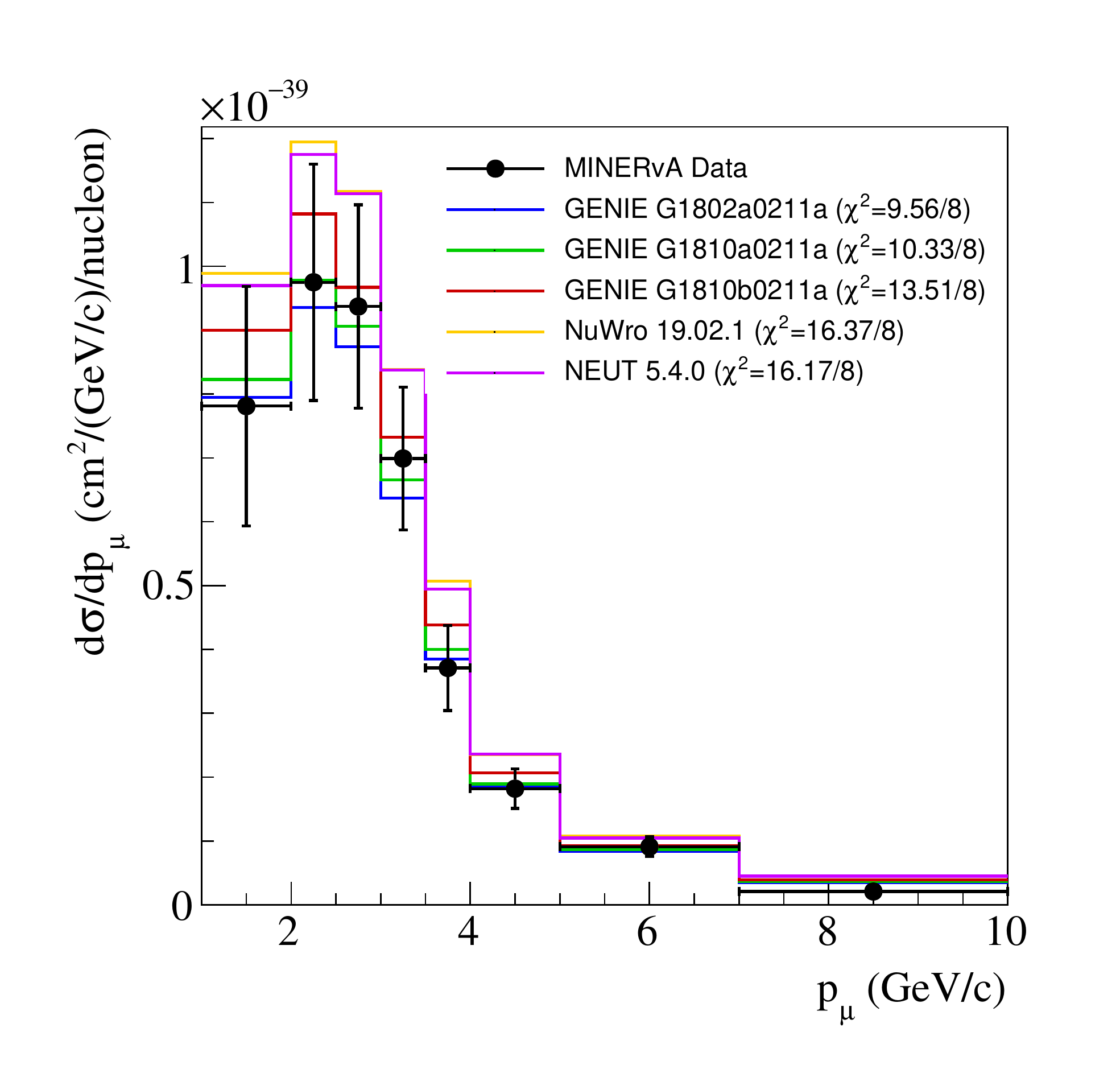}
    \includegraphics[width=0.48\linewidth]{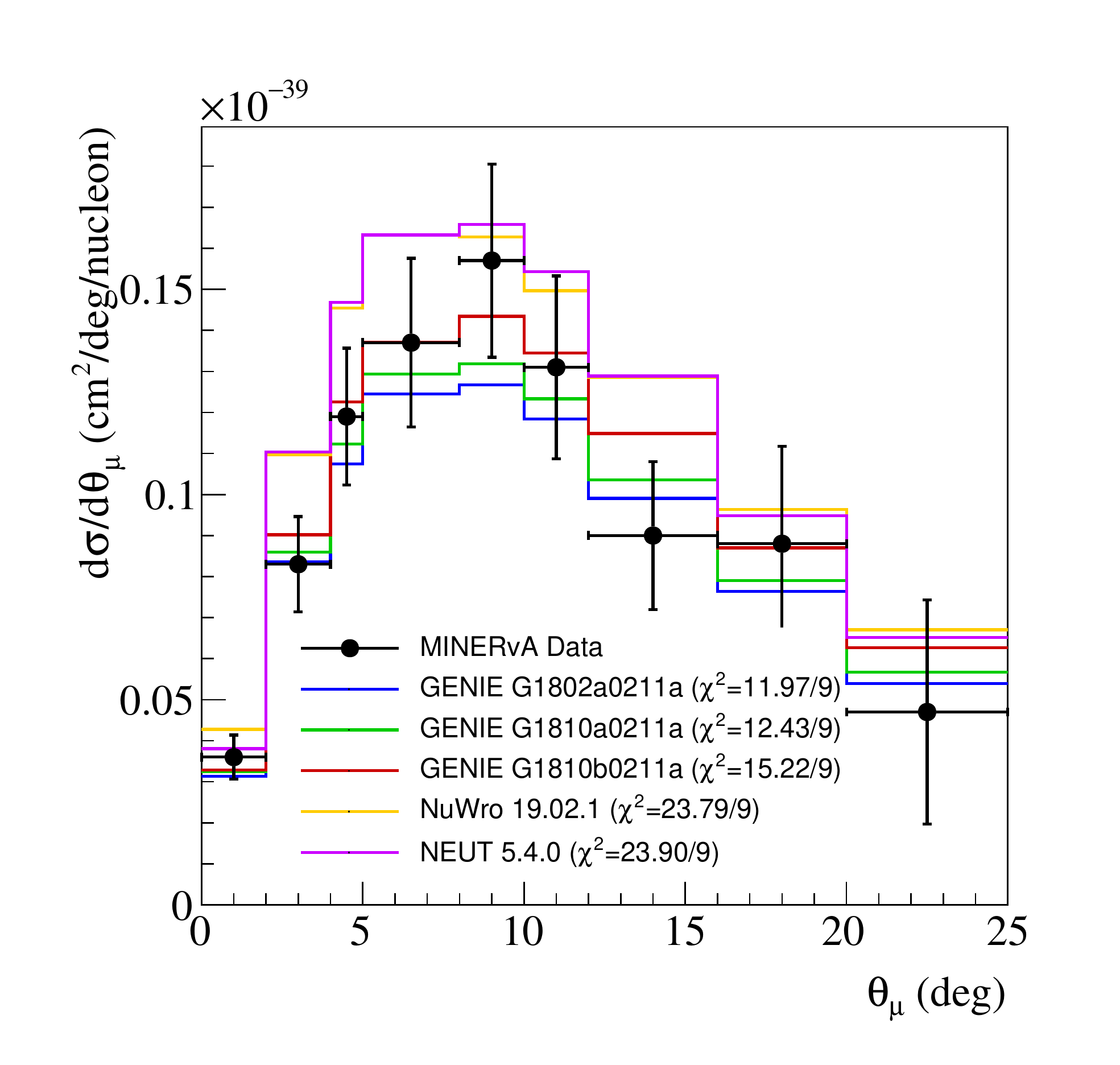}
    \includegraphics[width=0.48\linewidth]{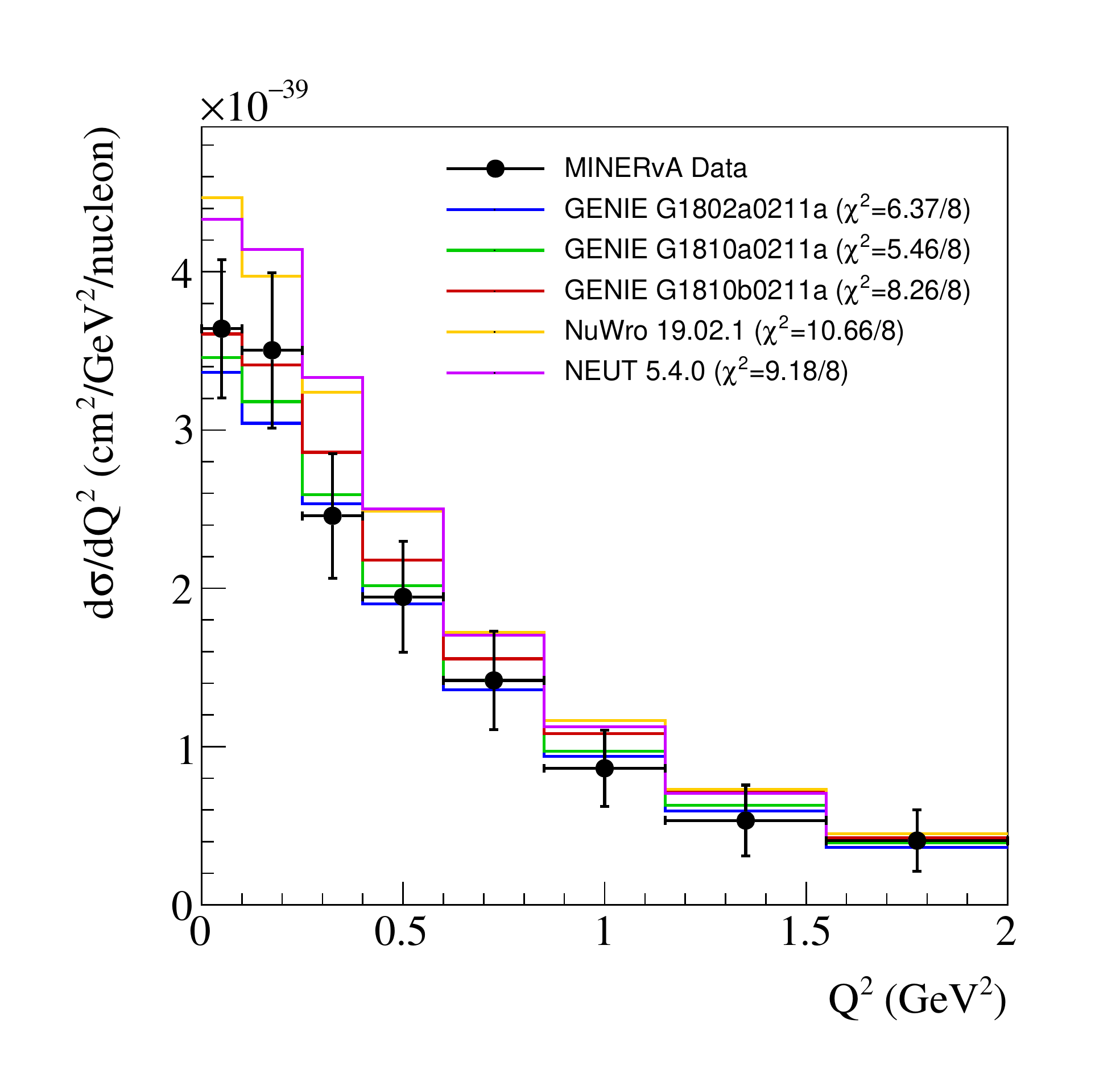}
    \includegraphics[width=0.48\linewidth]{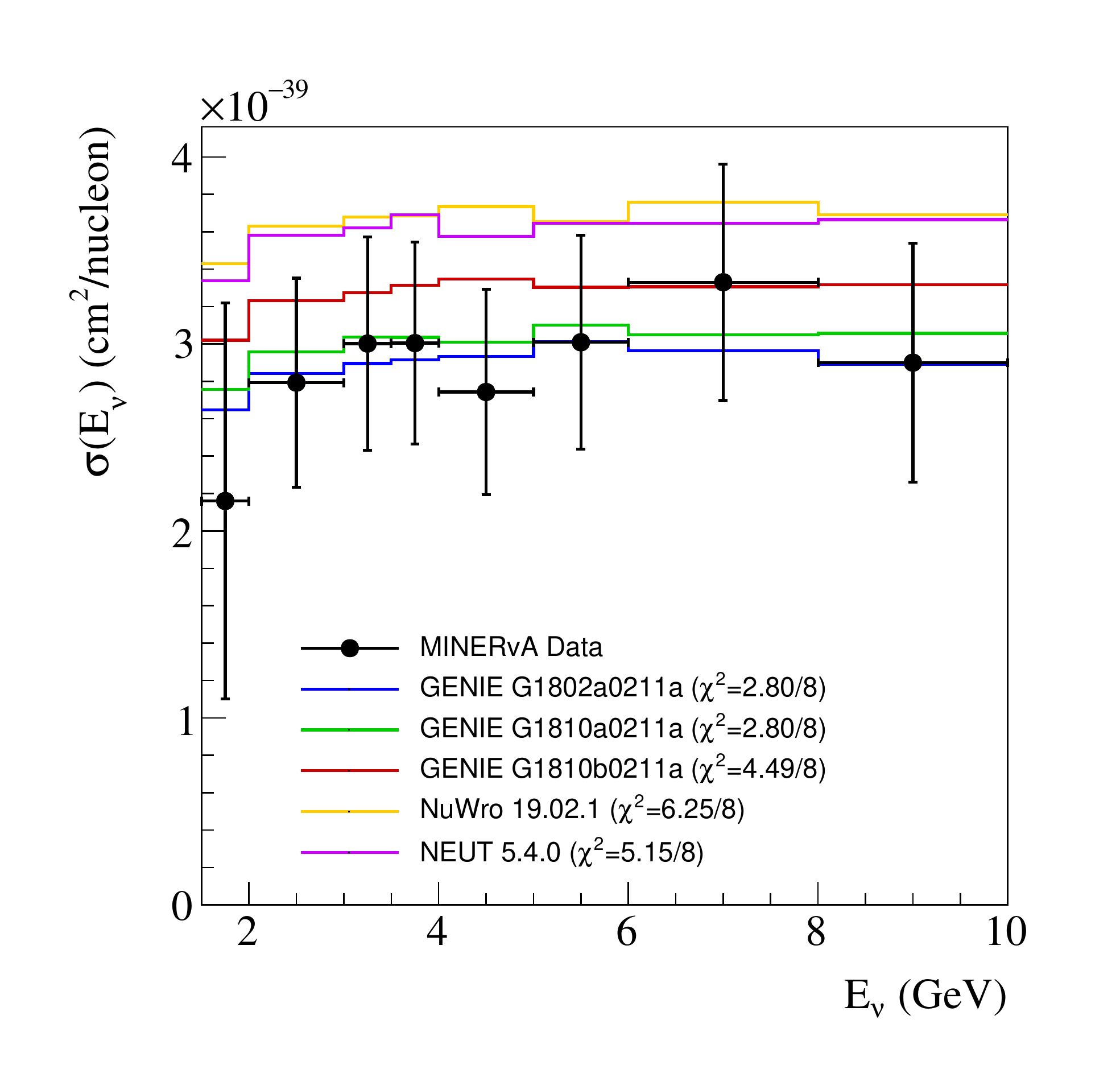}
    \caption{\minerva CC1$\pi^+$ cross section measurements and model comparisons in muon variables (top) and $Q^2$ and $E_{\nu}$ (bottom).}
    \label{fig:minerva_cc1pi_muon}
\end{figure*}

\begin{figure*}[ht!]
    \centering
    \includegraphics[width=0.48\linewidth]{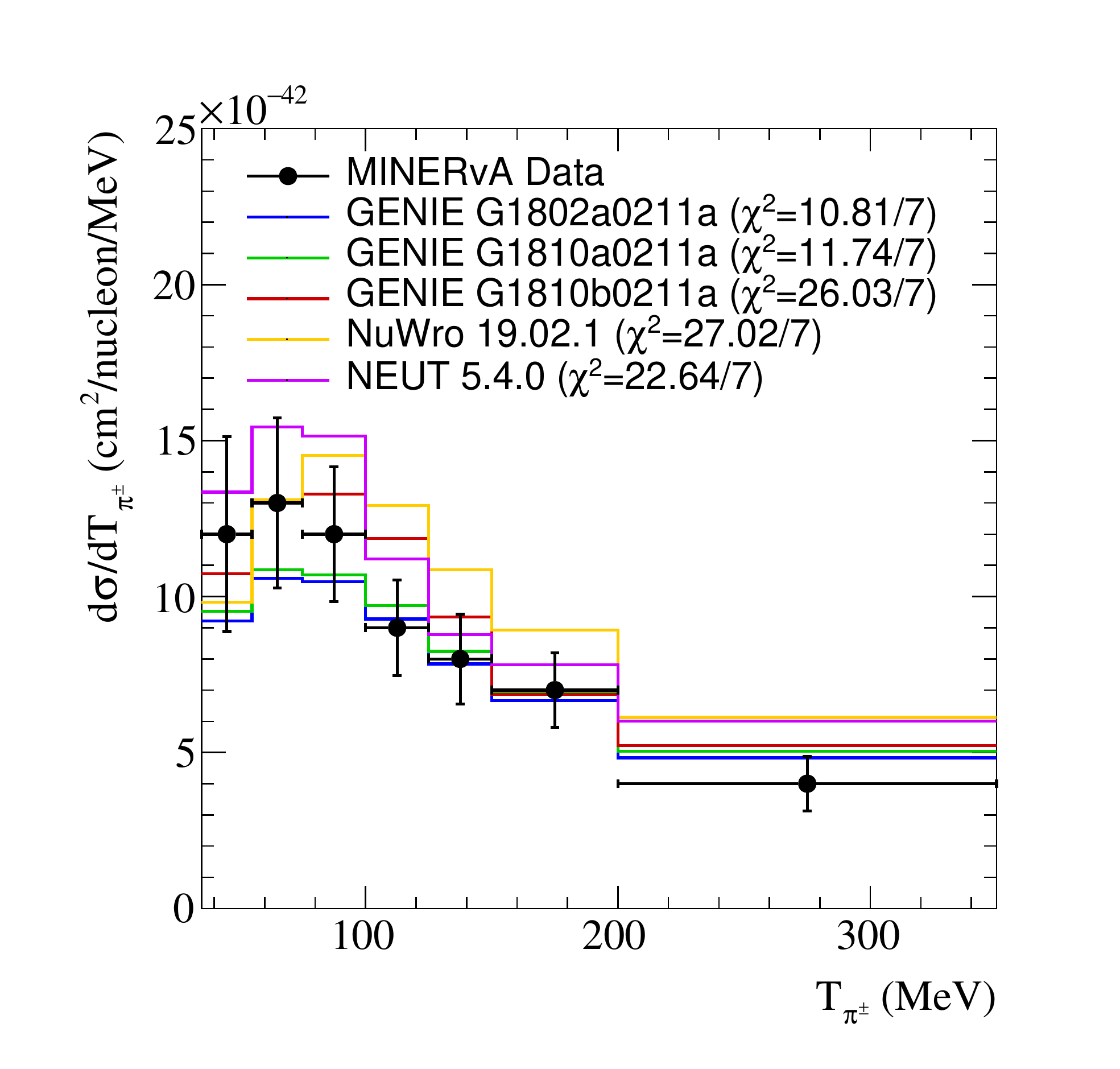}
    \includegraphics[width=0.48\linewidth]{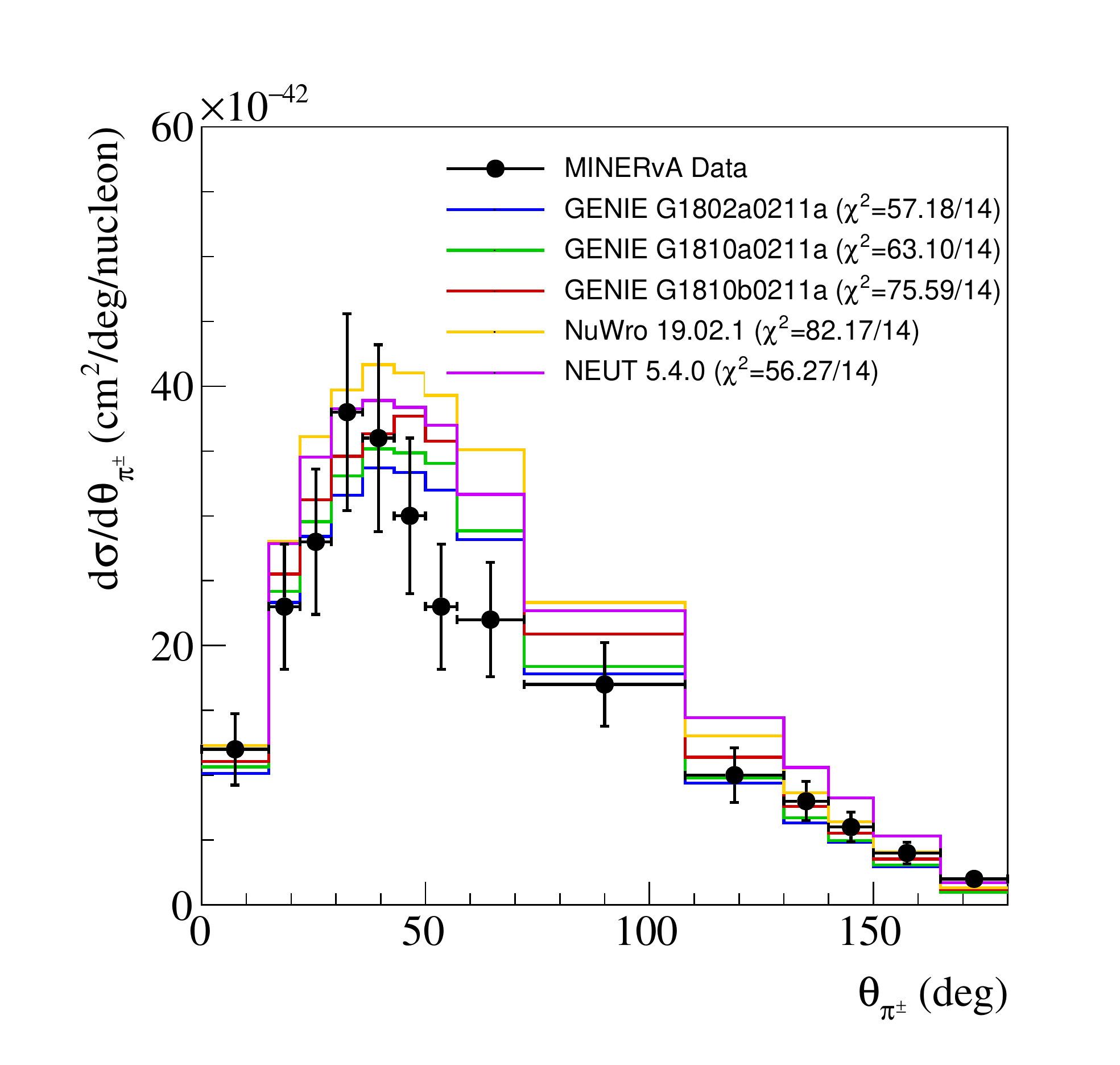}
    \caption{\minerva CC1$\pi^+$ cross section measurements and model comparisons in pion kinetic energy and angle.}
    \label{fig:minerva_cc1pi_pi}
\end{figure*}

One-dimensional differential cross sections were measured in muon and pion kinematic variables, four-momentum transfer squared, $Q^2$, and neutrino energy, $E_\nu$.   
Since all quantities were unfolded from the reconstructed quantities to the true quantities, theoretical calculations can be compared directly with them (acknowledging that this unfolding is model-dependent as it corrects for nuclear effects that can impact the relationship between $E_{\nu}^{\mathrm{true}}$ and $E_{\nu}^{\mathrm{reco}}$).
Results are shown in \autoref{fig:minerva_cc1pi_muon} and \autoref{fig:minerva_cc1pi_pi} with comparisons to the same calculations as for the previous T2K section. In the previous TENSIONS2016 paper~\cite{tensions2016}, the generators were 20-30\% above the MINERvA data~\cite{Eberly:2014mra}.  Since then, the data has been reanalyzed (\autoref{sec:minerva_pion_signaldef}) and the GENIE calculations have been improved with better fits to the $\nu_\mu$--D$_{2}$ data~\cite{GENIE_pitune,Wilkinson:2016wmz} and GENIE, NuWro and NEUT have also made improvements to their FSI models.

The lepton variables $p_\mu$ and $\theta_\mu$ can be used to gauge the overall cross-section magnitude. By-eye we see NEUT and NuWro have very similar predictions, and consistently over-predict the data below 5 GeV, and GENIE predicts the data better. However, when accounting for the strong bin-by-bin correlations, we see very large $\chi^2$ contributions from a single bin: the $7 < p_\mu < 10~\text{ GeV/c}$ bin, contributing 9 units of $\chi^2$ for NEUT, \Genietenb and NuWro, and 6 units for the other GENIE versions. Once this bin is excluded, NEUT and NuWro go from $\chi^2/N_\text{bins}\sim2$ to a very agreeable $\chi^2/N_\text{bins}\sim1$, and we see no definitively superior prediction. When comparing the shapes of the generators in $p_\mu$, they are remarkably similar, and the total predictions differ almost exclusively in normalisation. For the $\theta_\mu$ distribution, the predominant $\chi^2$ contribution is the $12^{\circ}<\theta_\mu<16^{\circ}$ bin, which contributes $\chi^2\sim10$ for NEUT and NuWro, and 7 units for GENIE. Again, if this bin is excluded all generators perform well in regards to the $\chi^2/N_\text{bins}$.

$Q^2$ can probe nuclear structure, and $E_\nu$ separates the interaction modes relatively well. However, the $Q^2$ and $E_\nu$ distributions have significantly more model dependence than the purely kinematic distributions observed in the detector. This is because they require corrections for initial and final state physics, coming from the input generator model. Comparing the predictions in \autoref{fig:minerva_cc1pi_muon}, the generators perform adequately in regards to the $\chi^2$, although we see relatively large differences between the predictions; at low $Q^2$ {\Genietwoa} predicts a 30\% smaller cross section than NuWro. NuWro and NEUT perform very similarly, likely due to similar form factors and tuning to data. They are somewhat higher than the data with comparable $\chi^2$, coming from the limited power of the data.
The CC$1\pi^\pm$ cross-section does not indicate a missing low-$Q^2$ drop-off and calculations appear to follow the data well, in common with CC$1\pi^-$~\cite{Le:2019jfy} production, but in contrast to \minerva's measurements of CC$1\pi^0$~\cite{Altinok:2017xua}. This is discussed more in comparison to T2K data in \autoref{sec:t2k_minerva_eff_comps}.

The pion variables are shown in \autoref{fig:minerva_cc1pi_pi}, which are more sensitive to resonance decay processes and FSI effects. We observe very similar behaviour between \Genietena and \Genietwoa, with a significantly lower cross-section than \Genietenb NEUT and NuWro, and with a significantly better $\chi^2$ in $T_\pi$; largely observed in the previous muon-based variables. Although the peak below $T_\pi<100\text{ MeV}$ is under-predicted, the higher region drives the $\chi^2$, with the last bin contributing 6.5 units of $\chi^2$ for \Genietwoa, and 16 units for NEUT. NuWro however receives its largest $\chi^2$ penalty in the 55--76 MeV and 100--125 MeV bins, with 6.5 and 7.5 units respectively.
Comparing \Genietena and \Genietenb, which differ only in FSI model, the empirical model produces an acceptable $\chi^2$ for the kinetic energy distribution, whereas the cascade based \Genietenb does not, and shares more features with NEUT and NuWro in prediction. 
Interestingly, the very similar predictions of NEUT and NuWro in the lepton variables in \autoref{fig:minerva_cc1pi_muon} are different in the pion variables; NEUT over-predicts almost the entire range but does well in the shape, whereas NuWro peaks at a higher $T_\pi$ than data, under-predicts the lowest bin, and over-predicts the remaining distribution at higher $T_\pi$, with a different shape to the other generators. 

In $\theta_\pi$ all generators produce an unsatisfactory $\chi^2$, with more than 4 units per bin on average. Similar to the T2K measurement, the rising shape of the peak is well-modeled, but the drop-off at higher angles is over-predicted. There is a large impact from the GENIE FSI model choice, and we see differences between NEUT and NuWro not observed in the muon kinematics. We note that the high-angle region is where the model dependence of the data may be the strongest, and note in the region of $\theta_\pi\sim100\degree$ that the data is almost perfectly reproduced by \Genietwoa and \Genietena predictions. 

\begin{table*}
    \caption{$\chi^2$ contributions for the generators against MINERvA CC1$\pi^+$ data, over a total of 54 bins.  The $\chi^2$ values for individual distributions are calculated using the provided covariance matrices. The summed $\chi^2$ for each generators does not account for correlations between measurements in different variables, so do not represent a proper global $\chi^2$.}
    \label{tab:minerva_1pi_chi2}
\begin{ruledtabular}
    \centering
    {\renewcommand{\arraystretch}{1.2}
    \begin{tabular}{l | c | c c c c c }
  
          Measurement & $N_\text{bins}$ & {\Genietwoa} & {\Genietena} & {\Genietenb} & NuWro & NEUT \\
          \hline
         $p_\mu$        & 8     & 9.56 & 10.33 & 13.51 & 16.37 & 16.17 \\
         $\theta_\mu$   & 9     & 11.97 & 12.43 & 15.22 & 23.79 & 23.90 \\
         $T_\pi$        & 7     & 10.81 & 11.74 & 26.03 & 27.02 & 22.64 \\
         $\theta_\pi$   & 14    & 57.18 & 63.10 & 75.59 & 82.17 & 56.27 \\
         $Q^2$          & 8     & 6.37 & 5.46 & 8.26 & 10.66 & 9.18 \\
         $E_\nu$        & 8     & 2.80 & 2.80 & 4.49 & 6.25 & 5.15 \\
        \hline
         All            & 54    & 98.69 & 105.86 & 143.10 & 166.26 & 133.31 \\
        
    \end{tabular}}
    \end{ruledtabular}
\end{table*}

It is clear from the $\chi^2$ table in \autoref{tab:minerva_1pi_chi2}, that {\Genietwoa} generally exhibits the best agreement with MINERvA data, with the exception of the $Q^2$ and $\theta_\pi$ distributions. We note \Genietwoa is most similar to the generator used in the analysis (GENIE 2.8.6). Comparing \Genietena and \Genietenb---which differ only by their pion FSI model---the empirical model used in the former consistently performs better, often similar to \Genietwoa.

NuWro generally has the worst $\chi^2$, and is very similar to NEUT in muon variables; similarities which disappear when pion variables are concerned. We note that the $Q^2$ and $E_\nu$ cross-section results holds very little statistical power, which is likely a result of large model uncertainties that come from correcting and unfolding to these variables in data. The $\theta_\pi$ predictions presents the largest $\chi^2/N_{\mathrm{bins}}$ with no generator performing well. This has also been seen in previous studies of this data~\cite{Stowell:2019zsh}.

\subsection{T2K and \minerva pion production phase-space comparisons}
\label{sec:t2k_minerva_eff_comps}

\subsubsection{Cross-section comparisons}
Direct comparison of the T2K and \minerva cross-section data is worthwhile because the targets are the same and the signal definitions are similar.  However, it is challenging given the differences in the neutrino fluxes: \minerva has a peak energy of 3.5 GeV, whereas the peak T2K energy is at 600 MeV, shown in \autoref{fig:expfluxes}. T2K's ND280 detector is at an off-axis angle of $2.5\degree$ which suppresses the high-energy tail as compared with the on-axis \minerva detector. In addition, the signal definitions and cuts (e.g. pion kinetic energy) tend to make overlap less common.
\minerva's cut on $W_{\mathrm{exp}}<1.4\text{ GeV}$ ensures that the cross section is dominated by the $\Delta(1232)$ as in T2K. However, the strength of the resonant and non-resonant contributions still depend on $E_\nu$, $W$ and $Q^2$, so will see different strengths at the two experiments. 

The small overlap region where a direct comparison is possible between \minerva and T2K is for events with $1<p_\mu <2~\text{GeV/c}$. \minerva's muon measurements are limited to a range in muon angle between roughly $0.9<\cos\theta_\mu<1.0$ where T2K also has coverage, and both  cover a nearly full angular range for pions. Thus the range where the two data sets can be compared is limited to the first bin of $p_\mu$ for \minerva (\autoref{fig:minerva_cc1pi_muon}), and the two highest momentum bins for T2K (lower right panel of \autoref{fig:t2k_cc1pi_mu}). In this limited region, the two results are largely consistent with each other, and the generator--data comparisons are consistent for the two, which is reassuring. 

\begin{figure*}[ht!]
    \centering
    \includegraphics[width=0.32\linewidth,page=1,trim=0cm 0.5cm 2cm 0cm,clip]{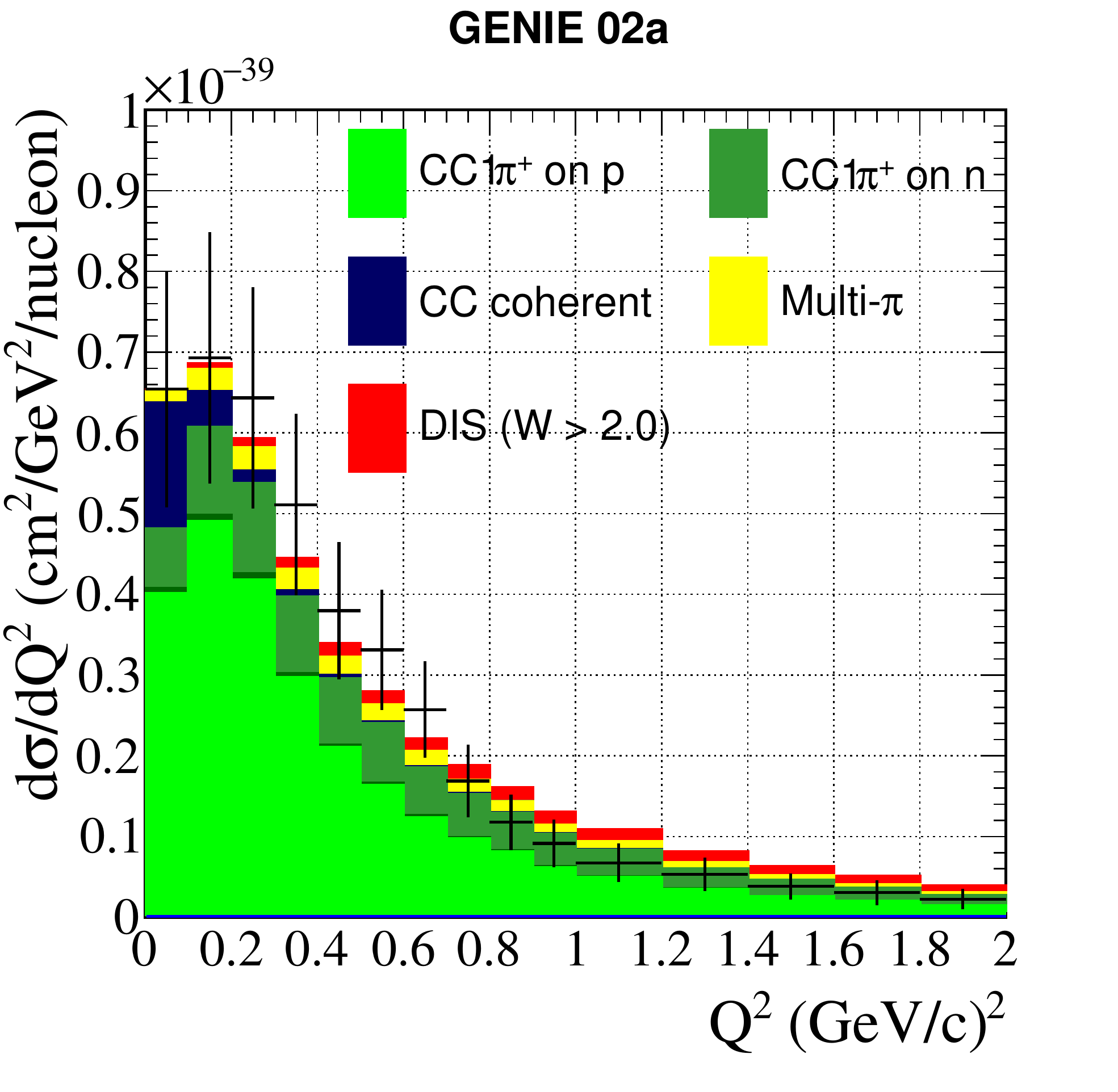}
    \includegraphics[width=0.32\linewidth,page=4,trim=0cm 0.5cm 2cm 0cm,clip]{T2K_CC1pip_CH_XSec_1DQ2_nu_modes}
    \includegraphics[width=0.32\linewidth,page=5,trim=0cm 0.5cm 2cm 0cm,clip]{T2K_CC1pip_CH_XSec_1DQ2_nu_modes}
    \caption{T2K CC1$\pi^+$ cross section in $Q^2_{\mathrm{rec}}$ for {\Genietwoa}, NuWro and NEUT, broken down by true interaction mode.}
    \label{fig:t2k_cc1pi_q2_mode}

    \includegraphics[width=0.32\linewidth,page=1,trim=0cm 1cm 1.8cm 0cm,clip]{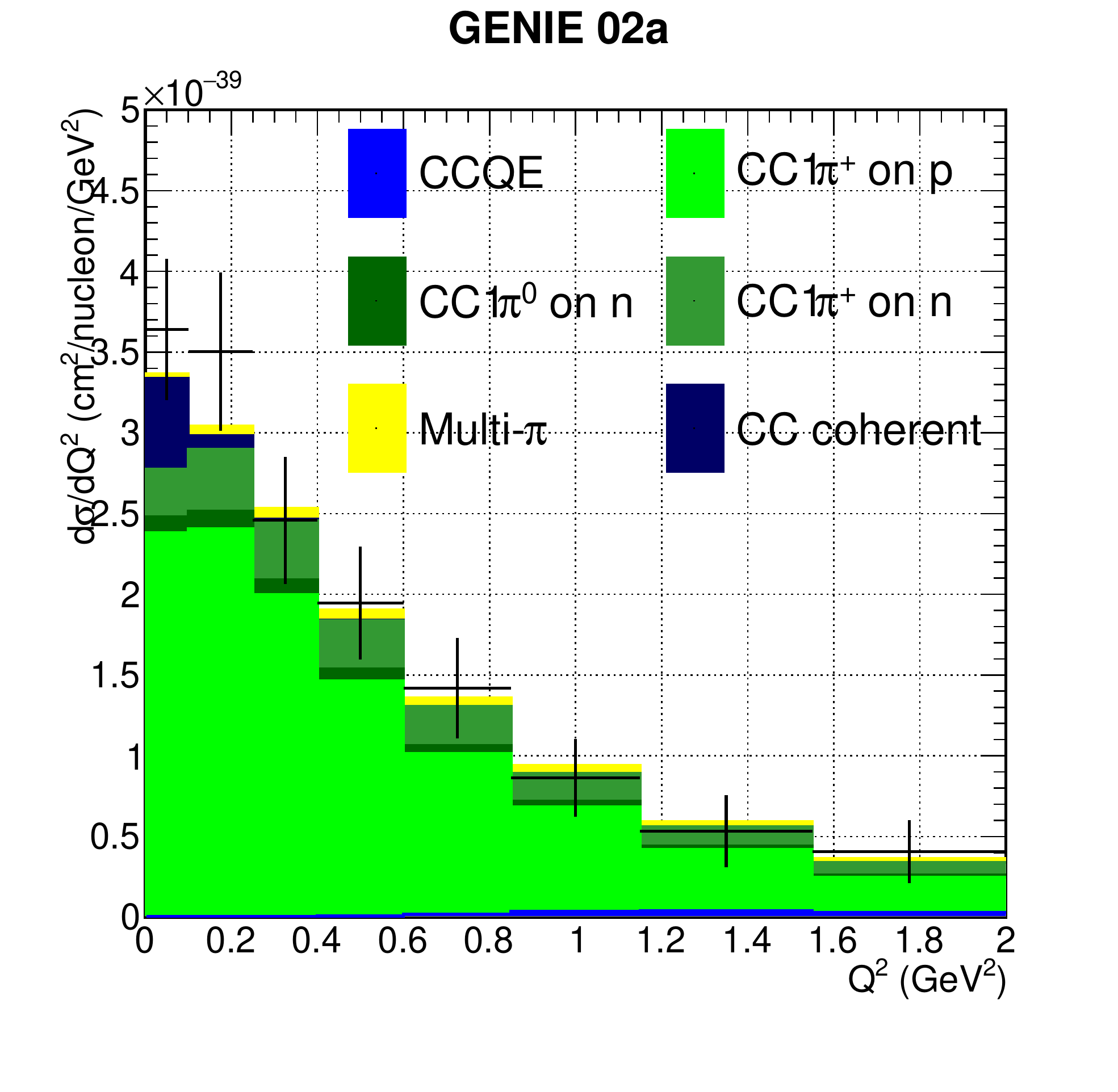}
    \includegraphics[width=0.32\linewidth,page=4,trim=0cm 1cm 1.8cm 0cm,clip]{MINERvA_CC1pip_XSec_1DQ2_nu_2017_modes}
    \includegraphics[width=0.32\linewidth,page=5,trim=0cm 1cm 1.8cm 0cm,clip]{MINERvA_CC1pip_XSec_1DQ2_nu_2017_modes}
    \caption{\minerva CC1$\pi^\pm$ cross section in $Q^2$ for {\Genietwoa}, NuWro and NEUT, broken down by true interaction mode. }
    \label{fig:min_cc1pi_q2_mode}
\end{figure*}

In Figs.~\autoref{fig:t2k_cc1pi_q2_mode} and \autoref{fig:min_cc1pi_q2_mode}, we show the breakdown of the predictions for $Q^2_{\mathrm{rec}}$ and $Q^2$ according to true interaction mode for the T2K and \minerva measurements, respectively. Separate panels are provided for {\Genietwoa}, NuWro and NEUT.
As expected from single nucleon cross sections, the composition is dominated by CC$1\pi^+$ on a struck proton target via resonances, with smaller contributions from CC$1\pi^+$ on a struck neutron target via resonances, and barely visible contribution from CC$1\pi^0$ via resonances, where the $\pi^0$ produces a $\pi^+$ in the final state via FSI. All generator codes include both non-resonant pion production and DIS processes. Although each code has different definitions of each process (see \autoref{sec:generator}), the sum is contained in the multi-$\pi$ category. There is no ambiguity for $W>2~\text{GeV}$ where true DIS models are most appropriate. 
We see approximately similar amounts of CC coherent at low $Q^2$ for all generators, which all use the Berger-Sehgal model.
Importantly, the $1\pi^+$ via resonance contributions are suppressed at lower $Q^2$ for all the generators. Interestingly, NuWro and NEUT have similar CC$1\pi^+1p$ contributions, whereas GENIE's prediction is significantly lower, even though the generators are tuned to similar bubble chamber data. This highlights the need for consistent high-statistics neutrino-nucleon data.  The strength of the CC$1\pi^+1n$ channel in the peak region is also different, and varies by a factor of 2 ranging from NuWro (smallest) to GENIE to NEUT.
Coherent and DIS contributions are larger for the T2K signal definition than \minerva, partially due to the $W_{\mathrm{rec}}<1.4$ GeV signal definition used by \minerva which cuts $\sim$50\% of coherent events below $Q^2<0.2\text{ GeV}^2$. The ratio of CC$1\pi^+$ from proton to CC$1\pi^+$ from neutron is very similar at both low and higher energy experiments. 

\begin{figure*}[ht!]
    \centering
    \includegraphics[width=0.32\linewidth,page=1,trim=0cm 0.5cm 1.8cm 0cm,clip]{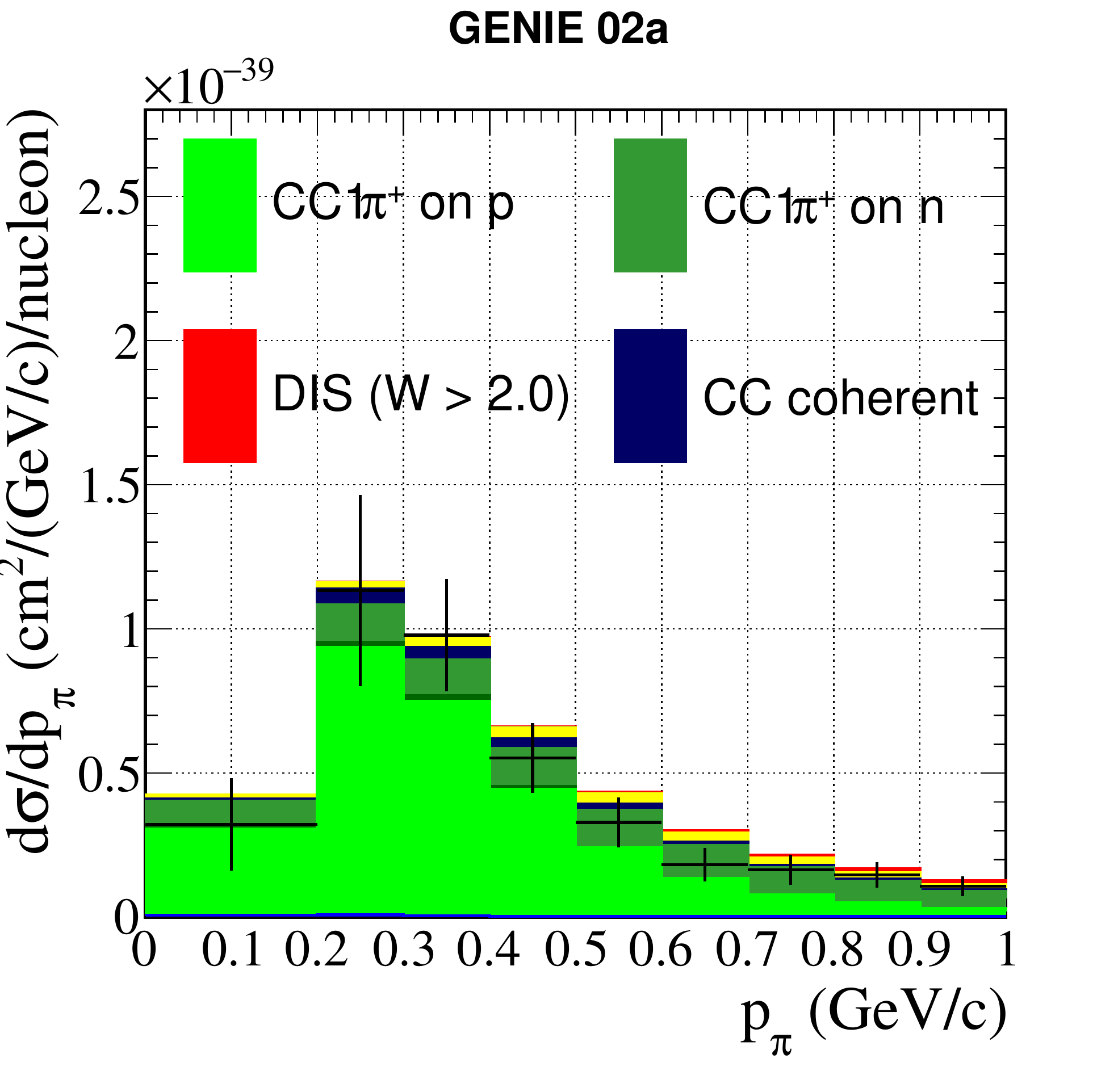}
    \includegraphics[width=0.32\linewidth,page=4,trim=0cm 0.5cm 1.8cm 0cm,clip]{T2K_CC1pip_CH_XSec_1Dppi_nu_modes}
    \includegraphics[width=0.32\linewidth,page=5,trim=0cm 0.5cm 1.8cm 0cm,clip]{T2K_CC1pip_CH_XSec_1Dppi_nu_modes}
    \caption{T2K CC1$\pi^+$ cross section in $p_\pi$ for {\Genietwoa}, NuWro and NEUT, broken down by true interaction mode, shown up to $p_\pi=1\text{ GeV/c}$.}
    \label{fig:t2k_cc1pi_ppi_mode}

    \includegraphics[width=0.32\linewidth,page=1,trim=0cm 1cm 1.5cm 0cm,clip]{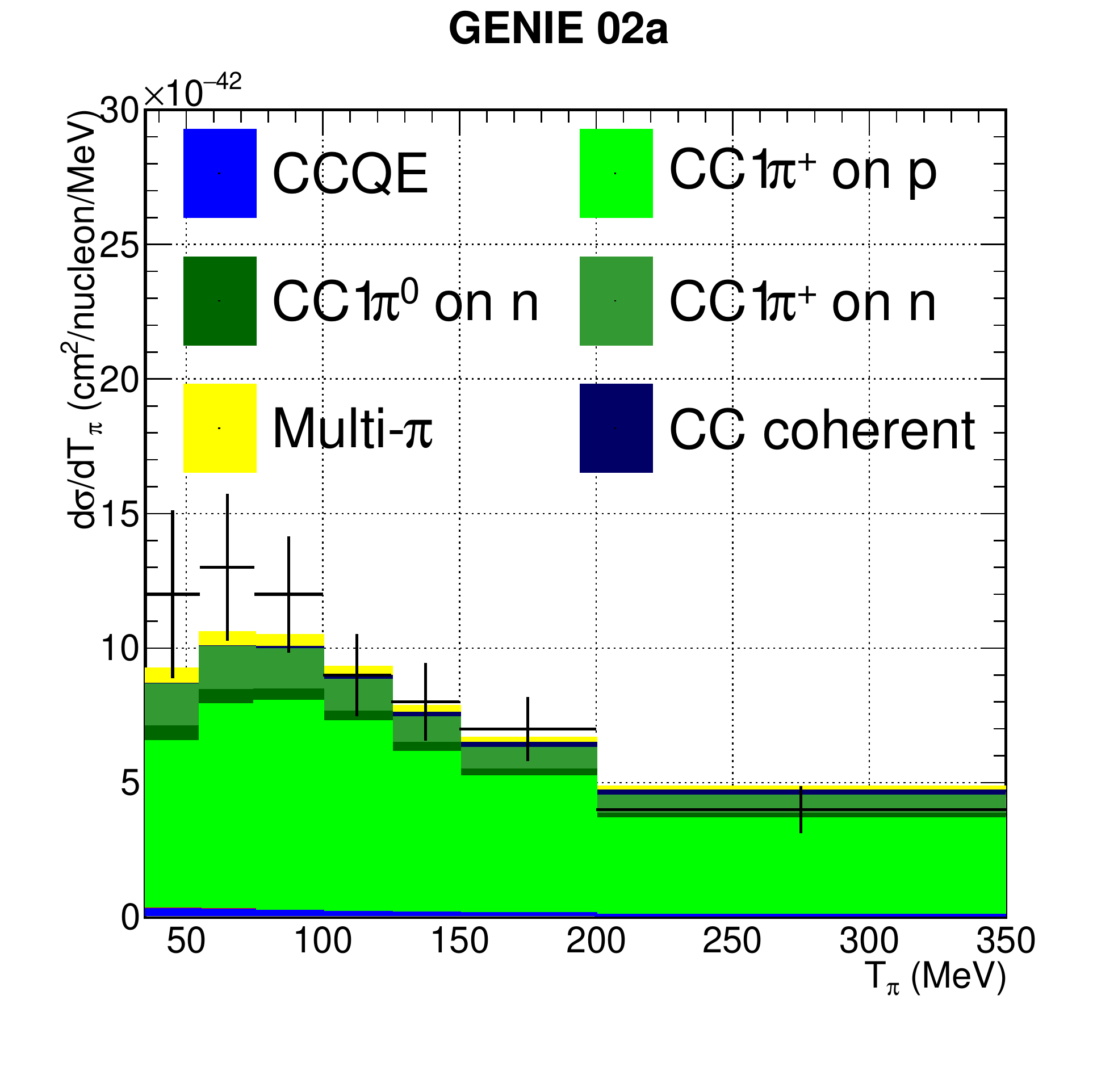}
    \includegraphics[width=0.32\linewidth,page=4,trim=0cm 1cm 1.5cm 0cm,clip]{MINERvA_CC1pip_XSec_1DTpi_nu_2017_modes}
    \includegraphics[width=0.32\linewidth,page=5,trim=0cm 1cm 1.5cm 0cm,clip]{MINERvA_CC1pip_XSec_1DTpi_nu_2017_modes}
    \caption{\minerva CC1$\pi^+$ cross section in $T_\pi$ for {\Genietwoa}, NuWro and NEUT, broken down by true interaction mode.}
    \label{fig:min_cc1pi_tpi_mode}
\end{figure*}

Low-$Q^2$ suppression in single pion production has seen some recent discussion in the community; NOvA introduced a low-$Q^2$ suppression to better match their charged-current resonant enhanced selections when using GENIE v2~\cite{Acero:2020eit}, as did MINERvA and NUISANCE collaborators when tuning GENIE v2 to their CC$1,N\pi^{+,0}$ data~\cite{Stowell:2019zsh}. When updating to GENIE v3, NOvA no longer requires such a suppression~\cite{alex_himmel_2020_4142045}, and neither does T2K when using their updated single pion model in NEUT. Our findings support these choices. All generators now include lepton mass effects in the Berger-Sehgal resonance model, and have updated the $\Delta$ form factors. It seems that these improvements to the nucleon-level interactions have improved the low-$Q^2$ discrepancies which were previously attributed to ``nuclear effects''. Additionally, all generators now also use an updated Berger-Sehgal coherent model, which MINERvA has found to better predict charged-current coherent data~\cite{PhysRevD.97.032014}.  

\subsubsection{Model dependence via signal definition and signal selection differences}
As earlier discussed, \minerva's CC$1\pi^\pm$ measurement specifies two separate kinematic limits in the signal definition: $1.5<E_\nu^\text{true}<10\text{ GeV}$ and $W_{\mathrm{rec}}<1.4\text{ GeV}$. The selection requires a back-exiting muon that is well-measured in MINOS, effectively limiting $p_\mu>1.5\text{ GeV}$ and $\theta_\mu<20\degree$. It also requires the pion to be contained within \minerva and be well-reconstructed, limiting the pion to $50<T_\pi<350\text{ MeV}$ and $\theta_\pi < 80\degree$ or $\theta_\pi>110\degree$. The cross-section is reported for all events with a single pion in the final state, although only single pion events with specific muon and pion kinematics were actually observed. The simulation was used to correct the data for the unmeasured region of phase space, at the cost of introducing a dependence on the simulation used to make the correction.

\begin{figure}[h]
    \centering
    \includegraphics[width=0.97\linewidth,page=2,trim=0cm 0cm 2cm 0.6cm,clip]{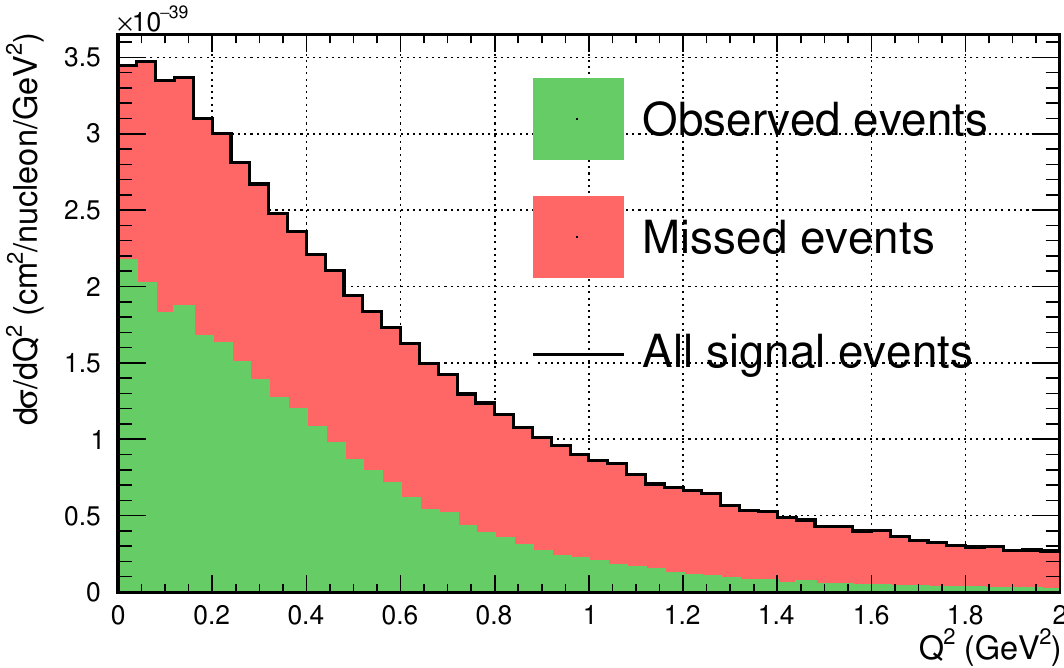}
    \caption{\minerva CC1$\pi^\pm$ cross section in $Q^2$ for {\Genietwoa}, looking at the effect of selection cuts on the generated events.} 
    \label{fig:min_q2_cuts}
\end{figure}

Here we investigate how each of the selection cuts applied to the true particles generated by {\Genietwoa} affects \minerva's CC1$\pi^\pm$ $d\sigma/dQ^2$ cross-section with $W_{exp} < 1.4\text{ GeV}$ and $1.5\text{ GeV}<E_\nu^{\mathrm{true}}<10\text{ GeV}$. Technically, the cuts should be applied on the reconstructed candidates' kinematics present in the full experiment simulation, but this was not available. 
Interestingly, a small coherent sample passes the signal definition in \autoref{fig:min_cc1pi_q2_mode}. However, these events are gone once the selection cuts are applied (see \autoref{sec:pion_discussion}).
When inspecting interaction mode contributions to the cross-section, there are no charged-current coherent events in {\Genietwoa} passing both the muon and pion kinematic cuts. This can be seen in the $T_\pi$ by-mode contribution in \autoref{fig:min_cc1pi_tpi_mode} which shows no CC coherent contribution, whereas the $Q^2$ distribution in \autoref{fig:min_cc1pi_q2_mode} shows a large contribution at low $Q^2$. Hence \minerva's phase space cuts in the signal selection removes the possibility to actually observe coherent events in the detector with these selection cuts, and the low $Q^2$ coherent contribution enters purely through the underlying simulation. 

\begin{figure}[h]
    \centering
    \includegraphics[width=0.97\linewidth,page=1,trim=0cm 0cm 1.8cm 0.6cm,clip]{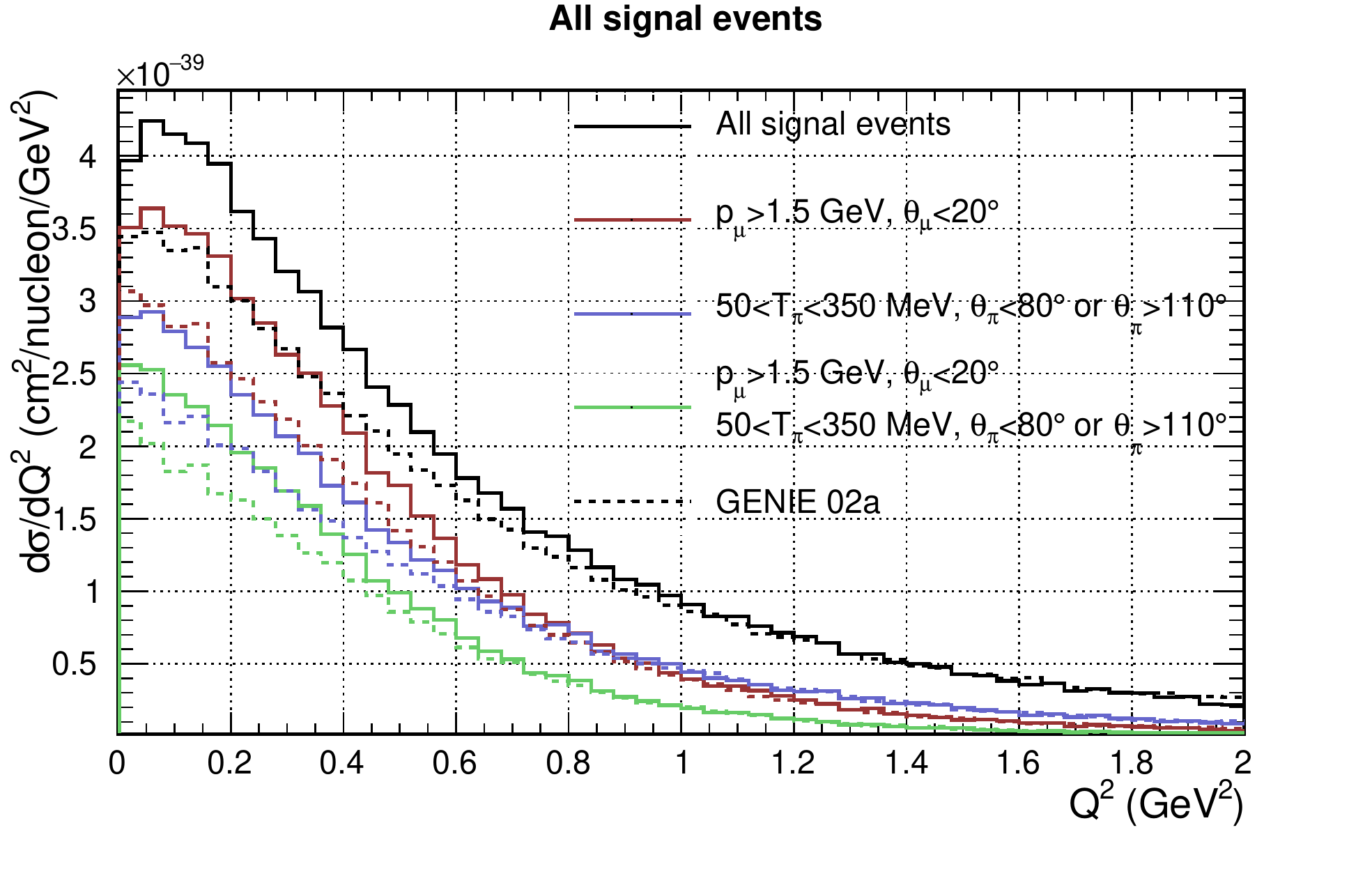}
    \caption{\minerva CC1$\pi^\pm$ cross section in $Q^2$ for NEUT 5.4.0.1 (solid) and {\Genietwoa} (dashed), looking at the effect of selection cuts on the generated events.}
    \label{fig:min_q2_cuts_neut}
\end{figure}

This is investigated in more detail in \autoref{fig:min_q2_cuts_neut}, where the cut sequence and signal definition is applied using GENIE (dashed lines) and NEUT (solid lines). GENIE cuts events at all $Q^2$ but each cut has a different effect on the shape.  With NEUT as the base model instead, the impact of the cuts change which events are observed in \minerva, especially when $Q^2<0.4\text{ GeV}^2$ where the physics content is not well known and the difference between GENIE and NEUT is 25\%. This implies that the extracted data may have been different if MINERvA had used NEUT as the model for the analysis, and therefore implies some degree of model bias. How significantly this affects the results is impossible to accurately quantify in this exercise.

\subsubsection{Efficiency comparisons}
\label{sec:pi_eff}
This section shows the T2K CC1$\pi^{+}$ and MINERvA CC1$\pi^{\pm}$ efficiencies, calculated using NEUT v5.3.6 and GENIE v2.6.2 
respectively, as a function of $p_{\mu}$, $\cos\theta_{\mu}$, $T_{\pi}$ and $\cos\theta_{\pi}$, using the method described in \autoref{sec:eff} to show the efficiencies as a function of a single variable.

\begin{figure*}[ht!]
    \centering
    \includegraphics[width=0.48\linewidth]{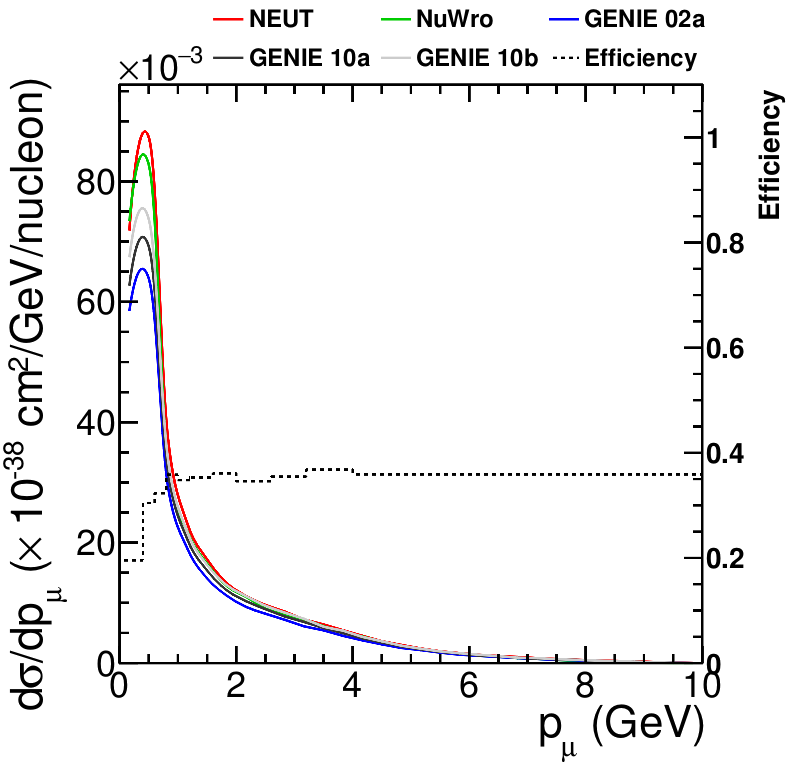}
    \includegraphics[width=0.48\linewidth]{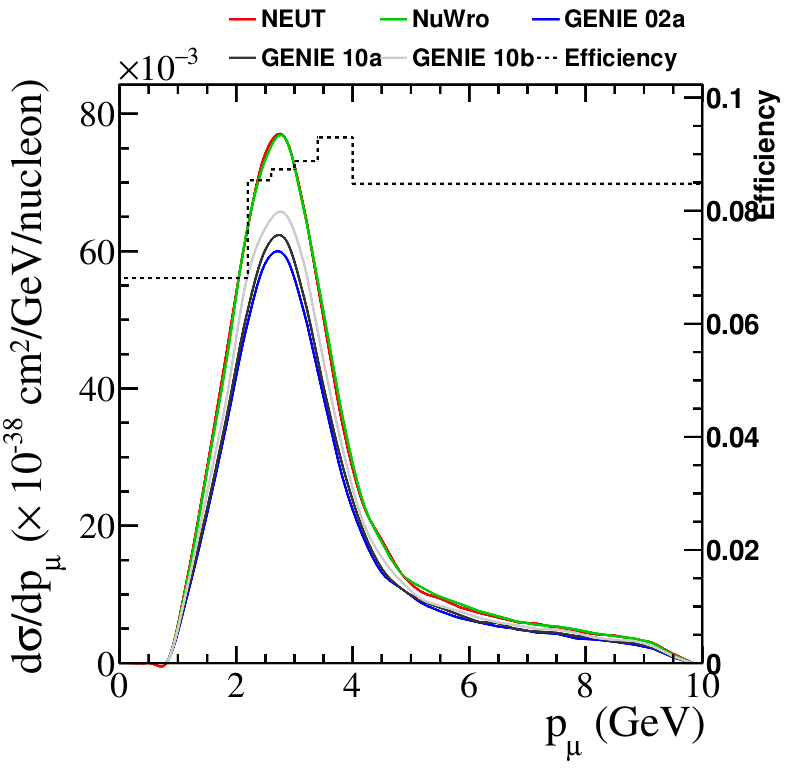}
    \caption{Muon momentum distributions for T2K (left) and \minerva (right). The dotted line is the efficiency for each experiment using the right axis scale and the solid lines are generator predictions for this variable using the left axis scale.}
    \label{fig:t2k_min_cc1pi_pmu}
\end{figure*}

The efficiency as a function of muon momentum is shown in \autoref{fig:t2k_min_cc1pi_pmu} for T2K and \minerva, overlaid with the respective model predictions from NEUT, NuWro, and GENIE. 
T2K has a flat muon momentum efficiency of $\sim$35\% above 1 GeV, dropping to $\sim$30\% at 200 MeV, and falling further to $\sim$20\% at lower momentum.  This was the basis for a cut on the muon momentum to avoid model dependence when extracting the result.  Still, the majority of T2K events fall in the region of changing efficiency where models disagree at the 15\% level. This could lead to model dependence in the efficiency correction.  We note that T2K does not extract a result solely as a function of the muon momentum and this effect is likely to be largely mitigated by extracting the cross section in $p_{\mu}$--$\cos\theta_{\mu}$. 
In contrast to T2K, \minerva's muon momentum efficiency is relatively flat at approximately 9\% across a broad momentum range but vanishes below 1.5 GeV.
Below $\sim$2 GeV, the various generator models agree very closely. Here, the efficiency drops below 7\%, averaged across a single bin. 
At the peak of the distribution, 30\% differences are observed and the agreement worsens with increasing momentum.  The shape of the efficiency is largely caused by the MINOS track matching requirements. 

\begin{figure*}[ht!]
    \centering
    \includegraphics[width=0.48\linewidth]{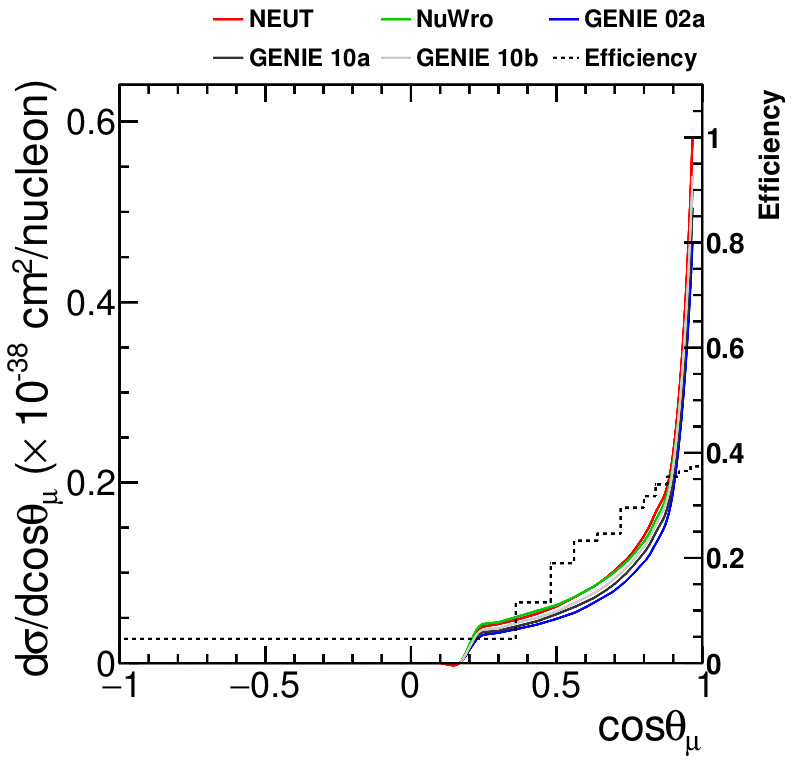}
    \includegraphics[width=0.48\linewidth]{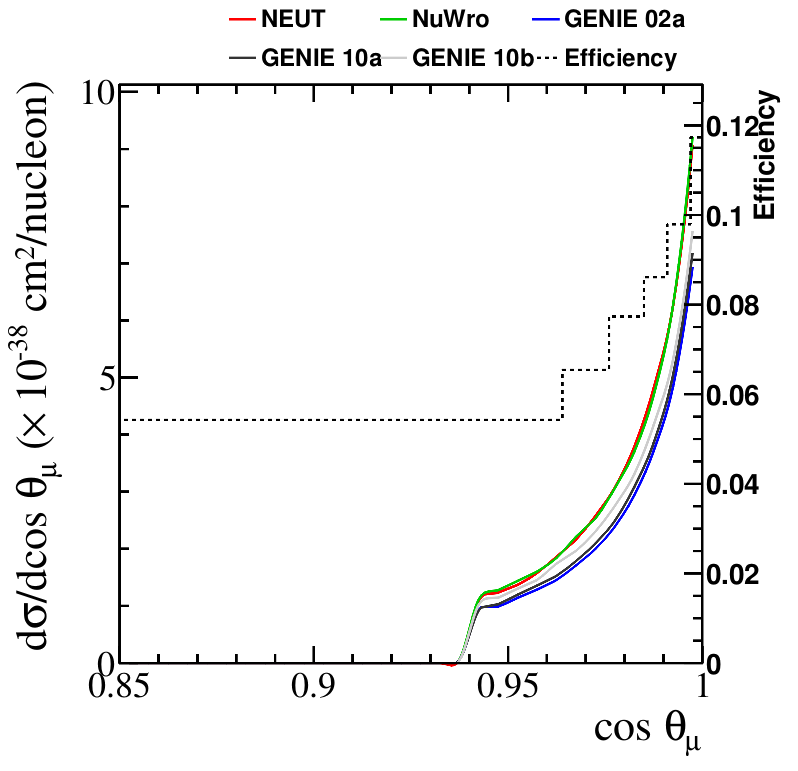}
    \caption{Muon angle for T2K (left) and \minerva (right).  Experimental efficiency is shown as a black dotted line using the right vertical scale and various generator prediction are shown using the left vertical axis.}
    \label{fig:t2k_min_cc1pi_qmu}
\end{figure*}

The efficiency as a function of muon angle is shown in \autoref{fig:t2k_min_cc1pi_qmu}.
Coverage in this variable is restricted for both T2K and \minerva.  However, the efficiency is largest at forward angles where the cross section is largest. \minerva's MINOS track matching requirements limits angular acceptance to less than $\cos\theta_\mu \sim0.93$ ($20\degree$), although the signal definition is not restricted to this region. The range for T2K extends much further, up to $\cos\theta_\mu \sim0.26$ ($75\degree$) for measurements that integrate over muon momenta (all forward angles are considered for the muon double differential cross section). \minerva falls off steadily from 12\% to 5\% over that range, while T2K falls off from 37\% to 10\%. In both cases the bulk of the events are forward-going, somewhat mitigating the effects of the efficiency decline on measurements that integrate over muon angle. T2K published a two-dimensional distribution in $p_\mu-\theta_\mu$, in part to decrease the model dependence shown here.  

\begin{figure*}[ht!]
    \centering
    \includegraphics[width=0.4\linewidth]{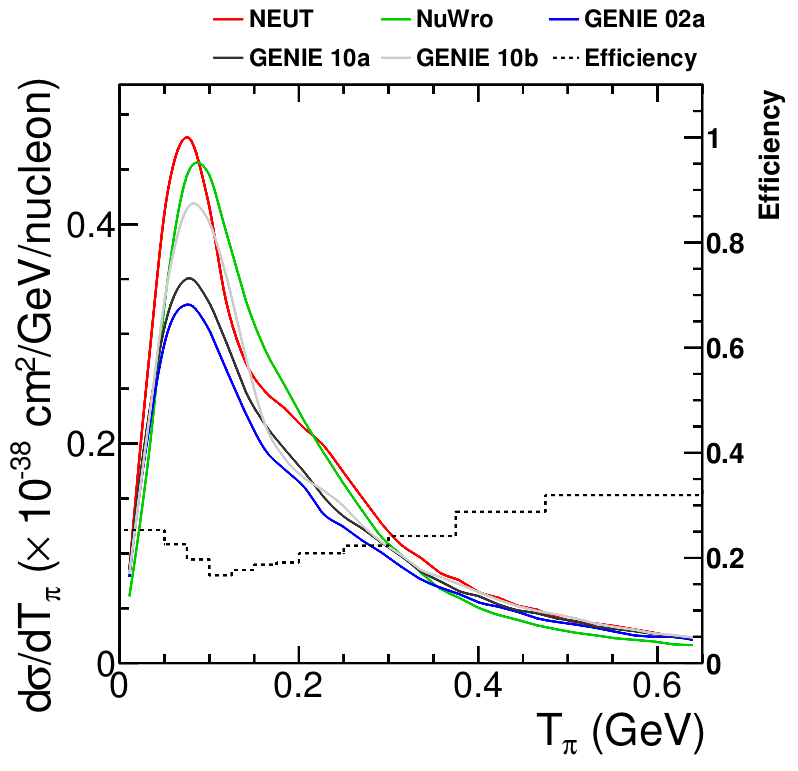}
    \includegraphics[width=0.4\linewidth]{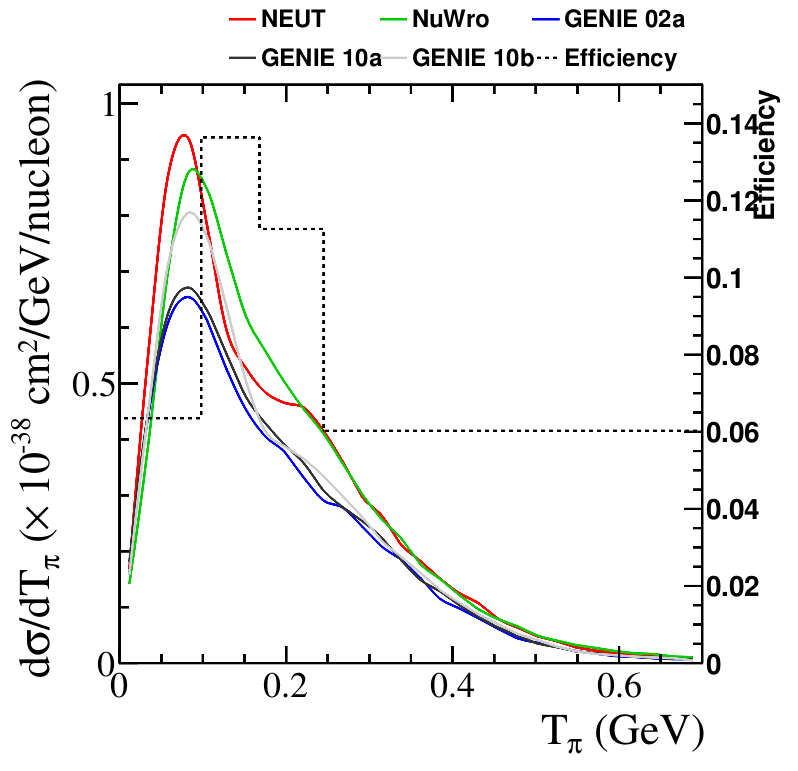}
    \caption{Pion kinetic energy for T2K (left) and \minerva (right).  Experimental efficiency is shown as a black dotted line using the right vertical scale and various generator prediction are shown using the left vertical axis.}
    \label{fig:t2k_min_cc1pi_Tpi}
\end{figure*}

The efficiency as a function of pion kinetic energy is shown in \autoref{fig:t2k_min_cc1pi_Tpi}.
T2K and \minerva have reasonable efficiency down to low $T_{\pi}$. The acceptance region extends up to high values of $T_{\pi}$ containing the full range of the cross section model predictions, although the MINERvA efficiency falls with higher $T_\pi$ whereas T2K's increases due to the two methods of tagging pions present in the analysis.

The variations in efficiency over the peak regions in $T_{\pi}$, coupled with the model disagreement in those regions, are troubling for measurements that integrate over the full pion phase space. For T2K, the efficiency drops from 25\% down to 20\%, and then rebounds up to 33\% in the $0<T_\pi<0.6~\text{GeV}$ region where the vast majority of the cross section lies. The cross section drops rapidly and generator predictions vary here by up to 25\%. The efficiency uncertainties are further complicated by nature of the event selection, which includes Michel tagged pions at the lower pion kinetic energies and tracked pions at higher energies. The details of the efficiency function shape over this region is sensitive to systematic errors in either of these two selections. However, the Michel tagged pion sample is only used in the muon and pion $T_{\pi}$ measurements: all other measurements include a $p_{\pi}>200$ MeV/c ($T_{\pi}>104$ MeV) phase space cut which intentionally eliminates most of the Michel sample's contributions. The issues with the increasing tracked pion efficiency through 600 MeV is still problematic.

\minerva suffers from similar issues, with the efficiency rising steadily from 6\% to 14\% in the region below the predicted cross section peak near $T_\pi=100~\text{ MeV}$. At the peak the generator models differ in their predictions by up to 30\%. Bins are large with respect to the change in efficiency on the high $T_{\pi}$ side of the peak as well, where the cross-section prediction continue to change quickly, and large model difference persist. In summary, there is potential for large model dependence across the full range of $T_{\pi}$, especially since no signal definition cut is placed on the variable.

\begin{figure*}[ht!]
    \centering
    \includegraphics[width=0.4\linewidth]{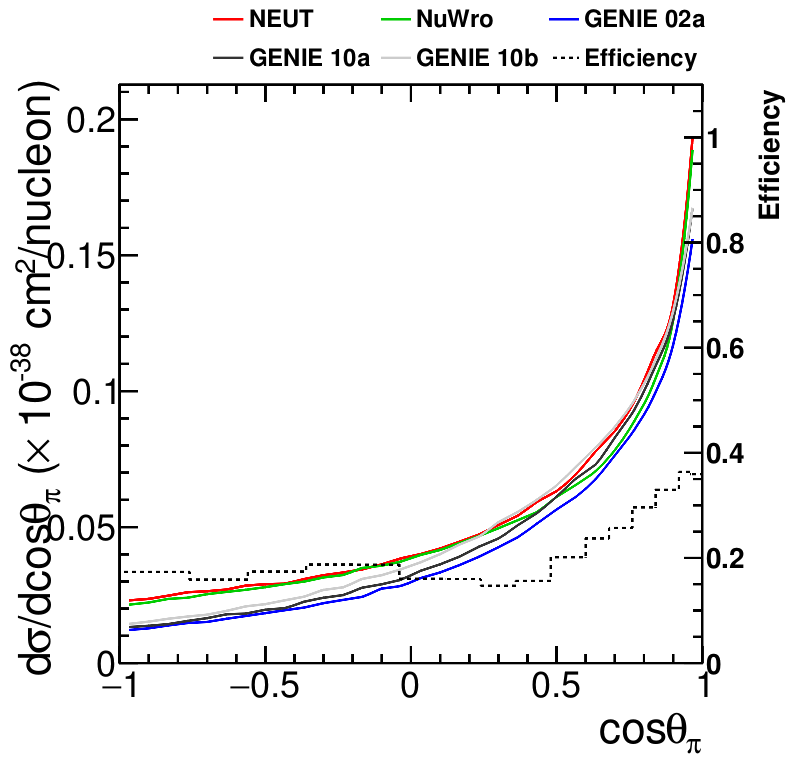}
    \includegraphics[width=0.4\linewidth]{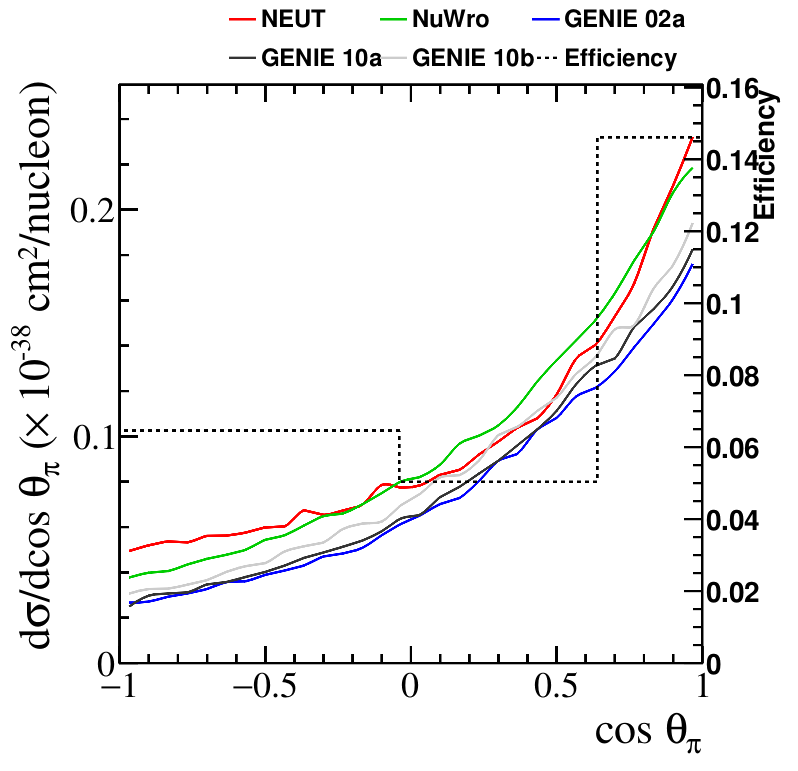}
    \caption{$\cos\theta_\pi$ for T2K (left) and \minerva (right).  Experimental efficiency is shown as a black dotted line using the right vertical scale and various generator prediction are shown using the left vertical axis.}
    \label{fig:t2k_min_cc1pi_qpi}
\end{figure*}

The efficiency as a function of the cosine of the pion angle is shown in \autoref{fig:t2k_min_cc1pi_qpi}. T2K's efficiency is flat at $\sim$20\% for backwards and perpendicular pions and rises gradually to $35\%$ for forward pions. \minerva has essentially no acceptance of perpendicular pions because of the scintillator strip orientation. Nevertheless, the \minerva CC1$\pi^{\pm}$ analysis shown here~\cite{Eberly:2014mra} covers the full range of pion angles. This may be partially mitigated by the smoothness of the cross section behaviour in this region, and the relative agreement between the generators as shown, but it clearly risks model dependence, which is impossible to accurately assess after the analysis. In particular, the \minerva $\theta_{\pi}$ bins which correspond to this low efficiency region in the published cross section measurement are not informative.

\subsection{Discussion and summary}
\label{sec:pion_discussion}
For this publication, the recent CC1$\pi^+$ data from T2K~\cite{t2k_cc1pi_fgd1} and the updated CC$1\pi^\pm$ MINERvA results~\cite{minerva_datarelease} were considered.  Both measurements studied $\nu_\mu$-CH interactions and report inclusive distributions of muon and pion angle and kinetic energy/momentum and other quantities of interest.  These measurements are at different neutrino energies and give the opportunity to examine energy dependence of the interaction.

One of the primary sources of tension at the 2016 workshop~\cite{tensions2016} came from comparison of the pion production results of MiniBooNE~\cite{Wilking} and \minerva~\cite{Eberly:2014mra}. The results had somewhat different signal definition (CC$1\pi^+$ vs CC1$\pi^\pm$, $E_\nu$ and $W$ cuts) and were at different beam energies.  It was also noted that the lack of covariance matrices in the MiniBooNE data release severely complicated interpretation of the results.  Although no direct comparison between data sets was possible due to differences in signal definition, there was clearly a more rapid evolution of the cross section in neutrino energy for the calculations than was observed in the data. Later calculations at the time~\cite{Sobczyk:2012zj} were unable to resolve the differences.

The first publication of pion production results from T2K~\cite{t2k_cc1pi_fgd1} provides significant new information with a more modern treatment of uncertainties, at a similar neutrino energy to MiniBooNE.  Attempts at comparing \minerva and T2K measurements run into similar issues as \minerva and MiniBooNE comparisons. Each measurement seeks a result that is objective and reproducible.
However, detector technology, geometry and selection sculpts the particle acceptances differently, imparting potential model dependence to the data.  
Although the process of minimizing bias from detector effects or model dependence is still underway, an attempt was made to compare data at the same kinematics to decrease the differences.  
For the small overlapping phase space of \minerva and T2K, we find good agreement, with NuWro and NEUT often describing T2K better, and GENIE performing better at \minerva energies. We also note that the generators used to extract the cross section at each experiment is often the one with the best description of the data.  This may be coincidence or come from biases towards the underlying MC program used in analysis.

In TENSIONS2016, an instance of model dependence related to the signal definition limitation on $W$ in the original \minerva analysis signal definition was identified and improved for the data release~\cite{minerva_datarelease} result studied here. However, no additional changes in the phase-space restrictions on particle kinematics in the signal definition were made. Separately, all the models changed due to more complete fitting to neutrino H/D data~\cite{Rodrigues:2016xjj,GENIEFreeNucleonTune}. The main effect is to make the change with beam energy smaller and all the models are now in better agreement with \minerva data.  
Agreement of generators with both data sets is reasonable for pion kinematics for the quoted uncertainties.  On the other hand, all generators have similar systematic problems describing the Adler angle variables presented by T2K.

The {\Genietwoa}, NuWro and NEUT predictions for the single pion production measurements are broken down by mode for T2K and \minerva $Q^2$ and pion momentum/kinetic energy signal definitions in~\autoref{fig:t2k_cc1pi_q2_mode}-\autoref{fig:min_cc1pi_tpi_mode}. Although T2K data extends to higher pion energies, both peak at the same energy within the resolutions provided as both are dominated by $\Delta$ production processes. This peak is produced by a balance of pion production and FSI processes.  Thus, both effects must be considered.  

Generator predictions have small-to-moderate  deviations from the data based on the $\chi^2$ values in Tables~\ref{tab:t2k_chi2} and \ref{tab:minerva_1pi_chi2}.  It is notable that NuWro has the lowest $\chi^2$ for T2K and the largest for \minerva.  The deviations aren't large and all calculations seem to have general agreement with the data.
The calculations tend to have better agreement for lepton variables. 
For variables involving pion kinematics the differences between generators grows and the agreement with data is worse, with the largest data--MC differences being observed in the Adler angle distributions. This is also largely reflected in the comparisons to \minerva data, where the $d\sigma/d\theta_\pi$ distribution is poorly described by all generators, followed by the $d\sigma/dp_\pi$ distribution.

It is difficult to make a quantitative assessment of the agreement of data and calculations with regard to energy dependence.
The calculations have similar agreement with both T2K and \minerva data according to $\chi^2$.  In addition, the average cross section of both data sets for overlapping kinematics (\autoref{sec:t2k_minerva_eff_comps})
have a ratio very similar to that of the calculations.      
The generators are also very similar in lepton kinematics, and differ mostly in hadron kinematics.
These observations all give confidence that the energy dependence of the calculations matches that of the data within the accuracy of the data.

Beyond comparisons with calculations, a more detailed examination of the T2K and MINERvA analyses was made.
Both measurements have restricted muon angle ranges, much more so for MINERvA. MINERvA also has a significant hole in the pion angle acceptance around 90$^{\circ}$, and for higher momentum pions. Low efficiency regions are excluded from the T2K analysis, but corrected for using the nominal generator in the MINERvA analysis. 
The MINERvA data for $\theta_\pi \sim 90^\circ$ and disagreements with simulations should be discounted.
The \minerva CC$1\pi^\pm$ measurement produces a full phase-space cross-section measurement in which the selection for observables $p_\mu,\cos\theta_\mu, T_\pi,\theta_\pi$ are selected to better match the detector's reconstruction capabilities.  Model dependence from these choices is demonstrated.  Although the analysis contains systematic errors to account for these effects, future analyzers should minimize the model dependence. 

Efficiency studies uncover problems in regions where both the efficiency changes rapidly and model predictions from different generators vary.  One example is for \minerva $T_\pi\lesssim~\text{100 MeV}$ which applies to the lowest bin in \autoref{fig:minerva_cc1pi_pi}.  The other case is $p_\mu\lesssim1~\text{ GeV/c}$ for T2K which applies to the lowest two bins in \autoref{fig:t2k_cc1pi_mu}, although the importance is less clear when the data is presented as a two dimensional distribution.
Finally, it should be noted that for T2K's $p_\mu \cos\theta_\mu$ measurement it uses both TPC and Michel tagged pions, leading to a more complex $T_\pi$ efficiency shape. However, this is mostly undone by introducing the $p_\pi>0.2~\text{ GeV/c}$ cut on the other distributions.

The importance of the choice of signal definition and analysis methods was discussed in \autoref{sec:xsec_methods}.  The pion production data available is all from early measurements and many of the cautions listed there apply to these data. The T2K analysis used a single iteration of d'Agostini unfolding and the MINERvA analysis used a signal definition with derived values.  Both introduce model dependence that could be avoided. Both collaborations have more advanced analyses in progress, but were not available for this work.

A useful strategy is to use a multi-generator approach.  The cross section is either extracted separately with different generators~\cite{Abe:2016tmq}, or a bias test using a different generator is performed (e.g. Ref.~\cite{Abratenko:2020sga}).  An example would be to measure the efficiency with two generators making different choices or varying models within a specific generator to their extremes. A simpler alternative is to use an alternate generator or model set as fake data.  However, these methods are necessarily limited to model dependencies based on the models implemented in generators. 

The theoretical treatment of the pion production process is also important.  It is complicated and clearly the current models implemented in generators are unable to explain all of the data, with pion kinematic variables a particular weakness. It is important to stress that these studies are only for $\Delta(1232)$ dominated interactions, and will be increasingly complicated at DUNE energies, due to higher resonances and soft inelastic scattering contributions, and a more complex $^{40}$Ar target, amongst others. This motivates the need to improve our understanding by including better models in the event generators. The description of the basic pion production process with nucleon targets is limited by the quality of data~\cite{Rodrigues:2016xjj,GENIEFreeNucleonTune}.  Better accounting for existing electron inelastic scattering from the nucleon~\cite{Tiator:2011pw} would be a significant advance because vector resonance form factor improvements since Ref.~\cite{Rein-Sehgal} have not been included in the event generators. 
Furthermore, much improved data for pion electroproduction using nuclear targets from the e4nu experiment~\cite{Papadopolou:2020zkd,CLAS:2021neh} at JLab are expected.  This will be important for testing pion production and FSI with high quality data. 
Additionally, none of the common event generators include nuclear medium corrections to the $\Delta$ production operator, but various models are available~\cite{Freedman:1982yp}.

\FloatBarrier

\section{MiniBooNE data sets}
\label{sec:minib}
MiniBooNE produced a series of key measurements: CCQE and CCQE-like~\cite{Katori-mbqe}, CC1$\pi^+$~\cite{Wilking}, CC1$\pi^0$~\cite{miniboone-ccpi0}, NC$\pi^0$~\cite{miniboone-ncpi0}, NC elastic (NCEL)~\cite{miniboone-ncelas}.
These data sets have had a long lasting impact on the field of neutrino cross-section physics, as they pioneered  important approaches in measuring neutrino cross sections with intense, modern beams. MiniBooNE developed the presentation of double-differential results with high statistics data sets, and a broad signal definition (only based on final state properties which gives reduced model dependence). Additionally, they set standards regarding the information made available to describe and support each analysis in data releases that have now become a regular part of a modern neutrino cross-section publication. The results spawned extensive theoretical development of models to explain the data, and experimental efforts to assess uncertainties on models.  
However, limitations of MiniBooNE data have been uncovered with time. In some of the early MiniBooNE cross section results, the off-diagonal correlations were not reported due to concerns about the influence of finite MC statistics in the generation of the detector model uncertainties.  For example, the lack of a complete correlation error matrix for some of the measurements (CCQE-like, CC1$\pi^+$) prevents the ability to compare data to updated models in a statistically valid way. 
This is exacerbated by the unfolding method used, which leads to strongly regularized and therefore correlated values between bins. In another example, flux and background model improvements developed after the measurements were completed made some of the earlier analyses more difficult to interpret. Finally, the measurements themselves are being superseded by other experiments (e.g. T2K, MicroBooNE) at a similar neutrino energy regime. One major appeal of the MiniBooNE results was the uniformity of acceptance of the detector, but improvements to T2K reconstruction and detectors will overcome this limitation.  In addition, the statistical precision of MINERvA has surpassed or is approaching that of MiniBooNE.

It is clear that the MiniBooNE neutral current elastic (NCEL) and $\pi^0$ (NC$\pi^0$) production measurements are of great value because of large statistics, excellent efficiency uniformity, and up-to-date error analysis.
NCEL is a forward-folded result, and they are published for neutrino and antineutrino enhanced running, each with a full covariance matrix.  Both channels are important and cover reactions that are otherwise very poorly known.
They should be employed with confidence for comparison with calculations and can be used with the same confidence as most other modern data sets.  Since these results are still unique, we have chosen not to include them in the present work where the emphasis is on comparing different data sets. 
Data sets such as CCQE-like and CC1$\pi^+$ should be used with care as the estimated uncertainties did not include the full correlation information.
Any comparison of calculation with theory should include at least one other data set if one of these MiniBooNE measurements is used; Appendix~\ref{sec:minibapp} includes plots of the MiniBooNE data which can be qualitatively compared to the other datasets discussed in earlier sections of this document.


\section{Generator studies}
\label{sec:gen_studies}

The opportunity to compare generator outputs from the samples produced for the data comparisons led to the following sections.  These are studies that have applicability to the field as a whole.

\subsection{Fraction of energy transfer imparted to neutral final-state particles}

In the challenging task of neutrino energy estimation, high-quality modeling of the production of neutral particles is critical due to the relative difficulty of reconstructing these particles in a detector~\cite{Acero:2019ksn,DUNE:2020ypp}. Large detector volumes are needed for good neutral particle reconstruction; unfortunately, this is not true for existing detectors and Monte Carlo programs are required to estimate neutral energy contributions.
Significant differences in neutral particle production modeling among the predictions of GENIE, NEUT, and NuWro can exist and they are examined briefly in this section.

To facilitate comparisons between the generator models, we define the observable $F$ to be the fraction of the total energy contained in the final state hadrons (equal to the leptonic energy transfer $q_0$) which is associated with the production of neutral final-state particles:
\begin{equation}
\label{eq:neutral_fraction}
F \equiv \frac{1}{q_0} \sum_j \big(E_j - \delta_j^B \, m_j \big)
\end{equation}
Here the sum runs over all neutral particles in the final state, and $E_j$ is the total energy of the $j$th neutral particle. The second term is subtracted to avoid counting the neutron mass $m_n$ for baryons in the final state: the symbol 
$\delta_j^B$ is one if the $j$th particle is a baryon and zero otherwise. For definiteness, the following particle species (and their antiparticles) are included in the sum in Eq.~\ref{eq:neutral_fraction}: $n$, $\pi^0$, $\gamma$, $D^0$, $\Lambda^0$, $\Sigma^0$, $K_L^0$, $K_S^0$, $\eta$, and $\omega$. The unstable members of the list decay into both charged and neutral particles.

\begin{figure*}
  \centering
  \includegraphics[width=0.49\textwidth]{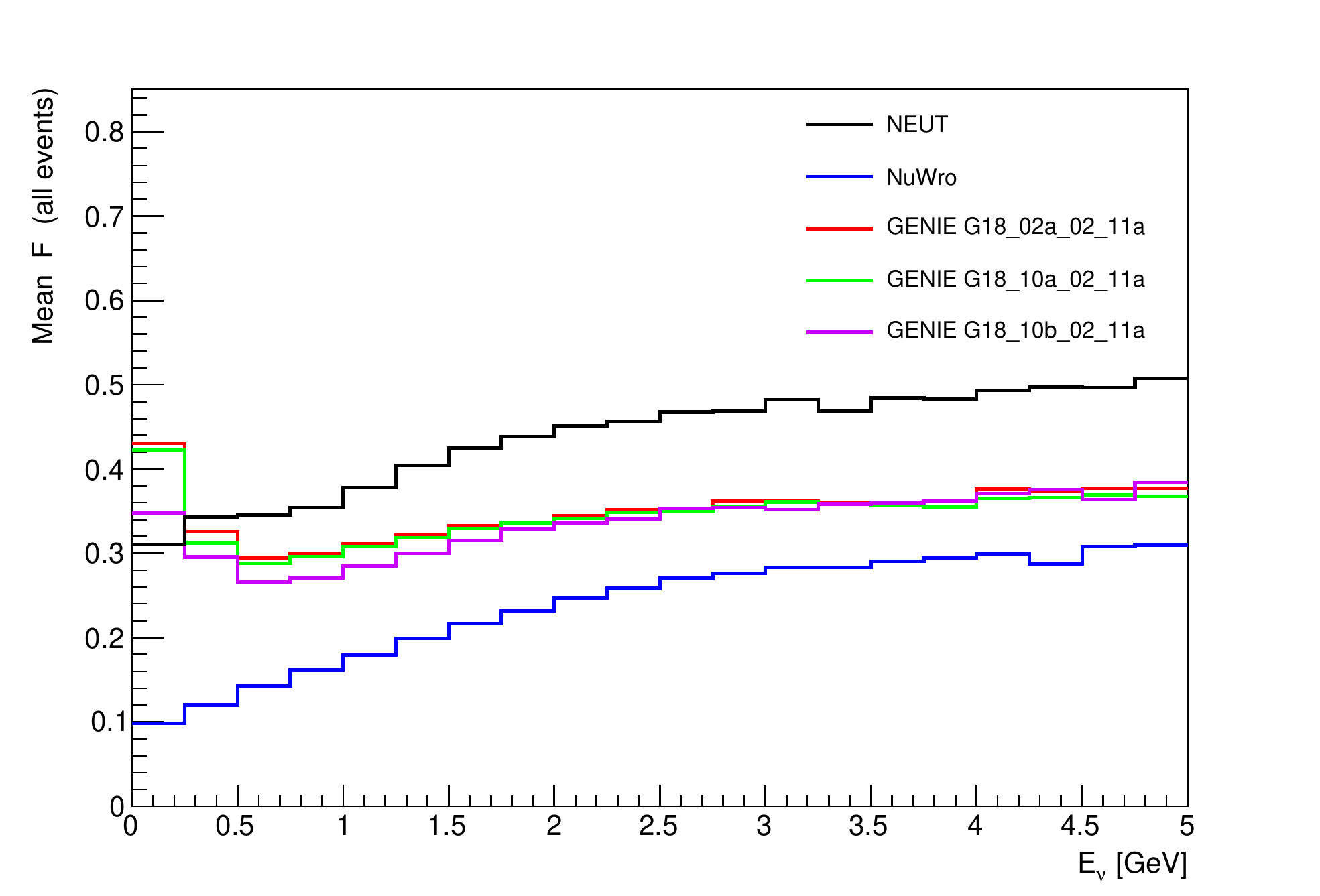} \hfill 
  \includegraphics[width=0.49\textwidth]{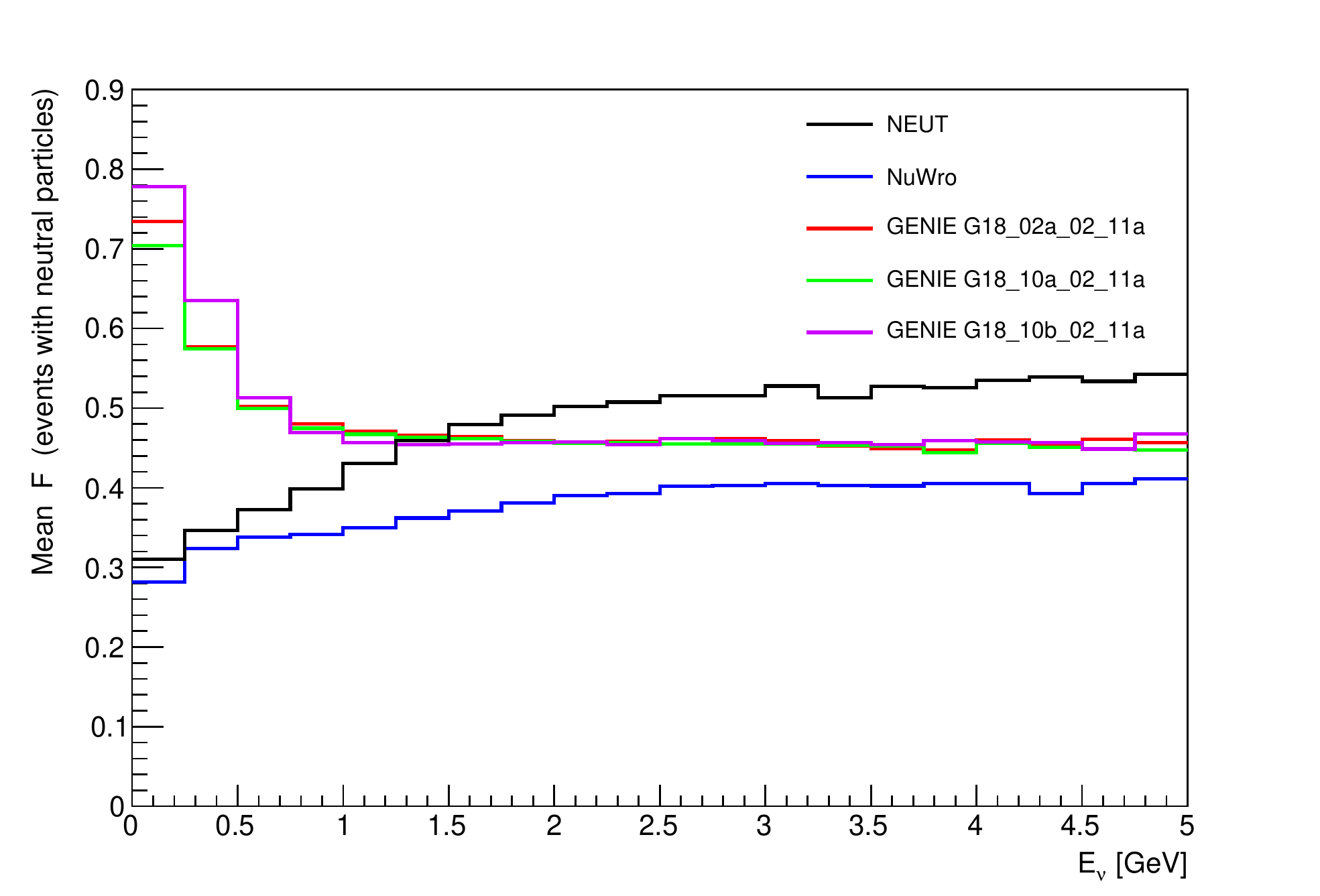} \\
  \includegraphics[width=0.49\textwidth]{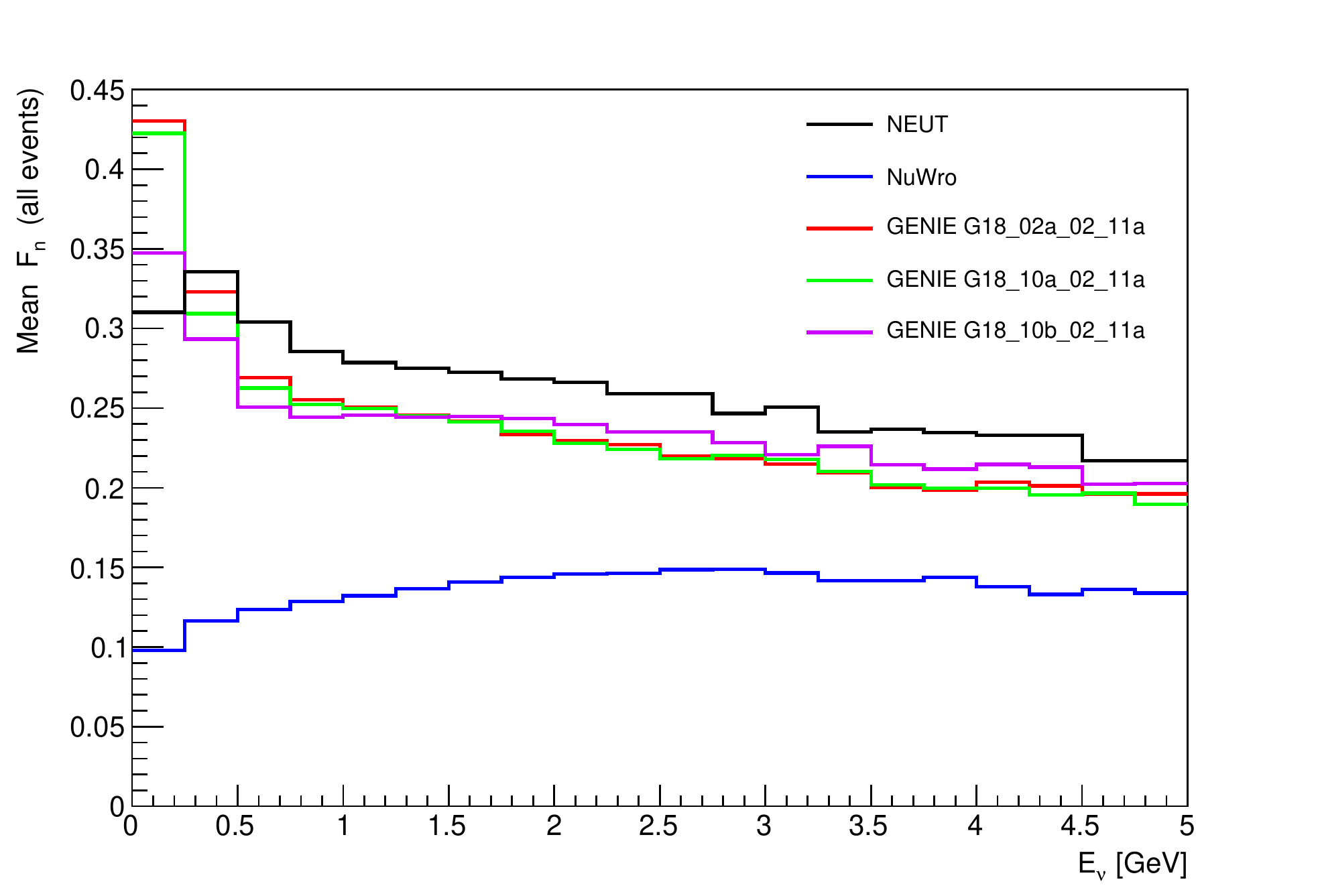} \hfill 
  \includegraphics[width=0.49\textwidth]{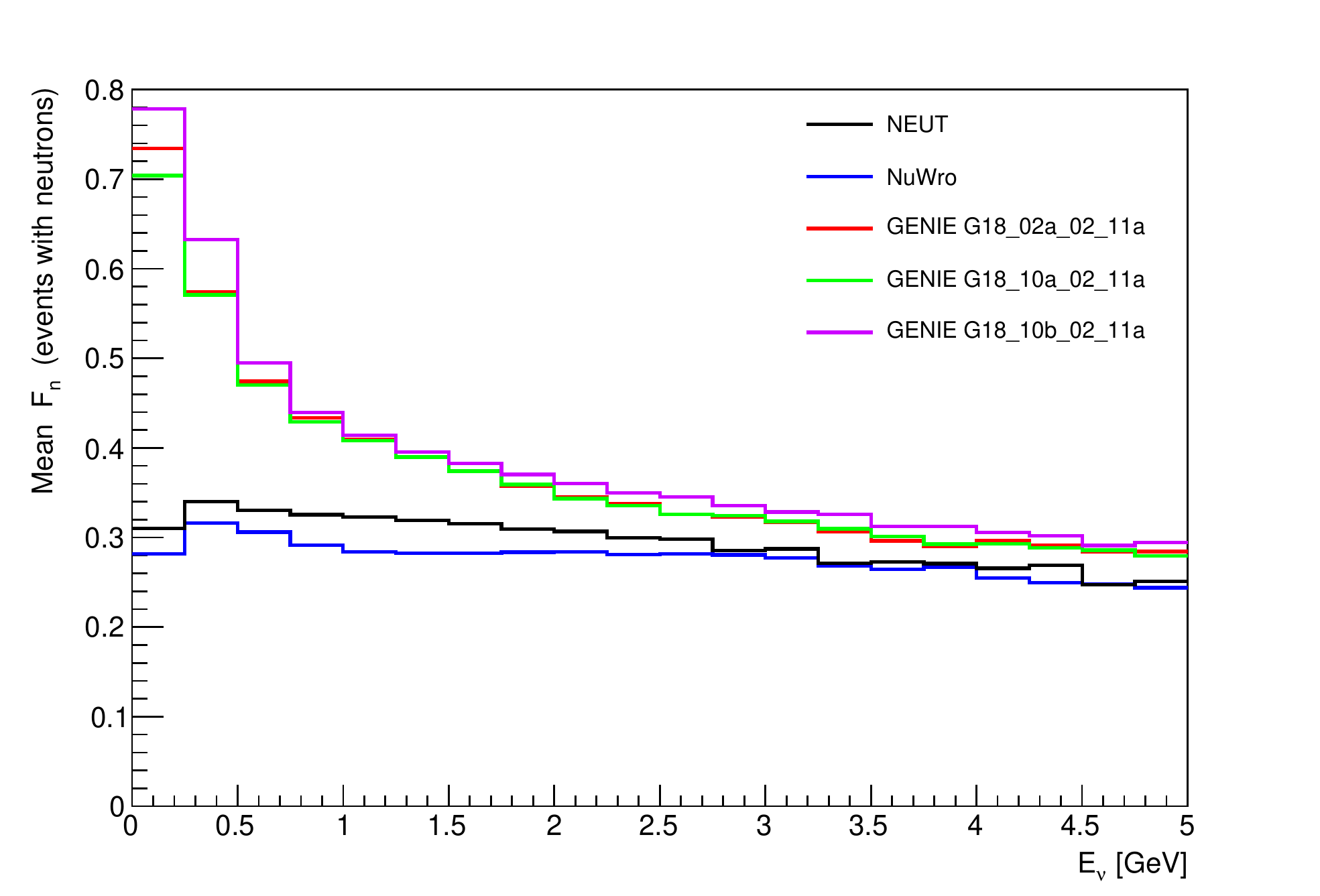} \\
  \caption{Mean fraction of the leptonic energy transfer imparted to final-state neutral particle species. All panels show distributions calculated for charged-current $\nu_\mu$ interactions on \isotope[40]{Ar}. TOP LEFT: Predictions including all neutral particles and all events in the sample. TOP RIGHT: Predictions including all neutral particles for events containing at least one final-state neutral particle. BOTTOM LEFT: Predictions including final-state neutrons only and all events in the sample. BOTTOM RIGHT: Predictions including final-state neutrons only for events containing at least one final-state neutron.}
  \label{fig:neutrals}
\end{figure*}

The mean value of the neutral energy transfer fraction $F$ is shown as a function of neutrino energy in the upper left panel of Fig.~\ref{fig:neutrals}. The simulated events used in the calculation, which are common to all panels of the figure, are for inclusive charged-current $\nu_\mu$ scattering on \isotope[40]{Ar} at energies most relevant for MicroBooNE (up to $E_\nu = \SI{5}{\GeV}$).
The upper right panel shows the generator predictions for the same distribution when at least one final-state neutral particle is required to be present. 
The bottom two panels repeat the same calculations using a modified definition of $F$ in which only final-state neutrons contribute to the sum. This observable, referred to as $F_n$, is equal to the sum of final-state neutron kinetic energies divided by the energy transfer $q_0$.

Variations in the GENIE model set have a significant impact on these distributions only at low neutrino energies, and typically only when comparing the two FSI models available in v3.0.6: hA2018 (red and green histograms) and hN2018 (purple histogram). Disagreements between GENIE and the other two generators are more pronounced, with the spread in the predicted mean values of $F$ reaching roughly a factor of two for an inclusive sample (upper left panel). For events containing at least one neutral particle (right-hand panels), a bifurcation between GENIE and NuWro/NEUT at low energies highlights an important physics difference: in both FSI models used by GENIE, an approximate treatment of compound nuclear decay is included which leads to an enhancement of low-energy neutron emission.

\subsection{Energy threshold dependence of proton and pion event yields}

Exclusive measurements with charged hadrons in the final state provide important insights into the physics of both the neutrino scattering and the nuclear environment. Most available data considers protons~\cite{Betancourt.119.082001,Abratenko:2020sga} and charged pions~\cite{Eberly:2014mra,t2k_cc1pi_fgd1}, both detectable through their ionization signatures. Each experiment makes choices consistent with their particle detection capabilities, including imposing a threshold for efficient tracking. MINERvA uses scintillator technology, with a threshold of $\sim100$~MeV (50~MeV) kinetic energy for protons (pions). T2K uses a combination of scintillator and TPC detectors, achieving a similar threshold to MINERvA. Liquid argon TPC detectors offer the promise of lower thresholds, with the 47 MeV proton threshold reported by MicroBooNE~\cite{Abratenko:2020sga} the lowest available in a fully automated reconstruction at this time.  

Generator models differ in their final state hadron kinematics, leading to differences in efficiency for a fixed threshold. To assess the impact of these effects among the generators used in this work, we determine the fraction of events above threshold as a function of the threshold value for specific final state topologies in T2K, MINERvA, and MicroBooNE, using the appropriate flux and target material in each case.

\begin{figure*}[ht]
  \centering
  \includegraphics[width=\textwidth]{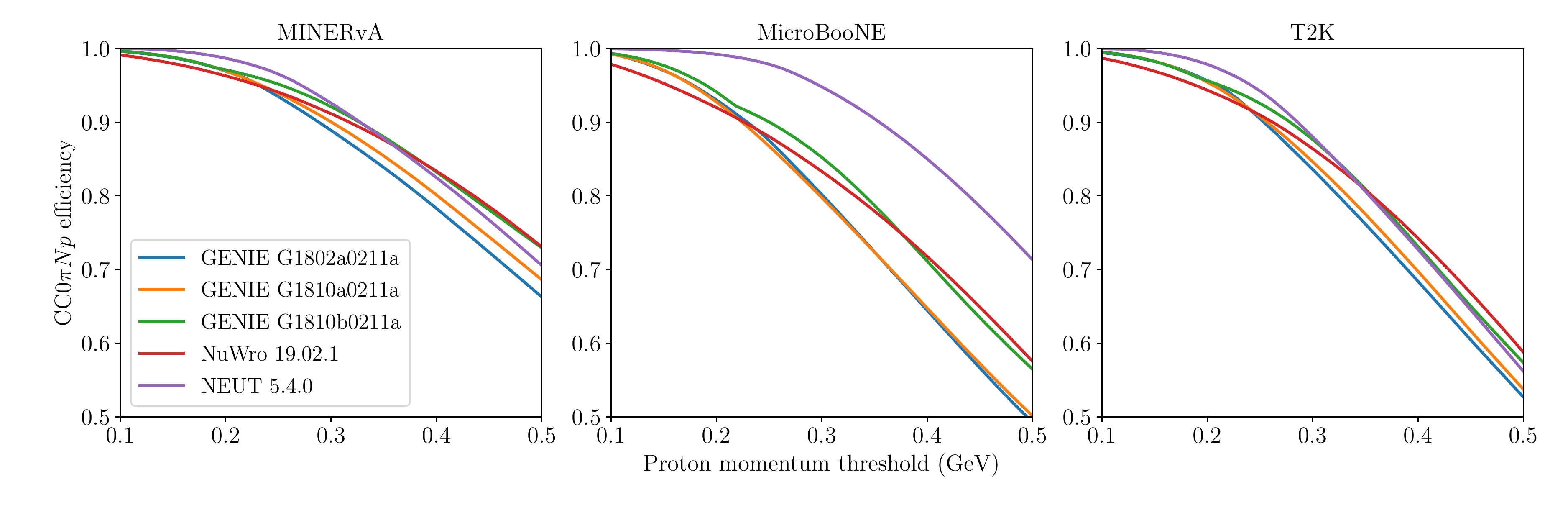}
  \caption{The fraction of true CC$0\pi Np$ events above threshold as a function of the true proton momentum threshold for different generators.}
  \label{fig:cc0pi-pthresh}
\end{figure*}

Figure~\ref{fig:cc0pi-pthresh} shows the effect of threshold on efficiency for protons detected for CC$0\pi$ final states. With several potential sources of low energy protons that are challenging to calculate, including varied assumptions in FSI models, this is an especially interesting issue. With lower-energy neutrinos, MicroBooNE and T2K are expected to be particularly sensitive to the threshold. While the codes generally show the same trends, some large discrepancies are evident. NEUT results are substantially different from the others due to its suppression of low-energy protons. At 300 MeV/c proton momentum (47 MeV kinetic energy) in MicroBooNE, the fraction of protons seen for NEUT is 13\% larger than the average of the others. {\Genietwoa} and {\Genietena}, which use an effective cascade FSI model, have about 5\% smaller fraction than NuWro or {\Genietenb}, which use a full cascade model. Experiments must carefully consider the effects of threshold choice.
 
\begin{figure*}[ht]
  \centering
  \includegraphics[width=\textwidth]{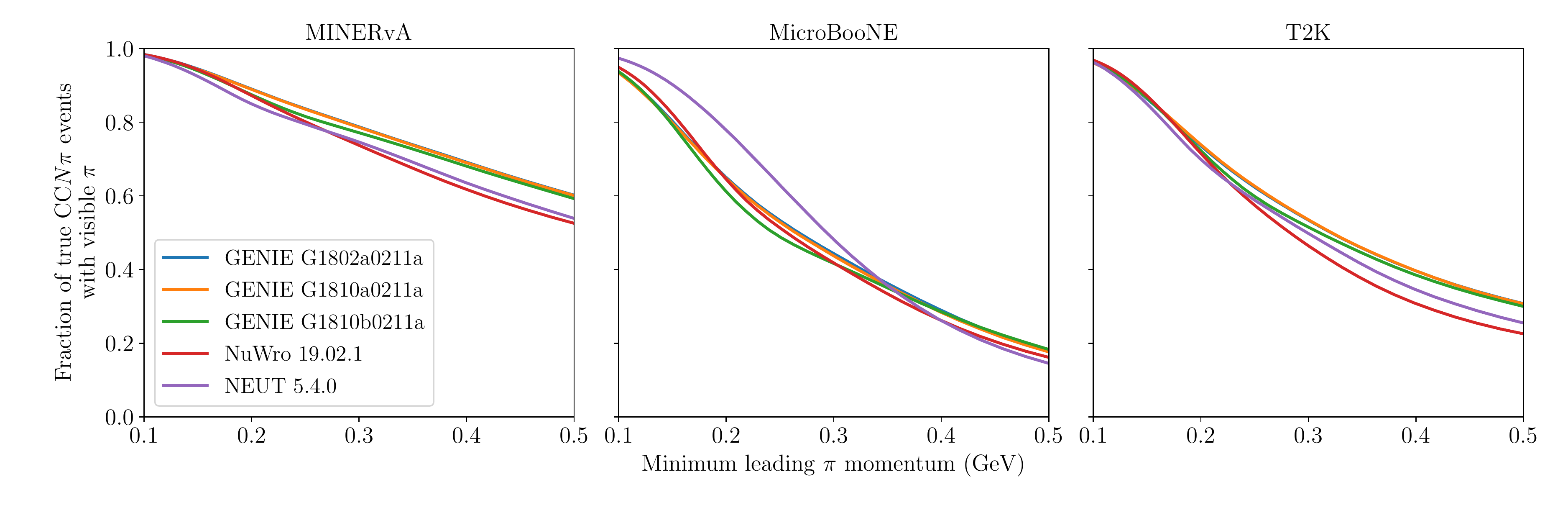}
  \caption{The efficiency for true CC$N\pi^\pm$ events as a function of the true leading $\pi^\pm$ momentum threshold, as compared among different generators.}
  \label{fig:ccpi-pithresh}
\end{figure*}

Figure~\ref{fig:ccpi-pithresh} shows the effect of threshold for pions detected in CC$0\pi^0 N\pi^\pm$ final states. It is notable that the largest deviations for protons are at low energy where the yield is growing rapidly. For positively-charged pions, the cross section is decreasing close to threshold and the largest deviations visible here come at larger energies. Since pions have lower ionization yields than protons at a given momentum, all experimental thresholds in use are in the region where there is very little dependence on generator. Effects of threshold are therefore relatively less important for charged pions than protons.

\section{Methodological limitations of the recent forward-folding cross-section results presented by the MicroBooNE experiment}
\label{sec:forwardfold}

Measured observables are always convoluted with detector effects. As described in
\autoref{sec:xsec_methods}, often, neutrino cross sections are presented after a deconvolution of these effects.
Such measurements are usually called ``unfolded''.
Unfolded measurements correct for smearing and other detector effects in order to get to the true underlying distributions.
If the observed neutrino interaction events in a certain observable bin $i$ is $N_i$, the cross section in true bin $j$ is usually calculated as
\begin{equation}
    \left( \frac{d\sigma}{dx} \right )_j = \sum_{i}\frac{U_{ji} (N_i - B_i)}{\Phi T \epsilon_j (\Delta x)_j}
\end{equation}
where $B_i$ are the number of background events in the observable bin, $\Phi$ is the total flux, $T$ number of target nucleons, $\epsilon_j$ is the efficiency in truth bin $j$, $\Delta x$ its width (or area) and $U_{ji}$ is the unsmearing matrix.
Unfolded measurements can be easily compared with theory predictions, and with unfolded results from other experiment.
The difficult part is to calculate the unsmearing matrix, and there exists different approaches to do it.

Unfortunately, unfolding is an ``ill-posed problem'' in which small statistical fluctuations can lead to large variations in the unfolded spectrum (with generally strongly (anti-)correlated unfolded data points).
Also they usually make some sort of assumption about the linearity of the statistical uncertainties (i.e. things are treated like normally distributed errors).
Another way of presenting cross section results is by using the ``forward-folding'' approach, in which the data is not deconvolved, but instead is compared with ``smeared'' (i.e. forward-folded) theoretical predictions.
Forward-folding measurements are currently being investigated with increased interest as they sidestep some issues of the unfolding methods and can thus be preferable from a statistical point of view (see e.g.~\cite{Cousins:2016ksu}).
To make these measurements comparable to other predictions, the method to smear the theoretical prediction has to be provided alongside with the measurement.

The MicroBooNE experiment has recently used the following method for their double-differential muon-neutrino charged-current inclusive cross-section measurement on argon~\cite{Abratenko2019} and their measurement with $N$ protons and no pions~\cite{Abratenko:2020sga}.
In the MicroBooNE method, a background-subtracted, efficiency-corrected cross-section measurement in reconstructed space is extracted from the measured data:
\begin{equation}
\label{eq:sigma}
    \sigma_i = \left(\frac{\dd^2\sigma }{\dd{p_\mu}\dd{\cos\theta_\mu}}\right)_i = \frac{N_i - B_i}{\tilde\epsilon_i  T  \Phi_{\nu_\mu}  (\Delta{p_\mu}  \Delta{\cos\theta_\mu})_i} \text{,}
\end{equation}
where $N_i$ and $B_i$ are the number of selected data events and the expected number of background events in reconstructed bin $i$, respectively.
$(\Delta{p_\mu} \cdot \Delta{\cos\theta_\mu})_i$ is $i^\text{th}$ bin area.
$T$ and $\Phi_{\nu_\mu}$ are the number of target nucleons and the integrated muon-neutrino flux, respectively.
$\tilde\epsilon_i$ is the average event selection efficiency:
\begin{equation}
\label{eq:eff}
    \tilde\epsilon_i = \frac{\sum_j S_{ij} N_j^\text{sel}}{\sum_j S_{ij} N_j^\text{gen}} \text{,}
\end{equation}
where $N_j^\text{sel}$ and $N_j^\text{gen}$ are the number of selected and generated events in truth bin $j$.
$S_{ij}$ is the smearing matrix describing the migration of events between kinematic bins, $S_{ij} =$$ P(\text{measured in bin}\, i \,|\, \text{generated in bin}\, j)$.

A covariance matrix $E$ describes the statistical and systematic uncertainties of the result.
The systematic part is generated by varying the input parameters of the simulation according to their respective uncertainties.
These variations can affect every single variable in Eq.~\eqref{eq:sigma} and~\eqref{eq:eff} except for the number of selected events, the bin area and the number of target nucleons.

To compare some model cross-section predictions $\sigma^\text{model}$ with the data, they are folded through the \emph{nominal} smearing matrix $S$.
Then the Mahalanobis distance~$d$ (i.e. ``chi-square'') between the predicted cross section and the measured data points is calculated using the covariance matrix:
\begin{widetext}
\begin{equation}
\label{eq:mahalanobis}
    d^2 = \sum_{i,k} \left[
    \left( \sigma_i - \sum_j{S_{ij}\sigma^\text{model}_j} \right)
    E^{-1}_{ik}
    \left( \sigma_k - \sum_l{S_{kl}\sigma^\text{model}_l} \right)
    \right] \text{.}
\end{equation}
\end{widetext}
If the measured data is a random variation of the model prediction,
$d^2$ should be $\chi^2$ distributed.

The idea behind providing the covariance matrix $E$ is that new generator predictions can be tested against the data using Eq.~\eqref{eq:mahalanobis}.
This requires all systematic uncertainties to be encoded in the covariance matrix.
Unfortunately, it seems like this does not hold true in the general case.
If the systematic uncertainties on the smearing itself are not negligible, the formalism described in~\eqref{eq:sigma} may underestimate these systematics. 

This can be easily shown when considering the extreme case of completely flat efficiencies (the ideal case in most scenarios).
When the selection efficiency $\epsilon$ is flat over all truth bins $j$,
Eq.~\eqref{eq:eff} becomes a constant expression:
\begin{equation}
    \tilde\epsilon_i
    = \frac{\sum_j S_{ij} N_j^\text{sel}}{\sum_j S_{ij} N_j^\text{gen}}
    = \frac{\sum_j S_{ij} N_j^\text{gen} \epsilon}{\sum_j S_{ij} N_j^\text{gen}}
    = \epsilon
    \text{.}
\end{equation}
No amount of variation in the smearing matrix will change this number, so in this case, the contribution of the smearing uncertainty to the covariance matrix is exactly~$0$.
For the measurement described in Ref.~\cite{Abratenko2019}, performance studies done by MicroBooNE~\cite{MicroBooNE:2018zgb} show that the effect of the smearing uncertainty is small compared to all other systematic uncertainties.
The interpretation of model comparisons using the covariance matrix $E$ are thus ``correct'', i.e. the partially ignored uncertainties have only a small effect.

This shows though that this method is not necessarily generalizable to an experiment in which the uncertainty on $S$ plays a bigger role.
To fully include the smearing uncertainty in the model comparisons, it must be taken into account when folding the truth-space model prediction to the reconstructed space.
One possibility of doing this is by using a set of response matrices, varied according to the detector uncertainties, as shown in Ref.~\cite{Koch2019}.

\section{Conclusion}
The field of neutrino interaction physics has expanded significantly in recent years, with numerous new measurements and calculations, and improved techniques. Importantly, analysis improvements now facilitate clearer connections among experimental results, and between those results and generator predictions. One notable shift is that nearly all experiments now use signal definitions expressed in terms of direct kinematic observables, e.g. final state lepton momentum and angle, and often make kinematic restrictions to better match experimental capabilities. This alignment to the topology of experimental measurements has helped to decrease the level of model dependence in extracted cross sections.  However, no direct comparison between results of different experiments is possible at this time.  Detailing these issues is the primary purpose of this paper.

Comparison of the results here with those in the first Tensions 2016 paper~\cite{tensions2016} shows the evolution.  At that time, both MiniBooNE and MINERvA used broad signal definitions and corrected for kinematic regions with low efficiency.  This has the advantage of being closer to theoretical results.  Since then, the dangers of model dependence have become much more apparent.  Nevertheless, some of the measurements here (particularly the MINERvA 1$\pi$ measurements) are done in the earlier style.  We use them as test cases showing the results of choices made in analysis.

Event generators continue to play a key role in comparing theory with measurements. Although model implementations necessarily lag behind most recent theory advances and certain assumptions are required to produce a composite model with complete phase space coverage, generator simulations can readily reproduce a broad array of signal definitions due to the availability of all variables in the calculation. Recent efforts have led to dramatic enhancements in the event generators, including a wider array of available models, inclusion of collective nuclear effects and processes involving nucleon-nucleon correlations, and improved tuning to data. These improvements are a result of a growing body of theoretical work, and strong collaboration among theorists and generator developers working toward implementation. We note that while generators tend to make similar advances, e.g. Valencia models~\cite{nieves_2011} for QE and $2p2h$ interactions and Salcedo-Oset~\cite{Salcedo-Oset} for pion FSI, full calculations to match experimental conditions involve many elements, and we find that agreement among calculations is good but not exact. The studies presented here include up-to-date configurations of the GENIE, NuWro, and NEUT generators, as well as a GENIE configuration using older model sets, as a point of comparison to earlier work. Notably, most of the more modern calculations were not in the original publications of the experimental measurements, and are presented here for the first time. 

The Tensions workshops have attempted to assess the overall status in the field of neutrino interactions in recent years. The first Tensions publication~\cite{tensions2016} was based on results available in 2016, focusing primarily on the final results from MiniBooNE and early T2K and MINERvA results. Comparisons made with generators available at the time revealed discrepancies among the experimental data, and highlighted the role of model dependence in cross-section measurements.  The results presented in this paper summarize the work conducted and initiated during a second workshop in 2019. In addition to an updated perspective including more recent and more detailed measurements alongside new and substantially improved event generators, we have presented new, detailed cross-experiment comparisons, studied key model differences among generators, and explored the limitations of forward-folding methods.
The comparisons in this document include a novel comparison of each interaction type in terms of the model dependence of the efficiency. 

Since the results are primarily for CH targets, the $A$ dependence cannot be studied in a detailed way.  However, the CH data from T2K ($\langle E_\nu \rangle \sim 1$~GeV) and MINERvA ($\langle E_\nu \rangle \sim 3.5$~GeV) allow examination of the energy dependence in a limited way. In some cases, bins with identical final state kinematics in each experiment were examined.
The following subsection summarizes those findings.

\subsection{Comparisons by interaction type}

First, we consider recent measurements of charged-current inclusive cross sections from T2K, MicroBooNE, and MINERvA, which provide a general view of neutrino interactions. The measurements show clearly that the modeling of the lepton kinematics works much better than the modeling of the hadronic part of the interactions. The T2K and MicroBooNE measurements using lepton kinematics yield $\chi^2$ per degree of freedom of order 2, while MINERvA's measurement including hadronic information is of order 10. (The larege $\chi^2$ values are likely to a strong effect from correlations.  This is still under study.)  Although clear differences exist in the relative performance of the generators among experiments, the poor fits overall make it difficult to draw definitive conclusions on particular sources of disagreement; all p-values are well below 0.01.

The different inclusive data sets seem to give contrasting conclusions when compared with various generators. For T2K, NEUT does a better job describing this data, NuWro predicts a smaller cross section in the forward region of the lepton compared to the other generators there. It also consistently predicts a larger peak at intermediate muon momenta, while these differences are not visible in the NuWro predictions for MicroBooNE.  GENIE $\chi^2$ results are intermediate between those of NEUT and NuWro.  For MicroBooNE, GENIE predictions have similar values of $\chi^2$ for all three models while the value from NEUT is the highest.  All generators seem to have equal understanding of A dependence in general.
For MINERvA measurements, NuWro gives the lowest $\chi^2$ compared with other predictions. However, all event generators show disagreement at low available energy and the peak of the distributions for the regions between $2p2h$ and RES. 

Concerning the $\nu_\mu$ CC-0$\pi$ cross-section, comparisons have been made between T2K and \minerva results and the generator predictions employed for this work. While it is difficult to draw an unambiguous conclusion, it is possible to underline some considerations. As described in \autoref{subsec:cc0piefficiency}, T2K and \minerva cover different $q_0$-$q_3$ regions and in order to directly compare results from the two experiments it is necessary to carefully select a common phase space. As reported in \autoref{subsec:t2kmincc0picomp}, the integrated $\nu_\mu$ CC-0$\pi$ cross sections in a phase space common to T2K and \minerva (p$_\mu>1.5$~GeV/c and $\theta_\mu<20^\circ$) are systematically higher than what is predicted by generators  under study (see Fig. \ref{fig:cc0pit2kvsminerva}). The only exception is for NEUT that seems to perfectly reproduce the \minerva integrated cross section. The other generator results tend to be $\sim$20\% below the data indicating that the energy dependence is better reproduced than the individual measurements.

For T2K, the systematic MC underestimation of the cross section is true also when considering the full muon momentum phase space in the same angular region. However, the detailed two-dimensional measurements shown in Fig. \ref{fig:t2knumucc0pich2dpcos}, \ref{fig:t2knumucc0pic2dpcos}, and \ref{fig:cc0piminerva2Dplot} reveal specific behaviors of the generators depending on the kinematic bins considered, as well as different degrees of agreement/disagreement with the data. In general, T2K carbon data seems to prefer NEUT and GENIE using LFG, while \minerva data does not show agreement with considered generators. Because of the different $q_0$-$q_3$ accessible to T2K and \minerva, the energy dependence of cross section is difficult to determine. Both experiments indicate an under-prediction in the forward lepton direction, for T2K is for high momentum ($>1$~GeV) bins, but the additional pion contribution via FSI at higher beam energies obscures the robustness of the predicted energy dependence. 

In a similar way, the different T2K and \minerva TKI variables prefer different generators as shown in \cref{fig:t2kcc0pinpstv}. Both T2K and \minerva data show poor $\chi^2$ agreement with respect to NuWro. Generators suffer in reproducing the low $\delta p_T$ region for \minerva, suggesting limits in the prediction of the initial state nucleons. On the other side, \minerva $\delta \alpha_T$ is well-reproduced by the generators other than NuWro, suggesting a more correct treatment of FSI at energies relevant for \minerva. T2K $\delta p_T$ shows lower $\chi^2$ values when compared to GENIE and NEUT with LFG, while GENIE with the \Genietwoa~tuning better reproduces $\delta \alpha_T$, which can be interpreted as a preference for GENIE's traditional effective cascade FSI model. Future measurements with additional statistics as well as further comparisons between T2K and \minerva data will certainly help in clarifying this complex picture.
Finally, generator comparisons with \minerva measurements in $Q^{2}_{QE}$ (see \autoref{subsec:minervaAscale}) underline the relevance of a correct FSI treatment for nuclei with increasing $A$. The comparisons suggest a preference for GENIE's traditional effective cascade FSI model (tuning {\Genietena} and {\Genietwoa}).

The pion production interaction remains a central concern in the field at the same time as its importance grows through the needs of DUNE~\cite{DUNE:nd2021}. It was the source of many issues with signal definition and data compatibility in TENSIONS2016~\cite{tensions2016}. Although \minerva has published a variety of measurements, the $1\pi^\pm$ data is of particular interest here since it can be compared to T2K.  While T2K has added measurements and \minerva has updated and expanded upon their first measurements, data remain insufficient to fully constrain models due to large uncertainties and limitations in data coverage. At the same time, improvements in modeling this interaction remain slow to implement.

The focus in this work is on a detailed comparison of the recent T2K CC1$\pi^+$~\cite{t2k_cc1pi_fgd1} and the updated \minerva  CC1$\pi^\pm$~\cite{minerva_datarelease} (which is heavily dominated by $\pi^+$) cross sections.
Both T2K and \minerva measurements stress inclusive measurements; this makes them sensitive to underlying efficiency evaluations.  However, generator predictions show that both are largely sensitive to $\pi^+$ production from bound protons.  The efficiency study (\autoref{sec:pi_eff}) show holes in both T2K and \minerva acceptance and the likelihood of resulting increased model dependence.  The overlap in kinematics is small but encouraging.  Within the estimated uncertainties, the measurements are in agreement, indicating that the energy dependence is handled correctly (in direct contrast to the MiniBooNE-\minerva comparisons of TENSIONS2016~\cite{tensions2016}). This provides a more detailed test than straight comparison with theory.

Both T2K and \minerva present numerous distributions that probe different parts of a complicated interaction.  Both qualitative and quantitative comparisons are described; see Tab. \ref{tab:t2k_chi2} and \ref{tab:minerva_1pi_chi2} for details. The muon distributions (Figs.~\ref{fig:t2k_cc1pi_mu} and \ref{fig:minerva_cc1pi_muon}) are largely insensitive to the details of FSI and  show features of the treatment of the initial state.  All generators do a reasonable job of describing these inclusive distributions with $\chi^2$ typically 1--2/bin.  Pion observables such as kinetic energy/momentum and polar angle  (Figs.~\ref{fig:t2k_cc1pi_pi} and \ref{fig:minerva_cc1pi_pi}) are more sensitive to the production mechanism and FSI. Features of the data in these observables are described poorly in general and values of $\chi^2$ are typically larger than 2 per bin. This is especially true for \minerva $\theta_\pi$, but it should be noted that \minerva $\theta_\pi$ shows some issues with model dependence in our studies.  The pion momentum distribution appears most sensitive to modeling.
Although the magnitude of various calculations is close to correct, the values of $\chi^2$ are large, indicating a problem with shape.  The most direct test of FSI is a comparison of GENIE $hA$ and $hN$ models with all other features held constant.  The more empirical $hA$ model describes the data better than the more theoretical $hN$ model.

Finally, the Adler angle distributions (angles of the decay pion in the parent resonance rest frame) for the T2K measurement show details of the $\Delta$(1232) decay.  This is the most detailed examination and shows the largest discrepancies between data and generator predictions.  Either the nuclear corrections are wrong or there is tension between older analyses for H/D targets~\cite{Radecky:1981fn} and the newer data for nuclear targets.

There is an interesting trade-off between \minerva and T2K 1$\pi^+$ measurements.  Both have the goal of providing accurate data that stands on its own with minimal model dependence and allows theory calculations to be compared directly with the data.  T2K puts more emphasis on decreasing the model dependence by restricting the phase space of the final state particles.  \minerva does this also, but to a smaller extent.
For example, \minerva uses a $W<1.4$~GeV constraint in the signal definition which allows more direct studies of the $\Delta$ resonance but creates significant potential for model dependence because events from the higher resonances become background.

MiniBooNE published a large body of neutrino cross-section data about 10 years ago. Although it has excellent statistics and broad kinematic coverage, some of those results have significant model dependence and incomplete systematic uncertainty analysis as compared to the more modern treatments. In particular, the lack of published correlation matrices for some MiniBooNE datasets presents a major difficulty. Quantitative model comparisons based on goodness-of-fit to these data should not be attempted unless great care is taken to address the statistical implications of the missing bin-to-bin correlations~\cite{Koch2021}. Provisional qualitative conclusions based on model comparisons to the MiniBooNE CC0$\pi$ and CC1$\pi$ measurements should be confirmed with more rigorous testing against modern measurements. The MiniBooNE NC elastic and NC1$\pi^0$ results are not subject to the same caveats: both provide high-quality measurements with full covariance matrices of otherwise poorly-known cross sections. Models may be compared to the NCEL and NC1$\pi^0$ data with the same level of confidence as typical modern measurements.

Only qualitative analysis of the MiniBooNE results was attempted in
Appendix \ref{sec:minibapp}. All modern calculations are in good visual agreement with the MiniBooNE CC0$\pi$ data considered in this publication with the most notable disagreement at very low $Q^2$. For CC1$\pi$, conclusions are harder to ascertain because the calculations are not in agreement with each other. The most notable problem is the significant normalization disagreement for NEUT, NuWro, and the older GENIE version with respect to the CC1$\pi^+$ MiniBooNE data. Better agreement of calculations with the MiniBooNE CC1$\pi^0$ data may indicate an issue with the CC1$\pi^+$ data.

A series of separate generator studies compared features of the models independent of comparison with data. The first considered the dependence of yield as a function of momentum threshold, and showed the sensitivity to choices the experiments make. Both proton and pion momentum yields vary significantly among the generators at the values typically chosen, indicating significant model sensitivity. The second study was of the neutral energy content which is an important contribution of the generators to all experiments. The variations among the generators are significant, both for the total neutral energy and for that contained in neutrons. Any experimental measurements in variables of this type are difficult but will be extremely valuable for constraining this energy which varies widely among generators.

The community continues to investigate and improve statistical methods to present cross-section measurements in mathematically correct and useful ways.
Disentangling the effects on the detector response from model-dependent nuisance parameters and from the actual parameters of interest remains a challenge.
The work within this group explored limitations of the forward-folding implementation employed by MicroBooNE in its first cross-section results~\cite{Abratenko2019}.
In this case, the method breaks down when the uncertainties of the detector smearing become large enough to have a notable effect on the result.
Other forward-folding strategies without this limitation exist, but have so far not been used in a published cross-section measurement, and introduce other challenges and complexities.
Statistical methods that have been used by other experiments, or even by the same experiment in the past, might not be sufficient for future experiments and ever more complicated analyses.
Additional effort from experiments to explore these techniques is essential ~\cite{Koch2019}.

\subsection{Looking forward}
We conclude with a general outlook on the field.  Progress since TENSIONS2016~\cite{tensions2016} is impressive.  The number and quality of experimental results for neutrino-nucleus cross sections has increased significantly and is still improving.  As a result, our studies expanded into inclusive interactions.  There have been a large number of mesonless (i.e. pionless at low neutrino energies such in T2K) quasielastic scattering results, both for neutrinos and antineutrinos.  The statistics of MINERvA results are approaching and in some cases exceeding
that of MiniBooNE (the previous standard) and have the ability to differentiate calculations better.  The additional detection of protons in the final state has allowed more focus on true QE interactions.  For pion production, MINERvA has published a variety of results.  The body of pion production results have expanded in the last few years with MINERvA results for all pion charges.  However, measurements are still limited in statistics and have potential problems with model dependence of various sources.  Of course, the pion production measurements remain challenging because of the difficulties identifying and measuring the energies of pions in tracking detectors or calorimeters. 

The measurements covered in this work are largely for CH targets because of the preponderance of scintillator target/detectors at the time. More recent results from MINERvA stress high atomic weight nuclei and MicroBooNE provides a liquid argon (LAr) target/detector.  Heavier nuclear targets and more accurate measurements with hadrons are clearly the frontier.  Recent MicroBooNE results~\cite{Abratenko:2020sga,MicroBooNE:2020fxd} have a proton tracking threshold of 47 MeV and liquid argon detectors should be able to go lower. This is especially interesting because the yield of protons and neutrons is expected to rise at low energy due to compound nucleus processes.  DUNE~\cite{DUNE:nd2021} will focus on LAr target/detectors.  Pion detection remains largely unexplored for LAr experiments because of the low fraction of events with a Michel electron.  The T2K near detector upgrade~\cite{Sgalaberna:2020xgq} will emphasize back angle particle detection and oxygen targets in preparation for upcoming measurements and Hyper-Kamiokande.  These are all very positive steps forward toward improved measurements and this document will hopefully aid that development.

\section{Acknowledgments}
The authors are grateful to the Univ. of Pittsburgh Center for High Energy Physics, Astronomy, and Cosmology (PITT PACC) for support during the workshop.

This manuscript has been authored by Fermi Research Alliance, LLC under Contract No. DE-AC02-07CH11359 with the U.S. Department of Energy, Office of Science , Office of High Energy Physics.

The authors acknowledge support from US Department of Energy under contract No.~DE-AC02-05CH11231, No.~DE-SC0007914, No.~DE-SC0008475, No.~DE-SC0021407 and No.~DE-SC0015903 and the National Science Foundation under Grant No. PHY-2047665.
This work was supported by a grant from the Science and Technology Facilities Council.
We acknowledge the support of CNRS/IN2P3, France.

The work was in part supported by a Cottrell Postdoctoral Fellowship, from the Research Corporation for Scientific Advancement award number 27467 and National Science Foundation Award CHE2039044.

Discussions about the MiniBooNE data with Rex Tayloe and Teppei Katori were valuable.


\appendix
\section{Additional MiniBooNE qualitative comparisons}
\label{sec:minibapp}

\begin{figure*} [ht]
	\centering
	\includegraphics[width=0.97\linewidth]{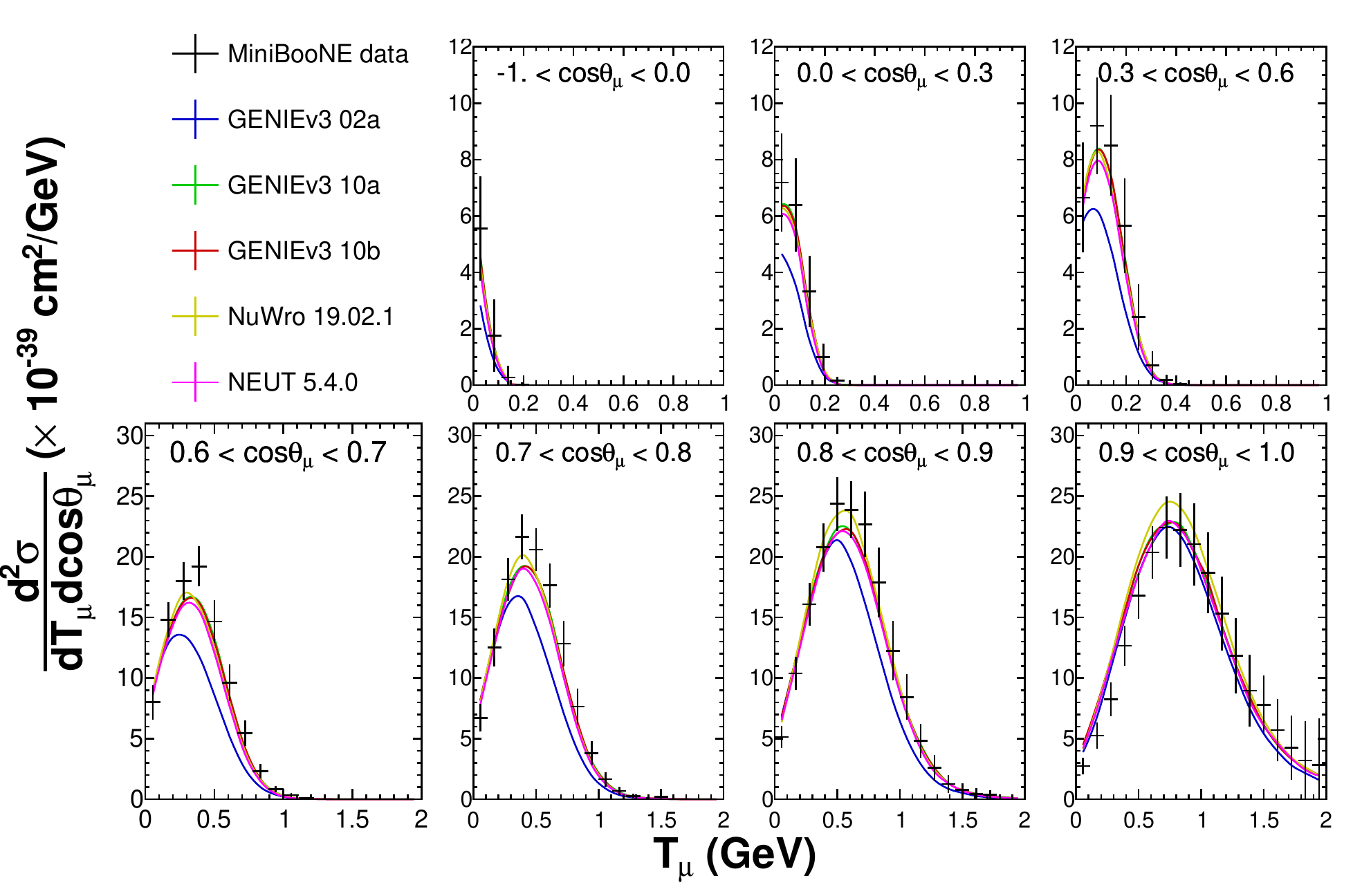}
	\caption{Comparison of MiniBooNE CCQE-like $T_{\mu}$, $\cos\theta_{\mu}$ data~\cite{Katori-mbqe} with GENIE, NEUT, and NuWro calculations.  Data is determined from the CCQE sample assuming true CCQE kinematics. Cross sections displayed come from the MiniBooNE data release~\cite{miniboone-data-release} for Ref.~\cite{Katori-mbqe} by adding the pion absorption correction back into the published CCQE values.  Each bin in the array is a $\theta_\mu$  angular distribution for the range in $\mu^-$ kinetic energy in the legend.The shown uncertainties are the shape-only uncertainties. There is an additional 10.7\% normalization uncertainty. }
	\label{fig:MBCCQElike2D}
\end{figure*}

Given the concerns raised in \autoref{sec:minib} we provide comparisons to MiniBooNE data in a limited way for this paper.
We note specifically that a lack of correlations may lead to incorrect conclusions being drawn from the data when a naive bin-by-bin comparison is made. But there is no established procedure to overcome this shortcoming for the currently available data. Only qualitative comparisons are provided, and a full interpretation should rely on other data sets presented (e.g. T2K, MINERvA, MicroBooNE).

\begin{figure}[ht]
	\centering
	\includegraphics[width=0.97\linewidth]{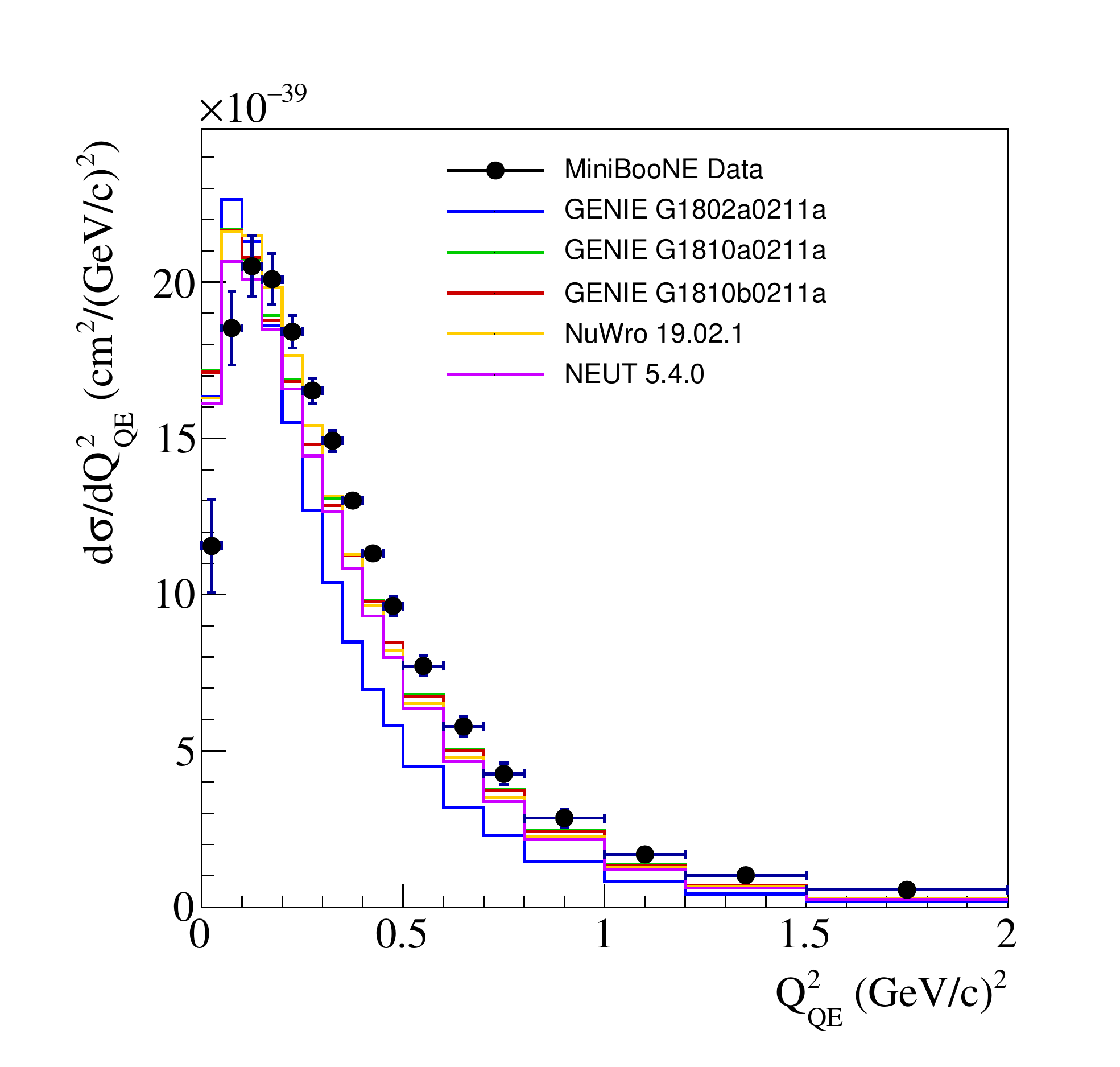}
	\caption{Comparison of MiniBooNE $Q^2_{\mathrm{QE}}$ data with GENIE, NEUT, and NuWro calculations.  The shown uncertainties are the shape-only uncertainties. There is an additional 10.7\% normalization uncertainty.}
	\label{fig:MBCCQElike}
\end{figure}

\begin{figure}[ht]
	\centering
	\includegraphics[width=0.97\linewidth]{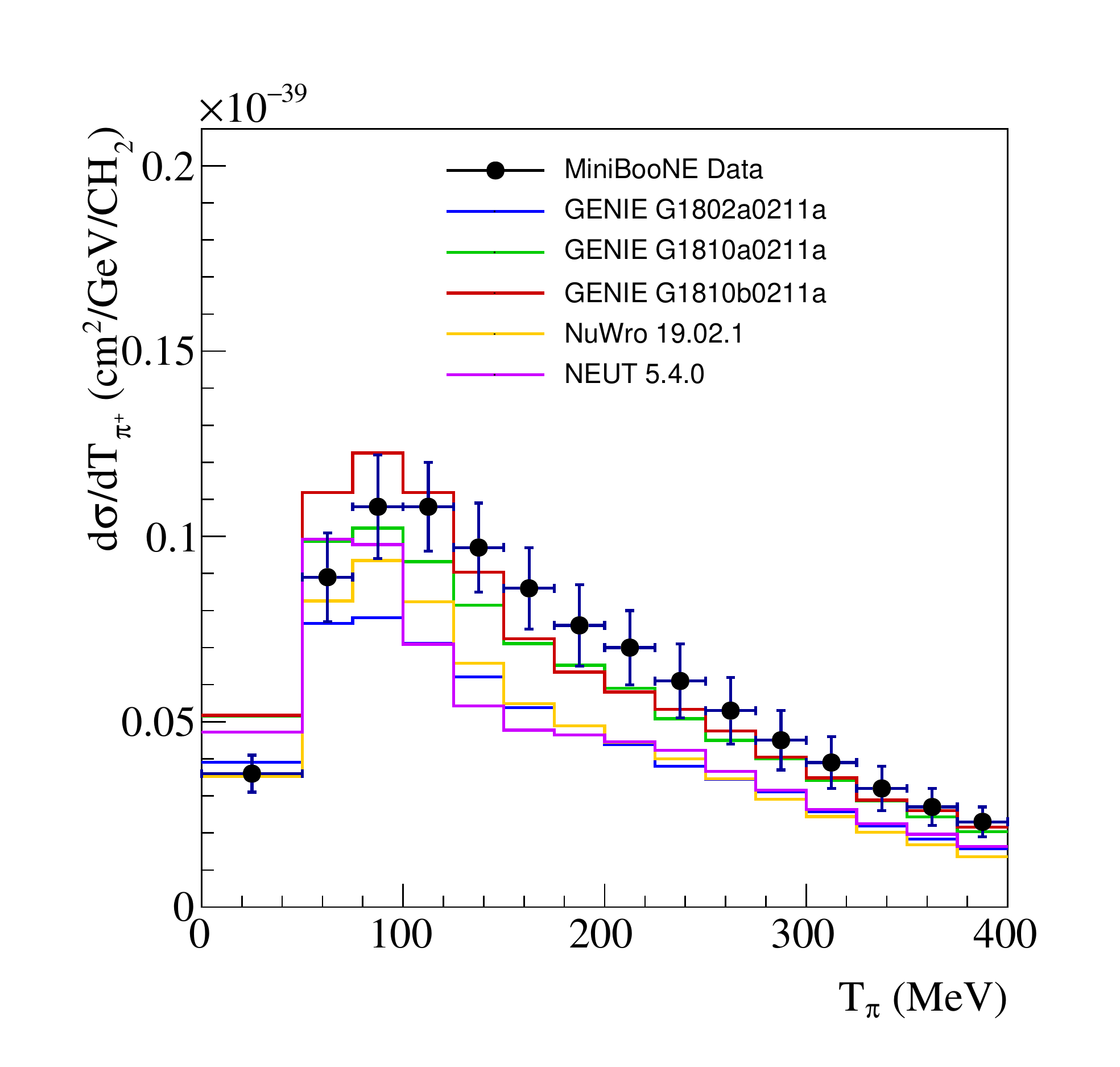}
	\caption{Comparison of MiniBooNE $T_{\pi^+}$ data~\cite{Wilking} with GENIE, NEUT, and NuWro calculations.}
	\label{fig:MBCC1pip}
\end{figure}

\begin{figure}[ht]
	\centering
	\includegraphics[width=0.97\linewidth]{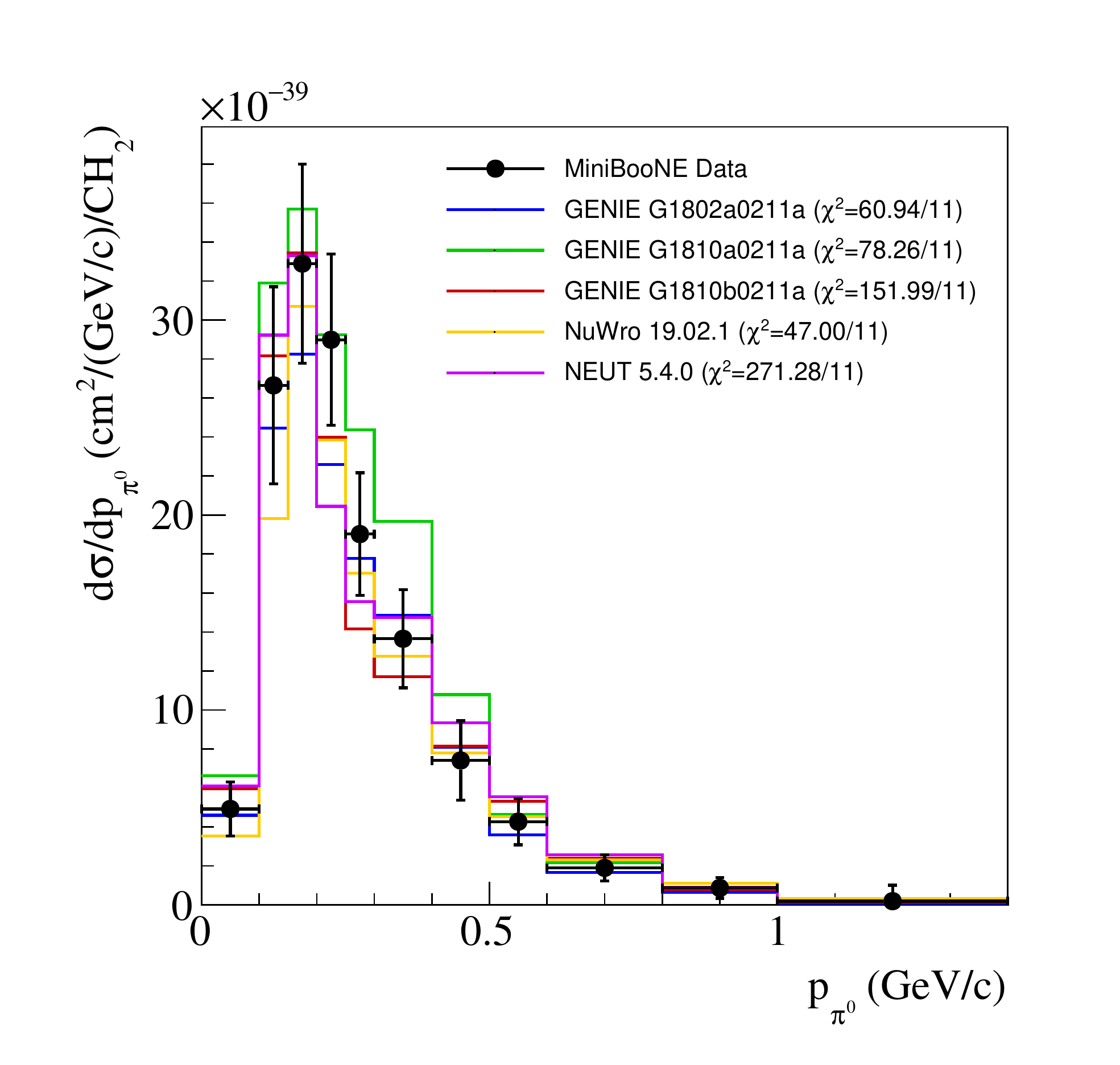}
	\caption{Comparison of MiniBooNE CC$1\pi^0$ $p_{\pi^0}$ data~\cite{miniboone-ccpi0} with GENIE, NEUT, and NuWro calculations.
	}
	\label{fig:MBCC1pi0}
\end{figure}

Here we discuss for this specific study CCQE-like and CC$1\pi^+$ results.  The first measurement, CCQE-like, included only the final state muon in the signal definition.  The CC$1\pi^+$ measurement introduced a novel way to identify positively charged pions in a Cerenkov detector through their inelastic scattering signature.  Comparisons of the CCQE-like $Q^2$ distribution, the CC$1\pi^+$ pion kinetic energy, and the CC$1\pi^0$ pion momentum  are shown in Figs.~\ref{fig:MBCCQElike}, \ref{fig:MBCC1pip}, and \ref{fig:MBCC1pi0} respectively.

While NEUT, G18\_10a, and G18\_10b are in qualitative agreement with the CCQE-like $Q^2$ distribution, GENIE G18\_02 has problems in describing the shape properly.  GENIE G18\_02 is significantly low at $Q^2>0.3$ GeV$^2$ and all calculations are above the data point at the lowest $Q^2$.  The corresponding plot for the MINERvA CCmeson-less measurement is shown in Fig~\ref{fig:cc0piminervaq2qetrackerl} for a higher energy neutrino beam. As a result, the $Q^2$ range is larger.  In addition, the statistical advantage at very low $Q^2$ is clear.  The agreement of the calculations with the MINERvA data is improved, but the problems at very low $Q^2$ persist.  A broader view is obtained with Fig.~\ref{fig:MBCCQElike2D} where the full range of  $p_\mu-\cos\theta_\mu$ can be seen.  This has the advantage of separating the components of the $Q^2$ calculation within the same sample.
All modern calculations are in good qualitative agreement with these data with the only problems seen for the older GENIE version at back angles and low momentum.

One of the central issues in TENSIONS2016~\cite{tensions2016} was the discrepancy in the $T_\pi$ normalization and shape between MINERvA~\cite{Eberly:2014mra} and MiniBooNE~\cite{Wilking} data.  As discussed in Ref.~\cite{Sobczyk:2014xza}, the normalization in the calculation is largely driven by the elementary $\nu_\mu p \rightarrow \mu^- \pi^+ p$ cross section and that makes a disagreement in normalization hard to understand from physics considerations.

The calculations for CC1$\pi^+$ in Fig.~\ref{fig:MBCC1pip} have a wide variation in magnitude and shape.  
Fig.~\ref{fig:minerva_cc1pi_pi}) shows the corresponding result for MINERvA (similar target and higher energy) and Fig.~\ref{fig:t2k_cc1pi_other} for T2K (similar energy and target).  
It is interesting to note that NuWro and NEUT tend to overpredict the T2K data but underpredict the MiniBooNE data.
The same tendency is seen with the GENIE calculations, esp. G18\_10a.
Since T2K and MiniBooNE have very similar target and energy distribution, the conclusion from this comparison is that the two experiments are inconsistent.
A proper error treatment would be required to make a more quantitative assessment.
On the other hand, NEUT and NuWro tend to overpredict MINERvA data and the GENIE calculations tend to be smaller than the data.
In contrast to the first comparisons~\cite{tensions2016}, GENIE v3 G18\_10a calculation is in agreement with both MiniBooNE and MINERvA.
Additional information is provided by the MiniBooNE CC$1\pi^0$ measurement~\cite{miniboone-ccpi0}.  The $\pi^0$ momentum distribution is shown in Fig.~\ref{fig:MBCC1pi0}; here, all calculations are in good qualitative agreement with the data, implying  differences in treatment of $\pi^+$ and $\pi^0$ in some calculations or in the data.  In detail, the $\chi^2/N_{bins}$ values are all large.

\end{document}